\documentclass[msc,deptreport,ai]{infthesis}      

\usepackage{listings}
\submityear{2019} 

\usepackage[many]{tcolorbox}
\usepackage{xcolor}

\lstdefinelanguage{JavaScript}{
  keywords={typeof, new, true, false, catch, function, return, null, catch, switch, var, if, in, while, do, else, case, break, handle, effect, fun, val, match, then, resume, control, for},
  keywordstyle=\color{blue}\bfseries,
  ndkeywords={class, export, boolean, throw, implements, import, this},
  ndkeywordstyle=\color{darkgray}\bfseries,
  morekeywords={[2]{weighted, yield_on_score, random_sampler, replay, finalize, advance}},
  keywordstyle={[2]{\color{olive}\bfseries}},
  morekeywords={[3]{score, sample, yield, ndet, total, div, console}},
  keywordstyle={[3]{\color{magenta}\bfseries}},
  identifierstyle=\color{black},
  sensitive=false,
  comment=[l]{//},
  morecomment=[s]{/*}{*/},
  commentstyle=\color{purple}\ttfamily,
  stringstyle=\color{red}\ttfamily,
  morestring=[b]',
  morestring=[b]"
}

\usepackage{amsmath}
\usepackage{commath}
\usepackage{dsfont}
\usepackage{wrapfig}
\usepackage{parcolumns}
\usepackage{xfrac}
\usepackage{courier}
\usepackage{changepage}
\usepackage{subcaption} 
\usepackage{tikz}
\usetikzlibrary{arrows}

\graphicspath{ {./images/} }

\begin{document}
\begin{preliminary}

\title{Modular probabilistic programming with algebraic effects (2019)}

\author{Oliver Goldstein supervised by Dr. Ohad Kammar\footnote{Email: oliver.goldstein@reuben.ox.ac.uk, ohad.kammar@ed.ac.uk}}

\abstract{
  Probabilistic programming languages, which exist in abundance, are languages that allow users to calculate probability distributions defined by probabilistic programs, by using inference algorithms. However, the underlying inference algorithms are not implemented in a modular fashion, though, the algorithms are presented as a composition of other inference components. This discordance between the theory and the practice of Bayesian machine learning, means that reasoning about the correctness of probabilistic programs is more difficult, and composing inference algorithms together in code may not necessarily produce correct compound inference algorithms. In this dissertation, I create a modular probabilistic programming library, already a nice property as its not a standalone language, called Koka Bayes, that is based off of both the modular design of Monad Bayes -- a probabilistic programming library developed in Haskell -- and its semantic validation. The library is embedded in a recently created programming language, Koka, that supports algebraic effect handlers and expressive effect types -- novel programming abstractions that support modular programming. Effects are generalizations of computational side-effects, and it turns out that fundamental operations in probabilistic programming such as probabilistic choice and conditioning are instances of effects. \footnote{See the GitHub from 2019: https://github.com/oliverjgoldstein/koka-bayes-writeup and see the language: https://github.com/oliverjgoldstein/koka-bayes}
}

\maketitle

\tableofcontents
\end{preliminary}

\chapter{Introduction}

\section*{Context}

The Bayesian approach to probabilistic reasoning is known as Bayesian Inference, and is an increasingly popular method to perform machine learning that enjoys strong mathematical foundations\footnote{Section 1.1, 1.2, 2.1 (intro) and 2.1.1, are loosely based on -- but not directly copied, from my IRR/IPP.}. Bayesian Inference has useful properties, compared to deep learning based machine learning, such as interpretability, explainability and expressiveness with respect to various forms of uncertainty \cite{yang2017explainable,barber2012bayesian}. At the heart of the Bayesian paradigm is the notion of a generative model. Generative models can account for the expressiveness of uncertainty and interpretability. In order to perform probabilistic modelling, knowledge about a domain and the uncertainty associated with that knowledge is encoded into a generative model, using expert knowledge. Inference algorithms estimate parameters of the model, that generate the observed data, by conditioning models on associated observed data. For instance, a climate researcher may create a basic model of a city when studying urban heat islands. The climate researcher would condition the model of the city with empirical data, such as the observed temperatures of the city, in order to infer values of interest such as latent city temperatures or factors that influence the temperature of a city. Probabilistic programming languages provide a generic framework to perform inference. \\

To highlight the importance of probabilistic programming languages, large multinational technology companies such as Microsoft, Google and Uber have recently started to invest in their own probabilistic programming languages (Infer.NET \cite{InferNET18}, Edward \cite{tran2018simple} and Pyro \cite{bingham2018pyro} respectively). Increasingly, machine learning use cases find their way into areas as diverse as critical defense national infrastructure \cite{li2018detection}\cite{branch2006imagery} or medical diagnostics \cite{bejnordi2017diagnostic}. Society is therefore increasingly in a position that relies on programmers to produce correct machine learning related code. As probabilistic programming becomes more important, the code that underlies probabilistic programming languages will face increasing scrutiny. Fortunately, there are two important programming language research goals which aim to ensure the correctness of code; modularity and type-safety. In particular, they ensure that programmers can be certain early on -- at compile time -- that composing code will not cause programs to crash or introduce unintended (inference) algorithmic bias. These research goals have two effects; as programming libraries become safer and more modular, machine learning models will become cleaner, simpler and easier to maintain, which will additionally make machine learning more trusted, and perhaps more widely adopted. \\

Algebraic effects are a recently discovered abstraction which enable modular programming \cite{plotkin2003algebraic}. However, algebraic effects do not use monads \cite{kammar2013handlers}; an abstraction derived from a branch of mathematics known as category theory. Monads are known to be both unintuitive as well as not being closed under composition. Monad Bayes is a library for modular probabilistic programming published only last year, that makes essential use of types. However, whilst it is based off a denotational validation \cite{scibior2017denotational}, it suffers from the issues associated with monads \cite{moore2018effect}. These issues motivate a previously unachieved construction, of a modular probabilistic programming language, that also uses this same denotational validation, to perform Bayesian inference using algebraic effects with expressive effect-types. 

\section*{Problem Statement}

Inference algorithms lie at the heart of probabilistic programming languages and are informally described as the combination of primitive blocks, for instance, Resample-Move Sequential Monte Carlo (RMSMC) \cite{gilks2001following} is described as Sequential Monte Carlo \cite{doucet2001introduction} (SMC) combined with one or more Markov Chain Monte Carlo steps (MCMC) \cite{neal1993probabilistic}. Whilst probabilistic programming languages have existed since 1994 \cite{gilks1994language}, most implementations do not program inference algorithms as compositions of other inference primitives \cite{scibior2018functional}. This poses a question which this project aims to answer: can probabilistic programming languages be created that build correctness and composability into their core? Monad Bayes \cite{scibior2018functional}, whilst modular, uses monads which have the previously stated problems with composability and intuitiveness. Currently, there is no probabilistic programming language, that demonstrates the use of algebraic effect handlers to produce correct, modular and composable inference. The demonstration of a language is important for several reasons. Firstly probabilistic programming is a new use case for Daan Leijen's language, Koka. Secondly, it affirms the theoretical link of the equivalence of monads and effects. Thirdly it provides industry with a modular approach to perform probabilistic programming.

\section*{Proposed solution and contribution}

I create a probabilistic programming library, Koka Bayes, with support for four inference algorithms, embedded in Koka, a contemporary functional language that supports algebraic effects. I construct the library iteratively, with the construction split into three vertical slices, proceeding in the order listed here:
\begin{enumerate}
    \item Library with a basic inference strategy, demonstrated over a basic model.
    \item Addition of four more inference algorithms to the library.
    \item Demonstration of inference, over a more complex model, conditioned on real world data.
\end{enumerate}
Koka Bayes is based off Monad Bayes's design and its semantic validation, and will be compared to it on scalability and execution time. I test and evaluate Koka Bayes with Monte Carlo inference, over a linear Gaussian model of climate.

\section*{Structure of dissertation}

Chapter 2 introduces the concepts that underpin probabilistic programming and effect handlers. Chapter 3 introduces and evaluates the theory and implementation of importance sampling -- a basic inference strategy. Chapter 4 presents four well known, and more advanced inference algorithms, which are motivated by the weaknesses of importance sampling. Chapter 5 demonstrates two of these inference algorithms over a model of climate change. Chapter 6 concludes with a summary of the contributions made and suggests future work.

\chapter{Background}

\section{Probabilistic programming}

Probabilistic programming languages support random sampling of variables, as well as the conditioning of programs as a function of both observed data and the sampled values of the random variables \cite{gordon2014probabilistic}. The goal of writing probabilistic programs is to perform inference over the probabilistic program in order to obtain the \emph{posterior} distribution, $p(\theta|D)$, specified by the program. The posterior distribution is a probability distribution over parameters $\theta$, conditioned on observed data $D$.
\begin{equation}
p(\theta|D) = \dfrac{p(\theta, D)}{p(D)} = \dfrac{p(D|\theta)p(\theta)}{p(D)} = \dfrac{p(D|\theta)p(\theta)}{\int p(D, \theta)\dif \theta} = \dfrac{p(D|\theta)p(\theta)}{\int p(D|\theta)p(\theta)\dif \theta}
\label{eqn:total-prob}
\end{equation}
Using programs to represent probabilistic models was conceived as an idea in 1978 \cite{saheb1978probabilistic} and programming languages for Bayesian modelling and inference have existed since at least 1994 \cite{gilks1994language}. Probabilistic programming languages have been used for various tasks from reasoning about theory of mind, to matching players who have similar skill levels on Xbox \cite{inferNET_applications,herbrich2007trueskill}\footnote{http://forestdb.org/}. There are two distinct kinds of probabilistic programming languages, both of which span a wide spectra of use cases\footnote{http://probabilistic-programming.org/wiki/Home}: 
\begin{itemize}
    \item Languages such as ProbLog 2 \cite{dries2015problog2} perform inference over logic programs, often called worlds, and concern themselves with computing the probability of a truth-related query about the world, given evidence or data. 
    \item Languages such as Edward \cite{tran2018simple} or Tabular \cite{gordon2014tabular}, infer, given data, the probability distribution of parameters that generated the given data.
\end{itemize}

Probabilistic programming aims to internalize general purpose inference mechanisms inside the compiler in order to allow the programmers to focus on writing models and have inference automatically done for them \cite{gordon2014probabilistic}. This specific goal has motivated the construction of many languages such as Gen \cite{cusumano2019gen}. However, Ackermann \& Freer \cite{ackerman2011noncomputable} show that, in general, computing continuous conditional probability distributions -- such as the posterior -- cannot be automated, as the halting problem can be reduced to the computation \cite{ackerman2011noncomputable}.

\subsection{Generative models}

Probabilistic programs represent Bayesian models. A Bayesian model is the combination of a statistical model with a prior distribution \cite{gutmann2018introduction}. A statistical model is a set of random variables $\textbf{x}_{\theta}$ parameterized by $\theta$. For each value of $\theta$, $x_{\theta}$ is a random variable with probability density function $p(x|\theta)$. Data is then used to pick out a particular member of the family $\{p(x|\theta)\}_{\theta\in\Theta}$. This process is called estimating or learning the parameters of a statistical model. Once a particular $\theta$ is chosen, a probabilistic model results. Probabilistic programs often do inference by conditioning Bayesian models on whether parameter samples drawn from the prior produce data equal to the observed data.

\begin{wrapfigure}{r}{0.39\textwidth}
  \begin{center}
    \includegraphics[width=0.4\textwidth]{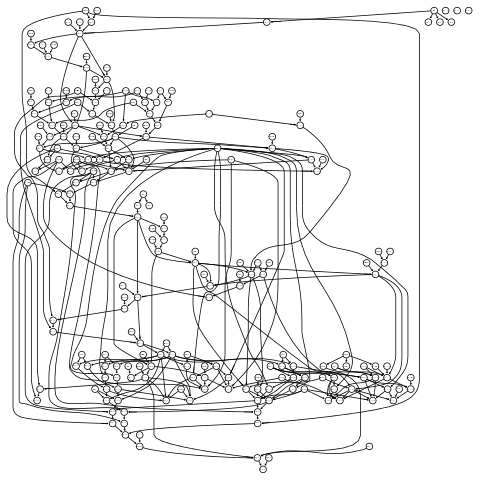}
  \end{center}
  \vspace{-20pt}
  \caption{Bayesian network from \cite{conati1997line}}
  \label{fig:andes-network}
\end{wrapfigure}

An example of a statistical model is a Bernoulli distribution $p(x | \theta) = \theta^{x}(1 - \theta)^{1 - x}$. Combining this statistical model with a probability distribution over $\theta$ results in a Bayesian model. An example of such a probability distribution is the beta distribution: $$p(\theta) = p(\theta | \alpha_{0}, \beta_{0}) = \dfrac{1}{Z(\alpha_{0}, \beta_{0})}\theta^{\alpha_{0} - 1}(1 - \theta)^{\beta_{0} - 1}$$Multiplying this prior with the statistical model results in the following Bayesian model: $$p(x, \theta | \alpha, \beta) = \dfrac{1}{Z(\alpha, \beta)} \theta^{x + \alpha + 1}(1 - \theta)^{\beta - x}$$. 

Probabilistic programming provides those familiar with generative models, three important benefits:
\begin{itemize}
    \item As the complexity of models increases, the mathematical language that represents them starts to matter. For example, other mathematical languages such as graphical models can be challenging to understand visually (see Figure \ref{fig:andes-network}), especially when models use complicated control flows. 
    \item Probabilistic programming separates inference from the modelling. This allows modellers to focus on modelling rather than the complex process of getting inference algorithms to work for their particular models.
    \item Many formalisms such as factor graphs or graphical models do not indicate the type of parameters to factor nodes or the distribution that underlie variable nodes. Probabilistic programs can express these types, which is useful information for modellers and inference algorithms.
\end{itemize}

\subsection{Inference}
There are three main forms of inference used by probabilistic programming languages; exact, Monte Carlo, and variational. 
\begin{itemize}
    \item \underline{Exact} inference algorithms calculate the posterior exactly, which means computing every term of the denominator in equation \ref{eqn:total-prob} (which corresponds to a sum in the discrete case). 
    \item \underline{Monte Carlo} inference algorithms approximate conditional probability distributions with samples generated from proposal distributions. Monte Carlo algorithms are general and guarantee asymptotically exact convergence. A prototypical example of a Monte Carlo inference algorithm, that I present in more detail in Chapter 4, is Sequential Monte Carlo (SMC) \cite{doucet2001introduction}. 
    \item \underline{Variational} inference algorithms approximate conditional probability distributions with functions chosen from a parametric class. These functions are then optimized to match the conditional probability distribution of interest by minimizing statistics that indicate the difference between two distributions -- typically the KL-divergence. As the parametric class may not in general include the true posterior distribution, variational methods are not guaranteed to be asymptotically exact. 
\end{itemize}

\section{Algebraic Effects}

Plotkin and Pretnar's \cite{plotkin2003algebraic} algebraic effects and handlers, are a language feature that generalize commonly occurring control flow constructs e.g. mutable state, non-determinism, exceptions, interactive input-output etc. Koka is a functional language that places algebraic effects at the core of its design. Koka also uses type signatures that not only describe the input and output value of a function, but also \emph{effect annotations} that indicate the side-effects that come from behaviours that a function may perform. \\

Listings \ref{lst:total} to \ref{lst:polymorphic} give simple examples of programming with effects. In Listing \ref{lst:total}, \texttt{add} is well defined for all values of \texttt{x} and \texttt{y}, and does not cause any effects, so the Koka type system allows its effect annotation to be \texttt{total}. By adding in a print statement, Listing \ref{lst:combined} may no longer be marked as total, and must be annotated with the \texttt{console} effect. Koka supports polymorphic effect types. In Listing \ref{lst:polymorphic}, the presence of \_e in the type signature means there is \emph{some} effect that can be unified with $\_e$, in addition to the already existing \texttt{total} effect type. In this case, $\_e$ will be unified with \texttt{console}. Specific effect operations can also be declared with the \texttt{effect} keyword. Listing \ref{lst:effects-used} details three effects used by Koka Bayes. The \texttt{sample} effect takes the empty type and returns a double, the \texttt{score} effect takes an \texttt{exp} data type and returns the unit type and the \texttt{yield} effect takes and returns the unit type. Adding the \texttt{sample} operation to a computation, requires its computation to include sample. In Koka, an effect type indicates that the computation may incur the particular effect, though it does not \underline{have} to incur this effect. However, if the effect type is not indicated in the signature then the computation will not incur the effect. Precise inference of the exact effect type is in general undecidable since Koka is Turing complete (consider a loop that may not terminate or cause an exception). \\

User defined effect operations cause a runtime exception unless the program provides an appropriate effect handler. Handlers provide the semantics for computations built out of effect operations \cite{pirog2018syntax}. In a computation, the presence of an effect acts as a signal, which propagates outwards until it reaches an effect handler of the computation with a matching clause \cite{kammar2017no}. The typing judgments for deep and shallow handlers \cite{kammar2013handlers} In Figure 2.2 and Figure 2.3 provide further insight on the workings of effect handlers. Handlers are specified via sets of mappings, which consist of \emph{return clauses} or \emph{effect operations clauses}. For each effect operation, both types of handler specify a continuation. A continuation represents the remaining computation which is currently being handled. The only difference between the types of deep and shallow handlers is the type of the continuation $k$. The continuations are similar in that they take the effect parameter $p$ with type $B_{i}$. However, the effects under which the continuation proceeds along with the return types of the continuations differ:
\begin{itemize}
    \item In the deep case, the effect operations $E$ are handled by the handler, leaving a new effect $E^{'}$. Handlers in Koka, default to deep handlers. 
    \item In the shallow case, because the continuation returns an un-handled computation, the continuation indicates that the effect $E$ is not discharged. 
\end{itemize}
\begin{adjustwidth}{-0em}{0em}
\noindent\begin{minipage}{0.5\textwidth}
\begin{lstlisting}[caption=Total addition function,frame=tlrb,label={lst:total}]{add1}
fun add(x : int, y : int) 
    : total int
    { return x + y }
\end{lstlisting}
\begin{lstlisting}[caption=Effect constructors,firstnumber=1,frame=tlrb,label={lst:effects-used}]{add1}
effect control yield() : ()
effect control sample() : double
effect control score( s : exp ):()

\end{lstlisting}
\end{minipage}\hfill
\begin{minipage}{.45\textwidth}
\begin{adjustwidth}{1em}{0em}
\begin{lstlisting}[caption=Addition with print,frame=tlrb, label={lst:combined}]{add2}
fun add(x : int, y : int) 
: <total,console> int {
    println("Hello World!")
    return x + y }
\end{lstlisting}
\begin{lstlisting}[caption=Polymorphic addition function,frame=tlrb,firstnumber=1,label={lst:polymorphic}]{add2}
fun add(x : int, y : int) 
: <total|_e> int {
    println("Hello World!")
    return x + y }
\end{lstlisting}
\end{adjustwidth}
\end{minipage}
\end{adjustwidth}

\begin{adjustwidth}{-5em}{0em}
\noindent\begin{minipage}{0.75\textwidth}
\begin{tcolorbox}[colback=gray!10!white,
                     colframe=white!20!black,
                     title= \textsc{\textbf{Figure 2.2}\\ Deep handler typing rules},  
                     center, 
                     valign=top, 
                     halign=left,
                     before skip=0.8cm, 
                     after skip=1.2cm,
                     center title, 
                     width=7.6cm]
 
$E = \{\text{op}_{i} : A_{i} \rightarrow B_{i} \}_{i}$ \\
$H = \{\textbf{return}\; x \mapsto M\} \uplus \{\text{op}_{i} \; p \; k \mapsto N_{i}\}_{i}$ \\
\begin{center}
$[\Gamma,\;p : A_{i},\; k : B_{i} \rightarrow_{E^{'}} C \vdash_{E^{'}} N_{i} : C]_{i}$
$\Gamma,\; x : A \vdash_{E^{'}} M : C$ \\
\par\noindent\rule{\textwidth}{0.4pt}
$\Gamma \vdash H : A\; {}^{E}\Rightarrow^{\;E^{'}} C$ \\
\end{center}
  \end{tcolorbox}
\end{minipage}
\begin{minipage}{.58\textwidth}
\begin{adjustwidth}{-9em}{0em}
\begin{tcolorbox}[colback=gray!10!white,
                     colframe=white!20!black,
                     title= \textsc{\textbf{Figure 2.3}\\ Shallow handler typing rules},  
                     center, 
                     valign=top, 
                     halign=left,
                     before skip=0.8cm, 
                     after skip=1.2cm,
                     center title, 
                     width=7.6cm]
 
$E = \{\text{op}_{i} : A_{i} \rightarrow B_{i} \}_{i}$ \\
$H = \{\textbf{return}\; x \mapsto M\} \uplus \{\text{op}_{i} \; p \; k \mapsto N_{i}\}_{i}$ \\
\begin{center}
$[\Gamma,\;p : A_{i},\; k : B_{i} \rightarrow_{E} A \vdash_{E^{'}} N_{i} : C]_{i}$
$\Gamma,\; x : A \vdash_{E^{'}} M : C$ \\
\par\noindent\rule{\textwidth}{0.4pt}
$\Gamma \vdash H : A\; {}^{E}\Rightarrow^{\;E^{'}} C$ \\
\end{center}
  \end{tcolorbox}
\end{adjustwidth}
\end{minipage}
\end{adjustwidth}

\begin{adjustwidth}{2em}{0em}
\noindent\begin{minipage}{0.43\textwidth}
\begin{lstlisting}[caption=Weighted effect handler,frame=tlrb,firstnumber=1,label={lst:weighted}]{weighted}
fun weighted(w : exp, 
action : () -> <score|e> a) 
            : e (exp, a) {
  var w_t := w
    handle(action) {
      return x -> (w_t, x)
      score(s) -> {
      w_t := mult_exp(w_t, s)
      resume(()) 
}}}
      
\end{lstlisting}
\end{minipage}\hfill
\begin{minipage}{.40\textwidth}
\begin{adjustwidth}{0em}{0em}
\begin{lstlisting}[caption=Random Sampler effect handler,frame=tlrb,firstnumber=1,label={lst:random-sampler}]{random-sampler}
fun random_sampler(action){
  handle(action) {
    sample() -> {
    resume(random())
}}}
\end{lstlisting}
\end{adjustwidth}
\end{minipage}
\end{adjustwidth}

The \texttt{weighted} handler in Listing \ref{lst:weighted} handles score effects. The handler (colored in \textcolor{olive}{\textbf{olive}}), is initially called with \texttt{wp}. Line 3 creates a reference to this weight, so its value can be updated as successive score effects are encountered. Line 4 handles the `action' using the \texttt{handle} keyword. The computation takes the unit type and returns an $a$ with with effect annotation \texttt{<\textcolor{magenta}{\textbf{score}}|e>}, i.e. it contains score effects (effects are colored in \textcolor{magenta}{\textbf{magenta}}) and any other effects $e$. As score effects are encountered, this handler multiplies the parameter to each score effect with the previously stored score, which, at the point in the program when the first score statement is encountered, will be the initial weight \texttt{wp}. This iterative multiplication progressively calculates and then assigns, a real valued score to a computation. After each handled score effect, the result of each multiplication is stored as the new score. The program is then resumed with the unit type. Return values are returned in a pair along with the stored weight. The \texttt{random\_sampler} handler in Listing \ref{lst:random-sampler} handles sample effects by resuming computations with random doubles, thus converting sample effects into Koka's built in ndet effects. \\

Algebraic effects have been used to perform probabilistic programming before in Pyro's Poutine library \cite{bingham2018pyro}. Moore and Gorinova et al. \cite{moore2018effect,gorinovaautomatic} realize that interceptors in Edward \cite{tran2018simple} and other operations such as Tracing, Conditioning or accumulating the log joint density of a program, are accidental implementations of effect handlers. However, none of the works use effect types and their inference algorithms are not based off of a denotational semantics and thus not necessarily correct.

\chapter{Modular inference: importance sampling}

\section{Implementation}

The source code of Koka Bayes is on Github\footnote{https://github.com/theneuroticnothing/koka-bayes} and was based off the design of Eff-bayes\footnote{https://github.com/ohad/eff-bayes}. However, there are two important differences between Koka Bayes and Eff-Bayes. Firstly, Eff-Bayes is in a different language that doesn't have effect annotations. Secondly, Eff-Bayes only implements Importance Sampling \& SMC. I created Unix/Linux based \texttt{make} commands to ensure straightforward installation and running. Importance Sampling is a simple inference algorithm which underlies three compound inference algorithms, and therefore is introduced in this Chapter.

\section{Models as factor graphs}

In the context of Koka Bayes, the input to inference algorithms are models. In Koka Bayes, following the design of Monad Bayes, a model is defined, as a \emph{thunk} or delayed computation that returns a value $x$ of polymorphic type $a$ accompanied by at least score and sample effects:

\begin{itemize}
\item The \texttt{sample} effect, when handled by the \texttt{random\_sampler} handler, equips models with an operation that returns a uniformly distributed probability floating point value from the unit interval, expressing probabilistic non-determinism thus the \texttt{ndet} effect.
\item The score effect, when handled by the \texttt{weighted} handler, provides a mechanism to successively multiply the arguments to score effects together. The weighted handler calculates the likelihood of model samples, given observed data.
\end{itemize}

\section{Importance Sampling}

Importance sampling approximates integrals with sample averages. Statistics of the posterior such as its expectation, or the partition function of a Bayesian model (denominator of Equation \ref{eqn:total-prob}). For example, the partition function can be computed as follows:

$$
    Z(\theta) = \int \tilde{p}(x;\theta) \dif x = \int \dfrac{\tilde{p}(x;\theta)}{q(x)} q(x) \approx \dfrac{1}{n} \sum_{i=1}^{n} \dfrac{\tilde{p}(x;\theta)}{q(x)} \dif x
$$
$$
\mathds{E}_{p}[x] = \int x p(x) \, \dif x \, \approx\dfrac{1}{n}\sum\limits^{n}_{i=1}x_{i}
$$
In general:
$$
\mathds{E}_{p}[g(x)] = \int g(x) p(x) \, \dif x \approx\dfrac{1}{n}\sum\limits^{n}_{i=1}g(x_{i})
$$
This generalization allows integrals to be approximated as follows:
$$
I = \int g(x)p(x) \dif x = \dfrac{\int g(x) \tilde{p}(x) \dif x}{\int \tilde{p}(x) \dif x} = \dfrac{\int g(x) \dfrac{\tilde{p}(x)}{q(x)} q(x) \dif x}{\int \dfrac{\tilde{p}(x)}{q(x)} q(x) \dif x}
$$
$$
= \dfrac{\mathds{E}_{q}[g(x)\dfrac{\tilde{p}(x)}{q(x)}]}{\mathds{E}_{q}[\dfrac{\tilde{p}(x)}{q(x)}]} = \dfrac{\mathds{E}_{q}[g(x)\dfrac{\tilde{p}(x)}{\tilde{q}(x)}]}{\mathds{E}_{q}[\dfrac{\tilde{p}(x)}{\tilde{q}(x)}]} \approx \dfrac{\dfrac{1}{n} \sum_{i=1}^{n}g(x_{i})\dfrac{\tilde{p}(x_{i})}{\tilde{q}(x_{i})}}{\dfrac{1}{n} \sum_{i=1}^{n} \dfrac{\tilde{p}(x_{i})}{\tilde{q}(x_{i})}} = \dfrac{\sum_{i=1}^{n}g(x_{i})w_{i}}{ \sum_{i=1}^{n}w_{i}}
$$ \\
where $q(x) = 0 \implies g(x) = 0$ and the importance weights ($w_{i}$) and the normalized importance weights ($niw$) are: \\
$$
    w_{i} = \dfrac{\tilde{p}(x_{i})}{\tilde{q}(x_{i})} \quad \quad \quad niw = \dfrac{w_{i}}{\sum_{i=1}^{n} w_{i}}
$$
In this formulation of importance sampling, neither the prior $\tilde{q}(x)$ nor the likelihood function $\tilde{p}(x)$ needs to be normalized. $w_{i}$ is referred to as an importance weight. 

\section{Importance Sampling implementation}

Importance sampling uses two steps. The first step constructs a representation of the posterior. This step uses both the \texttt{score} and \texttt{sample} effects, handled by two distinct handlers; \texttt{weighted} and \texttt{random\_sampler}, respectively. The second step normalizes the importance weights.

\subsection{Population}
Listing \ref{lst:importance-sampling} implements Importance Sampling as a function parameterized by a model, that returns a histogram approximating the posterior. We implement histograms by a list of pairs of values corresponding with their weights. Values may appear multiple times in the histogram. The \texttt{populate} function in Listing \ref{lst:populate} constructs the histogram, by repeatedly executing the model, following the design of Monad Bayes. Given the model terminates without causing an exception, each execution encounters a return statement, whereby the handled model returns a pair of, an \textit{importance weight} (via the weighted handler) and a returned model sample, which are added to the histogram. These two-tuples are referred to as particles \cite{doucet2001introduction}. The \texttt{weighted} handler constructs the score progressively as the running of the model encounters score statements. The weights that parameterize the weighted handler (importance weights) are initialized uniformly to $\sfrac{1}{k}$, where $k$ is the number of particles. Once the entire histogram has been filled, the histogram is normalized such that the weights sum to one. \\

\begin{adjustwidth}{2em}{0em}
\noindent\begin{minipage}{0.43\textwidth}
\begin{lstlisting}[caption=Populate,frame=tlrb,firstnumber=1,label={lst:populate}]{populate}
fun populate(k : int, 
model : () -> <score|e> b) : e histogram<b> {
  list(1, k) fun(i) {
    weighted(Exp(0.0)) {
      score(div_exp(Exp(0.0), 
      Exp(log(k.double)))) // score 1/k
      model()
}}}
\end{lstlisting}
\end{minipage}\hfill
\begin{minipage}{.40\textwidth}
\begin{adjustwidth}{0em}{0em}
\begin{lstlisting}[caption=First Order Markov Chain,frame=tlrb,firstnumber=1,label={lst:first-order}]{first-order}
val g = fun() {
    var x       := 0.0
    val vrs      = 1.0
    val scor-vrs = 0.2
    for(0, 5) fun(i) {
      x := normal(x, vrs)
      score(
      normal_pdf(x, scor-vrs, 3.0)
      )}
    x}
g
\end{lstlisting}
\end{adjustwidth}
\end{minipage}
\end{adjustwidth}

\begin{adjustwidth}{-0.5em}{0em}
\begin{lstlisting}[caption= Importance Sampling with 2000 particles,frame=tlrb,firstnumber=1,label={lst:importance-sampling}]{weighted}
fun importance_sampling(model : model<a,e>) : <ndet|e> histogram<a> {
  random_sampler{normalise(populate(2000, model))}}
\end{lstlisting}
\end{adjustwidth}

Given a model which only samples from the proposal density (prior distribution), then creating the histogram, handled with the \texttt{weighted} and \texttt{random\_sampler} handlers, corresponds to \textit{importance sampling}. However, models that sample variables of interest, in a sequential fashion, such that sample statements depend on previous sample statements via a transition probability distribution, parameterized by the value of the previously sampled value from the proposal density i.e. $p(x_{t-1}|x_{t})$, then, handling such a model corresponds to Sequential Importance Sampling  \cite{doucet2000sequential}. 

\section{Use case : First order Markov Chain}

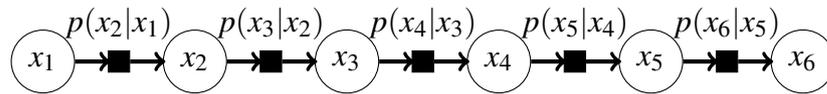
\begin{figure}
\begin{adjustwidth}{0em}{0em}
\centering
\begin{tikzpicture}[square/.style={regular polygon,regular polygon sides=4}]

\tikzset{vertex/.style = {shape=circle,draw,minimum size=1em}}
\tikzset{edge/.style = {->,> = latex'}}
\node[vertex] (x1) at  (1,0) {$x_{1}$};
\node[vertex] (x2) at  (3,0) {$x_{2}$};
\node[vertex] (x3) at  (5,0) {$x_{3}$};
\node[vertex] (x4) at  (7,0) {$x_{4}$};
\node[vertex] (x5) at  (9,0) {$x_{5}$};
\node[vertex] (x6) at  (11,0) {$x_{6}$};

\node (f1) at (2,0) [minimum size=0.2cm,draw,fill=black,label={$p(x_{2}|x_{1})$}] {};
\node (f2) at (4,0) [minimum size=0.2cm,draw,fill=black,label={$p(x_{3}|x_{2})$}] {};
\node (f3) at (6,0) [minimum size=0.2cm,draw,fill=black,label={$p(x_{4}|x_{3})$}] {};
\node (f4) at (8,0) [minimum size=0.2cm,draw,fill=black,label={$p(x_{5}|x_{4})$}] {};
\node (f5) at (10,0) [minimum size=0.2cm,draw,fill=black,label={$p(x_{6}|x_{5})$}] {};

\draw (x1) edge[-to,shorten >=-1pt,ultra thick] (f1);
\draw (f1) edge[-to,shorten >=-1pt,ultra thick] (x2);
\draw (x2) edge[-to,shorten >=-1pt,ultra thick] (f2);
\draw (f2) edge[-to,shorten >=-1pt,ultra thick] (x3);
\draw (x3) edge[-to,shorten >=-1pt,ultra thick] (f3);
\draw (f3) edge[-to,shorten >=-1pt,ultra thick] (x4);
\draw (x4) edge[-to,shorten >=-1pt,ultra thick] (f4);
\draw (f4) edge[-to,shorten >=-1pt,ultra thick] (x5);
\draw (x5) edge[-to,shorten >=-1pt,ultra thick] (f5);
\draw (f5) edge[-to,shorten >=-1pt,ultra thick] (x6);

\end{tikzpicture}
\end{adjustwidth}
\caption{Factor graph of first order Markov Chain.} \label{fig:gaussian-fg}
\end{figure}

Figure \ref{fig:gaussian-fg} demonstrates Importance Sampling on a Gaussian first order Markov Chain. The code corresponding to this model can be seen in Listing \ref{lst:first-order}. In this chain, all transition distributions are Gaussian. At each variable node, the sampled random value is scored with respect to a Gaussian function with the data acting as the mean -- modelling corruptions of the observed values by Gaussian noise. The parameters that account for the standard deviation of both the corruption and the transition distributions are chosen arbitrarily at constant values of 0.2 and 1.0 respectively. 

\section{Caveats}

Importance Sampling suffers from sensitivity to the choice of the proposal distribution. If the proposal distribution (the prior) assigns low probability mass in regions of high probability associated with the posterior distribution then few samples will be drawn from the posterior. This is reflected in Figure \ref{fig:is-prior} whereby the returned histogram returns samples from the Gaussian prior $\mathcal{N}(0,1)$. Examining the likelihood of each particle reveals the flaw of importance sampling. Figure \ref{fig:is-posterior} shows that most particles have zero probability and only a single particle from the prior, which was sampled from the posterior (a few points at $\Theta = 3$) have all of the probability mass. \\

Resample Move Sequential Monte Carlo (RMSMC) and Sequential Monte Carlo (SMC), are both able to overcome the limitations of Importance Sampling, through the use of a divide and conquer approach to posterior estimation \cite{buckland2007embedding}. Trace Markov Chain Monte Carlo (TMCMC), is, in the limit of samples, also able to overcome the limitations of a prior distribution that does not closely resemble the posterior, though the domain where it works efficiently is generally different to that of SMC. Other approaches to solving the issues of importance sampling exist in the literature, such as suggesting proposal distributions closer to the posterior \cite{liu1998sequential}. RMSMC, SMC and TMCMC are implemented in the next Chapter.

\begin{figure}
\label{fig:gaussian}
\begin{adjustwidth}{0em}{0em}
  \begin{subfigure}[b]{0.5\textwidth}
    \includegraphics[width=\textwidth]{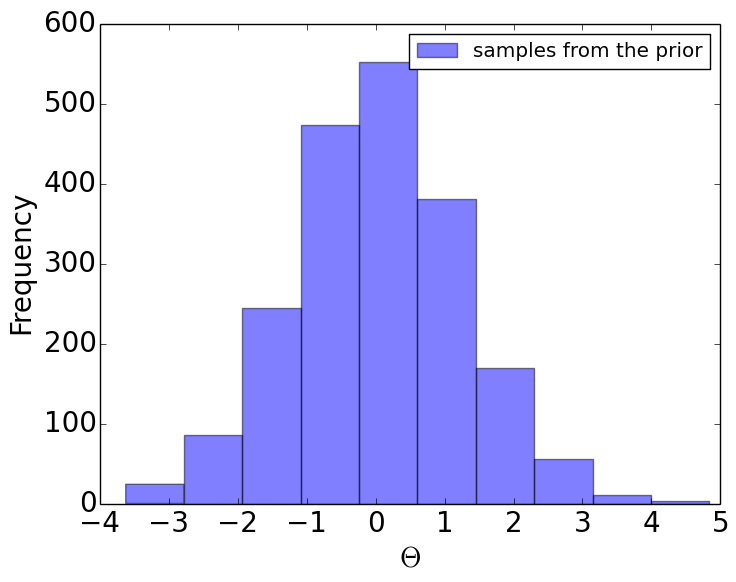}
    \caption{Importance sampling ; prior}
    \label{fig:is-prior}
  \end{subfigure}
  \begin{subfigure}[b]{0.5\textwidth}
    \includegraphics[width=\textwidth]{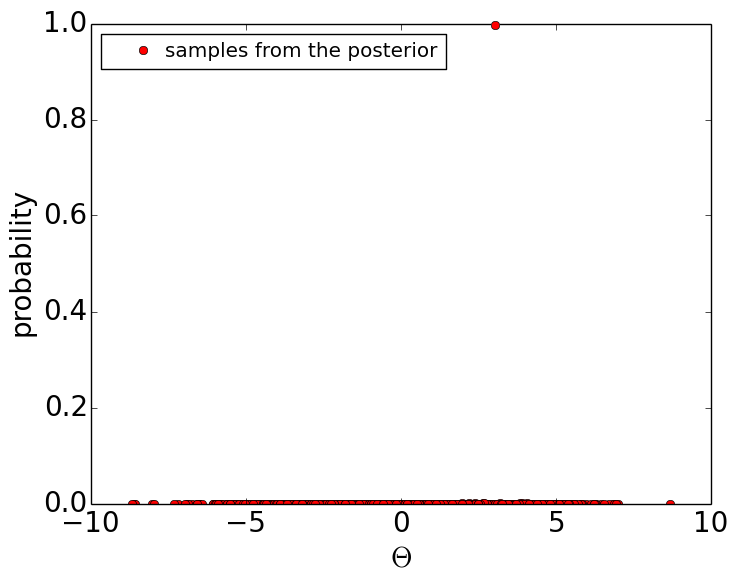}
    \caption{Importance sampling ; posterior}
    \label{fig:is-posterior}
  \end{subfigure}
\end{adjustwidth}
\begin{adjustwidth}{0em}{0em}
  \begin{subfigure}[b]{0.5\textwidth}
    \includegraphics[width=\textwidth]{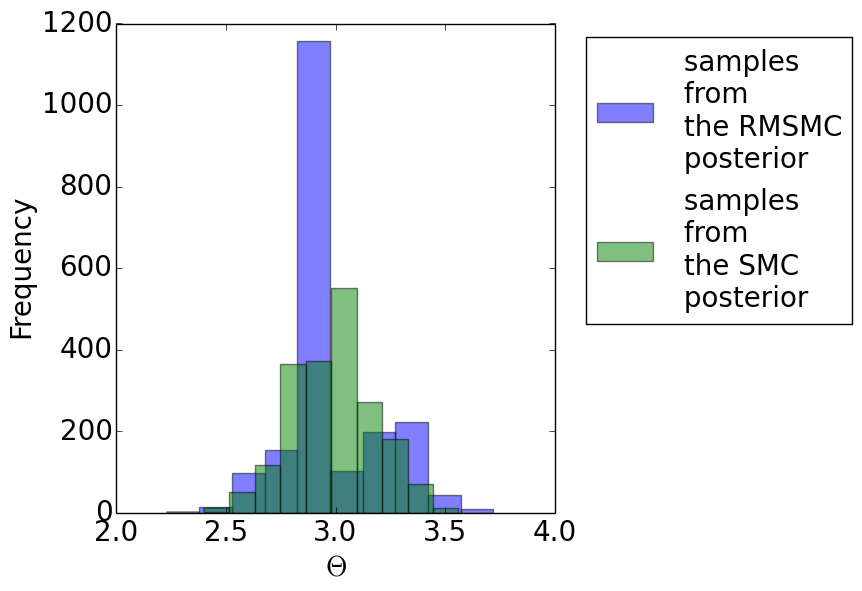}
    \caption{RMSMC / SMC ; posterior}
    \label{fig:is-prior}
  \end{subfigure}
  \begin{subfigure}[b]{0.5\textwidth}
    \includegraphics[width=\textwidth]{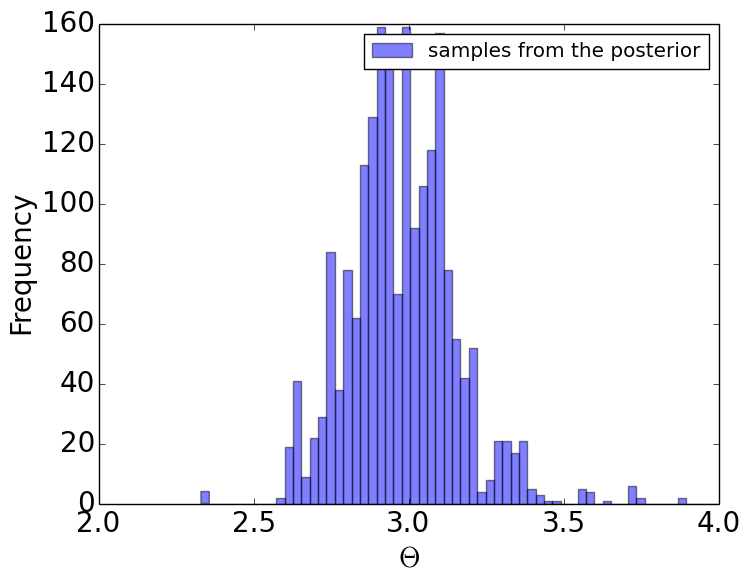}
    \caption{TMCMC ; posterior}
    \label{fig:is-posterior}
  \end{subfigure}
\caption{Results for inference over the first order Markov Chain model.}
\end{adjustwidth}
\end{figure}

\chapter{Inference algorithms}

This section describes four inference algorithms with operational descriptions of their implementations. In particular:

\begin{enumerate}
    \item I implement 2 building blocks from Monad Bayes: resampling and MH update.
    \item I combined these into: SMC, TMCMC, RMSMC and PMMH.
\end{enumerate}

\section{Resampling}

In order to solve the aforementioned limitations of importance sampling, Doucet et al. \cite{doucet2001introduction} implements a resampling mechanism that probabilistically removes particles with low importance weights, and probabilistically multiplies particles with high importance weights. Composing sequential importance sampling with this resampling procedure results in Sequential Monte Carlo \cite{scibior2018functional}. \\

Doucet et al. \cite{doucet2009tutorial} suggest three resampling procedures; multinomial, systematic and residual sampling. The multinomial sampling approach is adopted in the Sequential Monte Carlo implementation that I provide. In multinomial sampling, in order to generate $n$ samples from the target distribution, $n$ uniformly random samples are generated from the unit interval. Each of these randomly generated samples $u_{1},u_{2}...u_{n}$ are used as input to a generalized inverse cumulative probability distribution of the normalized weights $w$, i.e. $F^{-1}(u_{i})$. This will return the particle $x_{i}$ with $i$ such that $u_{k} \in [\sum_{s=1}^{i-1}w_{s},\sum_{s=1}^{i}w_{s})$ \cite{hol2006resampling}. This returns a particle with weight $w_{i}$ with probability $\dfrac{w_{i}}{\sum_{i=1}^{N}w_{i}}$. \\

From an operational point of view, the resample function in the code first calculates the total weight of all of the particles in the histogram that represents the posterior. A random sample from the uniform distribution is multiplied with this total weight, which results in a random proportion $R$ of the total weight. The histogram of all particles is then recursed over such that each particle subtracts its own weight, from $R$. Eventually the value is subtracted until it decreases to zero or below. At this point, the particle that is being considered at that point in the recursion is returned. In this way, the probability of a particle being returned is in direct proportion to its weight.

\section{Sequential Monte Carlo}

Sequential Monte Carlo (SMC) is an inference algorithm that aims to sample from high dimensional target probability densities $\pi(x_{1:n})$ (such as the posterior), which cannot be sampled from directly, by using samples from lower dimensional densities that increase in dimension i.e. $\pi(x_{1})$ then $\pi(x_{1:2})$ etc. Typically SMC is used for \textit{filtering} or \textit{smoothing} over models with outputs $y$ and latent variables $x$ \cite{doucet2001introduction,doucet2009tutorial}. In smoothing, the objective is; given data $\{y_{1}, ... y_{n}\}$, to infer the probability distribution $p(x_{t} | \{y_{1}, ... y_{n}\})$ s.t. $t < n$ and in the case of filtering to infer the probability distribution $p(x_{t} | \{y_{1}, ... y_{n}\})$ with $t = n$. More formally, the objective in filtering is to obtain the distribution $p(x_{1:n}|y_{1:n})$ where $p(x_{1:n})$ is the prior and $p(y_{1:n}|x_{1:n})$ is the likelihood \cite{doucet2001introduction}.
$$p(x_{1:n}) = \mu(x_{1})\prod_{k=2}^{n}f(x_{k}|x_{k-1})$$
$$p(y_{1:n}|x_{1:n}) = \prod_{k=1}^{n}g(y_{k}|x_{k})$$
In these equations, the function $f$ is the state-transition function and $g$ is the observation function. Using Bayes rule:
$$p(x_{1:n}|y_{1:n}) = \dfrac{p(y_{1:n}|x_{1:n})p(x_{1:n})}{p(y_{1:n})} = \dfrac{p(y_{1:n}|x_{1:n})p(x_{1:n})}{\int p(y_{1:n}|x_{1:n})p(x_{1:n}) \dif x_{1:n}}$$
and a simple proof provided by Doucet et al. \cite{doucet2000sequential} demonstrates a recursive relationship exists between posteriors at different steps. SMC exploits this recursive relationship between the posteriors at different time-steps in order to perform efficient inference \cite{doucet2000sequential}\cite{buckland2007embedding}:
$$p(x_{1:n}|y_{1:n}) = p(x_{1:n-1}|y_{1:n-1})\dfrac{f(x_{n}|x_{n-1})g(y_{n}|x_{n})}{p(y_{n}|y_{1:n-1})}$$
$$w_{n} = w_{n-1}\dfrac{f(x_{n}|x_{n-1})g(y_{n}|x_{n})}{p(y_{n}|y_{1:n-1})}$$
The SMC algorithm is presented in three steps \cite{doucet2009tutorial}. The first step samples particles, the second computes importance weights and the third resamples the particles.

\subsection{Sequential Monte Carlo implementation}

The Bootstrap Filter is an example of a well known, modular, SMC algorithm and is the combination of sequential importance sampling with a resampling procedure \cite{doucet2001introduction,buckland2007embedding}. The SMC algorithm implemented here is an instance of this algorithm. SMC first populates a histogram with particles that will represent the posterior distribution. The models that are used to populate the histogram are transformed into a series of functions, which when called, run the model by a specified number of score statements. This corresponds to advancing a computation by a number of suspension points. By partially evaluating a model, one is able to do a bit of inference followed by running the model and then resampling, before doing more inference. This corresponds to the layout of the generic SMC algorithm \cite{doucet2009tutorial} which has successive rounds of weight evaluation steps followed by resampling steps. This is achieved in code by running the model, through the \texttt{populate} function parameterized by the \texttt{populate\_func()} function on lines 1 -- 3 in Listing \ref{lst:func-smc}. The \texttt{populate\_func()} function composes the \texttt{advance} handler with the \texttt{yield\_on\_score} handler. \\

In Listing \ref{lst:yield}, the \texttt{yield\_on\_score} handler handles score effects by emitting a yield effect, before resuming the program. Yield effects allow the computation to be suspended. When handled by the \texttt{advance} handler, they can be positioned in such a way as to allow for the computation to proceed by a specific amount. The \texttt{advance} effect handler, upon encountering a yield statement, turns the handled computation into a function that takes a value $a$ of type int. This value $a$ is conditioned on by the handler, such that if it is greater than zero, the yield effect is discharged and the program is resumed with $a - 1$ placed on the stack, otherwise the yield effect persists and $0$ remains on the stack before program resumption. This corresponds to lines 6 - 11 in Listing \ref{lst:advance}. Calling these functions with an integer $n$, will mean at most $n$ score statements are handled in-between resampling steps, thus bringing the computation closer to returning a model sample. When this value reaches $0$, the computation will once again yield. When the advance handler resumes each computation, because of the deep nature of the advance handler, subsequent score statements will still be handled by this handler, which will now by default, emit a yield. This is because the parameter with which it is called, will stay at $0$ once it reaches $0$. The \texttt{advance} handler is used to handle yields which other advance handlers emit (i.e. they are composed). When the function induced by the advance handler is consistently passed a small integer $n$ and there are many score statements in the computation, then there will be a linear increase in the number of handlers used per score statement. \\

\begin{adjustwidth}{1em}{0em}
\noindent\begin{minipage}{0.45\textwidth}
\begin{lstlisting}[caption=Advance effect handler,frame=tlrb,firstnumber=1,label={lst:advance}]{weighted}
fun advance(action 
    : () -> <yield|e> b) 
    : e ((a : int) 
    -> <yield|e> b) {
  handle(action) {
    return x -> fun(a){x}
    yield() -> fun(a) {
      if (a > 0) then {
        (mask<yield>{
        resume(())})(a - 1)
      } else {
        yield()
        (mask<yield>{
        resume(())})(0)
}}}}
\end{lstlisting}
\end{minipage}\hfill
\begin{minipage}{.5\textwidth}
\begin{adjustwidth}{2.5em}{0em}
\begin{lstlisting}[caption=Yield on score effect handler,frame=tlrb,firstnumber=1,label={lst:yield}]{random-sampler}
fun yield_on_score(action) {
  handle(action) {
    return x -> x
    score(w) -> { 
    score(w); 
    yield(); 
    resume(())
}}}
\end{lstlisting}
\begin{lstlisting}[caption=Finalize effect handler,frame=tlrb,firstnumber=1,label={lst:finalize}]{finalize}
fun finalize(action) {
  handle(action) {
    return x -> x
    yield() -> resume(())
}}
\end{lstlisting}
\end{adjustwidth}
\end{minipage}
\end{adjustwidth}

\begin{adjustwidth}{1em}{0em}
\begin{lstlisting}[caption=Functions that underlie Sequential Monte Carlo,frame=tlrb,firstnumber=1,label={lst:func-smc},linewidth=14.2cm]{weighted}
fun populate_func() { 
    advance { yield_on_score { model() } } 
} // This function is use to initially populate the histogram.

fun adv_func(wm) {
  match(wm) { (w, m) -> { weighted(w) { advance { m(step_size)}}}} 
} // This function advances each particle by a certain step_size.

fun fin_func(wm) { 
    match(wm) { (w,m) -> { weighted(w) { finalize{ m(0) }}}}
} // This function handles remaining yields given no more steps.
\end{lstlisting}
\end{adjustwidth}
After only a few thousand handled score statements, each single score statement will be handled by many thousands of nested advance handlers, with thousands of separate parameters stored on the stack. This behaviour is the cause of a space leak, that is likely responsible for a dysfunctional behaviour of RMSMC and the poor time complexity of SMC. A solution to this issue is to use shallow effect handlers. Shallow effect handlers don't rehandle the computation and instead return a computation that may still incur effects which match those that have been handled. Shallow handlers can be called recursively on the handled computation, so that there is at most one handler per yield statement. I didn't use shallow handlers because Koka's current implementation does not support them. \\

After \texttt{populate\_func()} has populated the histogram, through repeatedly executing the model and handling it with the aforementioned handlers, the histogram is now a list of functions and is passed to the main loop that executes SMC. This main loop checks if the maximum number of steps has been reached. If there are still steps remaining -- a resampling step occurs and each model is advanced by \texttt{step\_size}. In order to advance the computations and handle yielding computations, the \texttt{adv\_func} function on lines 5 - 7 in Listing \ref{lst:func-smc}, is mapped over the histogram that represents the posterior. The \texttt{adv\_func} function has three components. The first is the weighted handler, which takes the existing particle weight and updates it by multiplying it with the parameters of successive score statements. In general, throughout the library, underflow issues are avoided when computing with large negative exponential powers, by encapsulating the powers within the \texttt{Exp} data-type and operating on the powers rather than the floating point result of evaluating the power. This means that a high degree of precision is likely to be preserved under multiplication, addition and division. The second is the \texttt{step\_size} parameter that is passed to each function in the histogram. This is passed to one of the models and is used to advance the model computation. The third is the advance handler that is used to handle yields from previous advance handlers. \\

The number of steps given as a parameter is an upper bound on the number of steps between resampling operations. This is because the advance handler handles return statements by simply discarding the parameter and returning the model sample. Resampling steps alone will reduce the diversity of the particles because particles with marginally lower weights than other particles will steadily be eliminated. \\

The other branch of the main loop is executed when there are no more steps remaining in the SMC algorithm. This branch, applies \texttt{fin\_func} to each particle in the histogram. There are also three key steps here. The first involves the weighted handler that keeps track of the model weight. The second is the \texttt{step\_size} parameter that is now zero. This can be understood by examining Listing \ref{lst:finalize}. Here, the finalize handler handles yield statements, by immediately resuming. The consequence of this is that if any other advance handler yields, the parameter $a$ will not affect the control flow. By immediately resuming each yield, we run the model to capacity and then we encounter a return statement. SMC returns the histogram as the output.

\section{Trace Markov Chain Monte Carlo}
\subsection{MCMC}
In the context of Bayesian Inference, Markov Chain Monte Carlo methods perform random walks over the posterior distribution by using transition kernels to generate samples based off current samples. Trace Markov Chain Monte Carlo (TMCMC), uses the space of program traces as the distribution and therefore uses two key components. The first key component is the concept of a program trace which is a list of random choices a program has made \cite{scibior2018functional}. The second key component is the Metropolis Hastings update procedure. \\

To describe MCMC, I follow the theoretical description of Metropolis Hastings provided by Mackay \cite{mackay2003information}. The Metropolis Hastings algorithm is a common instance of an MCMC method, and is comparable to importance sampling in that it is a technique to estimate target densities through the use of proposal densities. However, the denotational validation of inference, off of which Monad Bayes and therefore Koka Bayes is based, is not restricted to density functions \cite{scibior2017denotational}. The Metropolis Hastings algorithm assumes access to some multiplicative constant of the target density $P^{*}(x)$ s.t. $P(x) = \sfrac{P^{*}(x)}{Z}$ for any state $x$, as well as a proposal density $Q(x^{\prime};x^{i})$ that depends on the state at the current step $i$; $x^{i}$. This proposal density $Q$ can be \textit{any} fixed positive density ($Q(x^{\prime};x^{i}) > 0$ for all $x,x^{\prime}$) from which samples can be drawn, in order for the method to converge to the target density \cite{mackay2003information}. \\

MCMC uses a proposal distribution to construct a Markov Chain with the target distribution (the posterior), as the limiting distribution. In Koka Bayes, the target distribution is approximated with a list of samples. At the first step, an initial list is created and initialized with an initial state $x^{0}$ -- sampled arbitrarily. In general, at time-step $i$, new states $x^{i+1}$ are proposed, based off of the current state $x^{i}$. Whether or not new states are added to the list or not (the Metropolis Hastings update procedure), is conditional on the result of evaluating a Bernoulli random variable. This random variable is parameterized by an un-normalized ratio $a$:
\begin{equation}\label{eqn:mh-update-ratio}
a = \dfrac{P^{*}(x^{\prime})}{P^{*}(x^{i})} \dfrac{Q(x^{i};x^{\prime})}{Q(x^{\prime};x^{i})}
\end{equation}
such that if $a \geq 1$ then the new state is accepted, and if $a < 1$ then the new state is accepted with probability $a$. If the Bernoulli random variable evaluates to \texttt{False} then the new state is rejected, else the old state is added to the list once more. \\

The ratio $a$ is greater at higher probability areas of the target density, weighted by the symmetry of the proposal distribution. For example, if the proposal distribution is not symmetrical then the ratio will tend to move in a certain direction more than others. If the proposal distribution is symmetrical then the right hand fraction reduces to one. A symmetric proposal distribution $Q$ has the property: $ \forall i . Q(x^{i};x^{\prime}) = Q(x^{\prime};x^{i})$

\subsection{Program traces and the trace data structure}

The arbitrary initialization of the list that approximates the posterior, corresponds to the instantiation of a single list of all of the random choices made over a single run of the model. The posterior distribution in the space of traces is thus represented by a list of lists. This means that an initial list of \textit{trace values} (double-precision floating point randomly sampled values that correspond to random choices in the program) represents the starting point in the Metropolis Hastings algorithm. \\

The \texttt{Trace} data type is central to the implementation of TMCMC and is used as a parameter to iterations of the Metropolis Hastings update procedure. This data type consists of a four-tuple of: the model; the approximated probability of the model result given observed data (obtained via the \texttt{weighted} handler); the trace values, and the final model result itself. For example, for the first order Markov Chain in Listing \ref{fig:gaussian-fg}: the model is the function \texttt{g}, the approximated probability of the model is the weight that results from handling the score statements of the model, the trace values consist of twelve doubles resulting from the six normal distribution samples and the model result would consist of the final returned value \texttt{x}. 

\subsubsection{Perturbation of the program trace}

Perturbation of the program trace is the way in which new program traces are proposed. The perturbation occurs via the \texttt{perturb\_trace} function which randomly selects one of the trace values and replaces it with a new uniformly distributed sample. The MH update in Koka Bayes is based off the update procedure that Monad Bayes uses. The modification of the trace may induce a model which has more or less sample statements than there were previously (if the control flow is modified based on the outcome of a sample operation). If the model now encounters more sample statements, then these further samples are assigned uniformly random values from the unit interval. The approximated probability and the length of the traces of both the model with the perturbed trace, and without the unmodified trace, parameterizes the Bernoulli random variable in the MH update procedure, that is conditioned on when accepting proposed traces. This is a result from Monad Bayes, which generalizes the Metropolis Green Hastings theorem to program traces. The ratio that parameterizes the Bernoulli distribution is therefore different to the ratio $a$ in Eqn. \ref{eqn:mh-update-ratio}. The proposal distribution favours exploring programs with longer traces, which corresponds to proofs relating the Metropolis Green Theorem to program traces in \cite{scibior2017denotational}.

\subsection{Trace Markov Chain Monte Carlo implementation}
 The \texttt{tmcmc} function uses the \texttt{replay} effect handler over sample statements to initialize the list and when doing MH updates. The replay effect handler in Listing \ref{lst:replay} handles sample statements. It keeps track of the number of sample statements encountered during the action being handled and extends or truncates the trace where necessary. When a return statement is encountered, the modified trace is returned in a tuple along with the original return value.
 
 \newpage
\begin{adjustwidth}{1em}{0em}
\begin{lstlisting}[caption=Replay effect handler,frame=tlrb,firstnumber=1,label={lst:replay},linewidth=13cm]{weighted}
fun replay(trace, action) : <sample|e> (list<double>, a) {
  var new_trace = trace
  var index     := -1
  handle(action) {
    return x -> (split(new_trace, index+1).fst, x)
    sample() -> {
      index := index + 1
      match(new_trace[index]) {
        Nothing -> {
          val rnd = sample()
          new_trace := new_trace + [rnd]
          resume(rnd) }
        Just(random_value) -> resume(random_value)
}}}}
\end{lstlisting}
\end{adjustwidth}
\begin{adjustwidth}{1em}{0em}
\begin{lstlisting}[caption=MH Update ,frame=tlrb,firstnumber=1,label={lst:mh-step},linewidth=13cm]{random-sampler}
fun mh_step(trace) {
  val n_trace = perturb_trace(trace)
  match(trace) {Trace(model, p, old_tr, _) -> {
      val p2b = with_randomness(model, n_trace)
      match(p2b){ (new_tr, (q, b)) -> {
        val ratio = min(1.0, (q * old_tr.length) 
        / p * new_tr.length)))
        val accept = bernoulli(ratio)
        if(accept) {  Trace(model, q, new_tr, b) } else {trace}
}}}}}
\end{lstlisting}
\end{adjustwidth}
The \texttt{mh\_step} function implements the Metropolis Hastings update procedure. This function is parameterized by a \texttt{Trace} data structure. The update procedure modifies the trace of the data structure on line 2 in Listing \ref{lst:mh-step} using \texttt{perturb\_trace}. After the trace has been modified, the model is run with the \texttt{with\_randomness} function on line 6 in Listing \ref{lst:mh-step} that uses two handlers; the \texttt{replay} effect handler parameterized by the perturbed trace, and the \texttt{weighted} effect handler. This function returns a tuple of three things; the new modified trace, the new probability of the model and the final sample that results from the model execution. The result of the \texttt{accept} Bernoulli random variable dictates which \texttt{Trace} data structure, \texttt{mh\_step} returns. \\

The \texttt{tmcmc} function takes a model, the number of tmcmc steps, an initial probability of the model and a burnin parameter (which controls for the number of initial samples that are discarded) and returns a pair. The burnin is used to control for the highly dependent nature of the simulation on initial program traces. The first element of the returned pair is another list of pairs. Each pair in this list contains trace values paired with the corresponding return values. The reason that both the trace and the result are stored are due to efficiency reasons, in that once the posterior trace list is returned, the model does not need to be run again. The second element is the final \texttt{Trace} data structure which results when the algorithm terminates. The first element of the returned pair is enough to be able to approximate the posterior.

\section{Resample Move Sequential Monte Carlo}

Resample Move Sequential Monte Carlo (RMSMC) \cite{gilks2001following} uses a move step in addition to resampling steps in SMC. Move steps perform one or more iterations of MCMC after each resampling step. This solves an important issue of SMC which is that particles with high importance weights at early time-steps, may induce poor random choices later on, which degenerates particles, that are otherwise good approximations of the posterior.

\subsection{Resample Move Monte Carlo implementation}

In-between resampling and advance steps, each particle has Markov Chain Monte Carlo Metropolis Hastings updates step applied to it. This is achieved in Listing \ref{lst:rmsmc}. Specifically this formula is placed into the \texttt{adv\_func} function from Listing \ref{lst:func-smc}. Four parameters are passed to the tmcmc function: a model, \texttt{t\_steps} is the number of Metropolis Hastings steps to perform, $w$ -- the weight of the model sample and the number of burnin samples which is set to $0$.

\begin{adjustwidth}{0em}{0em}
\begin{lstlisting}[caption= Resample Move Sequential Monte Carlo Update. ,frame=tlrb,firstnumber=1,label={lst:rmsmc},linewidth=13cm]{weighted}
tmcmc( f() { m(step_size) }, t_steps, w, 0 ).snd.trace_m
\end{lstlisting}
\end{adjustwidth}

\section{Particle Marginal Metropolis Hastings} 

Resample-Move uses MH steps inside SMC. Particle Marginal Metropolis Hastings  (PMMH) reverses this pattern by using SMC as an MH proposal distribution \cite{andrieu2010particle}. PMMH is defined over models that decompose to a prior over parameters $p(\theta)$ coupled with a state space model $p_{\theta}(x_{1:t}|y_{1:t})$ that features latent variables. PMMH calculates the posterior distribution $p(\theta | y_{1:t}) \propto p_{\theta}(y_{1:t})p(\theta)$, by using SMC to obtain an estimate of $p_{\theta}(y_{1:t})$ which is then factored into the MH acceptance ratio. PMMH is primarily used for parameter estimation in time-series models \cite{scibior2018functional}.

\subsection{Particle Marginal Metropolis Hastings implementation}

The \texttt{parameter\_model} represents the prior distribution over parameters, the result of which is used to parameterize the \texttt{main\_model}. The main model has SMC (without weight normalization) applied to it and the sum of the importance weights is used as the score that approximates $p_{\theta}(y_{1:t})$. The TMCMC algorithm is then applied to this combined model. As all sample and score statements inside the main model are handled by SMC, TMCMC will only perform inference over un-handled samples inside the parameter model.

\begin{adjustwidth}{1em}{0em}
\begin{lstlisting}[caption=PMMH algorithm ,frame=tlrb,firstnumber=1,label={lst:rmsmc},linewidth=11cm]{pmmh}
fun new_model(parameter_model : model<b,_e>, 
main_model : b -> model<a,_e1>,
particle_num : int, step_num : int) 
: model<b, _e> {
  val g = fun() {
    val params = parameter_model()
    val smc_hist = smc(particle_num, step_num, 1, 
        main_model(params), False)
    score(sum_weights(smc_hist))
    params
  }
  g
}
fun pmmh(parameter_model, main_model) {
  val pmmh_model = new_model(parameter_model, 
    main_model, 10, 10)
  tmcmc(pmmh_model, 10, Exp(0.0), 0)}
\end{lstlisting}
\end{adjustwidth}
\section{Evaluation}
\subsection{Quantitative Evaluation}

\begin{figure}
\begin{adjustwidth}{0em}{0em}
  \begin{subfigure}[b]{0.5\textwidth}
    \includegraphics[width=\textwidth]{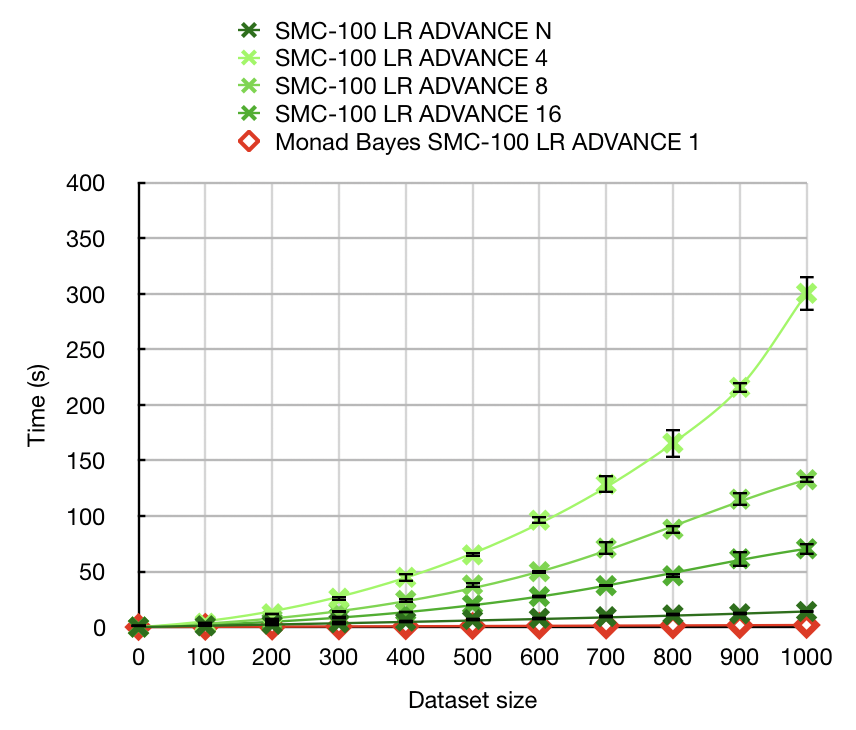}
    \caption{SMC performance : increasing dataset.}
    \label{fig:smc-dataset}
  \end{subfigure}
  \begin{subfigure}[b]{0.5\textwidth}
    \includegraphics[width=\textwidth]{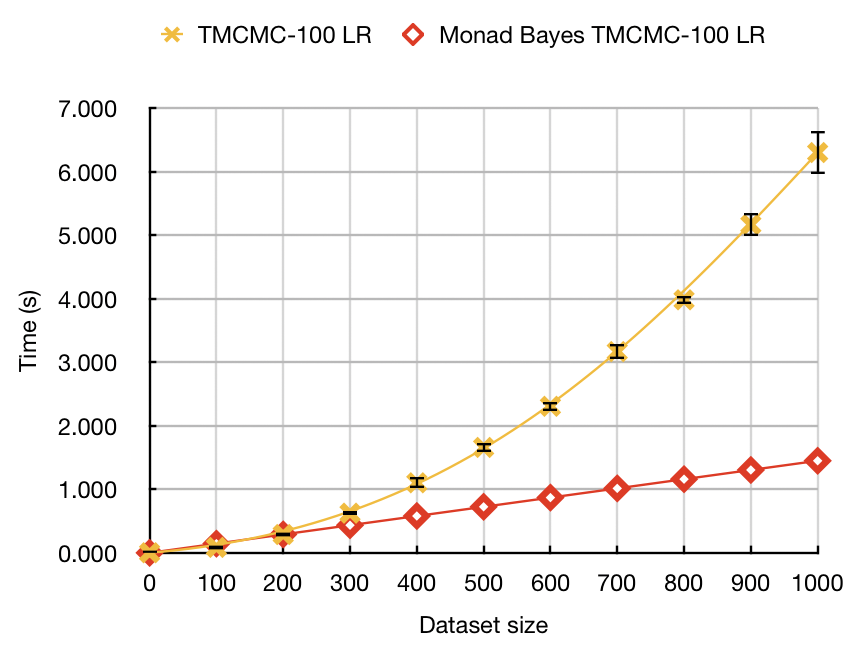}
    \caption{TMCMC performance : increasing dataset.}
    \label{fig:tmcmc-dataset}
  \end{subfigure}
\end{adjustwidth}
\begin{adjustwidth}{0em}{0em}
  \begin{subfigure}[b]{0.5\textwidth}
    \includegraphics[width=\textwidth]{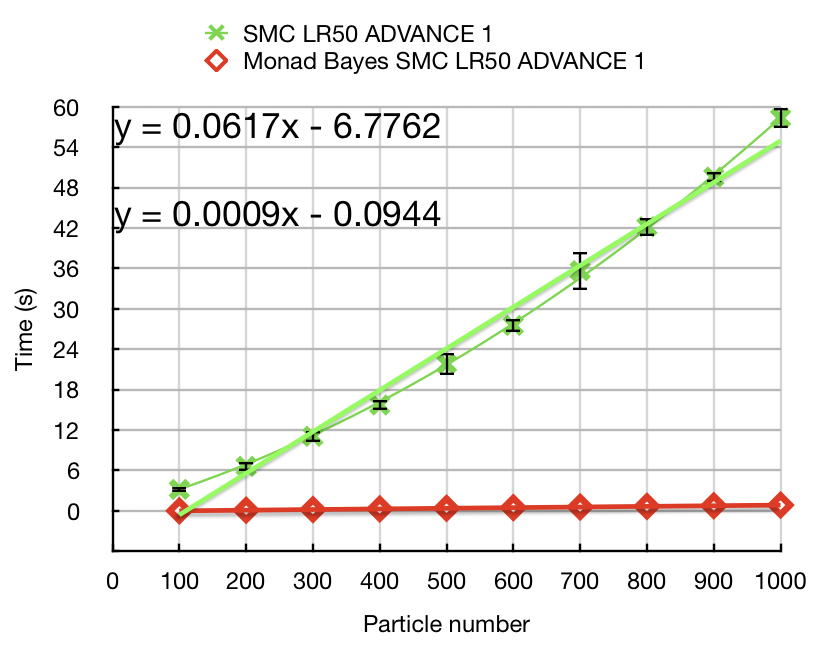}
    \caption{SMC performance : increasing particles.}
    \label{fig:smc-particle}
  \end{subfigure}
  \begin{subfigure}[b]{0.5\textwidth}
    \includegraphics[width=\textwidth]{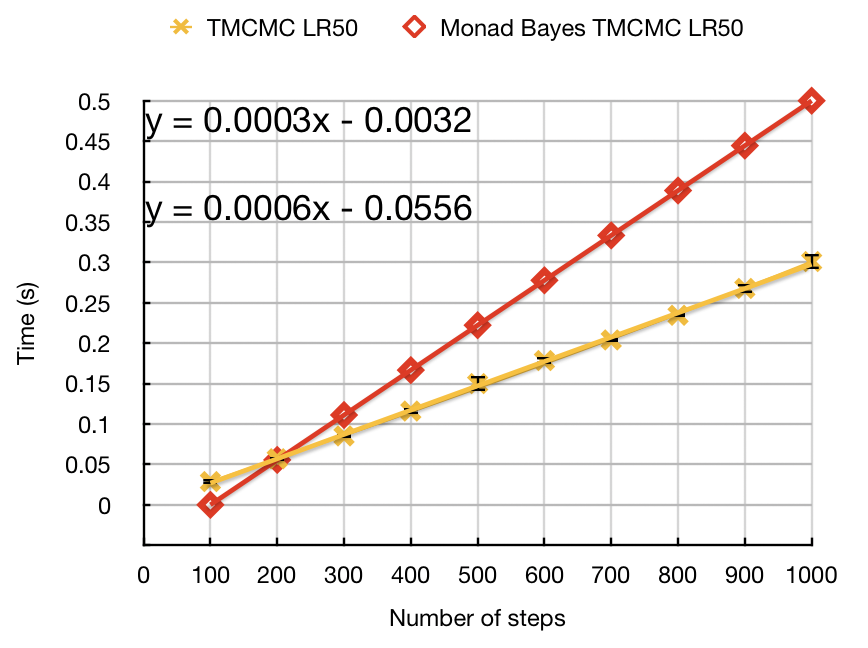}
    \caption{TMCMC performance : increasing steps.}
    \label{fig:tmcmc-step}
  \end{subfigure}
\end{adjustwidth}
\caption{The top row uses the dataset size as an independent variable. The bottom row uses the number of steps / particle number as an independent variable. LR50 denotes a logistic regression model with 50 data points. ADVANCE N in general corresponds to the number of score statements advanced through before resampling steps occur. The algorithm name followed by a number denotes the number of particles or steps of the algorithm. The error bars indicate the standard-deviation of the time over four runs of the algorithms. The top equation in \ref{fig:smc-particle} refers to Koka Bayes, whilst in \ref{fig:tmcmc-step} it refers to that of Monad Bayes.}
\label{fig:timing}
\end{figure}

Figure \ref{fig:timing} details timing experiments, run over a Logistic Regression model ported from Monad Bayes to Koka Bayes, on a Macbook Pro with a 2.7 GHz Intel Core i7 processor (16 GB 2133 MHz LPDDR3 RAM). The Logistic Regression model was chosen to ensure comparability between the libraries. However, there are considerable issues in comparing the performance of Koka Bayes to that of Monad Bayes. In particular, I did not run Monad Bayes over the same processor architecture that I tested Koka Bayes on, and Monad Bayes uses suspension points after every score statement in the Logistic Regression model. This means that after every score statement, resampling steps occur. Performing resampling steps after every score statement becomes intractable for models with many score statements for Koka Bayes's implementation of RMSMC and (to some extent) SMC. I expect that this intractability derives from the space leak of the deep advance handler. RMSMC has poor overall performance in that whilst it copes with arbitrary rejuvenation steps -- it tolerates only one resampling step and therefore is not displayed in this section. Improvement in the efficiency of RMSMC is left as future work, because Koka needs to fix an identified bug with shallow handlers. This would then allow me to implement SMC$^{2}$, which is a compound inference algorithm based off of RMSMC.  \\

Figure \ref{fig:smc-dataset} shows the exponential ($R^{2}>0.985$ -- this coefficient of determination means that 98.5\% of the variation in time is described by the variation of an exponential function fitted via maximum likelihood estimation over parameters A and B from the parametric class : $Ae^{Bx}$) scaling of the SMC algorithm as the \texttt{step\_size} parameter is decreased -- which controls how many score statements are handled before resampling steps occur -- is changed from N (ADVANCE N) to 4 (ADVANCE 4). Further profiling would be warranted to investigate the reasons underlying the increase in time complexity, i.e. whether it is due to the deep advance handler or due to the resampling mechanism interacting with handlers or some other reason. However, the remarkably quick computation corresponding to advancing the computation all the way (ADVANCE N) before performing resampling steps, suggests that running the model and the resampling procedure without many nested handlers, consists of a negligible fraction of the total computation time. Monad Bayes has a linear scaling, whilst doing one resampling step after every score statement (ADVANCE 1) and takes under two seconds rather than over three hundred. Figure \ref{fig:smc-particle} shows that with an increasing number of particles, Koka Bayes achieves an approximately linear scaling ($R^{2}>0.98$) for the SMC algorithm. However, the gradient of the scaling is approximately 68 times steeper than Monad Bayes. \\

I did not control for the processor architecture, in comparing Koka Bayes to Monad Bayes. However the relative performance between TMCMC and SMC, indicates that TMCMC tends to be faster than SMC. Figure \ref{fig:tmcmc-dataset} shows a polynomial of order two increase, in the time complexity of the Koka Bayes TMCMC algorithm, as the dataset increases in size. Monad Bayes only has a linear scaling over increasing dataset sizes. Unfortunately, without Koka based profiling tools, it is hard to identify where this change in performance stems from. Whilst Koka Bayes performs worse over an increasing dataset size, Figure \ref{fig:tmcmc-step} shows that it is almost twice as fast as Monad Bayes as the number of MH steps in the algorithm increases, even though Koka Bayes has a higher startup time.

\subsection{Qualititative Evaluation}

Table \ref{tab:loc} details the number of lines of code (LoC) in the inference algorithms implemented in Koka Bayes. The LoC reflect the implementation complexity and implementation effort of algorithm implementations and provide some indication of the ease, with which reasoning about the correctness of the algorithms can be done. Koka Bayes performs competitively with respect to both WebPPL and Anglican. The modular design of Koka Bayes reflects that of Monad Bayes and therefore I expect that the number of lines of code to implement further compound inference algorithms such as SMC${^2}$ \cite{scibior2018functional} will be smaller than the number of lines required to build the algorithms in monolithic designs such as WebPPL and Anglican. In addition, the type-safety of Koka Bayes prevents many different forms of runtime errors that occur in languages such as Pyro, WebPPL and Anglican. \\

Many of the same design principles that Monad Bayes uses are also found in Koka Bayes. The similarity of the libraries is supported by the equivalence expressivity of algebraic effects and monads \cite{forster2017expressive}. For instance, as previously mentioned, Monad Bayes uses the MonadSample, MonadCond and MonadInfer monads. In Koka Bayes, these correspond to the sample, score effects and the model type, respectively. Monad Bayes also uses the principle of suspension points in a program that can be advanced or finished. Koka Bayes's similarity to Monad Bayes, crucially means that it is close to Scibiors semantic validation of inference \cite{scibior2017denotational}. This means that reasoning about the code with respect to the presentation of the inference algorithms is easier.
\begin{table}
\begin{center}
\begin{tabular}{ |c|c|c|c|c| }
 \hline
 Library & MH & SMC & RMSMC & PMMH \\ 
 \hline
 Koka Bayes & 124 & 104 & 30 & 10 \\ 
 Monad Bayes & 67 & 70 & 11 & 4 \\ 
 WebPPL & 314 & 334 & 0 & N/A \\ 
 Anglican & 100 & 87 & N/A & N/A \\
 \hline
\end{tabular}
\caption{\label{tab:loc} Lines of code excluding comments and import statements. To deal with excessive newlines caused by additional curly brackets, I compacted curly brackets into single lines. This makes for a fairer comparison with Monad Bayes's Haskell code base which doesn't use curly brackets. N/A indicates that the algorithm is not available in a given language.}
\end{center}
\end{table}

\chapter{Use case : climate change modelling}

The 2015 Paris Agreement aims to keep temperature increases to ``well below 2 $^\circ$C above pre-industrial levels and pursuing efforts to limit the temperature increase to 1.5 $^\circ$C  above pre-industrial levels" \cite{rogelj2016paris}. Remarkably, no temporal or spatial demarcation as corresponding to ``pre-industrial levels" \cite{schurer2017importance} has been officiated. The consequence of this, is that the notion of climate change since ``pre-industrial levels" is ambiguous. Hawkins et al. \cite{hawkins2017estimating} suggest a period of 1720-1800 as the period corresponding to pre-industrial levels. They arrive at this estimate by controlling for macroscopic climate events such as volcanic events, solar activity and changes in the Earth's orbit. Fortunately, there is a free, publicly usable Kaggle dataset\footnote{https://www.kaggle.com/berkeleyearth/climate-change-earth-surface-temperature-data}, released by Berkeley Earth, of average global land temperatures dating back to 1750. With this dataset, I demonstrate the use of Sequential Monte Carlo and Trace Markov Chain Monte Carlo based inference over a simple extension of a linear Gaussian state-space model. The goal of this Chapter is not to draw new conclusions about climate change, but to demonstrate the potential of Koka Bayes.

\section{The Kalman Filter}

Linear Gaussian state-space models are known as Kalman Filters \cite{roweis1999unifying}. The linear Gaussian model builds on a vector of latent unobserved states $\textbf{x}$ which produce an observable vector $\textbf{y}$. The evolution of unobserved states follows a first order Markov Chain. In particular, later latent states are related to previous latent states via multiplication with a fixed \textit{state-transition} matrix $\textbf{A}$ and addition with a time-dependent vector $\textbf{w}_{t} \sim \mathcal{N}(0, Q)$ where Q is a co-variance matrix.
$$
\textbf{x}_{t+1} = \textbf{A}\textbf{x}_{t} + \textbf{w}_{t}
$$
At discrete time intervals $t$, the corresponding unobserved state $\textbf{x}_{t}$ produces an observable output $\textbf{y}_{t}$. The observable vectors $\textbf{y}$ are evaluated by transforming the unobserved state $\textbf{x}$ by a \textit{observation} matrix $C$ and then adding a vector $\textbf{v} \sim \mathcal{N}(0,R)$ where R is a co-variance matrix.
$$
\textbf{y}_{t} = \textbf{C}\textbf{x}_{t} + \textbf{v}_{t}
$$

\subsection{Extended linear Gaussian state-space model}

The extended linear Gaussian model I create is both a restriction and a generalization of the Kalman Filter. I restrict the model by assuming that the unobserved states $\textbf{x}$ are one-dimensional floating point values. A corollary of this assumption is that the state-transition matrices $\textbf{A}$ are one-dimensional floating point values. I generalize the model by having a time dependent state transition matrix $\textbf{A}_{t}$ as well as a time dependent observation matrix $\textbf{C}_{t}$. In addition, the covariance matrix $R$ is also time dependent. With this formulation, the observable vectors $\textbf{y}$ are still conditionally independent of each other, which can easily be verified by d-separation \cite{geiger1990d}. The joint probability distribution of the model therefore factorizes into the same form as a Hidden Markov Model:

$$
p(\{\textbf{x}_{1}... \textbf{x}_{n}\}, \{\textbf{y}_{1}... \textbf{y}_{n}\}, \theta) = p(\theta) p(\textbf{x}_{1} | \theta) \prod_{t=1}^{n-1}p(\textbf{x}_{t+1}|\textbf{x}_{t}, \theta) \prod_{t=1}^{n}p(\textbf{y}_{t}|\textbf{x}_{t}, \theta)
$$

$$
\textbf{y}_{t} = \textbf{C}_{t}\textbf{x}_{t} + \textbf{v}_{t}
$$
\begin{equation}\label{eqn:extended-transition}
\textbf{x}_{t+1} = \textbf{A}_{t}\textbf{x}_{t} + \textbf{w}_{t} \quad \textbf{w}_{t} \sim \mathcal{N}(0, 1)
\end{equation}

There are two important machine learning tasks that are typically performed when considering Kalman Filter models \cite{roweis1999unifying}. The first is parameter \textit{learning} or \textit{system identification}, which involves being given a series of outputs $\{\textbf{y}_{1}, ... \textbf{y}_{n}\}$ and to subsequently infer the values of the parameters $\theta = \{\textbf{A},\textbf{C},Q,R\}$. The other task is \textit{filtering} mentioned in Section 4.2.1, which involves obtaining the probability distribution $p(\textbf{x}_{t} | \textbf{y}_{1}, ... , \textbf{y}_{n})$ with $t = n$. The inference methods perform both tasks over the climate model presented.

\section{Experiments}

\subsection{Berkeley Earth Dataset}

The dataset consists of single data points that represent average land temperatures for the entire Earth, for each month from the period of 1756 until 2016 (260 years). There are no data points missing from the dataset. Each datapoint is paired with a corresponding 95\% confidence interval. These uncertainty intervals take into account both spatial incompleteness of the temperature readings and statistical uncertainty, which arises because the temperature readings themselves may not be true reflections of the Earth's temperature \cite{rohde2013berkeley}. Whilst no data is missing from the dataset provided to me, in early years only two weather stations are used to calculate the entire Earth's temperature and so the corresponding uncertainty values for those years with fewer weather stations are much higher. In order to convert the confidence intervals to standard deviations such that they can be parameterized by a Gaussian random variable, I sought the advice of both Dr.~Schurer and Dr.~Zeke Hausfather -- the curator of the Berkeley dataset. They advised that in order to convert the confidence intervals to a standard deviation, divide each value by 3.92 (the number of standard errors that fit into a 95\% confidence interval) to obtain the standard error and using that as the standard deviation -- corresponding to a sample size of one. 

\subsubsection{Setting model hyper-parameters}

I split the Berkeley Earth data chronologically, into thirteen vertical slices of twenty years, giving a balance between resolution and sample size. There need to be enough slices so that important climatic trends are represented, yet not too many slices -- because there would then be few data points per block, resulting in both less interesting inference and higher variance in the temperature values which may not indicate the true latent temperature of the Earth. This partition results in twenty year slices of climate readings which, for each slice, I slice horizontally into months. This leaves thirteen twenty year slices and $\textbf{x}_{t}$ represents the latent temperature of a particular twenty year month block. I assume that every twenty years, the Earths monthly climate changes in accordance with Eqn. \ref{eqn:extended-transition}. For each twenty year month block the model detailed in Listing \ref{lst:climate-model} is used. \\

\newpage

\begin{adjustwidth}{1em}{0em}
\begin{lstlisting}[caption=Linear Gaussian Model applied to climate data for only one month,frame=tlrb,firstnumber=1,label={lst:climate-model},linewidth=14cm]{weighted}
var x := [] // Reference to list that stores latent variables.
for(0, 12) fun(i) {
  if(i == 0) {
    x := x + [(sample() * diff) + subtract] // Initial state dist.
  } else {
    val a_val = normal(a_mean,a_std_dev) // state-transition
    val w_val = normal(w_mean,w_std_dev) // transition noise
    x := x + [exn-get(x, i - 1) * a_val + w_val] // Transition.
  }
  val month_ys = exn-get(ys,i) // Read in temperatures.
  val month_vs = exn-get(vs,i) // Read in uncertainty.
  val c_row = multivariate_gaussian(month_ys.length, 1.0 + mult_bias_of_thermometer, mult_bias_of_thermometer_std_dev)
  val v_row = convert_uncertainty_to_rand(month_vs)
  val predictions = plus(mult(exn-get(x, i)
                        , c_row), v_row) // Predictions
  score_predictions(month_ys, predictions, score_var)
  ()
}
\end{lstlisting}
\end{adjustwidth}

I chose parameters based on a consultation and subsequent email follow-ups with Dr. Andrew Schurer -- an expert climate researcher based in Edinburgh -- who has personal experience with the Berkeley dataset. I used this consultation to try and elicit priors for global temperature values in 1750. I used a prepared script, an interactive email dialogue and discussions about likely percentiles in order to elicit priors, as per recommendations in the literature \cite{chaloner1996elicitation}. I found prior elicitation to be challenging, because the expert was unable to provide an absolute estimate in degrees Celsius of the temperature, and he largely depended on estimates of relative climate values from other empirical climate simulations to obtain such `priors'. In order to avoid being biased from such estimates, I decided to use an uninformative uniform prior over $\textbf{x}_{0}$ parameterized by the minimum and maximum temperatures in the first twenty months, for each month. The uniform distribution was parameterized in this manner, to allow the Monte Carlo based inference strategies to remain tractable, yet realistic. This choice is realistic because it excludes inferred initial temperatures below and above the minimum and maximum recorded temperature during the first block. This choice creates tractability because the search space is reduced relative to an alternative choice; using the lowest and highest temperatures ever recorded on Earth. In the end, I did not receive a useful prior from the expert. \\

The observation matrices $\textbf{C}_{t}$ reflect the systematic multiplicative bias of the thermometers that are used to measure the temperatures \cite{tian2013modeling}. In order to reflect the uncertainty of the expert on this particular bias, and to reflect the high accuracy of mercury as well as electronic thermometers, this parameter was set to $\mathcal{N}(1,0.05)$ throughout the slices. However, experimenting with different suggestions made by domain experts in thermometer calibration would be warranted to improve the model. The additive noise parameter $\textbf{v}_{t}$ is parameterized by uncertainty values given in the original dataset. The state-transition values $\textbf{A}_{t}$, are parameterized as $\mathcal{N}(1, 0.4)$, in line with empirical observations of the given Berkeley data. The score statements were parameterized by Gaussian probability density functions with $\mu$ chosen to be the observed data and $\sigma$ chosen at a value of $4.7$. This corresponds to assuming the observable data are corrupted by Gaussian noise. The standard deviation was small enough to ensure relatively accurate inference, yet large enough to accommodate both numerical stability issues in the inference process and prevent particle degeneracy in SMC.

\subsection{Inference hyper-parameters}

The parameters I chose to control for the scale of the inference, as well as the time taken for the inference, are detailed in Table \ref{tab:time}. I chose the largest number of particles for SMC and the highest step number for TMCMC, that Koka managed without throwing a runtime exception. In particular, Koka is unable to write or read files with more than a few thousand lines and Koka is unable to perform key functions such as map and split over lists that have many thousands of elements. I ran both inference algorithms twice to get an idea of the variance of the inference procedures when inferring temperatures. As Koka can only handle 2,000 particles -- a value small enough that it fails to provide high resolution of the posterior distribution at earlier latent states, I combined four runs of SMC for each month that I analyzed. I used a burnin of 80,000 samples for TMCMC.

\begin{table}
\begin{center}
\begin{tabular}{ |c|c|c|c|c| }
 \hline
 Algorithm & Time (s) & Steps/particles & Posterior representation & Runs \\ 
 \hline
 TMCMC & 1900 $\pm$ 320 & 100,000 & 20,000 samples & 2 \\ 
 SMC & 1100 $\pm$ 40 & 2,000 & 8,000 samples & 2 \\ 
 \hline
\end{tabular}
\caption{\label{tab:time} Inference hyper-parameters and also time taken. Time was recorded on a Macbook Pro (2016).}
\end{center}
\end{table}

\section{Results}
\begin{figure}
\begin{adjustwidth}{-3em}{0em}
\centering
\begin{tabular}{ccc}
\subcaptionbox{Global temperature change over time\label{fig:exp1}}{\includegraphics[width = 2.0in]{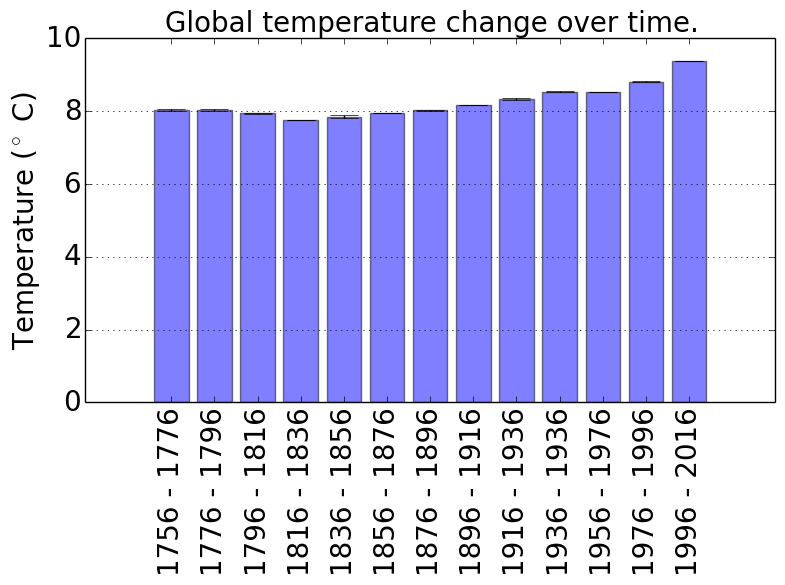}} &
\subcaptionbox{March temperature change SMC\label{fig:exp2}}{\includegraphics[width = 2.0in]{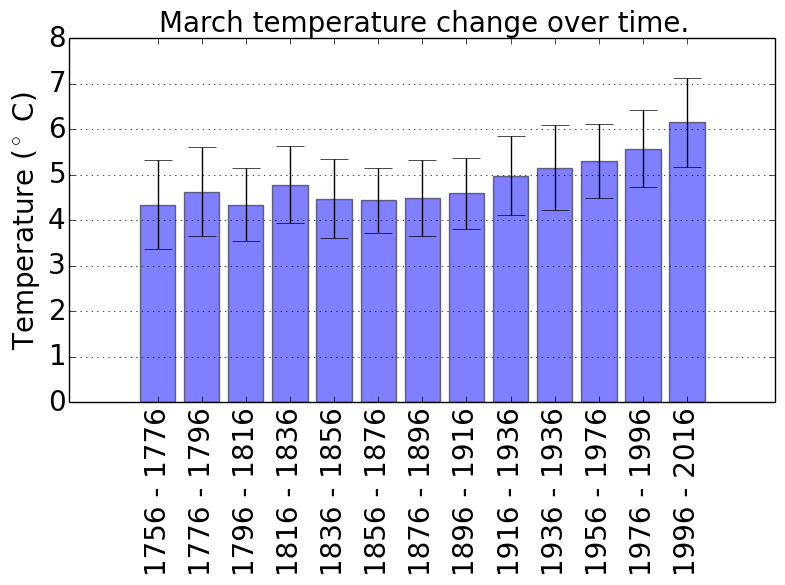}} &
\subcaptionbox{March temperature 1756-1776\label{fig:exp3}}{\includegraphics[width = 2.0in]{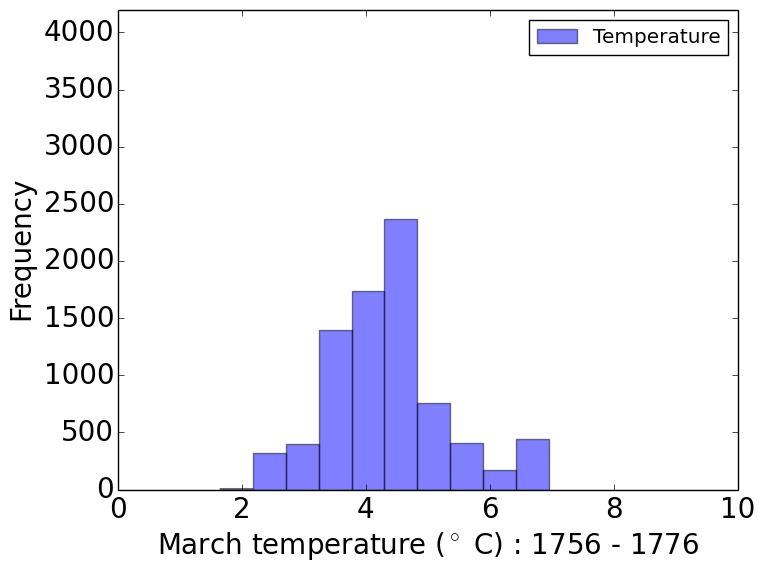}}\\
\subcaptionbox{March temperature 1836-1856\label{fig:exp4}}{\includegraphics[width = 2.0in]{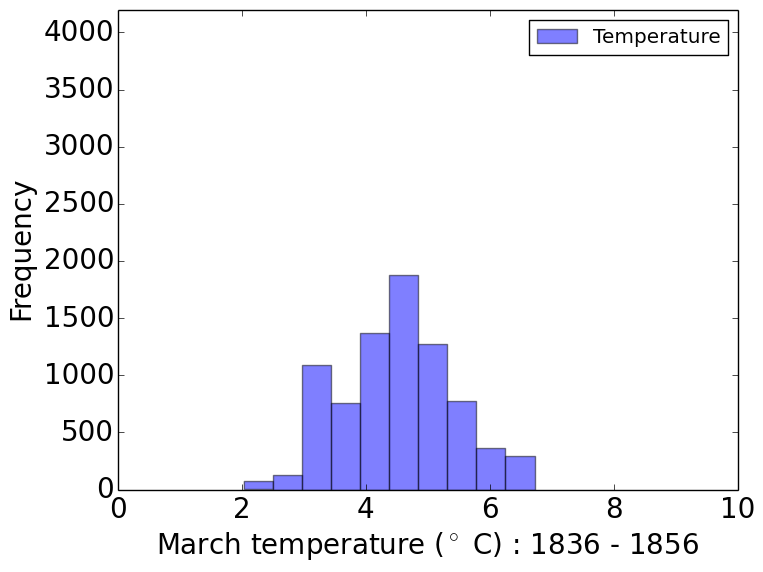}} &
\subcaptionbox{March temperature 1916-1936\label{fig:exp5}}{\includegraphics[width = 2.0in]{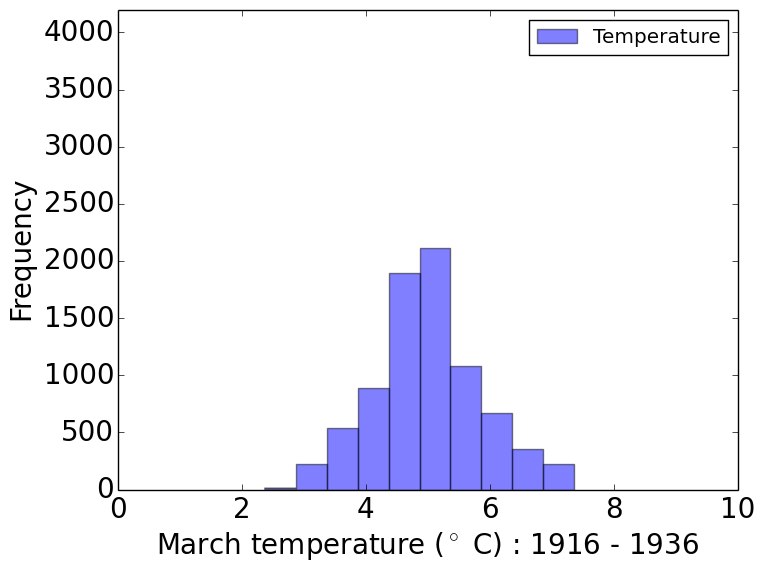}} &
\subcaptionbox{March temperature 1996-2016\label{fig:exp6}}{\includegraphics[width = 2.0in]{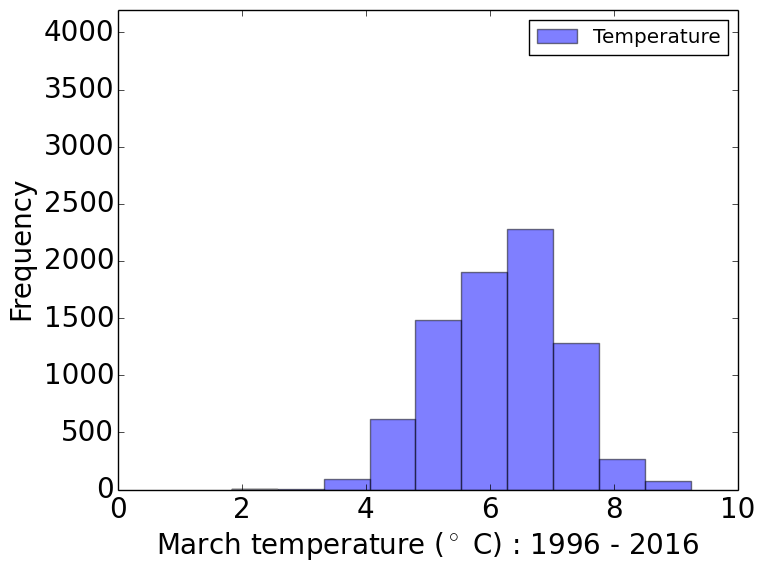}}\\

\subcaptionbox{Global temperature change over time\label{fig:exp7}}{\includegraphics[width = 2.0in]{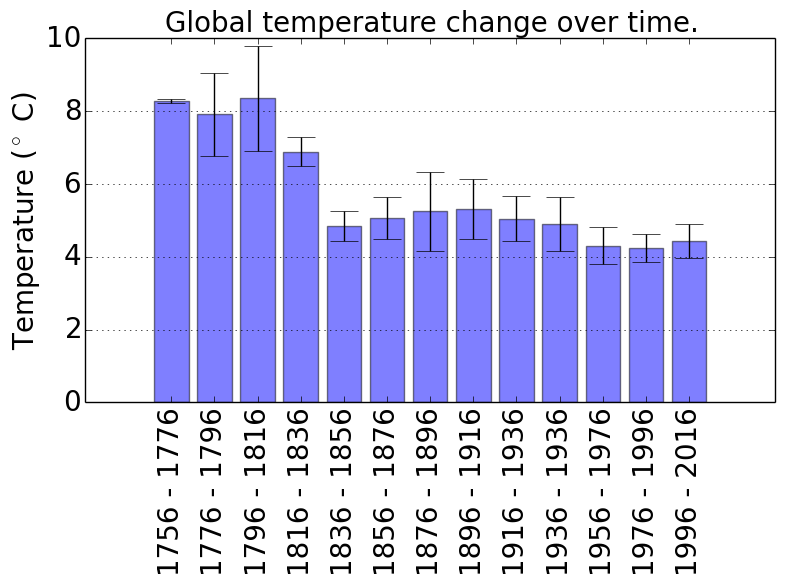}} &
\subcaptionbox{March temperature change TMCMC\label{fig:exp8}}{\includegraphics[width = 2.0in]{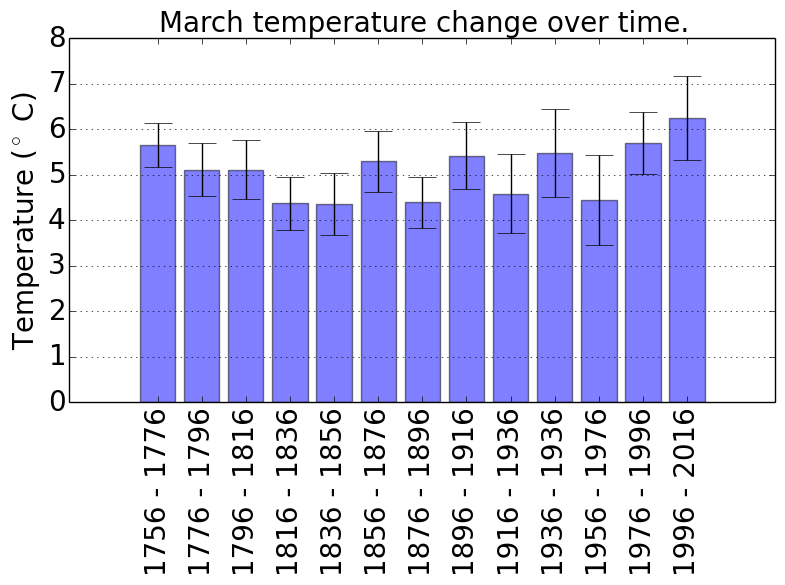}} &
\subcaptionbox{March temperature 1756-1776\label{fig:exp9}}{\includegraphics[width = 2.0in]{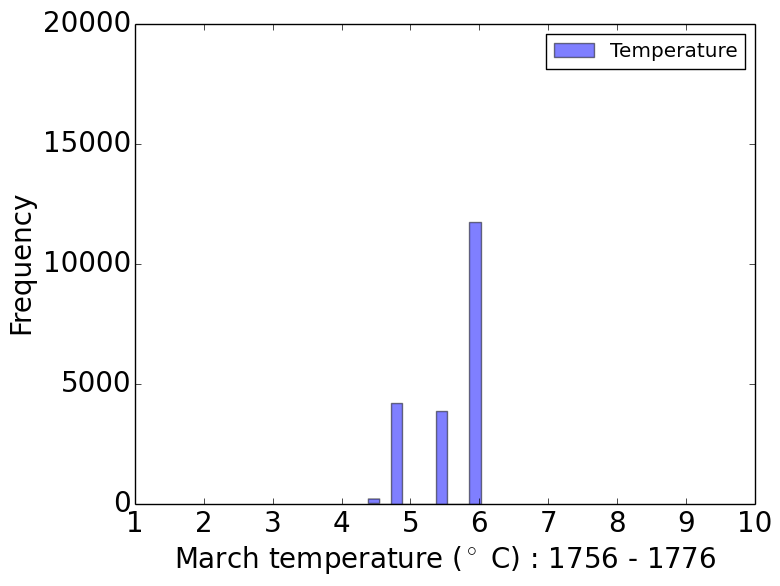}}\\
\subcaptionbox{March temperature 1836-1856\label{fig:exp10}}{\includegraphics[width = 2.0in]{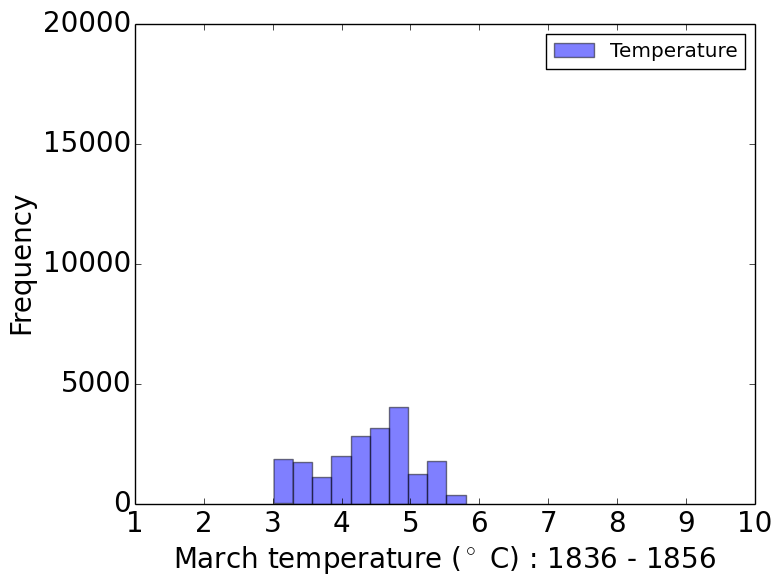}} &
\subcaptionbox{March temperature 1916-1936\label{fig:exp11}}{\includegraphics[width = 2.0in]{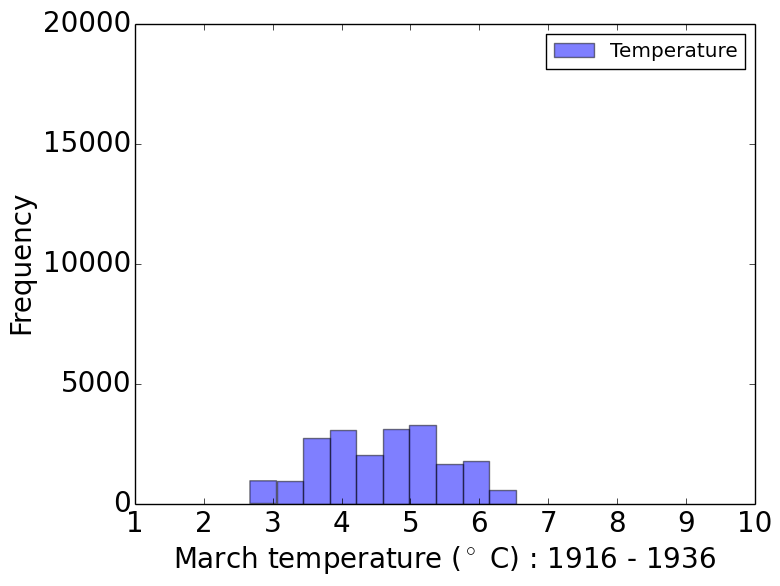}} &
\subcaptionbox{March temperature 1996-2016\label{fig:exp12}}{\includegraphics[width = 2.0in]{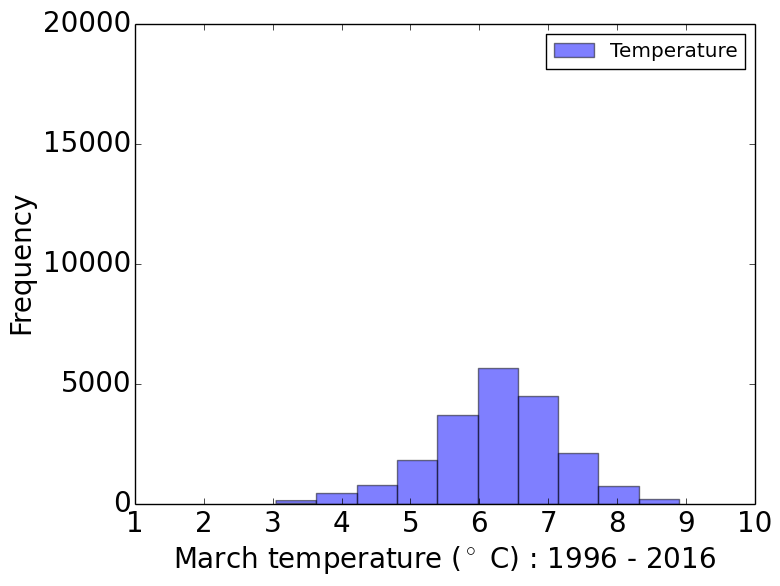}}\\
\end{tabular}
 \caption{The top two rows show the result of inference using SMC and the bottom two rows show the result of inference using TMCMC. The confidence intervals of the temperature change over March for both SMC and TMCMC represent the standard deviation of all model samples that represent the posterior. The confidence intervals of the global temperature change refer to the standard deviation of running the inference twice. }
\label{fig:experimental-results1}
\end{adjustwidth}
\end{figure}

\begin{figure}
\begin{adjustwidth}{-2em}{0em}
\centering
\begin{tabular}{cc}
\subcaptionbox{August temperature 1756 - 1776\label{exp:13}}{\includegraphics[width = 2.45in]{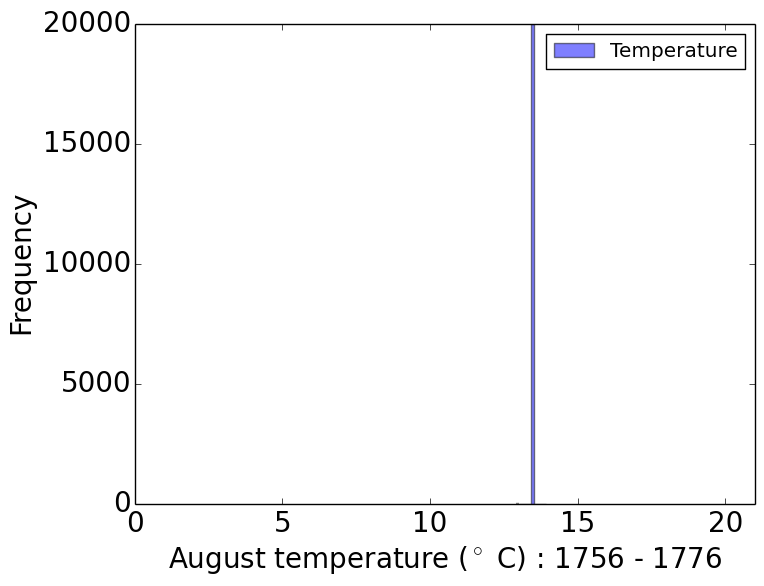}} &
\subcaptionbox{Source : Berkeley Earth \cite{berkeleyearth}\label{exp:15}}{\includegraphics[width = 2.45in]{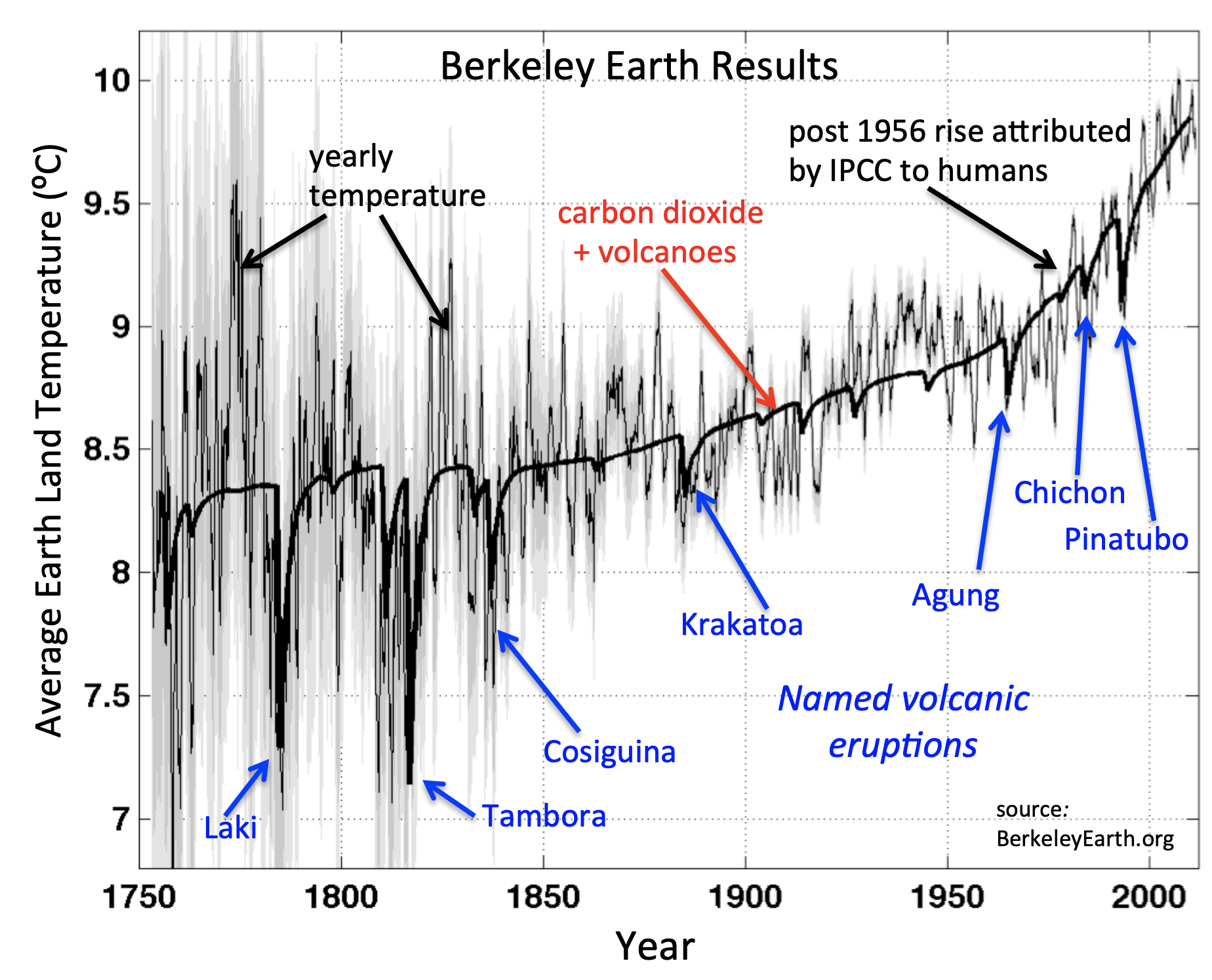}}\\
\end{tabular}
\caption{The leftmost image depicts the tendency of the TMCMC algorithm to become trapped in local optima, in this case for August, with only single samples representing the posterior. The rightmost image depicts the results of a press release in 2012 by Berkeley Earth depicting just over 250 years of global warming.}
\label{fig:experimental-results2}
\end{adjustwidth}
\end{figure}

Figure \ref{fig:exp1} and \ref{fig:exp7} demonstrate that overall, SMC and TMCMC differ over the inferred global temperature change. SMC infers an \textit{increase} in global temperatures of approximately 1.6 degrees Celsius over the entire time period. TMCMC infers a \textit{decrease} in global temperatures of approximately three degrees Celsius. However, there are similarities; for example, during the month of December, the temperatures of both algorithms show a similar increase between 1756 and 2016, of 1.5 degrees in the case of TMCMC and of about 1.8 degrees in the case of SMC. TMCMC and SMC both infer Gaussian distributed temperatures for the month of March. Often, TMCMC ends up in local optima which results in a single sample representing the entire posterior for all time periods. Examples of this phenomena are depicted in Figure \ref{fig:experimental-results2}, for two twenty year blocks at the beginning and end of the 260 year period studied.

\section{Evaluation and future work}

The global temperature change inferred by SMC corresponds to the findings of Berkeley Earth researchers \cite{rohde2013new}. SMC performs well here because it is an algorithm designed with linear state-space models in mind \cite{doucet2001introduction}. However, even with the use of SMC, the compatibility intervals for monthly temperatures in 1756 - 177 often overlap with those of 1996 - 2016 (see Figure \ref{fig:exp2} for an example). This overlap means that whether climate change is occurring, can, to some extent be doubted. Future work would investigate inference algorithms that achieve more accurate compatibility intervals. The reason the compatibility intervals for SMC based inference are large is related to the parameter that controls for corruption of the noise, which has to be large enough to prevent particle degeneracy in SMC. RMSMC reduces particle degeneracy and would likely allow for a smaller noise corruption parameter thus providing more accurate inference. \\

TMCMC infers the wrong temperature change. I suggest a few reasons for this. Firstly, TMCMC works by optimizing over a real valued vector space with many dimensions -- equal to the number of random choices that a model performs. In the case of the climate model there are a little under 1100 sample statements, though one could imagine a large model with many millions of sample statements -- with some samples having less impact on the inference process than others. Even performing one hundred thousand TMCMC steps, in expectation, each sample statement will only be perturbed 110 times. This is a vanishingly small fraction of the total number of sample statements needed to explore every possible trace and so it may only explore a local optima. Secondly, because the model has a sequential structure, perturbing sample statements that correspond to early latent states of the model, later on in the inference process, carries a risk. This is because the change of sample affects later model parameters, and so if all other later samples had already been perfectly optimized for -- then the perturbation will cause these samples to be corrupted. Whilst the probability of perturbing early states is low, if early states are rarely perturbed then the diversity of samples of earlier latent states will also be low, corresponding to a posterior distribution for early and sometimes later samples that consists of only one value, as per Figure \ref{exp:13}. \\

Some parameters themselves may be more important than others, so attaching a notion of importance (which could affect how many perturbations occur) to certain sample statements may be a useful direction of research. Further work could investigate the use of TMCMC that stages inference of different parameters, based on dependencies in the model. Providing users of the library with a method to visualize Trace plots -- diagrams that provide a visual cue as to whether Markov Chains are well mixed, would also make the library more transparent. In addition to such visualization tools, the library could provide useful statistical tests such as the Kolmogorov-Smirnov test, to indicate whether two distributions are equivalent -- thus allowing combinations of parallel computed independent Markov Chains. \\

Future work could improve both the model and its dataset. For example, ameliorating the na\"ivety of the Extended Linear Gaussian model using a model that describes the locations of the temperature recordings and potentially the physical dynamics of the thermometers. The simplicity and paucity of the dataset also creates issues. In earlier years, few weather stations were used to estimate the temperature of the whole planet, which makes those estimates unreliable. Other sources of data, such as tree ring or ice core data coupled with the underlying data generating processes could be factored in to the model to improve estimates of climate change. In this way, a better generative model of global warming can be created which incorporates aspects such as the temperature capacity of mercury or urban heat islands. One could investigate statistical model comparison measures such as the Bayesian Information Criterion or Bayes factors to compare different models. For the scope of this project, creating complex models with corresponding inference algorithms that can cope with complex models, would likely be intractable both in terms of development time and inference time. Therefore, I left it as future work. To test further inference algorithms, a simplified simulator of the Earth could be created, that yields noisy measurements of ground truth latent states which could be compared to inferred states. 

\chapter{Conclusions}

\section{Summary}

I have implemented a basic modular probabilistic programming library in Koka based off the design of Monad Bayes. The library supports Importance Sampling and four compound inference algorithms (SMC, TMCMC, RMSMC \& PMMH). I have compared the temporal scaling of TMCMC and SMC on larger datasets and inference, against that of Monad Bayes, finding that TMCMC performs competitively, but SMC does not. I have demonstrated the library on an extended linear Gaussian model with a dataset consisting of Earth's average land temperatures with uncertainties over the 260 year period starting in 1756 and ending in 2016. \\

During this experimental programming project, I encountered a number of challenges. I learned that developing with Koka is difficult, because a number of unforeseen internal compiler bugs and runtime exceptions occurred, which required time expensive workarounds.  This is because Koka is a pre-alpha research language. Access to Daan Leijen -- the author of the Koka language -- was therefore crucial. At a few points in the development process (notably when finalizing RMSMC and creating a Gamma distribution sampling function from a a handbook on Monte Carlo methods \cite{kroese2013handbook}) I collaborated with Daan Leijen in order to overcome a number of internal compiler errors. The lack of Koka profiling tools and documentation, also hindered development efforts, especially when attempting to develop a shallow handler to replace the inefficient deep handler. The Bayesian inference library is not yet ready for deployment because of the readiness level of Koka and the library's current performance. Whilst the library will be of use to programming language researchers, the model and library is unlikely to be of use to statistical and climate researchers due to the limitations of Koka as well as the model. The model is not a serious contender for climate change modelling due to the naïve simplicity of the underlying generative model and the limited datasets used in the inference process. Koka has issues in dealing with even moderately sized data sets, far smaller than those used in climate related institutions, such as the MET office. Koka is only able to read and write (only overwrite) files with at most a few thousand, limited length lines. In order to overcome this issue, I staggered the reading and writing into separate stages. In light of such challenges, it is clear that modular probabilistic programming based off of the design of Monad Bayes can be implemented with algebraic effects.

\section{Future / further work}

Future work would focus on making Koka and the library more stable, reliable and performant before adding more sophisticated inference algorithms and models. Of particular interest would be the implementation of variational inference through the use of guide programs \cite{cusumano2019gen}, profiling tools to assess inference convergence, as well as exact inference through disintegration \cite{shan2017exact}. Implementing back-propagation would help bring inference to a level competitive with deep learning \cite{pfeffer2017learning}. Improving the computational efficiency of RMSMC \& SMC, through programming language constructs such as shallow handlers, and the statistical efficiency, through the use of d-separation to calculate conditional independence of variables that assist in inference, would hopefully allow the speed of the library to scale similarly to other probabilistic programming languages. It would be interesting to see if probabilistic programming with effects could be implemented in a dependently typed programming language such as Agda \cite{norell2008dependently}, where the type system could offer even more guarantees than the type system of Koka. Whilst the development process was challenging, I enjoyed both constructing the library and learning from obstacles overcome. I believe that this work provides a solid foundation for further exploration into the exciting interface of algebraic effects and probabilistic programming.

\bibliographystyle{unsrt}
\bibliography{mybibfile}

\appendix

\chapter{Additional figures for SMC / TMCMC based inference over the extended linear Gaussian climate model.}

I am including these Figures to demonstrate the inference results for SMC and TMCMC over all months of the year. For each inference type, for each month, and for each twenty year block, a histogram is displayed. Each histogram uses the parameters found in Table 5.1. The histogram represents the values of the particles for SMC and the model values corresponding to each trace for TMCMC. In addition, the overall change for the months are plotted with uncertainty intervals that result from the standard deviation of the inferred temperature when running the algorithm twice. SMC performs well for all of the months (it corresponds to temperature increases that other researchers have found), in part because SMC was designed with linear Gaussian state-space models in mind. TMCMC sometimes performs reasonably well, but often becomes stuck in local optima for some months (e.g. August) -- suffering from low diversity in these cases.

\begin{figure}
\begin{adjustwidth}{-6em}{0em}
\centering
\begin{tabular}{ccc}
\subcaptionbox{January -- SMC: 1756 - 1776\label{fig:jan1}}{\includegraphics[width = 2.0in]{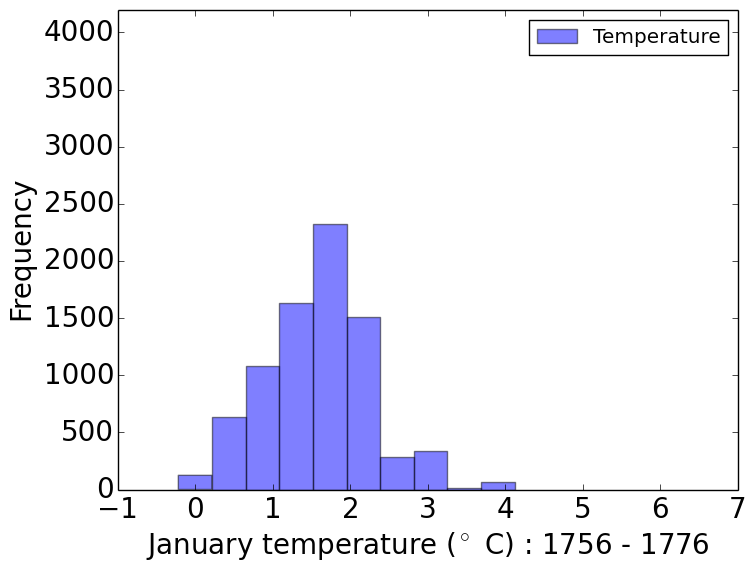}} &
\subcaptionbox{January -- SMC: 1776 - 1796\label{fig:jan2}}{\includegraphics[width = 2.0in]{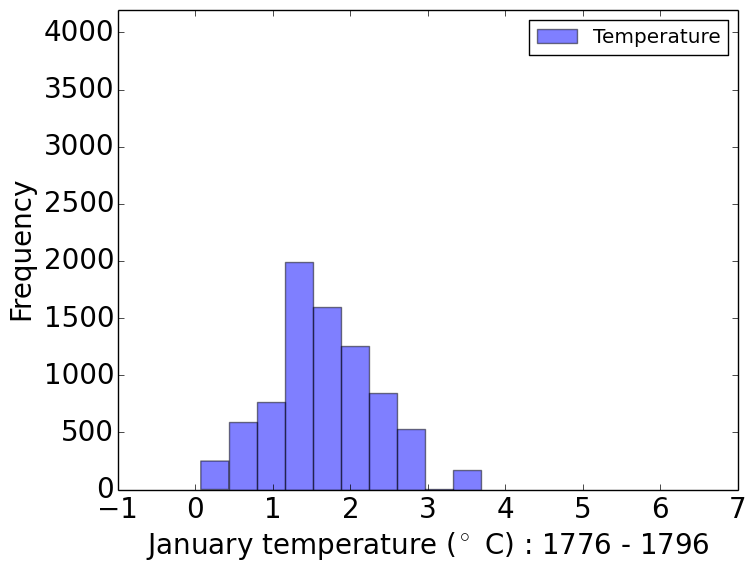}} &
\subcaptionbox{January -- SMC: 1796 - 1816\label{fig:jan3}}{\includegraphics[width = 2.0in]{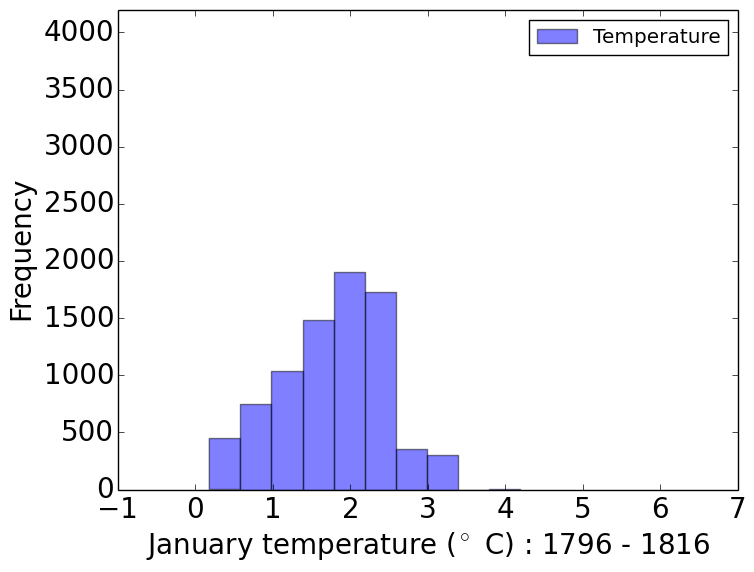}}\\
\subcaptionbox{January -- SMC: 1816 - 1836\label{fig:jan4}}{\includegraphics[width = 2.0in]{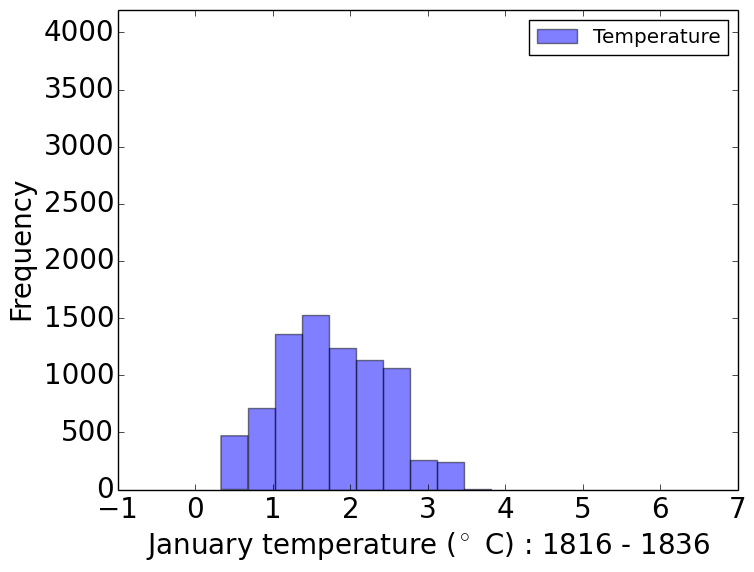}} &
\subcaptionbox{January -- SMC: 1836 - 1856\label{fig:jan5}}{\includegraphics[width = 2.0in]{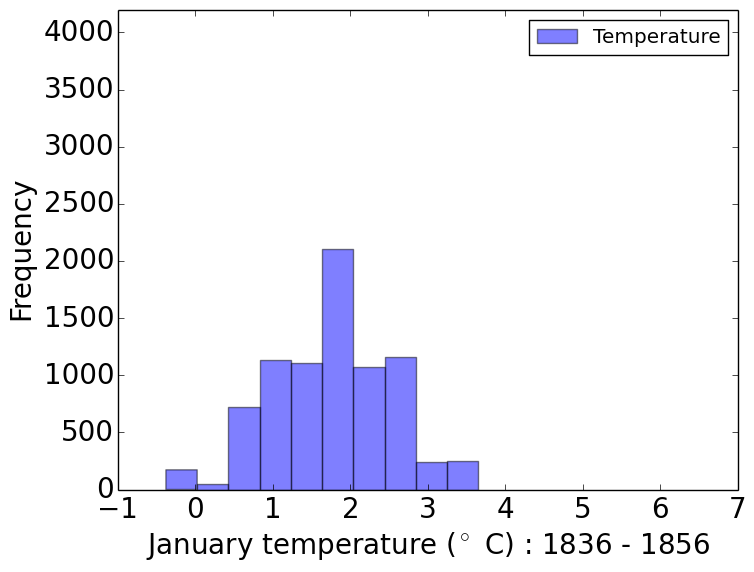}} &
\subcaptionbox{January -- SMC: 1856 - 1876\label{fig:jan6}}{\includegraphics[width = 2.0in]{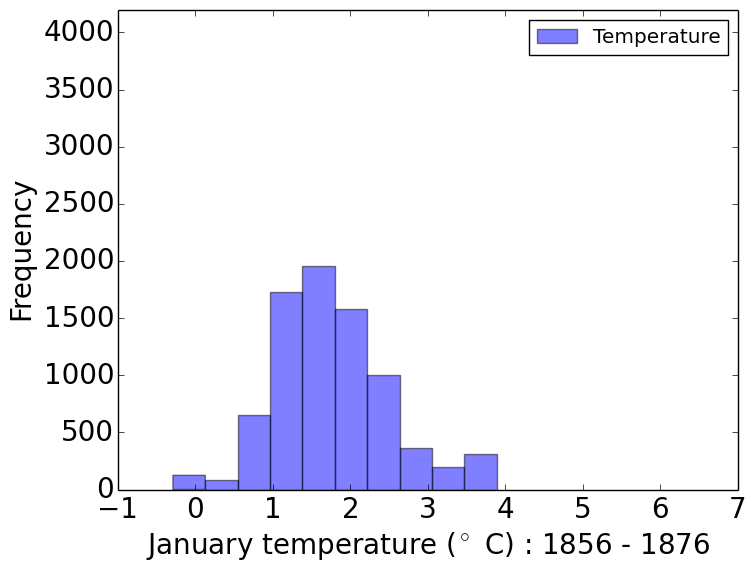}}\\
\subcaptionbox{January -- SMC: 1876 - 1896\label{fig:jan7}}{\includegraphics[width = 2.0in]{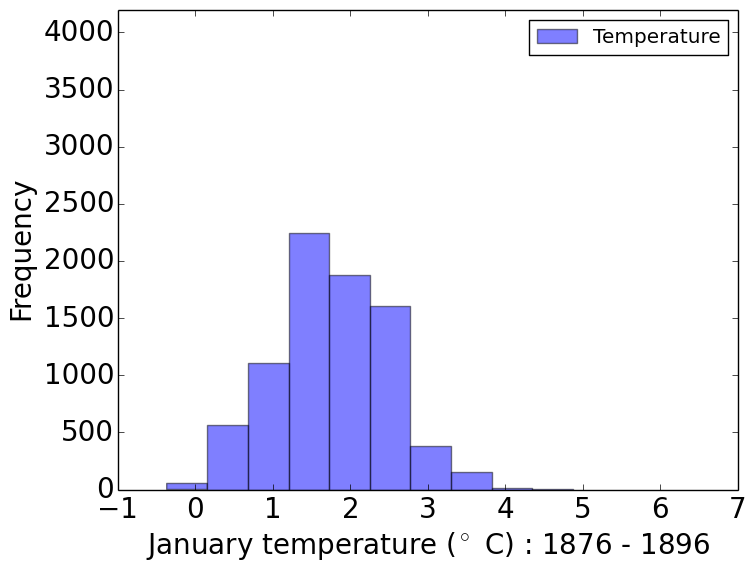}} &
\subcaptionbox{January -- SMC: 1896 - 1916\label{fig:jan8}}{\includegraphics[width = 2.0in]{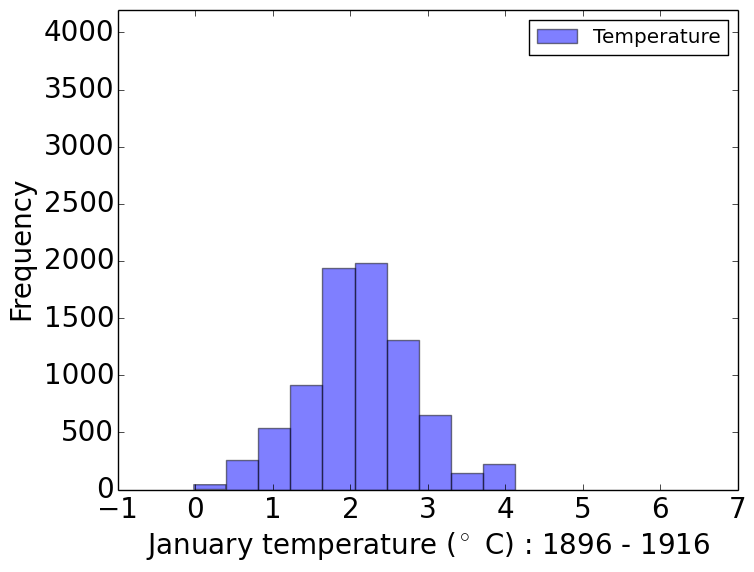}} &
\subcaptionbox{January -- SMC: 1916 - 1936\label{fig:jan9}}{\includegraphics[width = 2.0in]{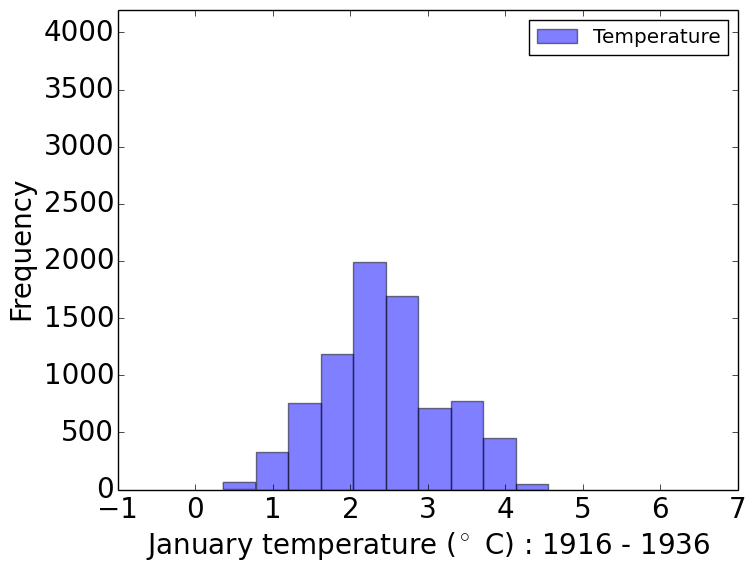}}\\
\subcaptionbox{January -- SMC: 1936 - 1936\label{fig:jan10}}{\includegraphics[width = 2.0in]{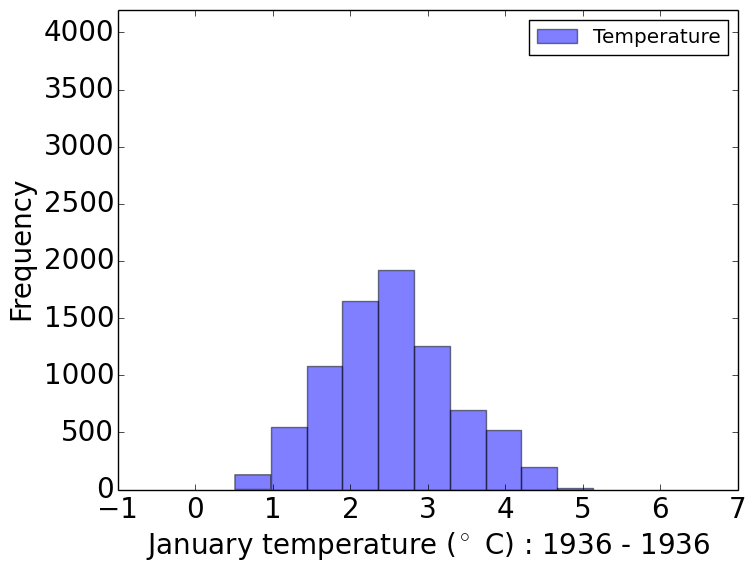}} &
\subcaptionbox{January -- SMC: 1956 - 1976\label{fig:jan11}}{\includegraphics[width = 2.0in]{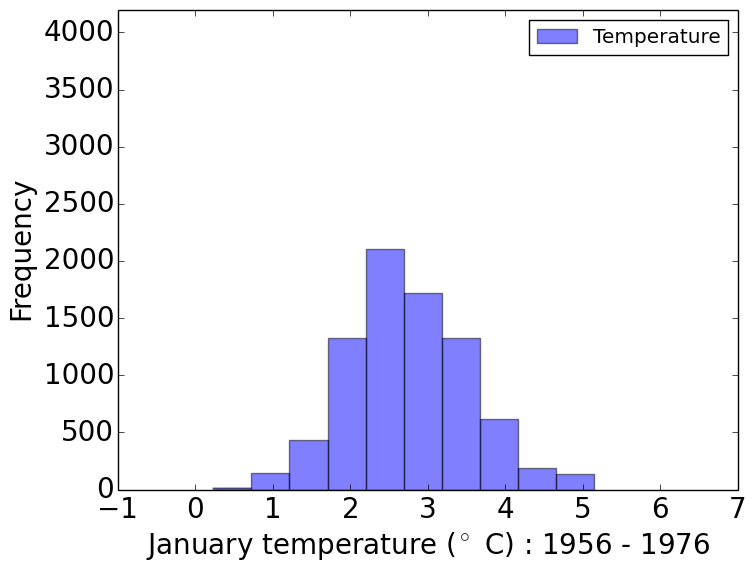}} &
\subcaptionbox{January -- SMC: 1976 - 1996\label{fig:jan12}}{\includegraphics[width = 2.0in]{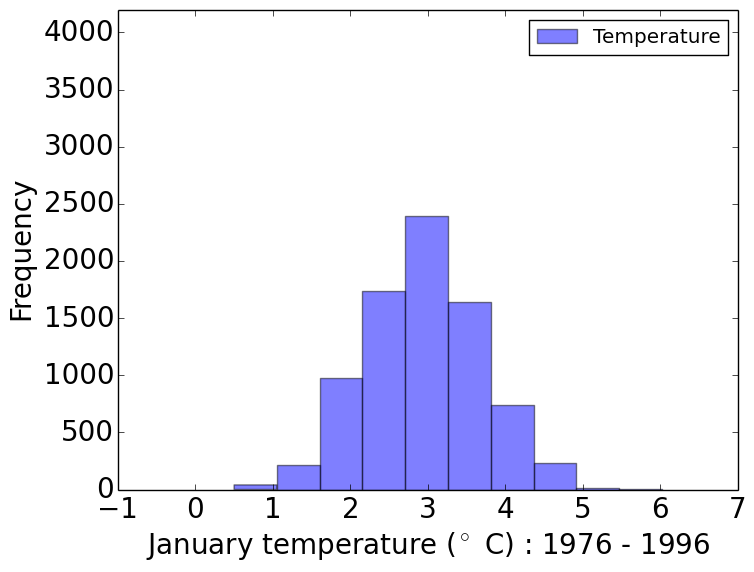}}\\
\subcaptionbox{January -- SMC: 1996 - 2016\label{fig:jan13}}{\includegraphics[width = 2.0in]{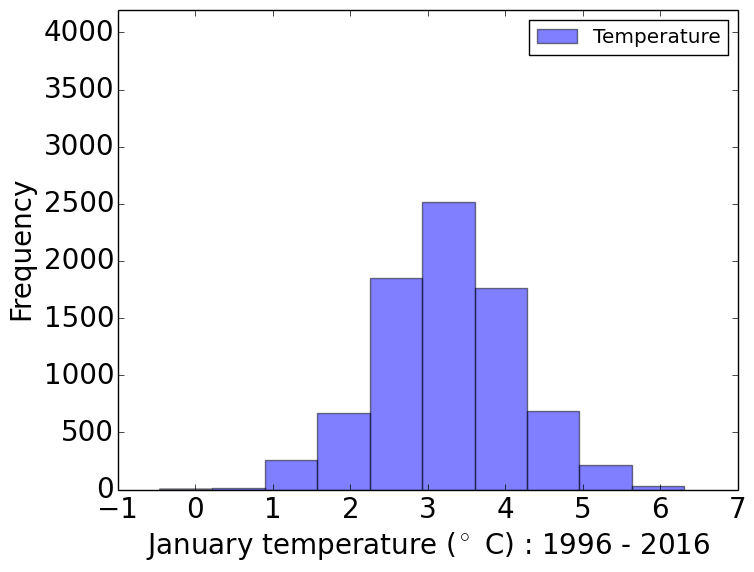}} &
\subcaptionbox{January -- SMC: Total\label{fig:jan14}}{\includegraphics[width = 2.0in]{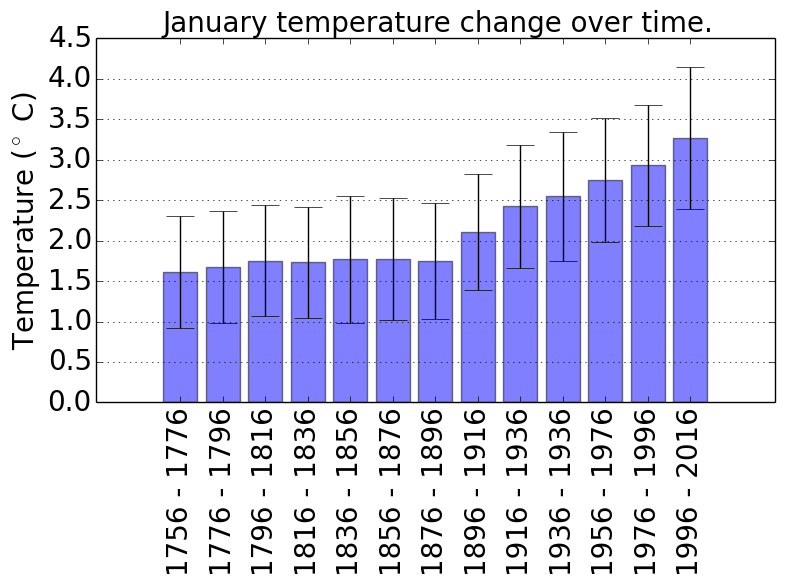}} &
\end{tabular}
\label{fig:jan1}
\end{adjustwidth}
\end{figure}

\begin{figure}
\begin{adjustwidth}{-6em}{0em}
\centering
\begin{tabular}{ccc}
\subcaptionbox{February -- SMC: 1756 - 1776\label{fig:feb1}}{\includegraphics[width = 2.0in]{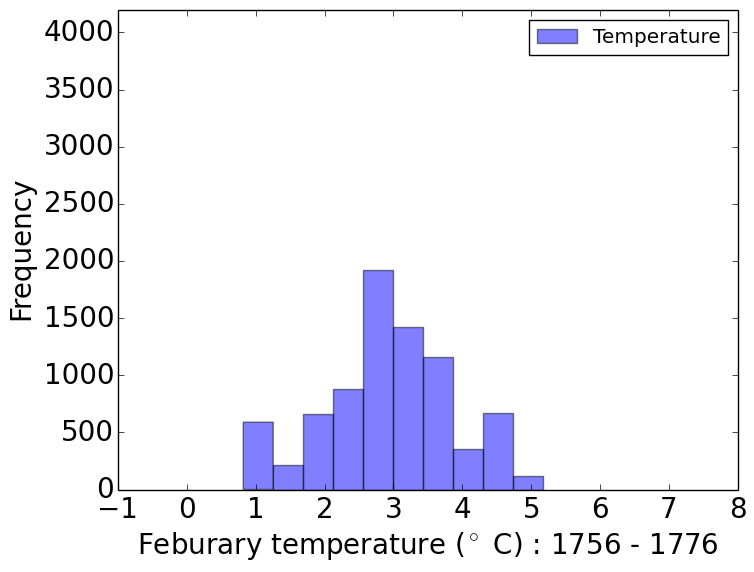}} &
\subcaptionbox{February -- SMC: 1776 - 1796\label{fig:feb2}}{\includegraphics[width = 2.0in]{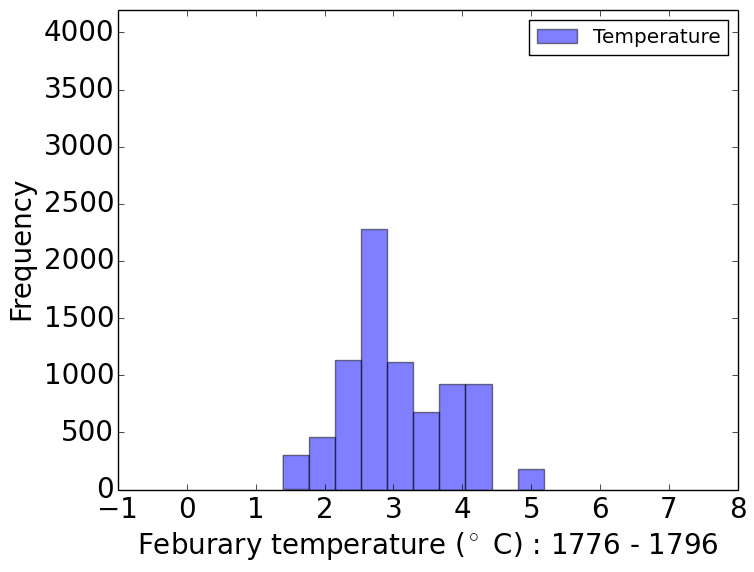}} &
\subcaptionbox{February -- SMC: 1796 - 1816\label{fig:feb3}}{\includegraphics[width = 2.0in]{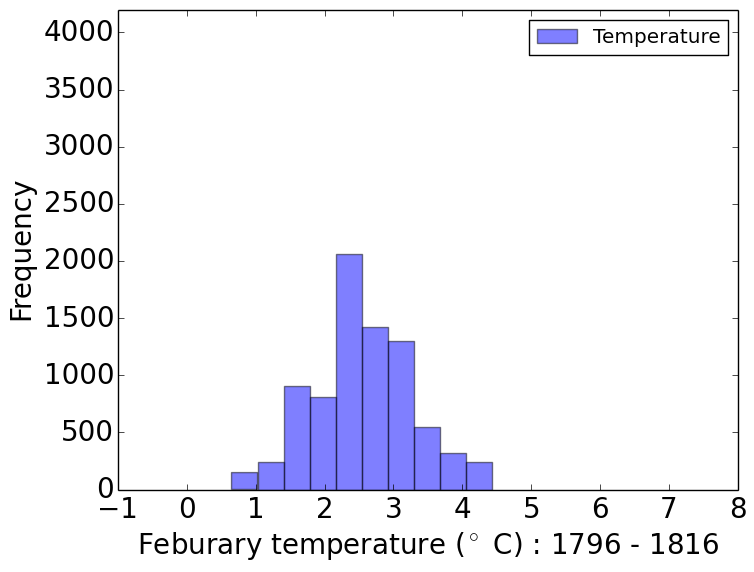}}\\
\subcaptionbox{February -- SMC: 1816 - 1836\label{fig:feb4}}{\includegraphics[width = 2.0in]{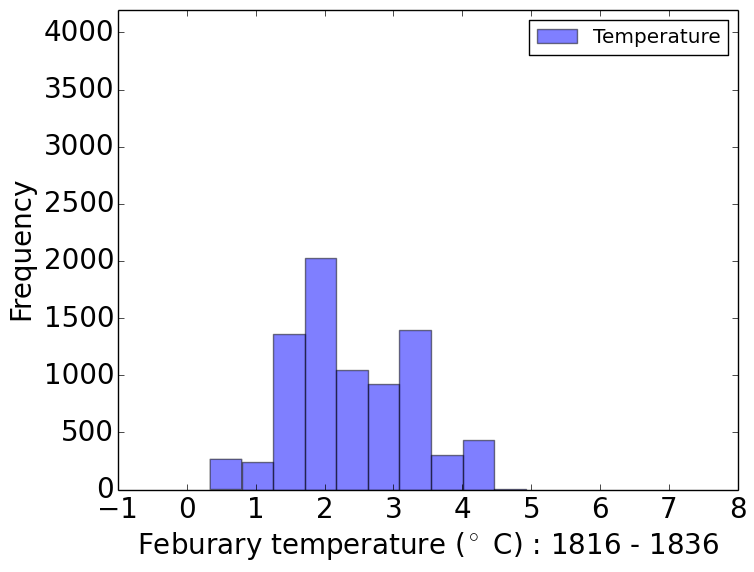}} &
\subcaptionbox{February -- SMC: 1836 - 1856\label{fig:feb5}}{\includegraphics[width = 2.0in]{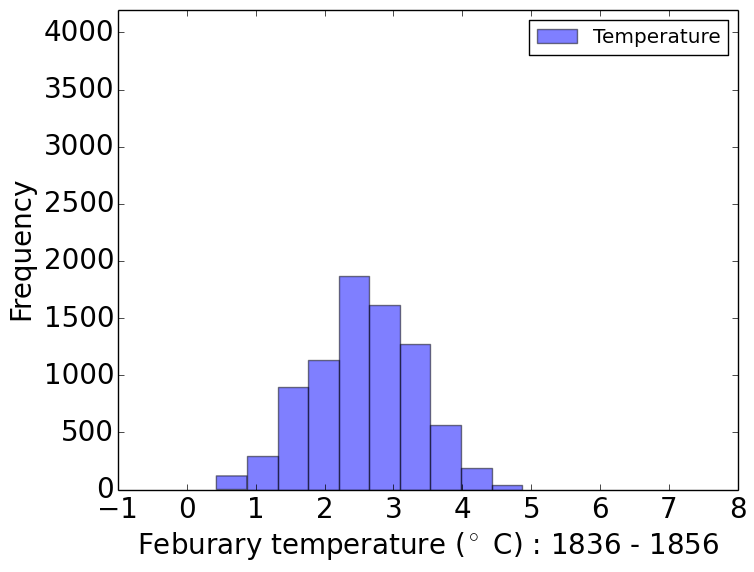}} &
\subcaptionbox{February -- SMC: 1856 - 1876\label{fig:feb6}}{\includegraphics[width = 2.0in]{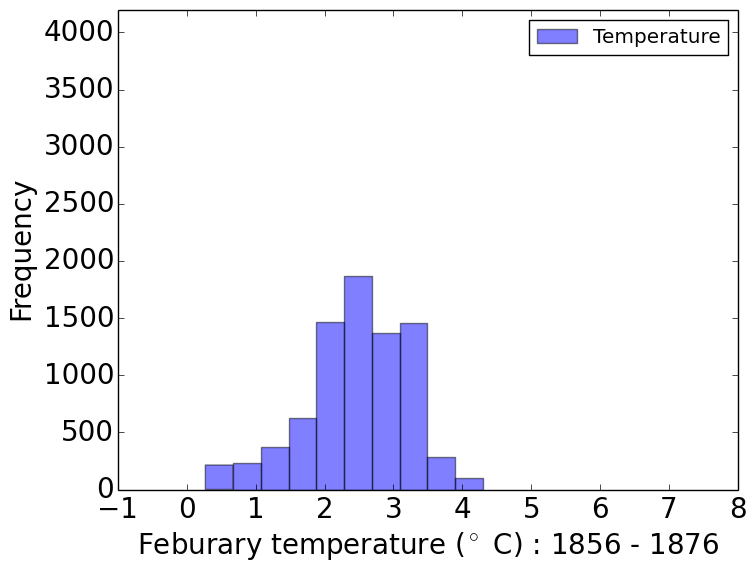}}\\
\subcaptionbox{February -- SMC: 1876 - 1896\label{fig:feb7}}{\includegraphics[width = 2.0in]{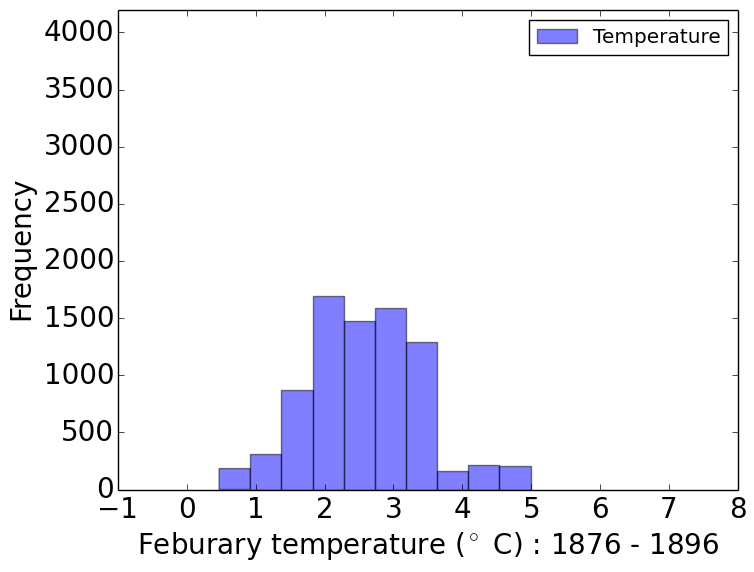}} &
\subcaptionbox{February -- SMC: 1896 - 1916\label{fig:feb8}}{\includegraphics[width = 2.0in]{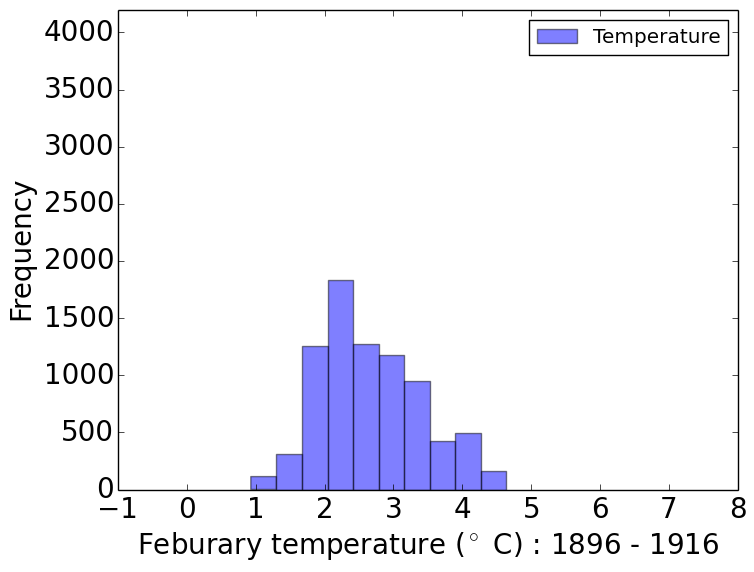}} &
\subcaptionbox{February -- SMC: 1916 - 1936\label{fig:feb9}}{\includegraphics[width = 2.0in]{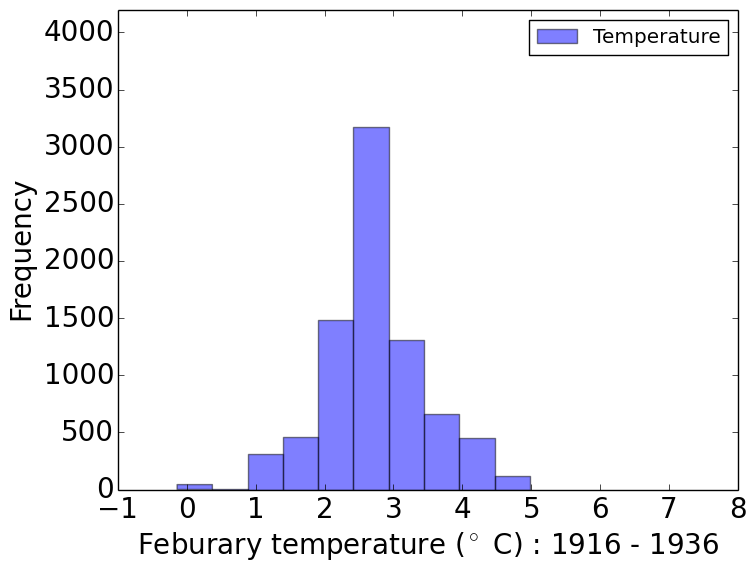}}\\
\subcaptionbox{February -- SMC: 1936 - 1936\label{fig:feb10}}{\includegraphics[width = 2.0in]{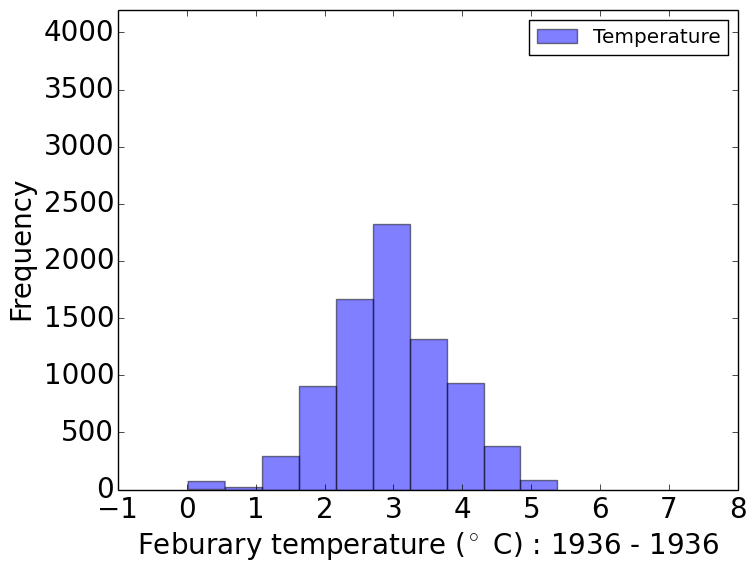}} &
\subcaptionbox{February -- SMC: 1956 - 1976\label{fig:feb11}}{\includegraphics[width = 2.0in]{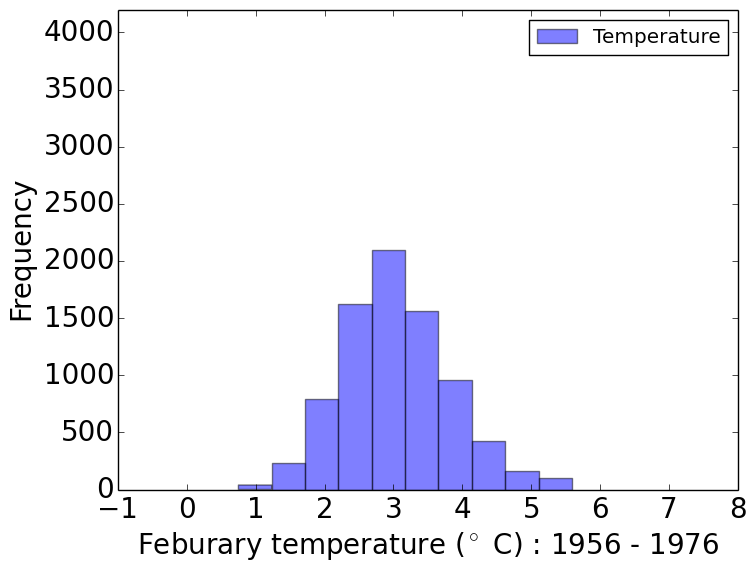}} &
\subcaptionbox{February -- SMC: 1976 - 1996\label{fig:feb12}}{\includegraphics[width = 2.0in]{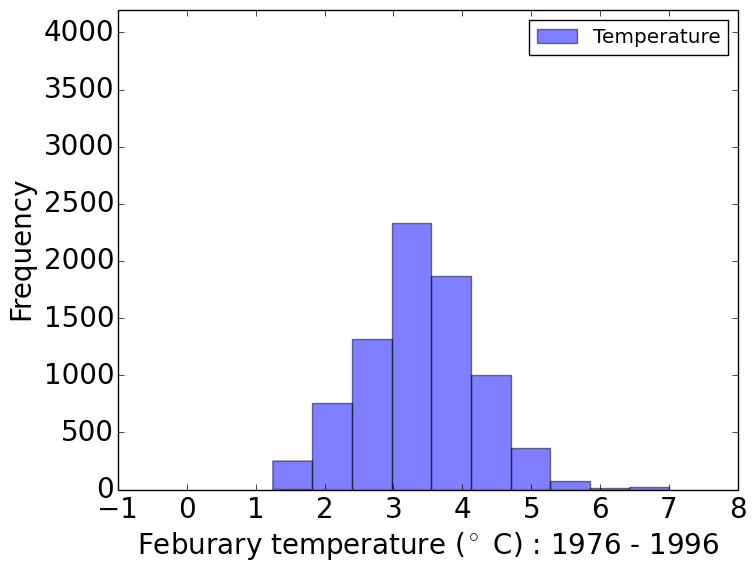}}\\
\subcaptionbox{February -- SMC: 1996 - 2016\label{fig:feb13}}{\includegraphics[width = 2.0in]{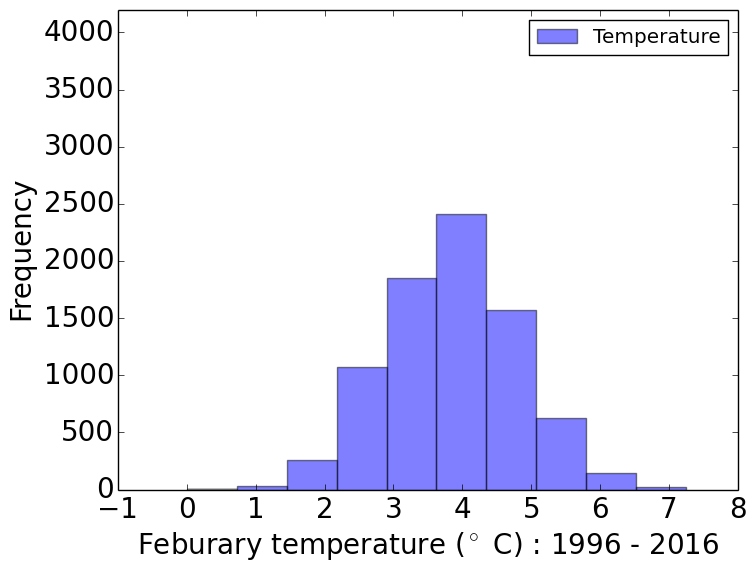}} &
\subcaptionbox{February -- SMC: Total\label{fig:feb14}}{\includegraphics[width = 2.0in]{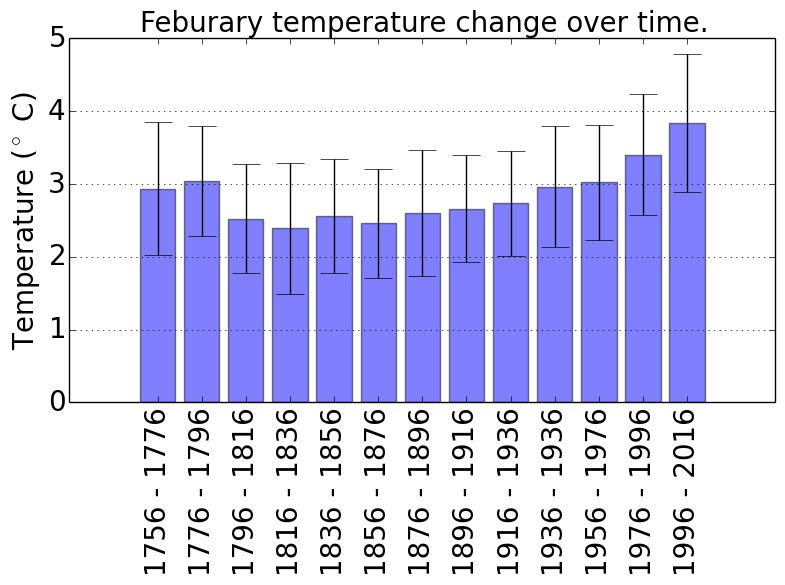}} &
\end{tabular}
\label{fig:feb1}
\end{adjustwidth}
\end{figure}

\begin{figure}
\begin{adjustwidth}{-6em}{0em}
\centering
\begin{tabular}{ccc}
\subcaptionbox{March -- SMC: 1756 - 1776\label{fig:mar1}}{\includegraphics[width = 2.0in]{images/smc/mar/comb_smc_hist_1.png}} &
\subcaptionbox{March -- SMC: 1776 - 1796\label{fig:mar2}}{\includegraphics[width = 2.0in]{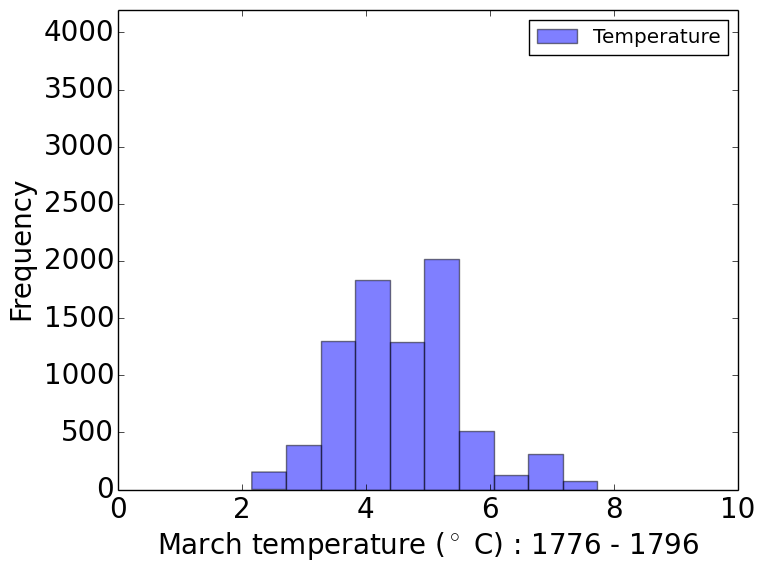}} &
\subcaptionbox{March -- SMC: 1796 - 1816\label{fig:mar3}}{\includegraphics[width = 2.0in]{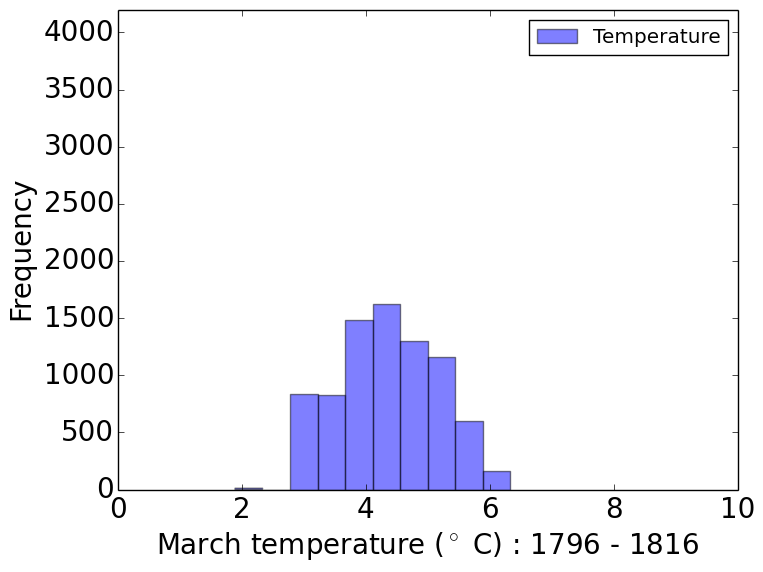}}\\
\subcaptionbox{March -- SMC: 1816 - 1836\label{fig:mar4}}{\includegraphics[width = 2.0in]{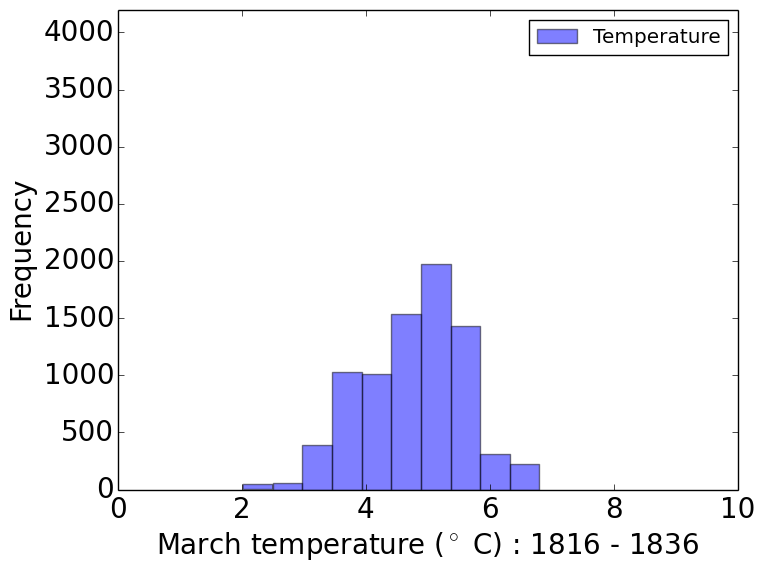}} &
\subcaptionbox{March -- SMC: 1836 - 1856\label{fig:mar5}}{\includegraphics[width = 2.0in]{images/smc/mar/comb_smc_hist_5.png}} &
\subcaptionbox{March -- SMC: 1856 - 1876\label{fig:mar6}}{\includegraphics[width = 2.0in]{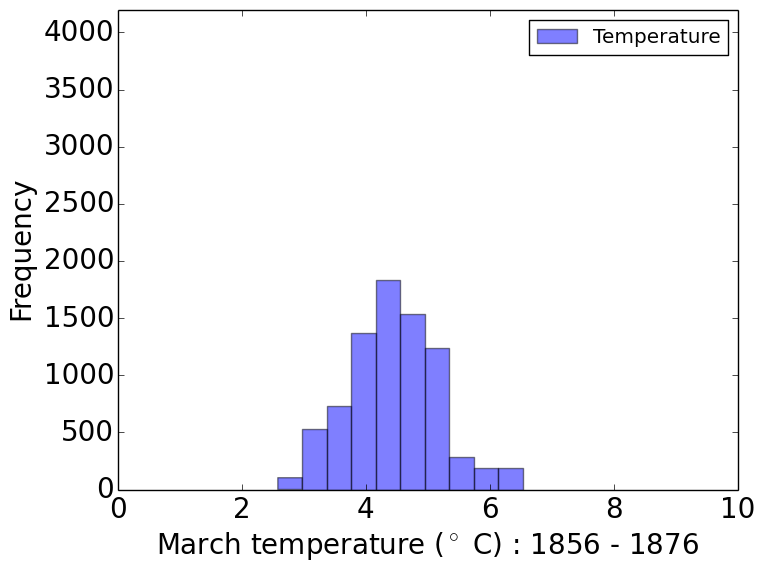}}\\
\subcaptionbox{March -- SMC: 1876 - 1896\label{fig:mar7}}{\includegraphics[width = 2.0in]{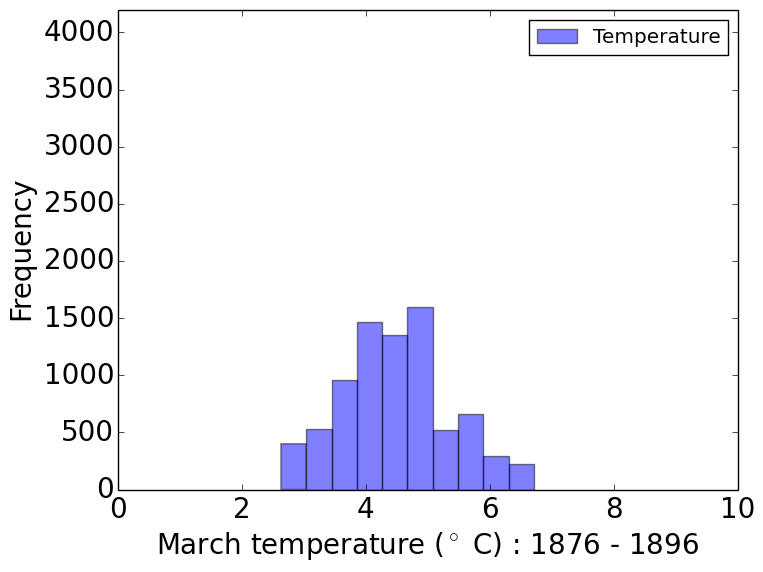}} &
\subcaptionbox{March -- SMC: 1896 - 1916\label{fig:mar8}}{\includegraphics[width = 2.0in]{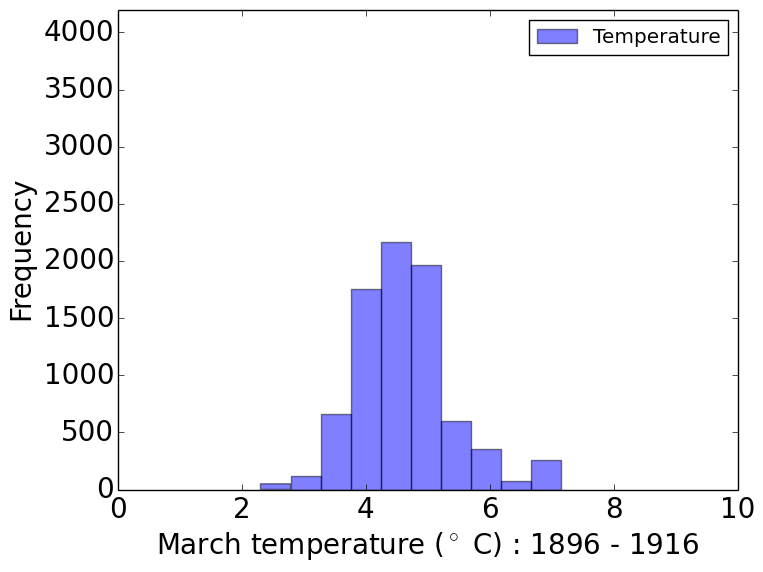}} &
\subcaptionbox{March -- SMC: 1916 - 1936\label{fig:mar9}}{\includegraphics[width = 2.0in]{images/smc/mar/comb_smc_hist_9.png}}\\
\subcaptionbox{March -- SMC: 1936 - 1936\label{fig:mar10}}{\includegraphics[width = 2.0in]{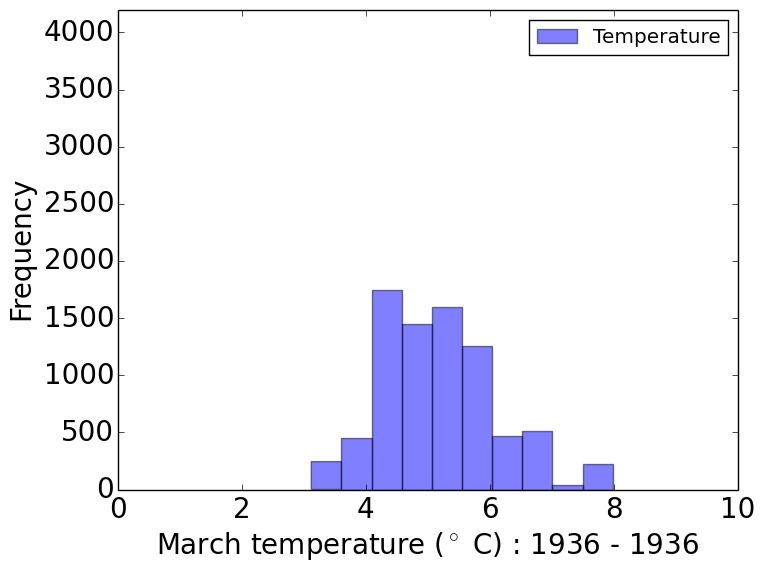}} &
\subcaptionbox{March -- SMC: 1956 - 1976\label{fig:mar11}}{\includegraphics[width = 2.0in]{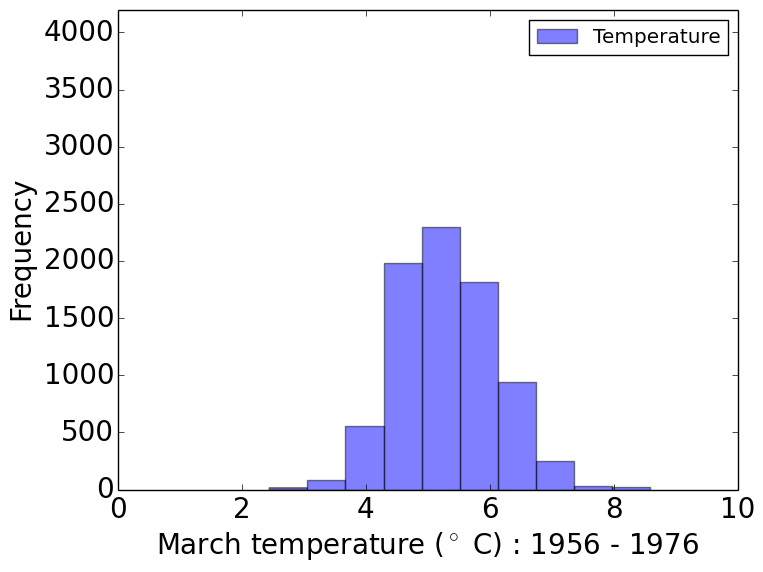}} &
\subcaptionbox{March -- SMC: 1976 - 1996\label{fig:mar12}}{\includegraphics[width = 2.0in]{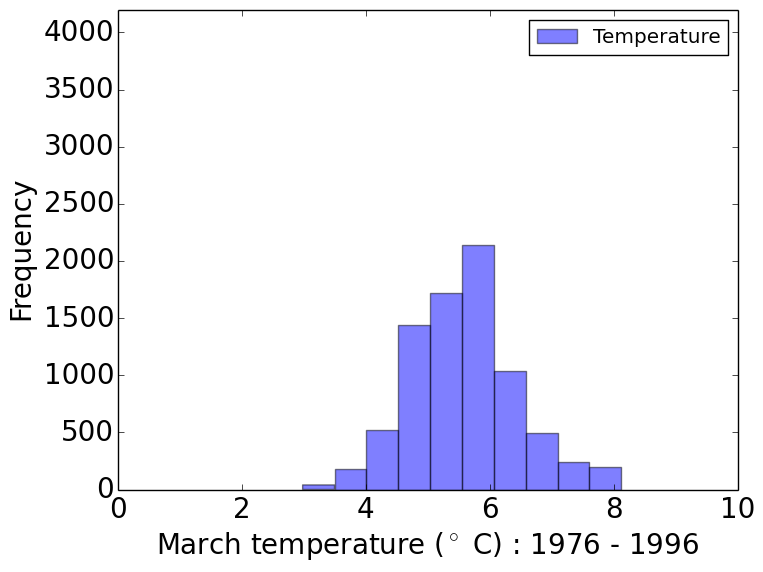}}\\
\subcaptionbox{March -- SMC: 1996 - 2016\label{fig:mar13}}{\includegraphics[width = 2.0in]{images/smc/mar/comb_smc_hist_13.png}} &
\subcaptionbox{March -- SMC: Total\label{fig:mar14}}{\includegraphics[width = 2.0in]{images/smc/mar/aggregate.png}} &
\end{tabular}
\label{fig:mar1}
\end{adjustwidth}
\end{figure}

\begin{figure}
\begin{adjustwidth}{-6em}{0em}
\centering
\begin{tabular}{ccc}
\subcaptionbox{April -- SMC: 1756 - 1776\label{fig:apr1}}{\includegraphics[width = 2.0in]{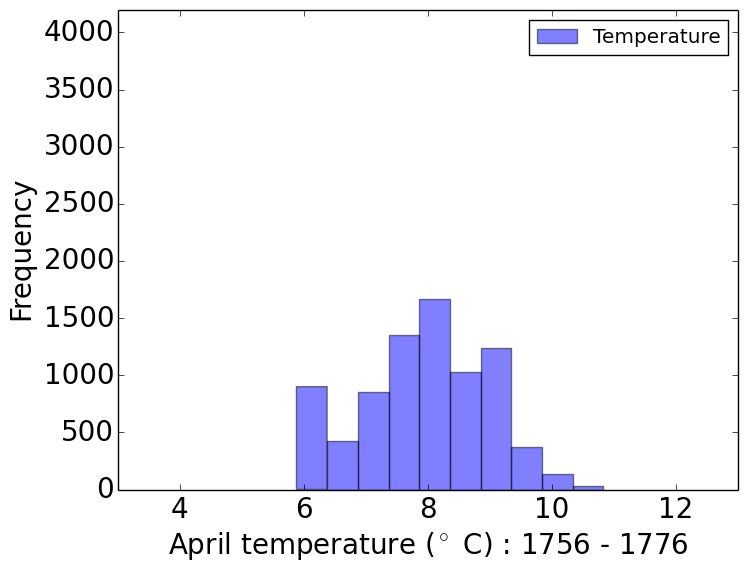}} &
\subcaptionbox{April -- SMC: 1776 - 1796\label{fig:apr2}}{\includegraphics[width = 2.0in]{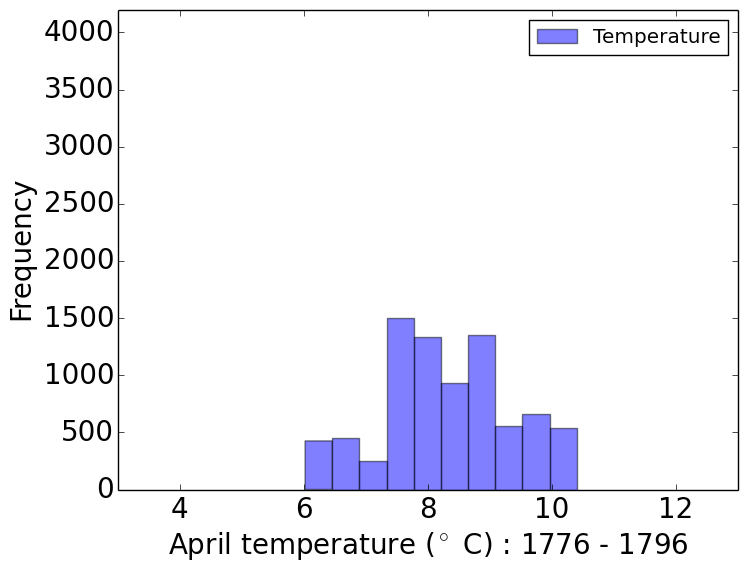}} &
\subcaptionbox{April -- SMC: 1796 - 1816\label{fig:apr3}}{\includegraphics[width = 2.0in]{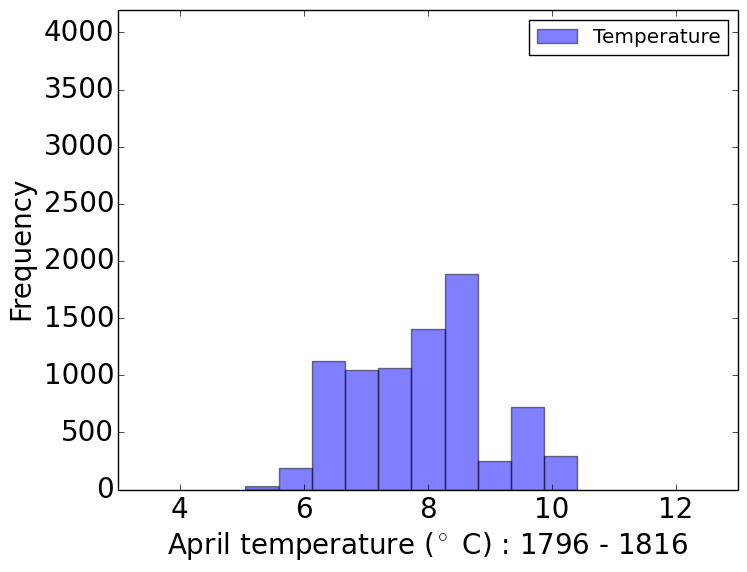}}\\
\subcaptionbox{April -- SMC: 1816 - 1836\label{fig:apr4}}{\includegraphics[width = 2.0in]{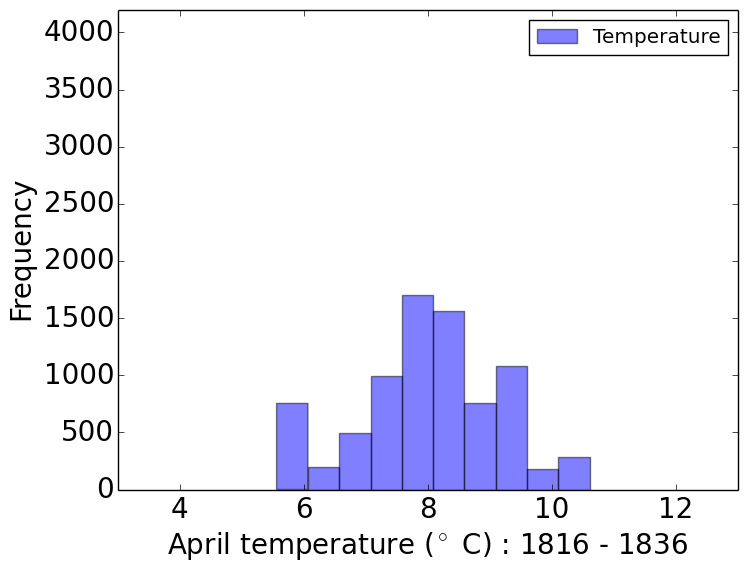}} &
\subcaptionbox{April -- SMC: 1836 - 1856\label{fig:apr5}}{\includegraphics[width = 2.0in]{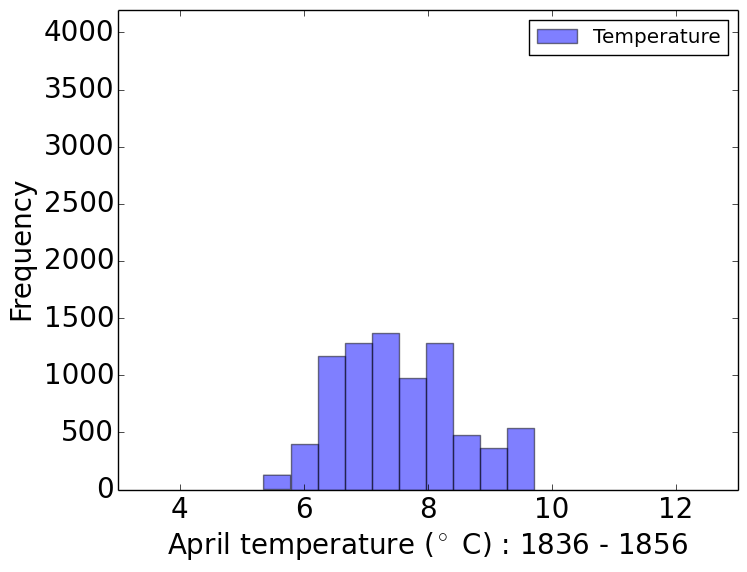}} &
\subcaptionbox{April -- SMC: 1856 - 1876\label{fig:apr6}}{\includegraphics[width = 2.0in]{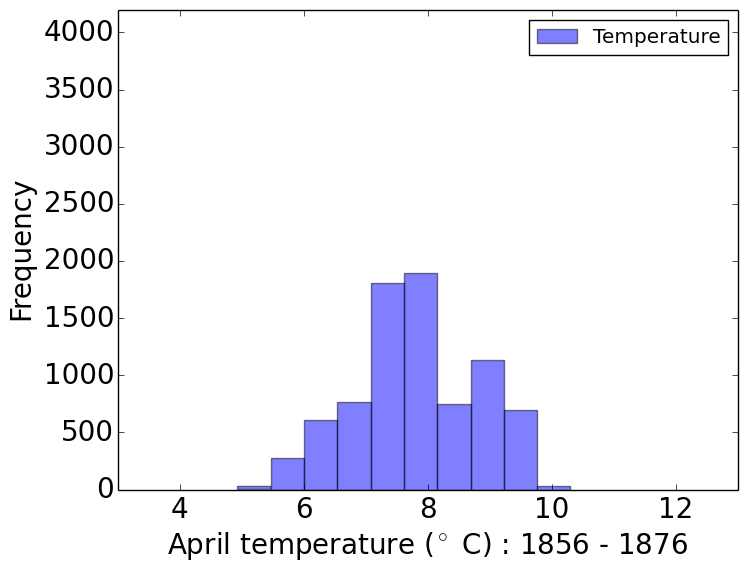}}\\
\subcaptionbox{April -- SMC: 1876 - 1896\label{fig:apr7}}{\includegraphics[width = 2.0in]{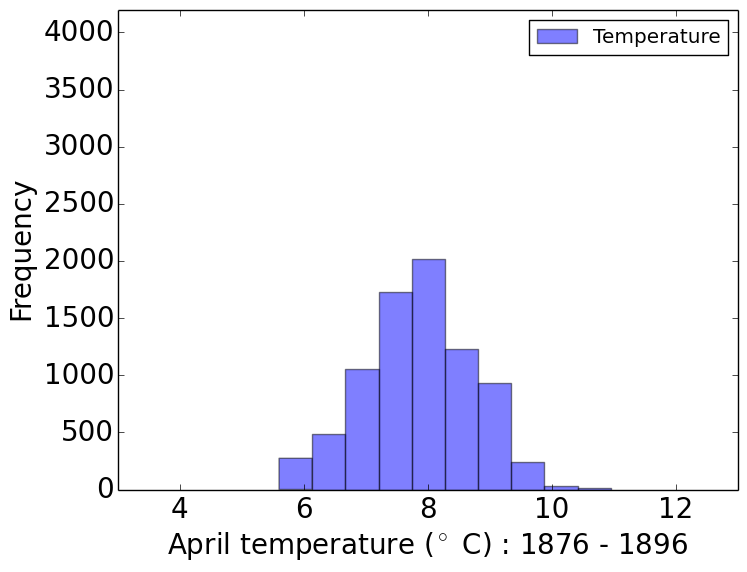}} &
\subcaptionbox{April -- SMC: 1896 - 1916\label{fig:apr8}}{\includegraphics[width = 2.0in]{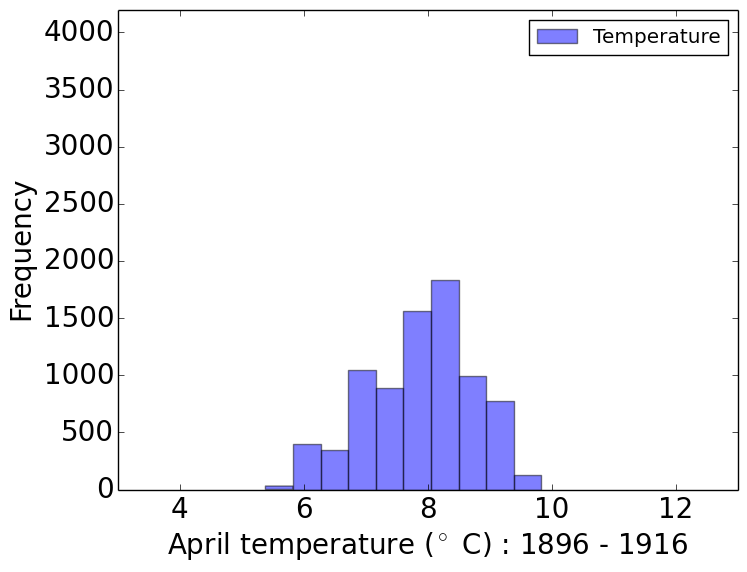}} &
\subcaptionbox{April -- SMC: 1916 - 1936\label{fig:apr9}}{\includegraphics[width = 2.0in]{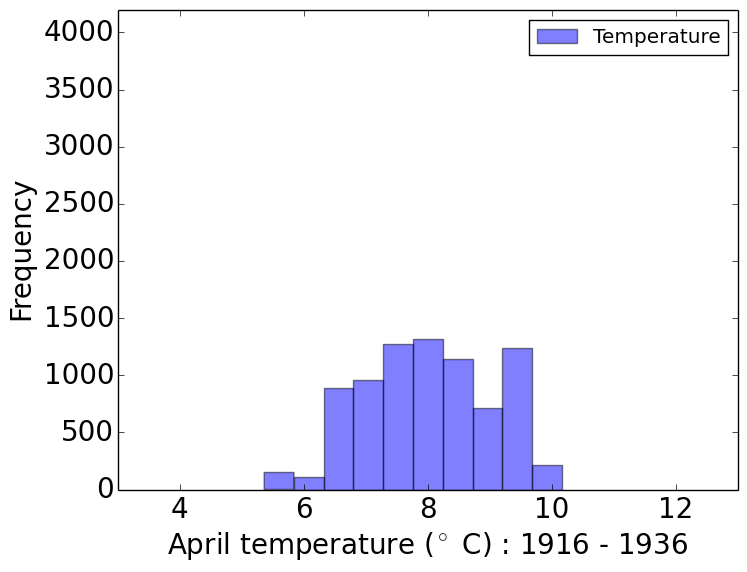}}\\
\subcaptionbox{April -- SMC: 1936 - 1936\label{fig:apr10}}{\includegraphics[width = 2.0in]{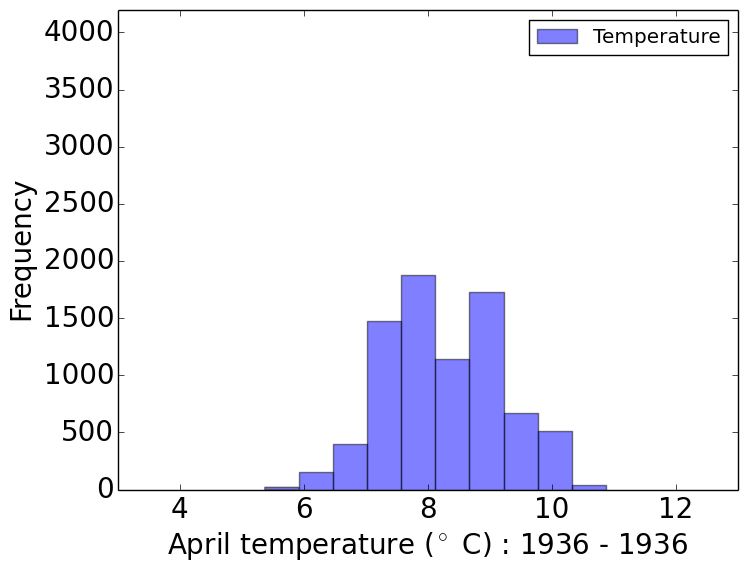}} &
\subcaptionbox{April -- SMC: 1956 - 1976\label{fig:apr11}}{\includegraphics[width = 2.0in]{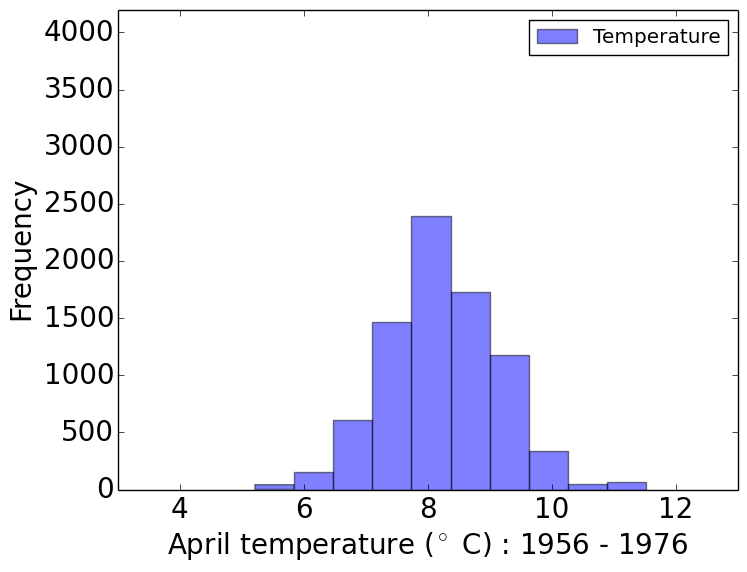}} &
\subcaptionbox{April -- SMC: 1976 - 1996\label{fig:apr12}}{\includegraphics[width = 2.0in]{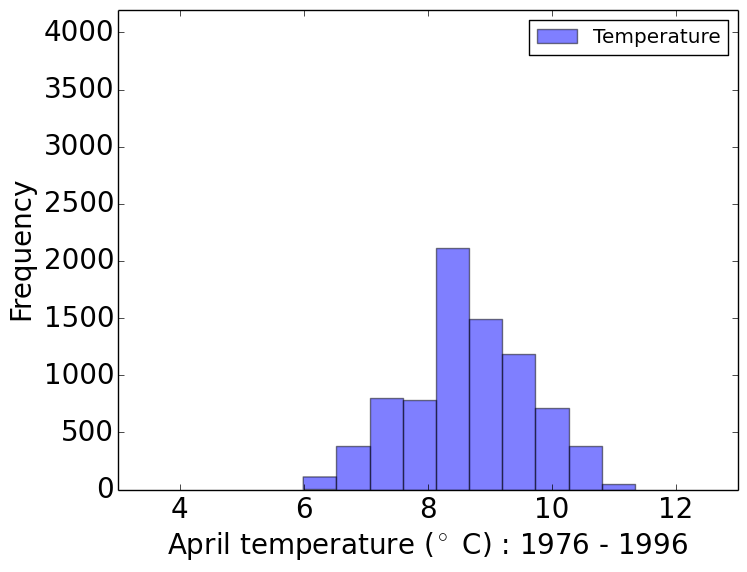}}\\
\subcaptionbox{April -- SMC: 1996 - 2016\label{fig:apr13}}{\includegraphics[width = 2.0in]{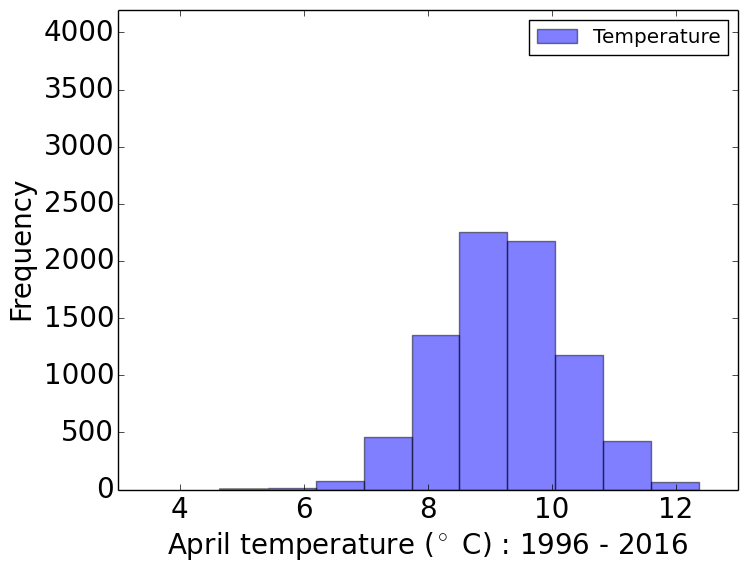}} &
\subcaptionbox{April -- SMC: Total\label{fig:apr14}}{\includegraphics[width = 2.0in]{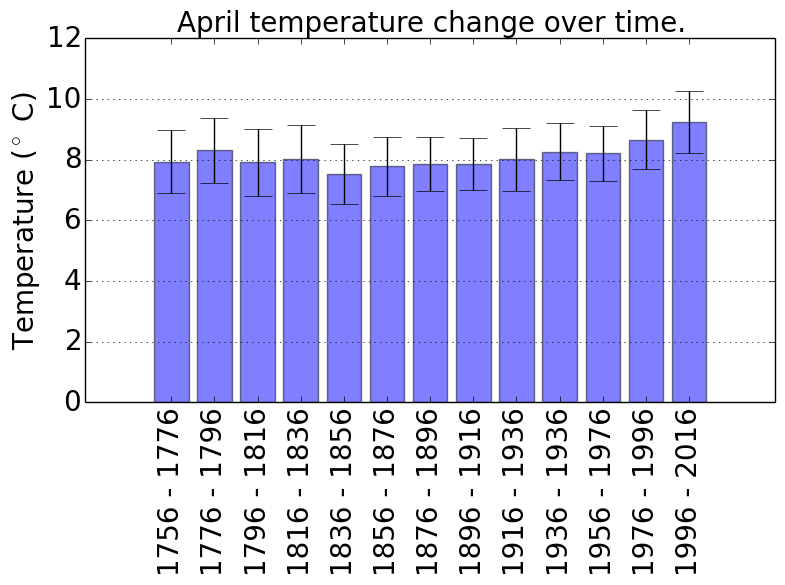}} &
\end{tabular}
\label{fig:apr1}
\end{adjustwidth}
\end{figure}

\begin{figure}
\begin{adjustwidth}{-6em}{0em}
\centering
\begin{tabular}{ccc}
\subcaptionbox{May -- SMC: 1756 - 1776\label{fig:may1}}{\includegraphics[width = 2.0in]{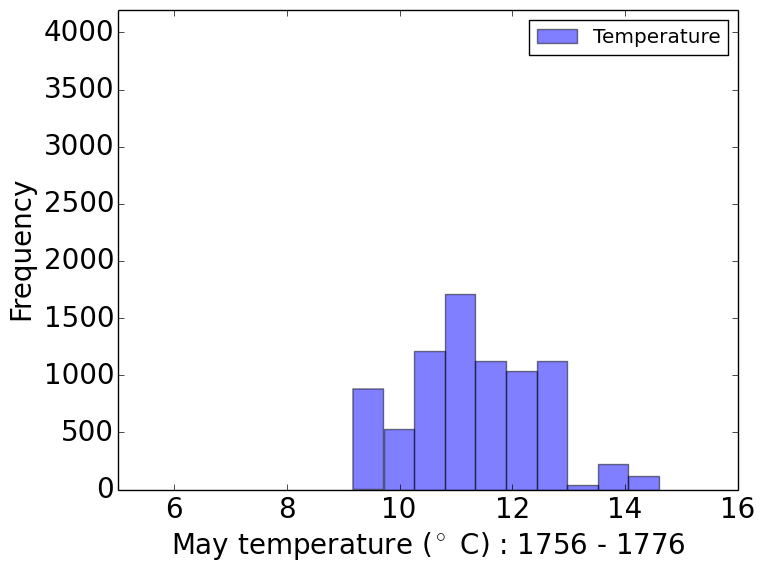}} &
\subcaptionbox{May -- SMC: 1776 - 1796\label{fig:may2}}{\includegraphics[width = 2.0in]{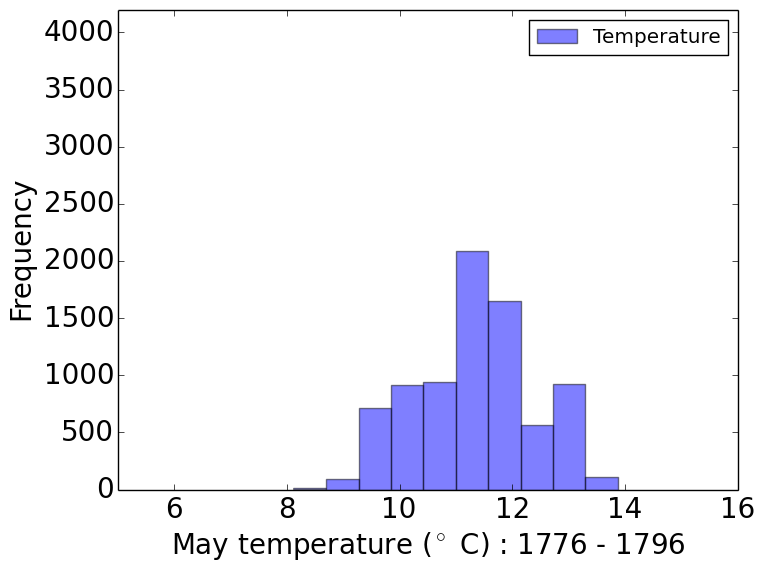}} &
\subcaptionbox{May -- SMC: 1796 - 1816\label{fig:may3}}{\includegraphics[width = 2.0in]{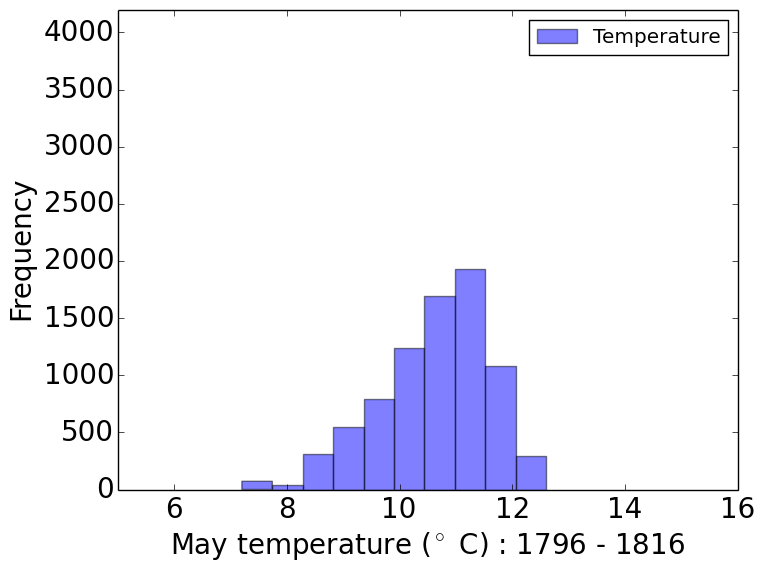}}\\
\subcaptionbox{May -- SMC: 1816 - 1836\label{fig:may4}}{\includegraphics[width = 2.0in]{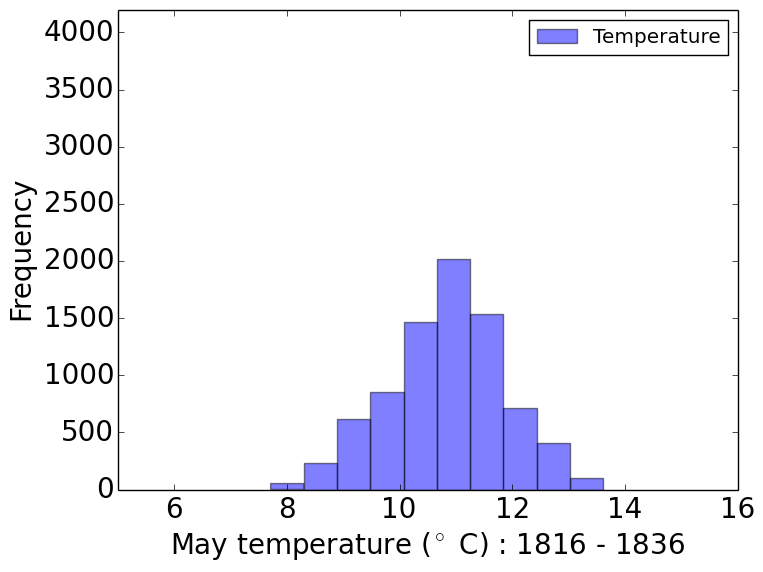}} &
\subcaptionbox{May -- SMC: 1836 - 1856\label{fig:may5}}{\includegraphics[width = 2.0in]{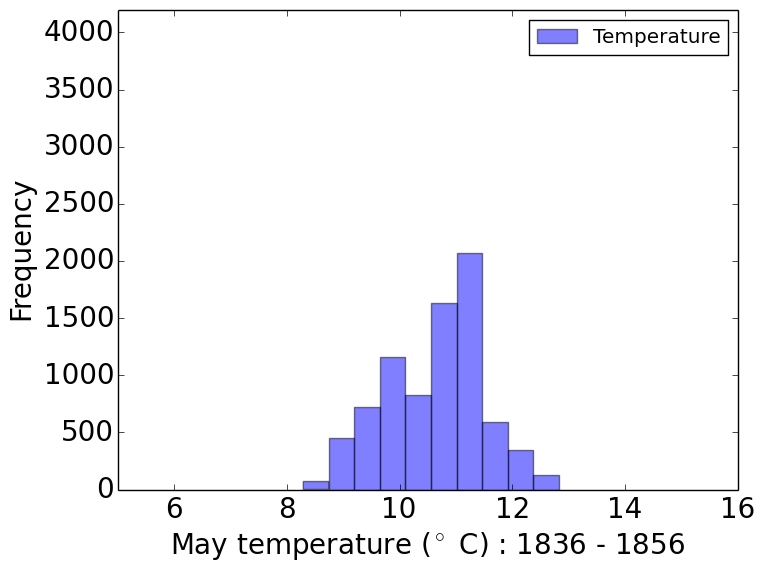}} &
\subcaptionbox{May -- SMC: 1856 - 1876\label{fig:may6}}{\includegraphics[width = 2.0in]{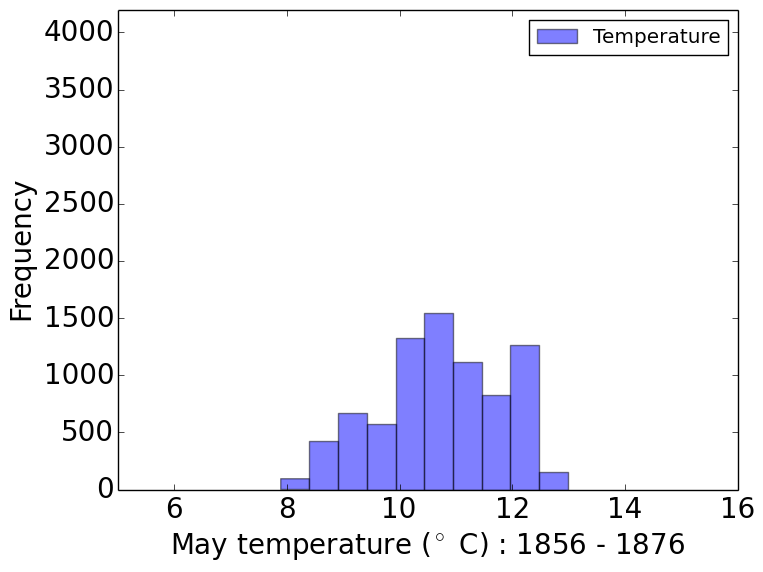}}\\
\subcaptionbox{May -- SMC: 1876 - 1896\label{fig:may7}}{\includegraphics[width = 2.0in]{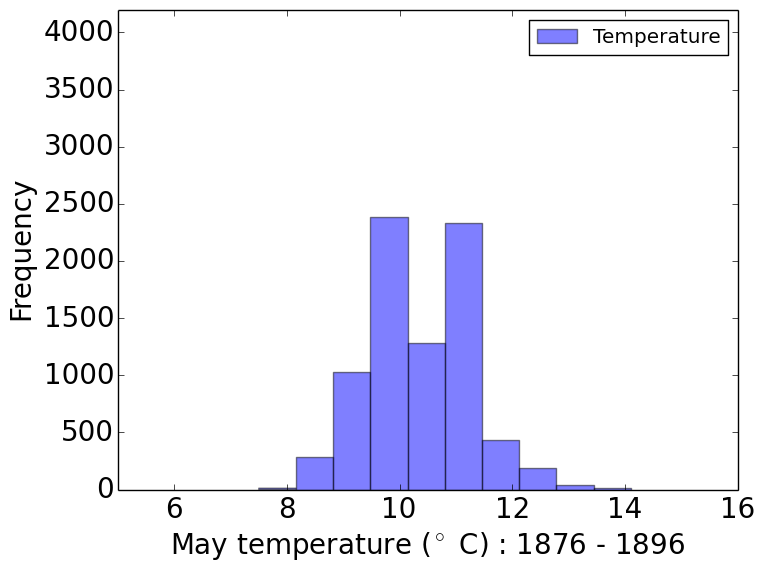}} &
\subcaptionbox{May -- SMC: 1896 - 1916\label{fig:may8}}{\includegraphics[width = 2.0in]{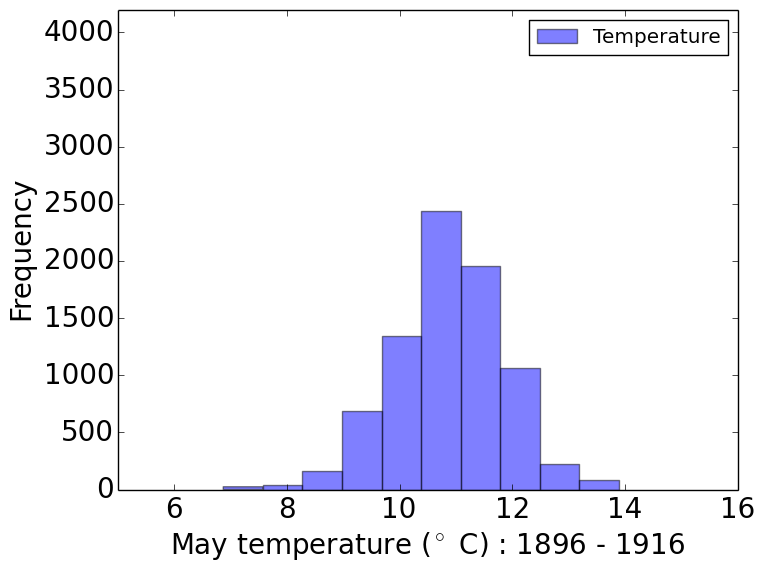}} &
\subcaptionbox{May -- SMC: 1916 - 1936\label{fig:may9}}{\includegraphics[width = 2.0in]{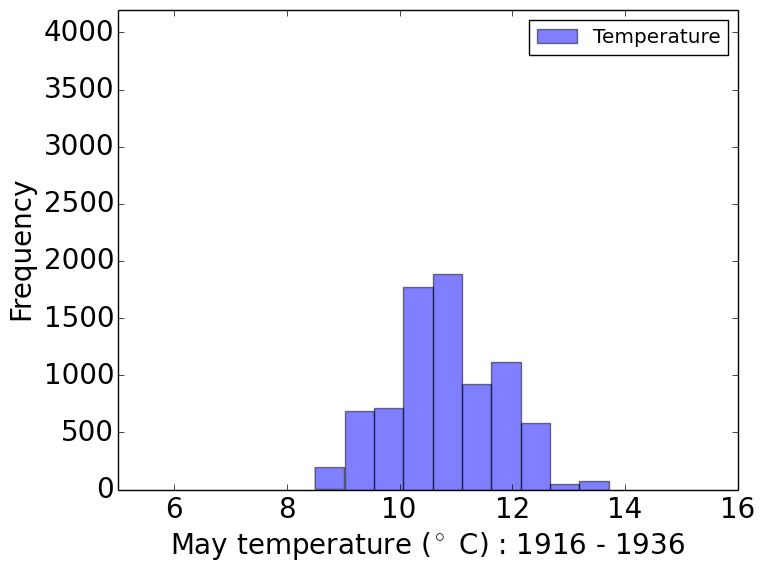}}\\
\subcaptionbox{May -- SMC: 1936 - 1936\label{fig:may10}}{\includegraphics[width = 2.0in]{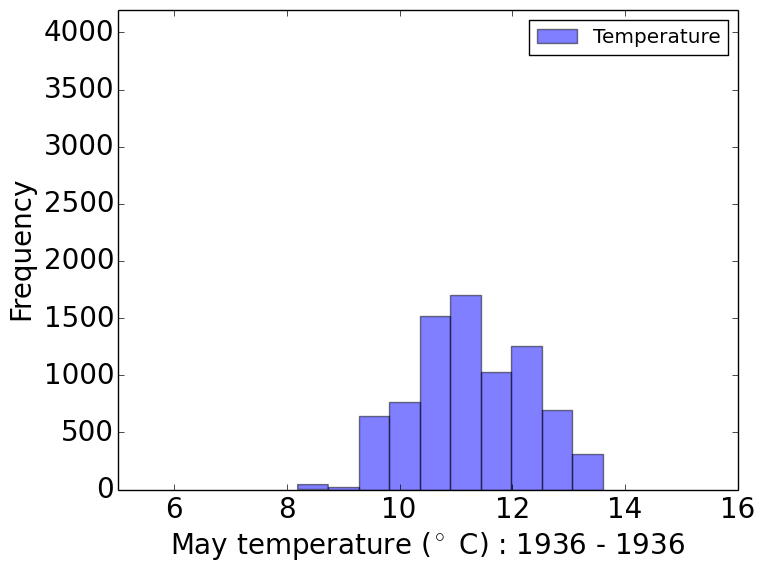}} &
\subcaptionbox{May -- SMC: 1956 - 1976\label{fig:may11}}{\includegraphics[width = 2.0in]{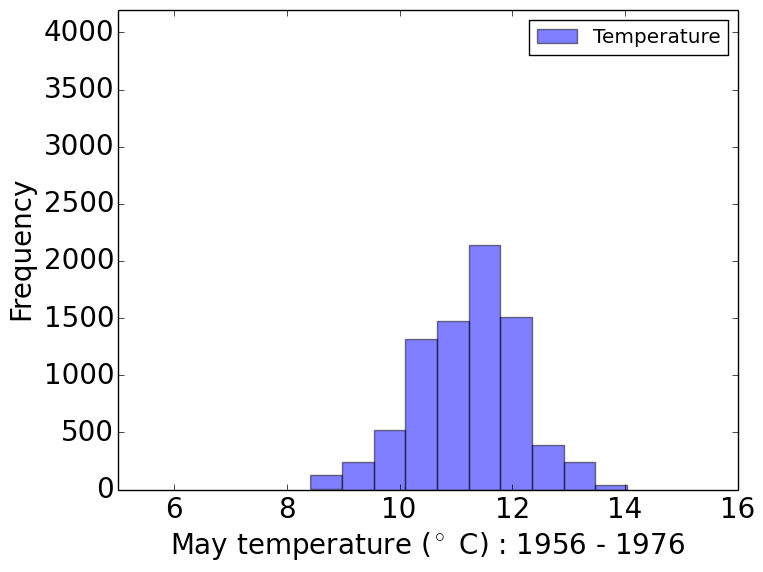}} &
\subcaptionbox{May -- SMC: 1976 - 1996\label{fig:may12}}{\includegraphics[width = 2.0in]{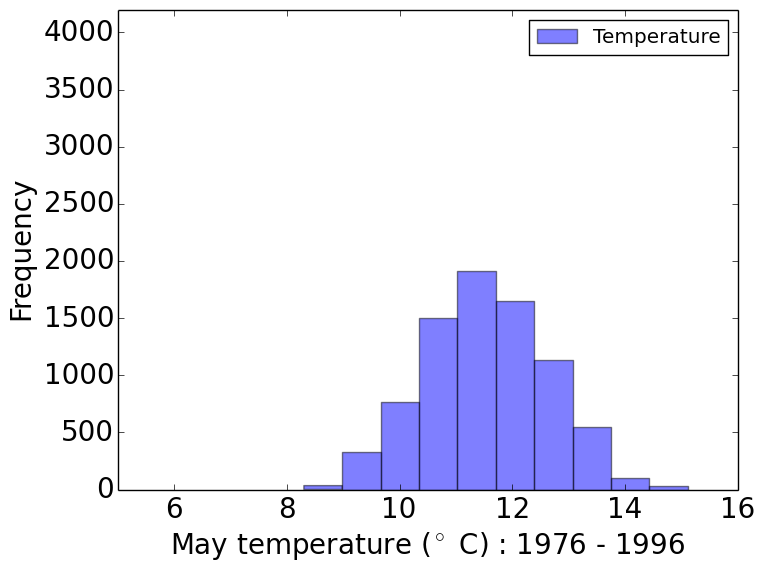}}\\
\subcaptionbox{May -- SMC: 1996 - 2016\label{fig:may13}}{\includegraphics[width = 2.0in]{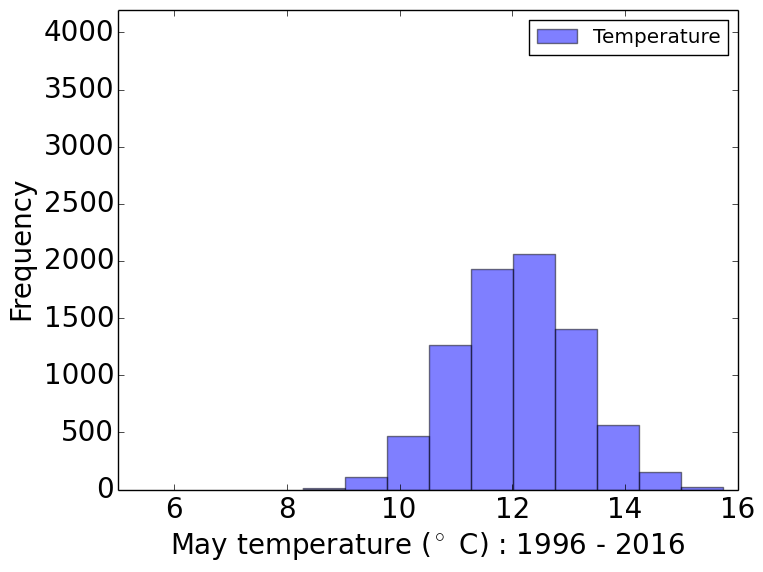}} &
\subcaptionbox{May -- SMC: Total\label{fig:may14}}{\includegraphics[width = 2.0in]{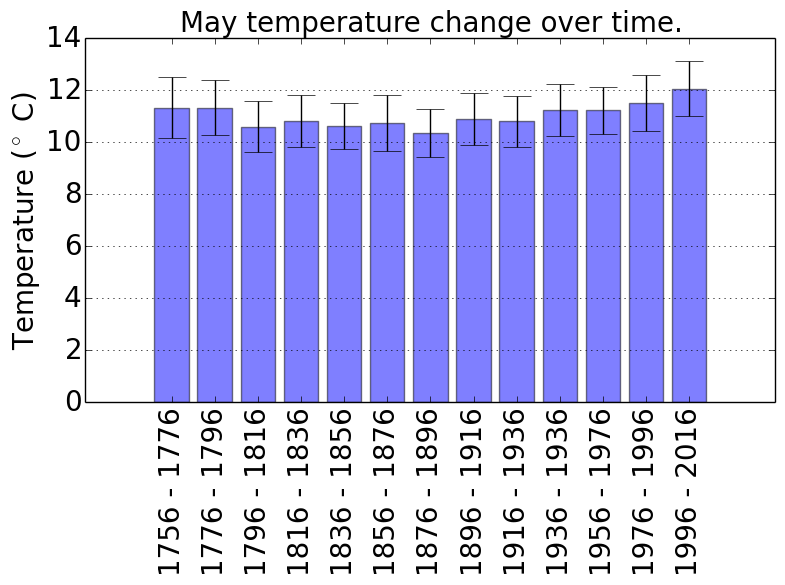}} &
\end{tabular}
\label{fig:may1}
\end{adjustwidth}
\end{figure}

\begin{figure}
\begin{adjustwidth}{-6em}{0em}
\centering
\begin{tabular}{ccc}
\subcaptionbox{June -- SMC: 1756 - 1776\label{fig:jun1}}{\includegraphics[width = 2.0in]{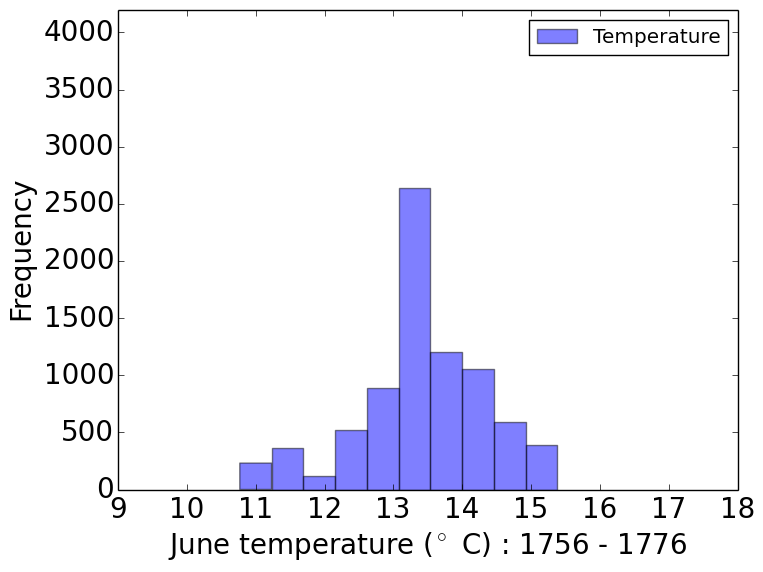}} &
\subcaptionbox{June -- SMC: 1776 - 1796\label{fig:jun2}}{\includegraphics[width = 2.0in]{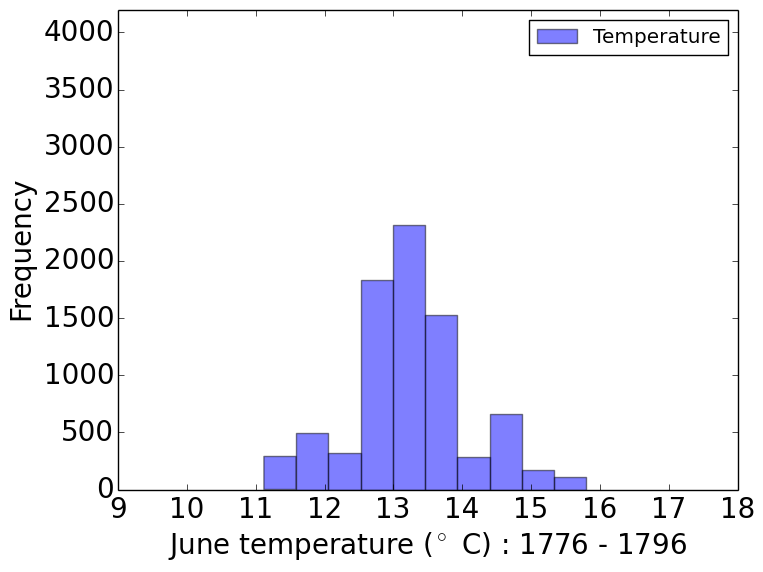}} &
\subcaptionbox{June -- SMC: 1796 - 1816\label{fig:jun3}}{\includegraphics[width = 2.0in]{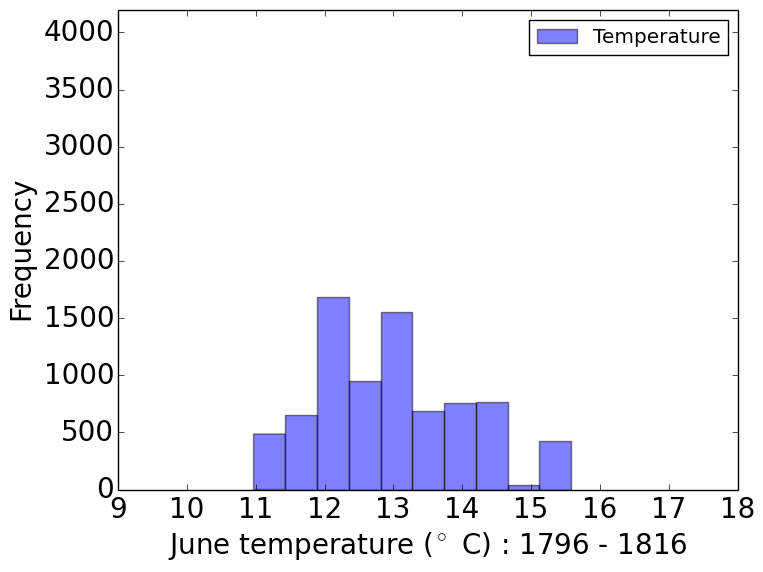}}\\
\subcaptionbox{June -- SMC: 1816 - 1836\label{fig:jun4}}{\includegraphics[width = 2.0in]{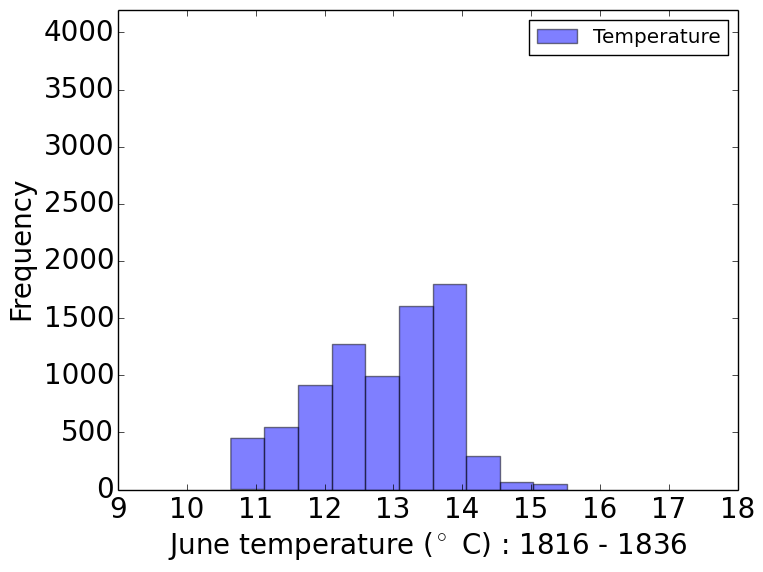}} &
\subcaptionbox{June -- SMC: 1836 - 1856\label{fig:jun5}}{\includegraphics[width = 2.0in]{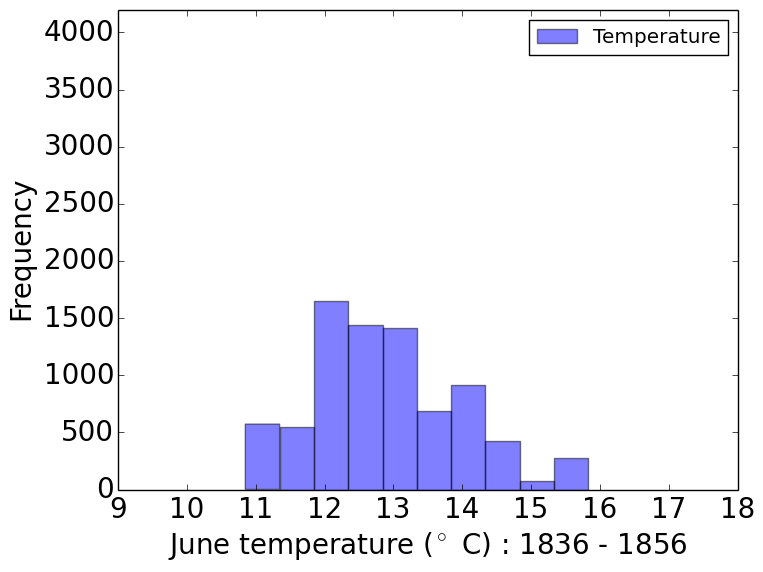}} &
\subcaptionbox{June -- SMC: 1856 - 1876\label{fig:jun6}}{\includegraphics[width = 2.0in]{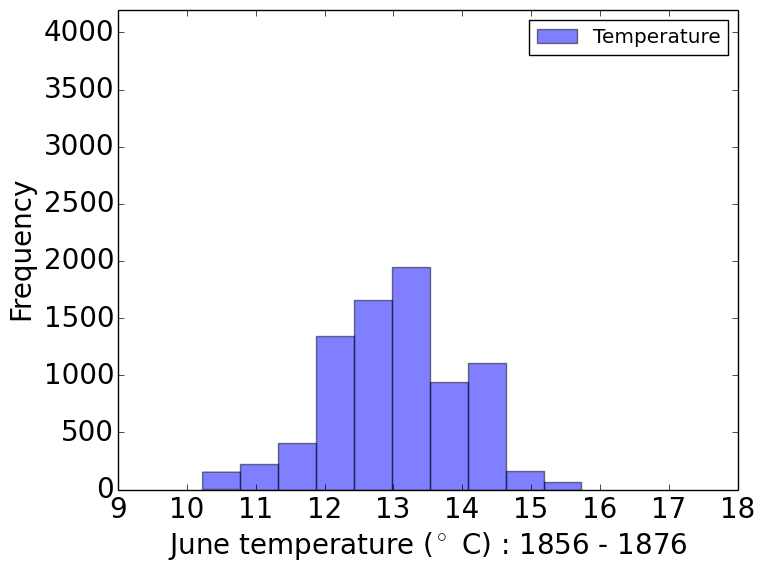}}\\
\subcaptionbox{June -- SMC: 1876 - 1896\label{fig:jun7}}{\includegraphics[width = 2.0in]{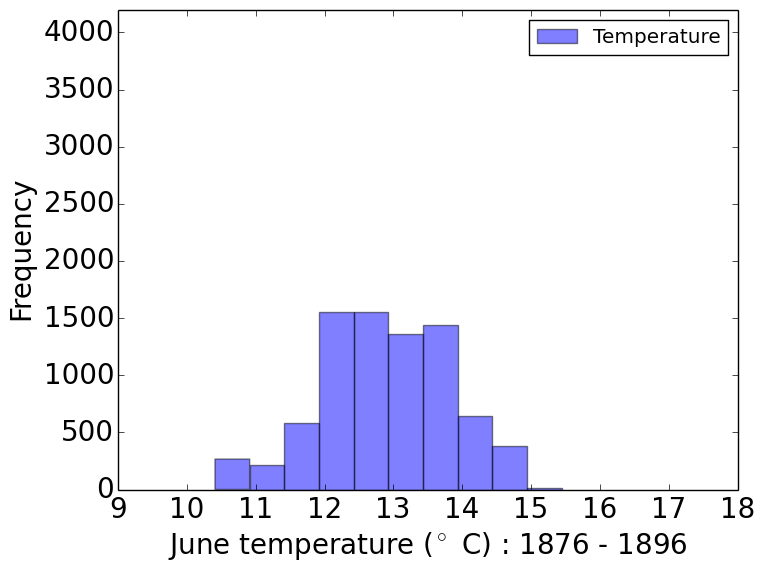}} &
\subcaptionbox{June -- SMC: 1896 - 1916\label{fig:jun8}}{\includegraphics[width = 2.0in]{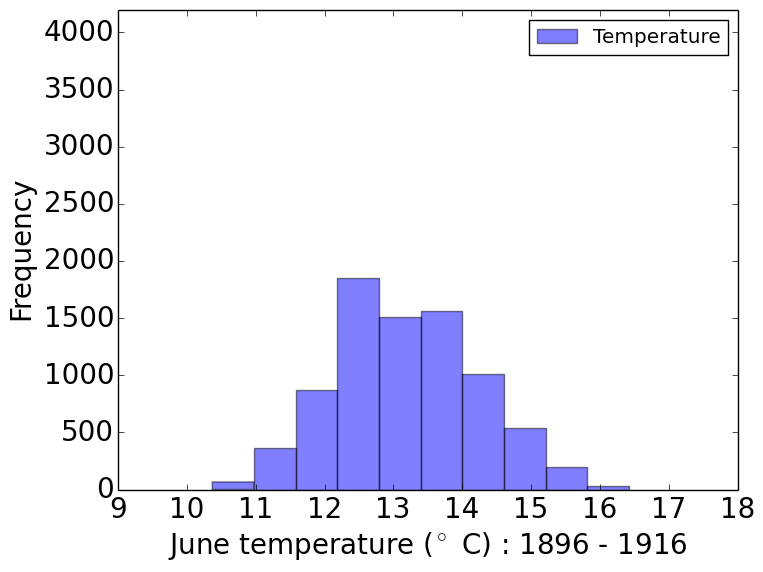}} &
\subcaptionbox{June -- SMC: 1916 - 1936\label{fig:jun9}}{\includegraphics[width = 2.0in]{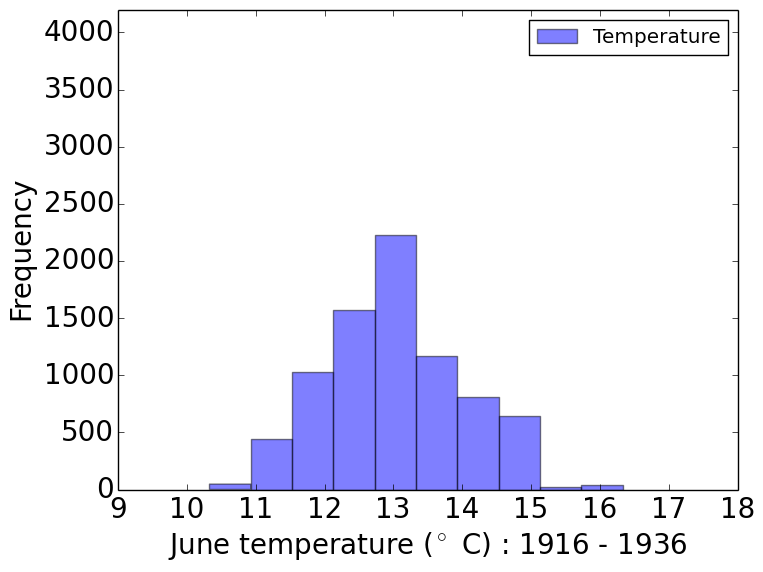}}\\
\subcaptionbox{June -- SMC: 1936 - 1936\label{fig:jun10}}{\includegraphics[width = 2.0in]{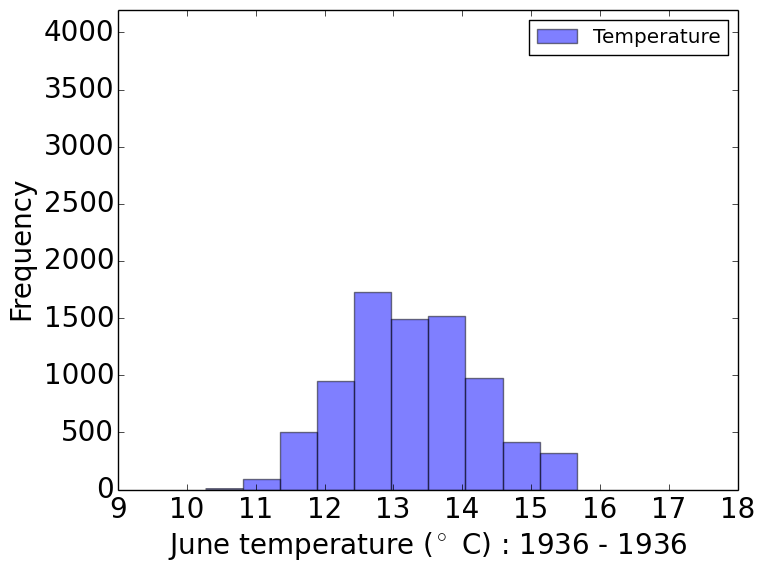}} &
\subcaptionbox{June -- SMC: 1956 - 1976\label{fig:jun11}}{\includegraphics[width = 2.0in]{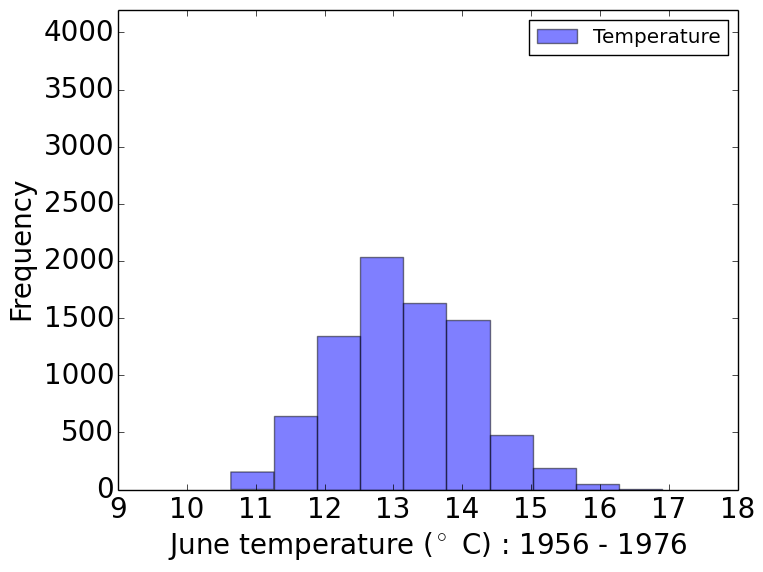}} &
\subcaptionbox{June -- SMC: 1976 - 1996\label{fig:jun12}}{\includegraphics[width = 2.0in]{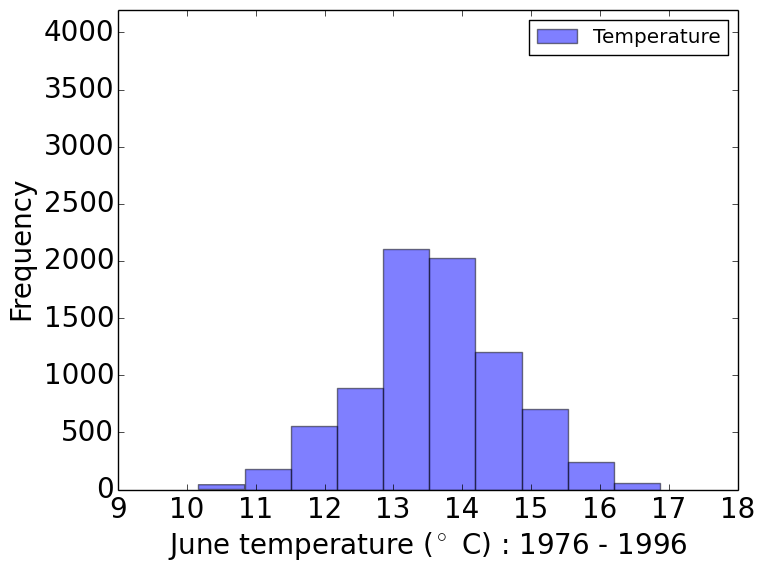}}\\
\subcaptionbox{June -- SMC: 1996 - 2016\label{fig:jun13}}{\includegraphics[width = 2.0in]{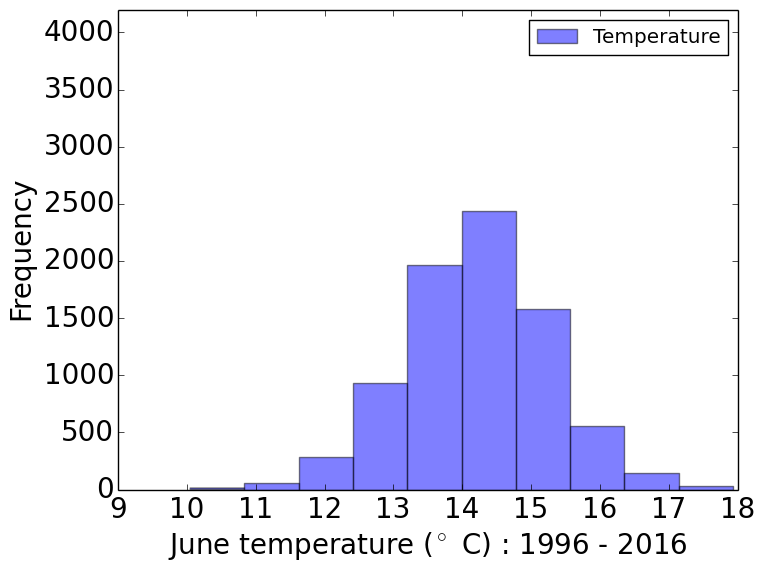}} &
\subcaptionbox{June -- SMC: Total\label{fig:jun14}}{\includegraphics[width = 2.0in]{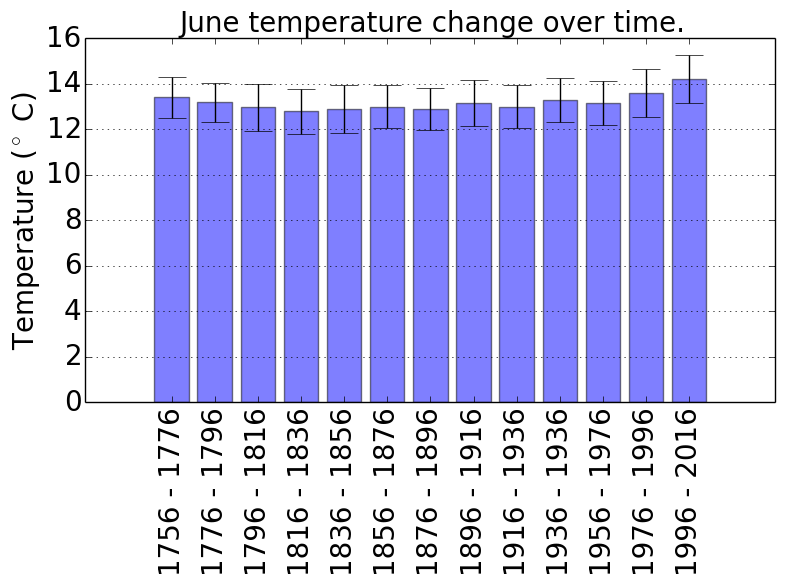}} &
\end{tabular}
\label{fig:jun1}
\end{adjustwidth}
\end{figure}

\begin{figure}
\begin{adjustwidth}{-6em}{0em}
\centering
\begin{tabular}{ccc}
\subcaptionbox{July -- SMC: 1756 - 1776\label{fig:jul1}}{\includegraphics[width = 2.0in]{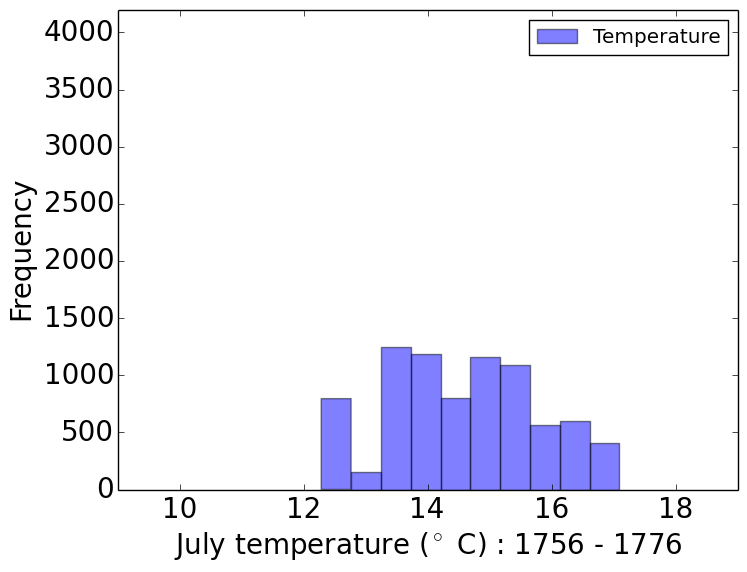}} &
\subcaptionbox{July -- SMC: 1776 - 1796\label{fig:jul2}}{\includegraphics[width = 2.0in]{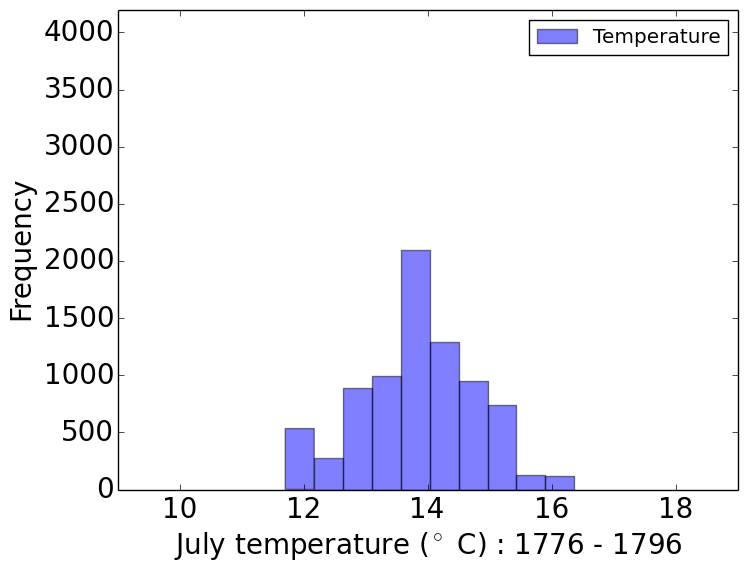}} &
\subcaptionbox{July -- SMC: 1796 - 1816\label{fig:jul3}}{\includegraphics[width = 2.0in]{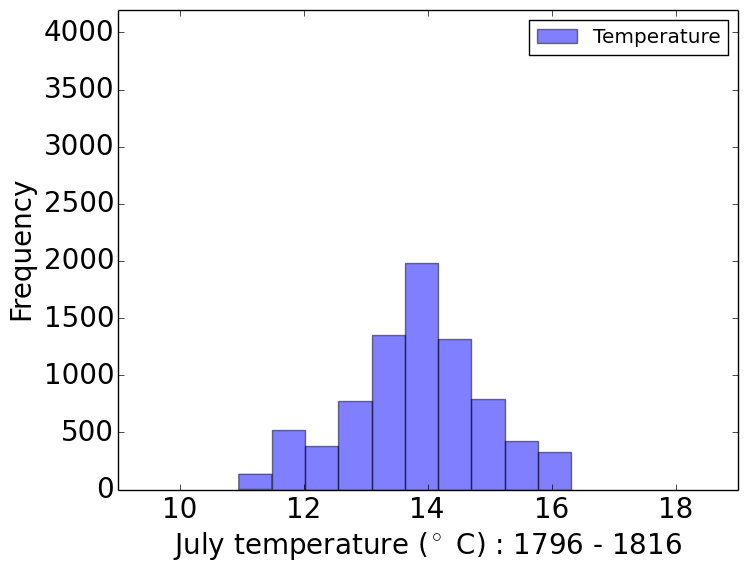}}\\
\subcaptionbox{July -- SMC: 1816 - 1836\label{fig:jul4}}{\includegraphics[width = 2.0in]{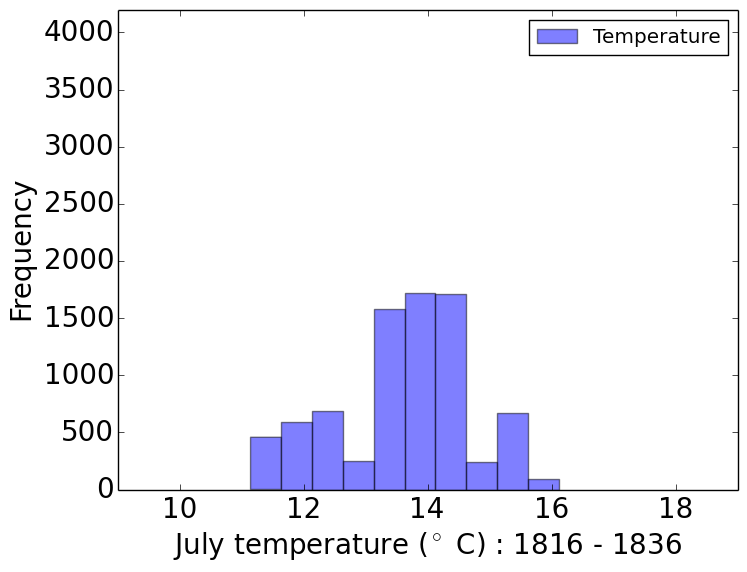}} &
\subcaptionbox{July -- SMC: 1836 - 1856\label{fig:jul5}}{\includegraphics[width = 2.0in]{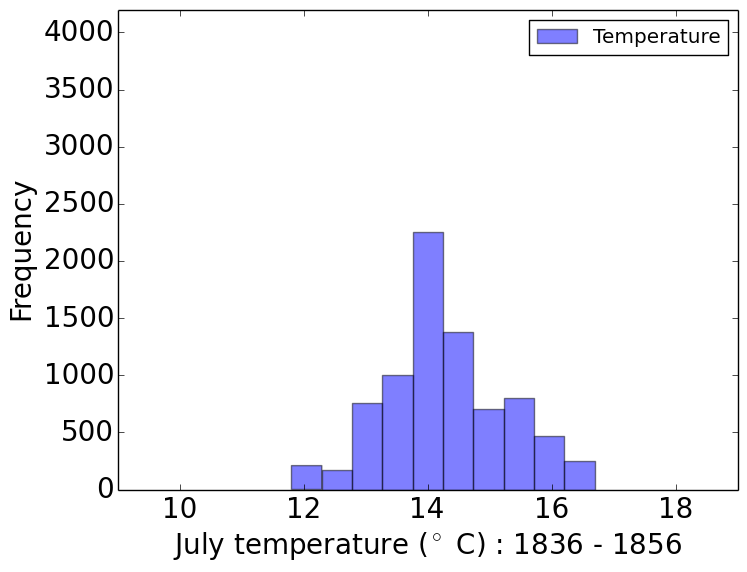}} &
\subcaptionbox{July -- SMC: 1856 - 1876\label{fig:jul6}}{\includegraphics[width = 2.0in]{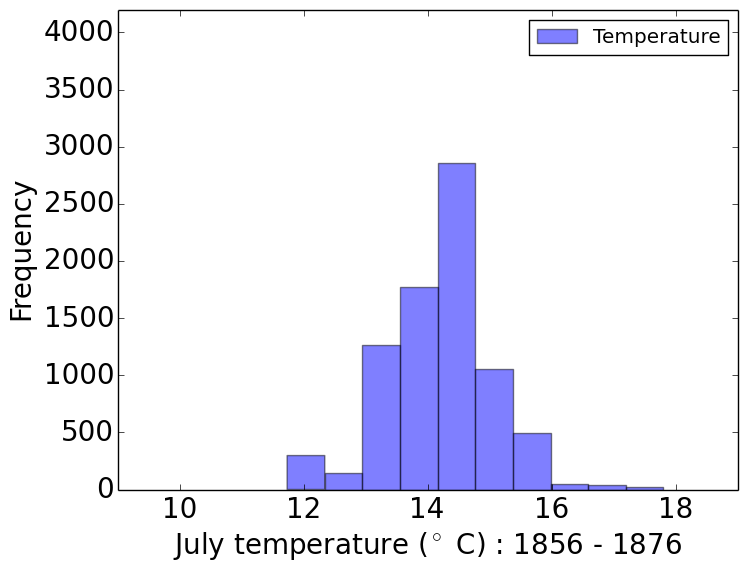}}\\
\subcaptionbox{July -- SMC: 1876 - 1896\label{fig:jul7}}{\includegraphics[width = 2.0in]{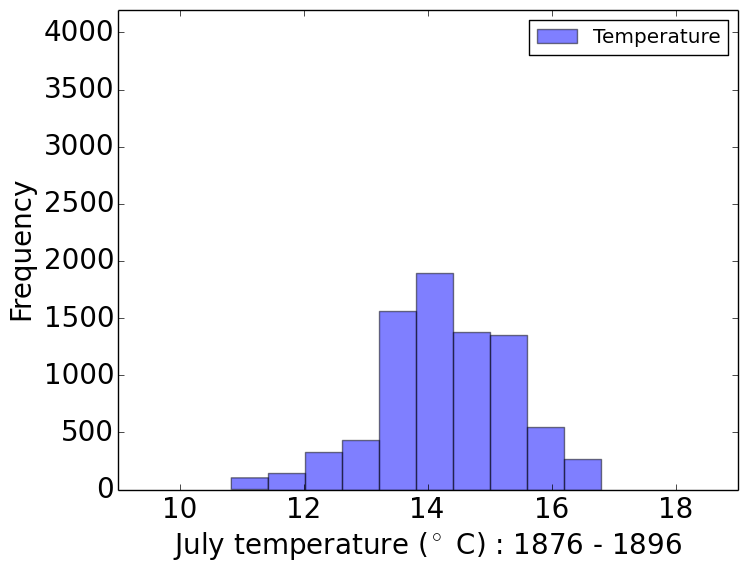}} &
\subcaptionbox{July -- SMC: 1896 - 1916\label{fig:jul8}}{\includegraphics[width = 2.0in]{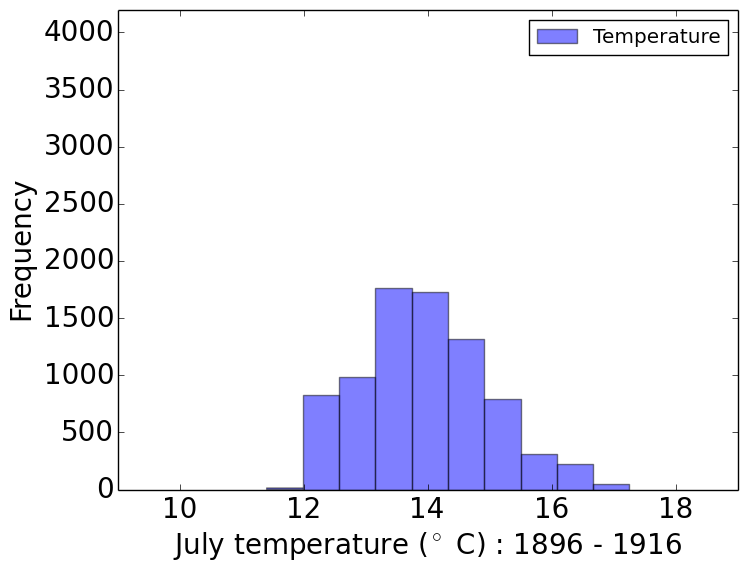}} &
\subcaptionbox{July -- SMC: 1916 - 1936\label{fig:jul9}}{\includegraphics[width = 2.0in]{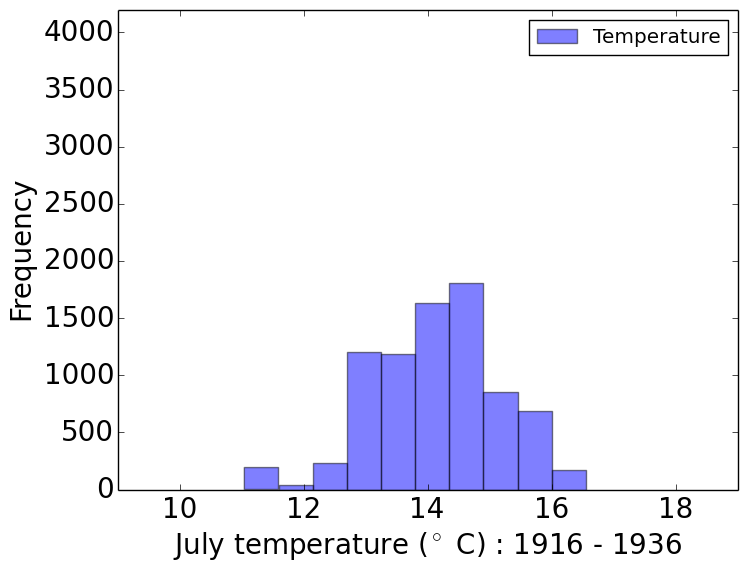}}\\
\subcaptionbox{July -- SMC: 1936 - 1936\label{fig:jul10}}{\includegraphics[width = 2.0in]{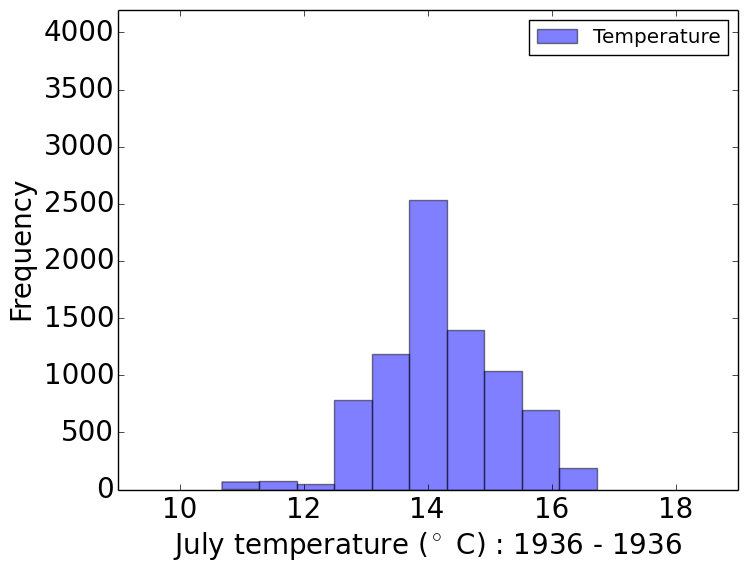}} &
\subcaptionbox{July -- SMC: 1956 - 1976\label{fig:jul11}}{\includegraphics[width = 2.0in]{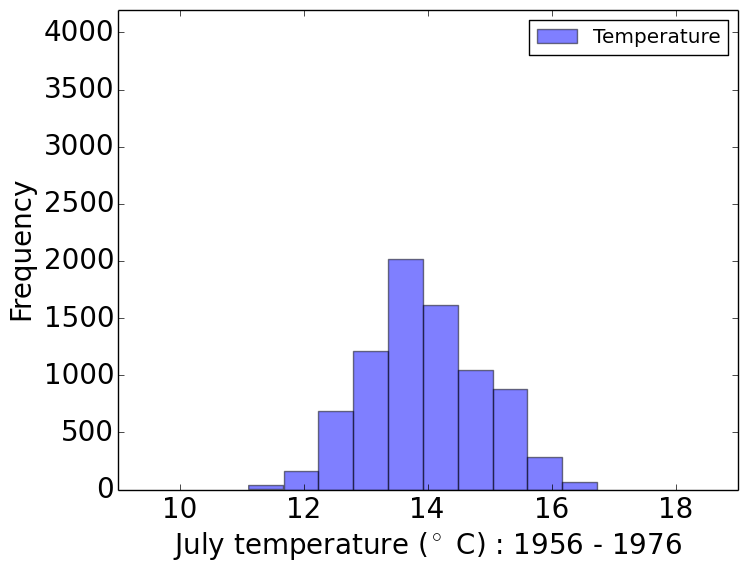}} &
\subcaptionbox{July -- SMC: 1976 - 1996\label{fig:jul12}}{\includegraphics[width = 2.0in]{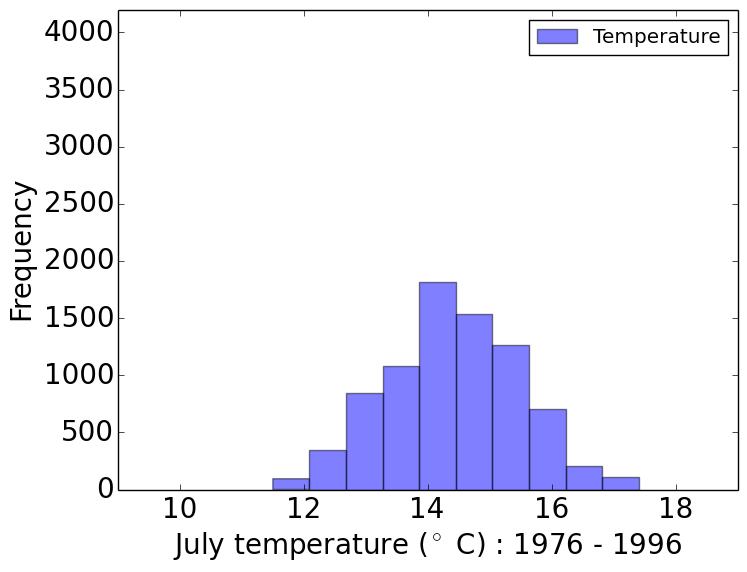}}\\
\subcaptionbox{July -- SMC: 1996 - 2016\label{fig:jul13}}{\includegraphics[width = 2.0in]{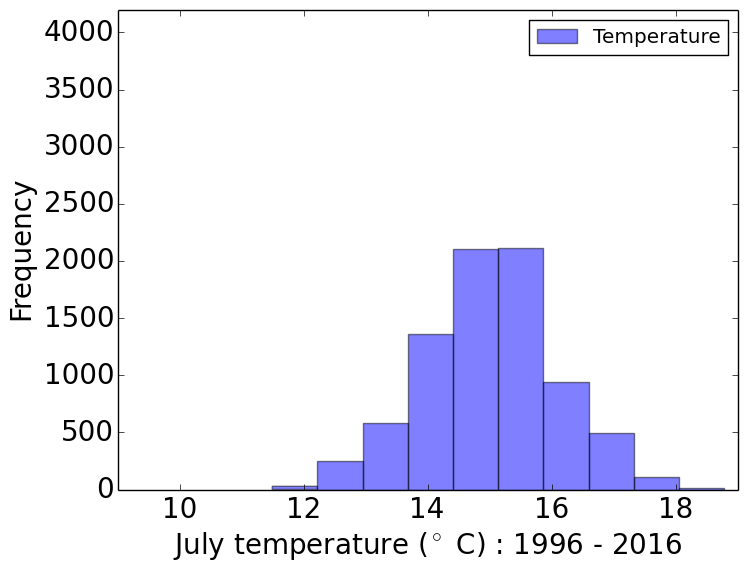}} &
\subcaptionbox{July -- SMC: Total\label{fig:jul14}}{\includegraphics[width = 2.0in]{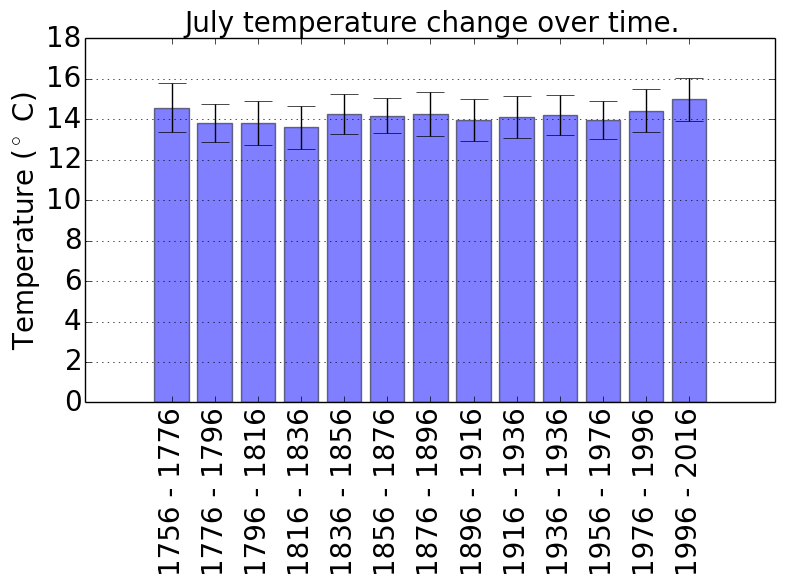}} &
\end{tabular}
\label{fig:jul1}
\end{adjustwidth}
\end{figure}

\begin{figure}
\begin{adjustwidth}{-6em}{0em}
\centering
\begin{tabular}{ccc}
\subcaptionbox{August -- SMC: 1756 - 1776\label{fig:aug1}}{\includegraphics[width = 2.0in]{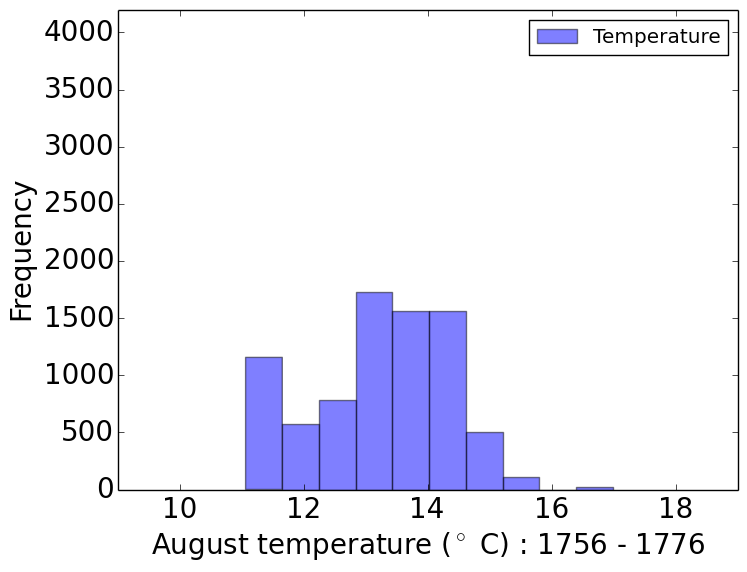}} &
\subcaptionbox{August -- SMC: 1776 - 1796\label{fig:aug2}}{\includegraphics[width = 2.0in]{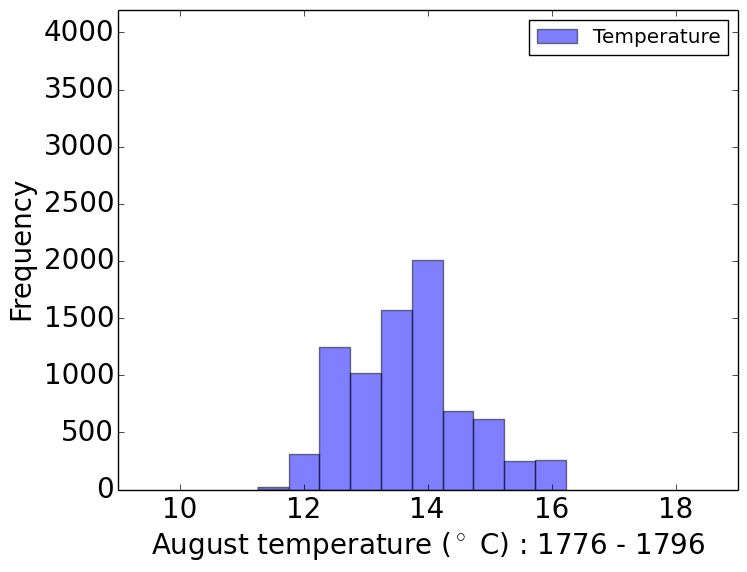}} &
\subcaptionbox{August -- SMC: 1796 - 1816\label{fig:aug3}}{\includegraphics[width = 2.0in]{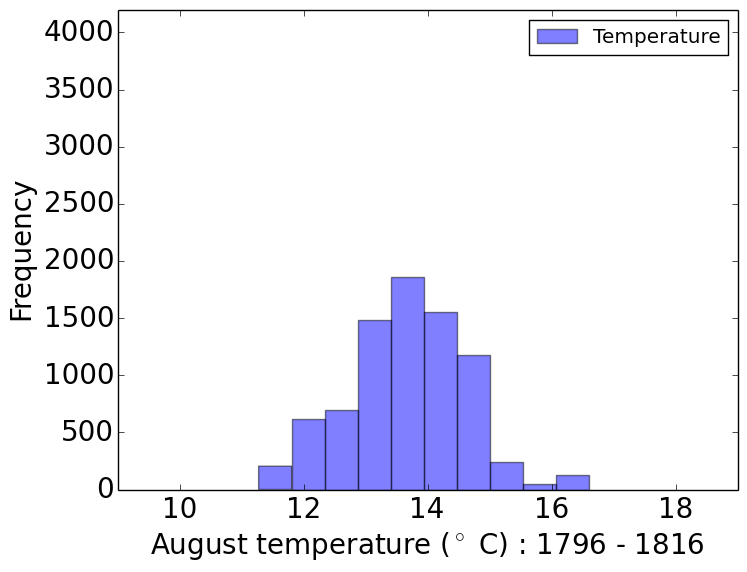}}\\
\subcaptionbox{August -- SMC: 1816 - 1836\label{fig:aug4}}{\includegraphics[width = 2.0in]{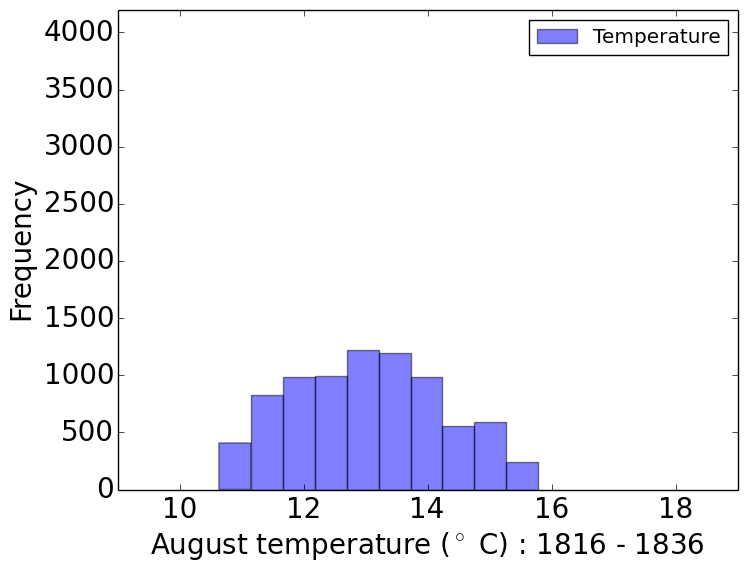}} &
\subcaptionbox{August -- SMC: 1836 - 1856\label{fig:aug5}}{\includegraphics[width = 2.0in]{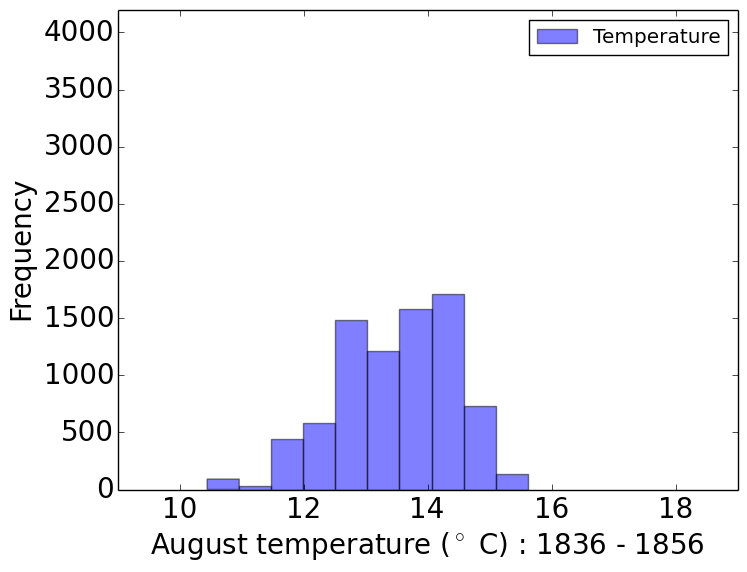}} &
\subcaptionbox{August -- SMC: 1856 - 1876\label{fig:aug6}}{\includegraphics[width = 2.0in]{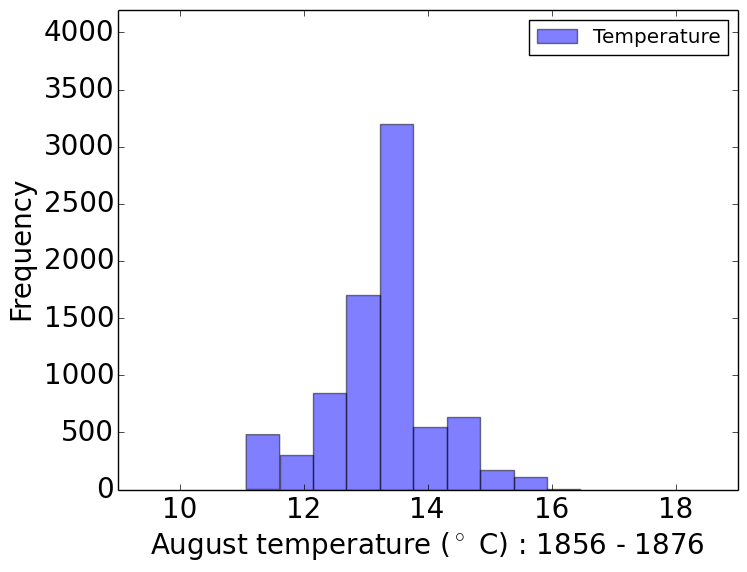}}\\
\subcaptionbox{August -- SMC: 1876 - 1896\label{fig:aug7}}{\includegraphics[width = 2.0in]{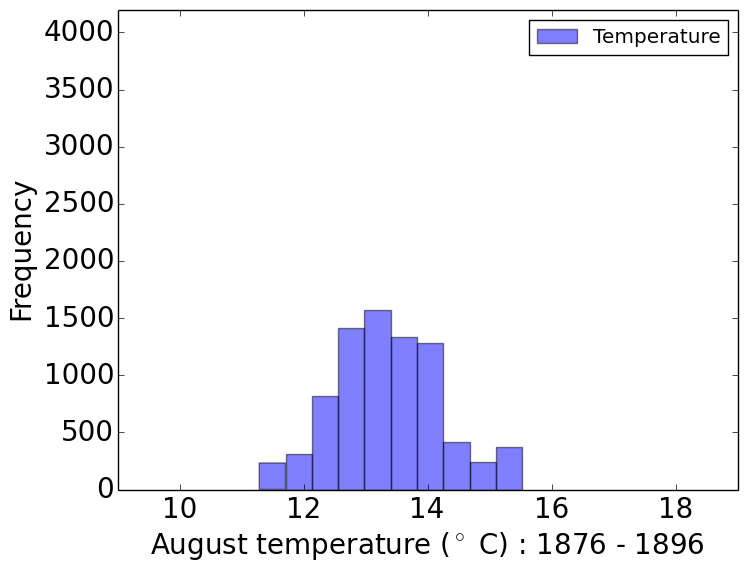}} &
\subcaptionbox{August -- SMC: 1896 - 1916\label{fig:aug8}}{\includegraphics[width = 2.0in]{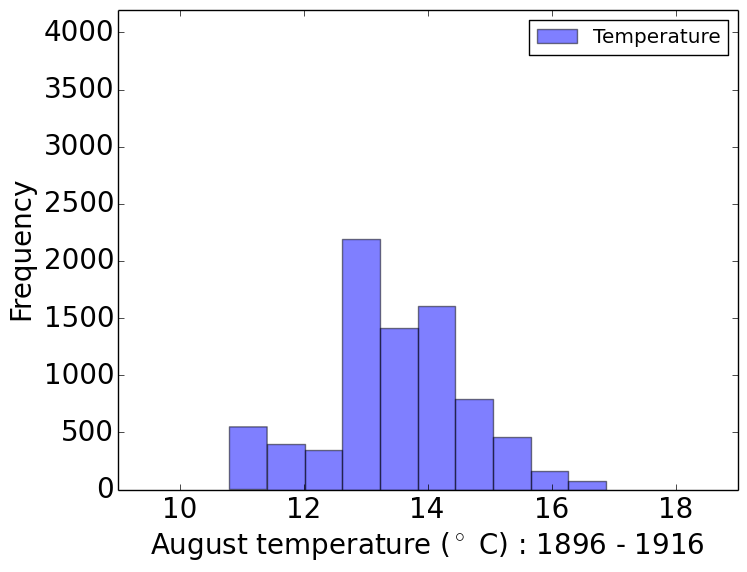}} &
\subcaptionbox{August -- SMC: 1916 - 1936\label{fig:aug9}}{\includegraphics[width = 2.0in]{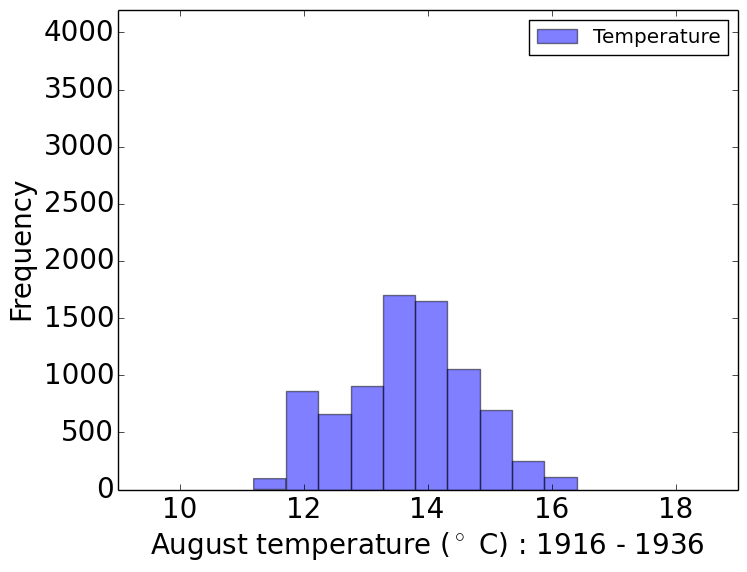}}\\
\subcaptionbox{August -- SMC: 1936 - 1936\label{fig:aug10}}{\includegraphics[width = 2.0in]{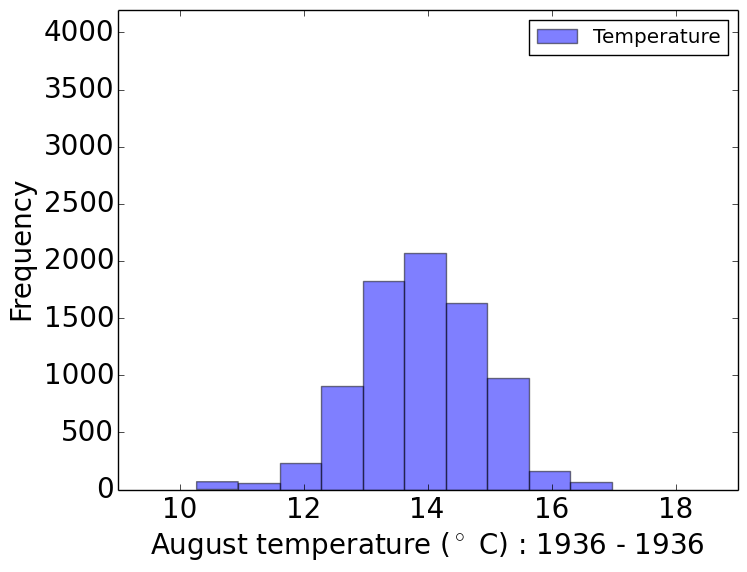}} &
\subcaptionbox{August -- SMC: 1956 - 1976\label{fig:aug11}}{\includegraphics[width = 2.0in]{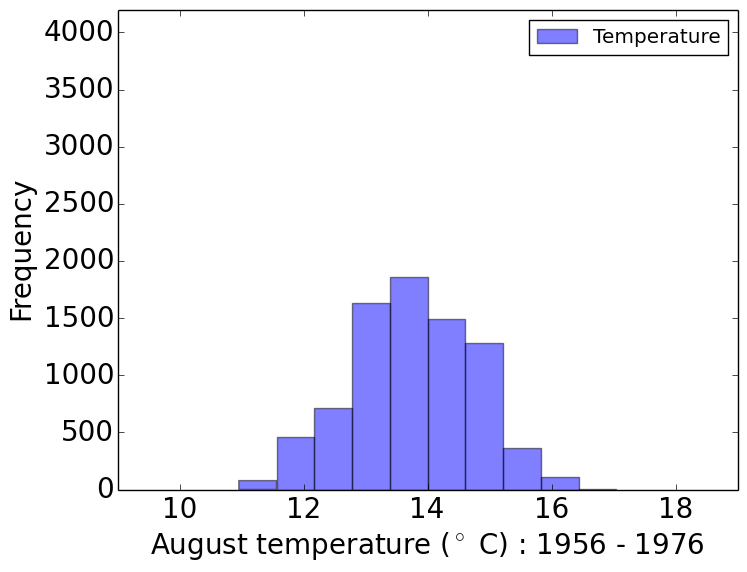}} &
\subcaptionbox{August -- SMC: 1976 - 1996\label{fig:aug12}}{\includegraphics[width = 2.0in]{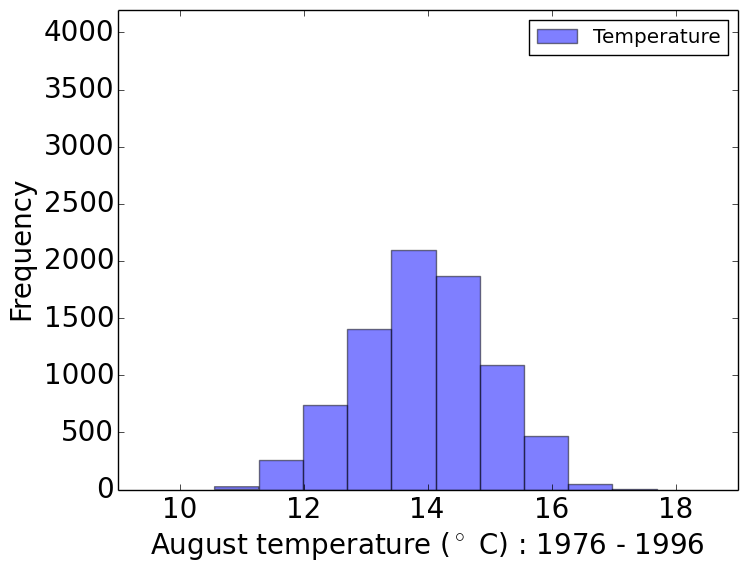}}\\
\subcaptionbox{August -- SMC: 1996 - 2016\label{fig:aug13}}{\includegraphics[width = 2.0in]{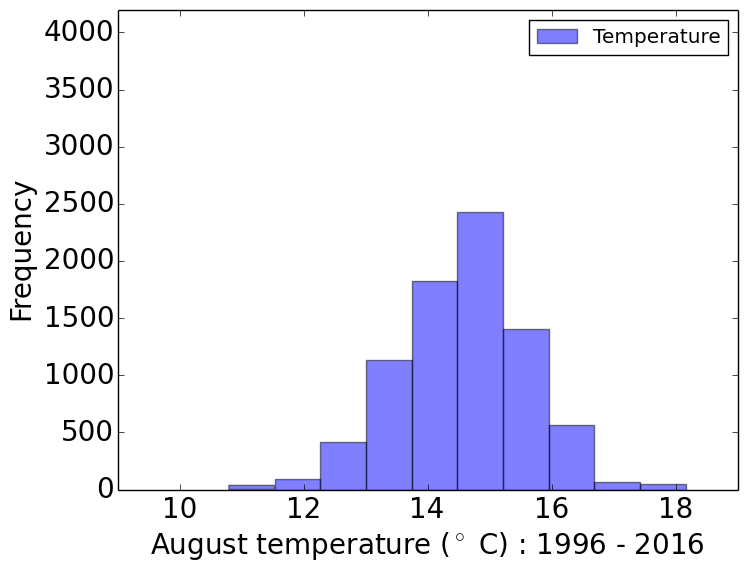}} &
\subcaptionbox{August -- SMC: Total\label{fig:aug14}}{\includegraphics[width = 2.0in]{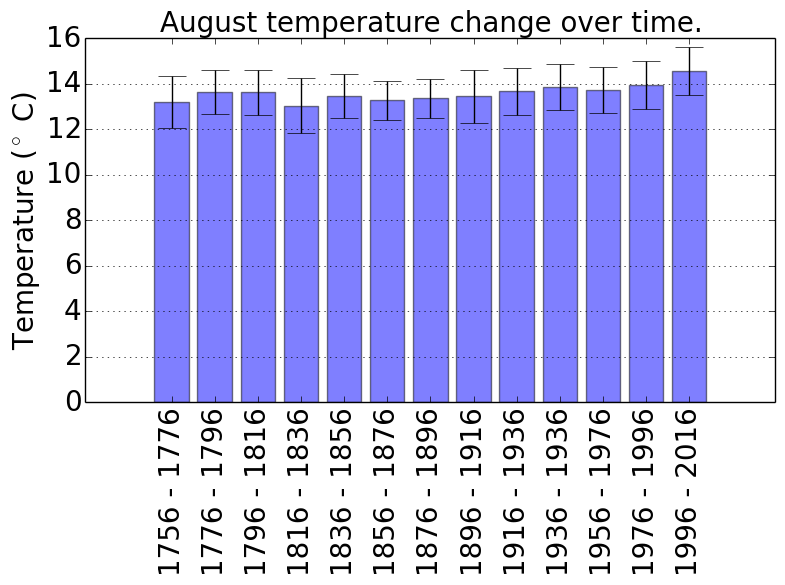}} &
\end{tabular}
\label{fig:aug1}
\end{adjustwidth}
\end{figure}

\begin{figure}
\begin{adjustwidth}{-6em}{0em}
\centering
\begin{tabular}{ccc}
\subcaptionbox{September -- SMC: 1756 - 1776\label{fig:sep1}}{\includegraphics[width = 2.0in]{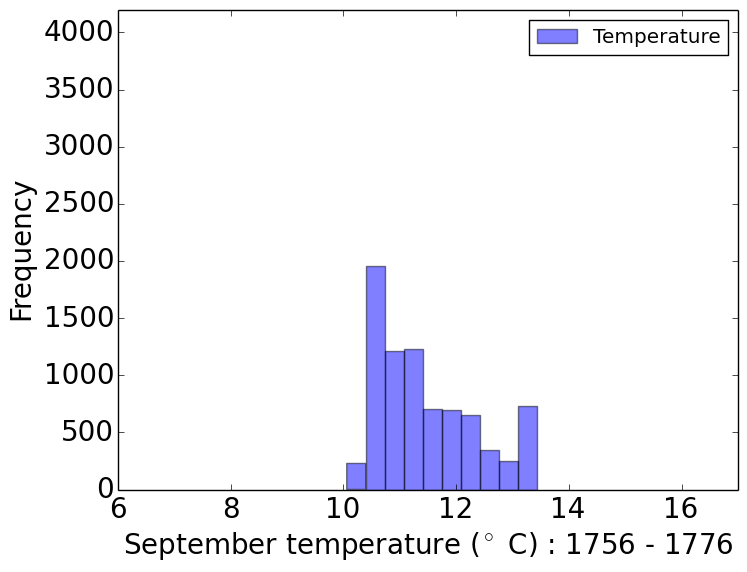}} &
\subcaptionbox{September -- SMC: 1776 - 1796\label{fig:sep2}}{\includegraphics[width = 2.0in]{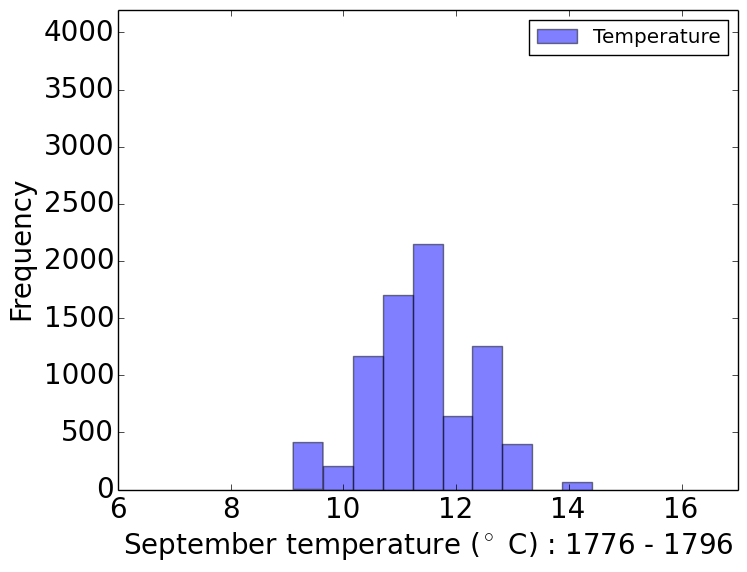}} &
\subcaptionbox{September -- SMC: 1796 - 1816\label{fig:sep3}}{\includegraphics[width = 2.0in]{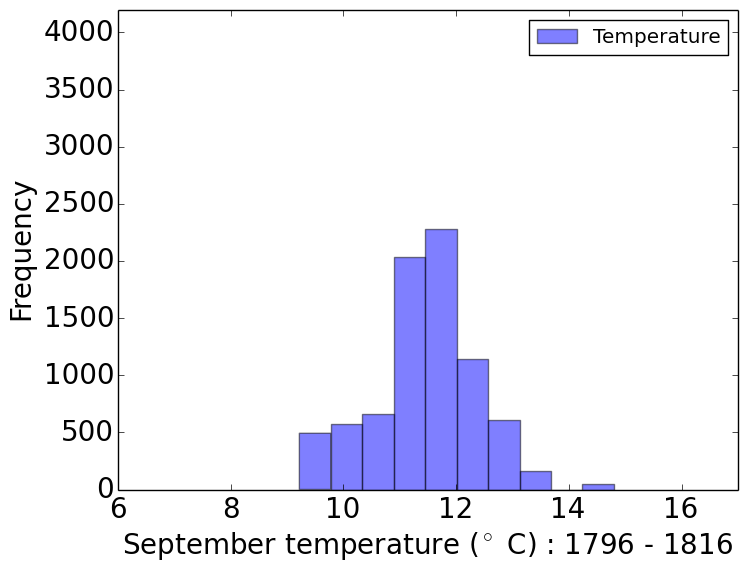}}\\
\subcaptionbox{September -- SMC: 1816 - 1836\label{fig:sep4}}{\includegraphics[width = 2.0in]{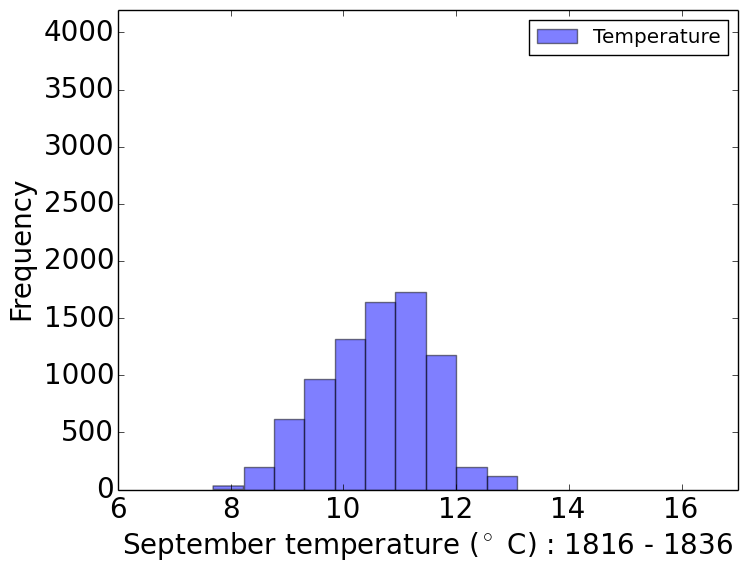}} &
\subcaptionbox{September -- SMC: 1836 - 1856\label{fig:sep5}}{\includegraphics[width = 2.0in]{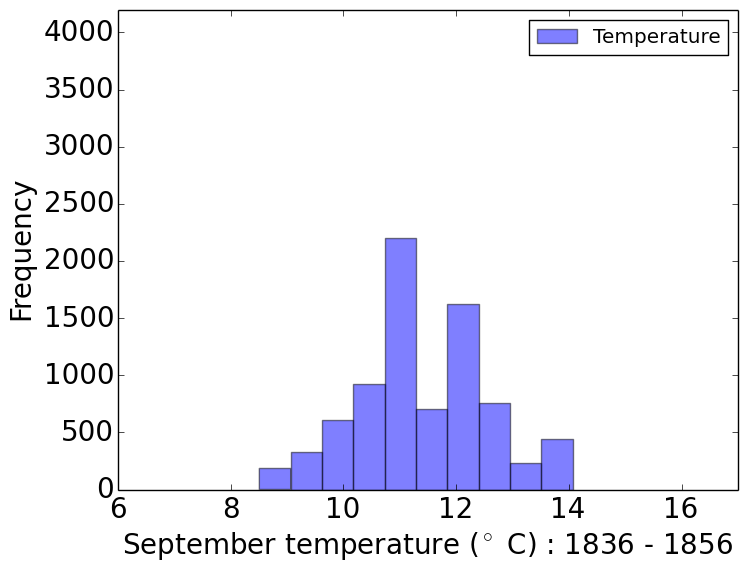}} &
\subcaptionbox{September -- SMC: 1856 - 1876\label{fig:sep6}}{\includegraphics[width = 2.0in]{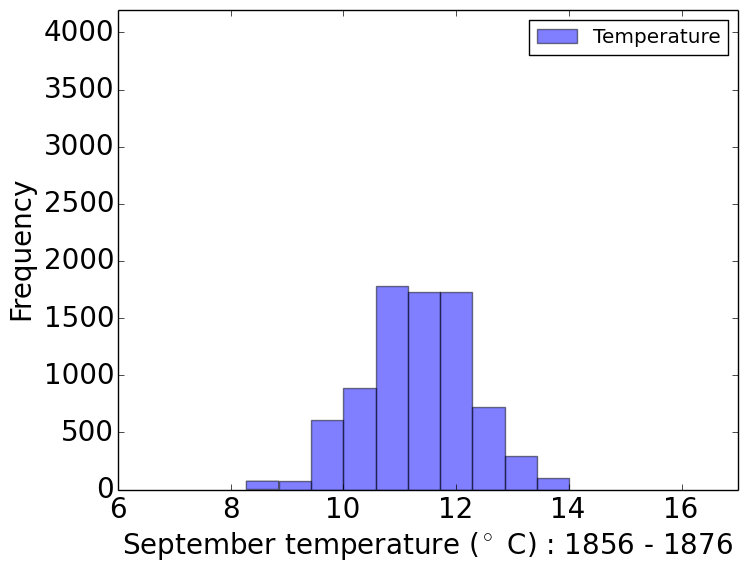}}\\
\subcaptionbox{September -- SMC: 1876 - 1896\label{fig:sep7}}{\includegraphics[width = 2.0in]{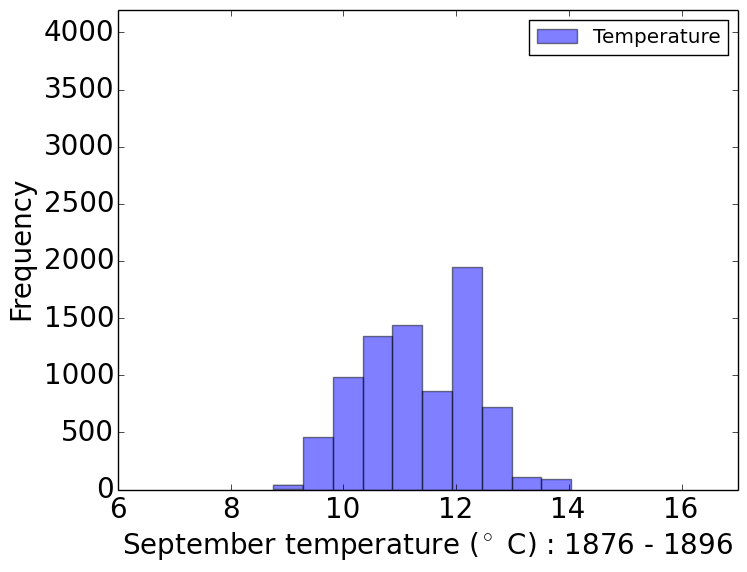}} &
\subcaptionbox{September -- SMC: 1896 - 1916\label{fig:sep8}}{\includegraphics[width = 2.0in]{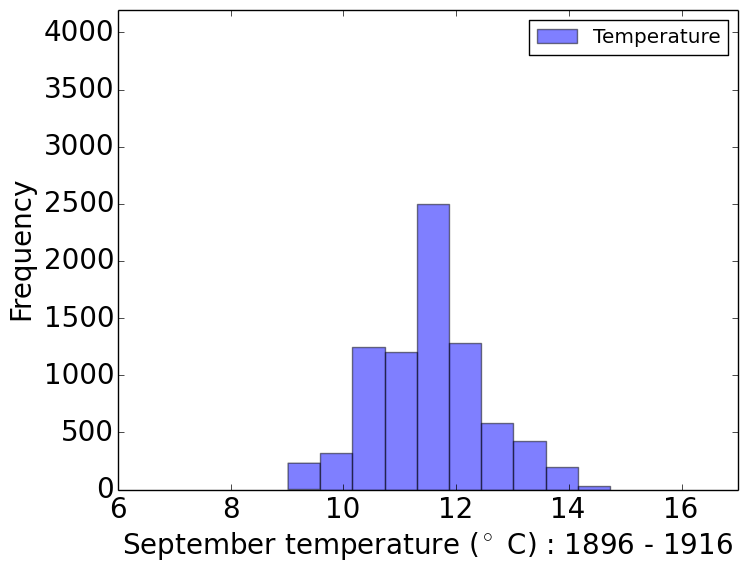}} &
\subcaptionbox{September -- SMC: 1916 - 1936\label{fig:sep9}}{\includegraphics[width = 2.0in]{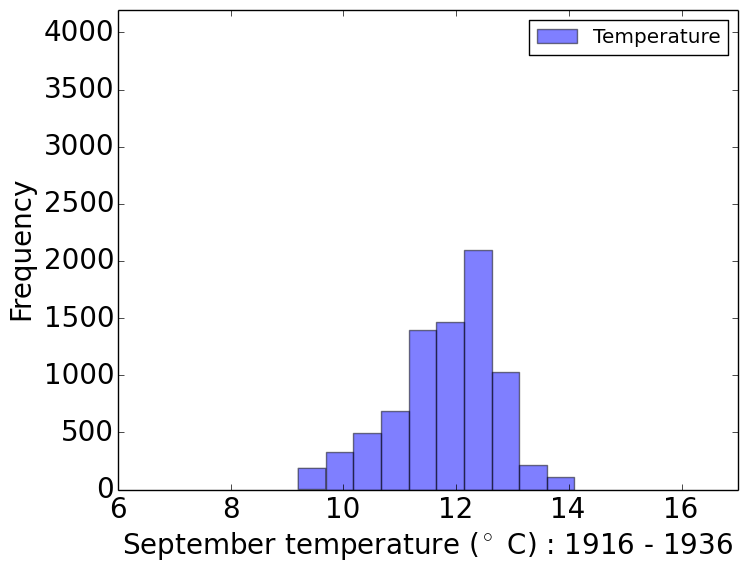}}\\
\subcaptionbox{September -- SMC: 1936 - 1936\label{fig:sep10}}{\includegraphics[width = 2.0in]{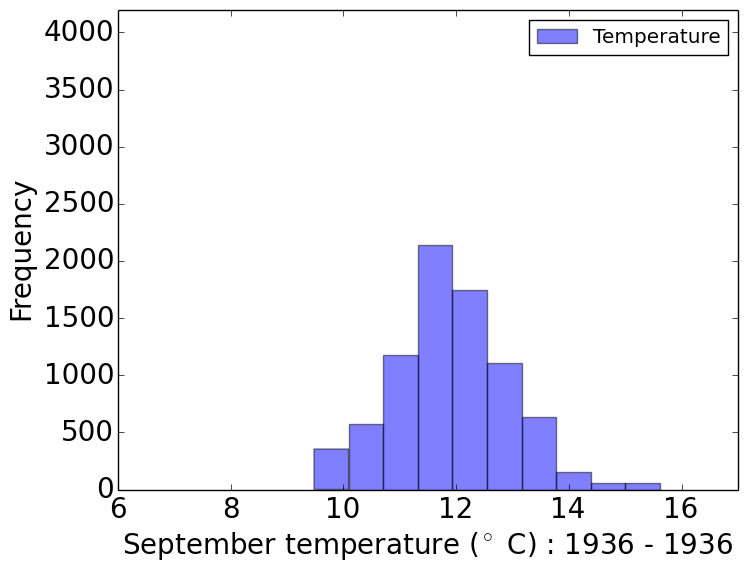}} &
\subcaptionbox{September -- SMC: 1956 - 1976\label{fig:sep11}}{\includegraphics[width = 2.0in]{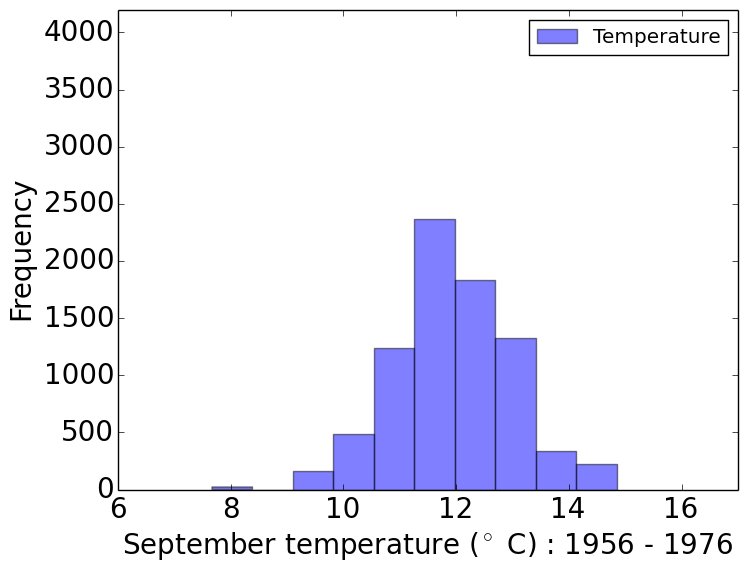}} &
\subcaptionbox{September -- SMC: 1976 - 1996\label{fig:sep12}}{\includegraphics[width = 2.0in]{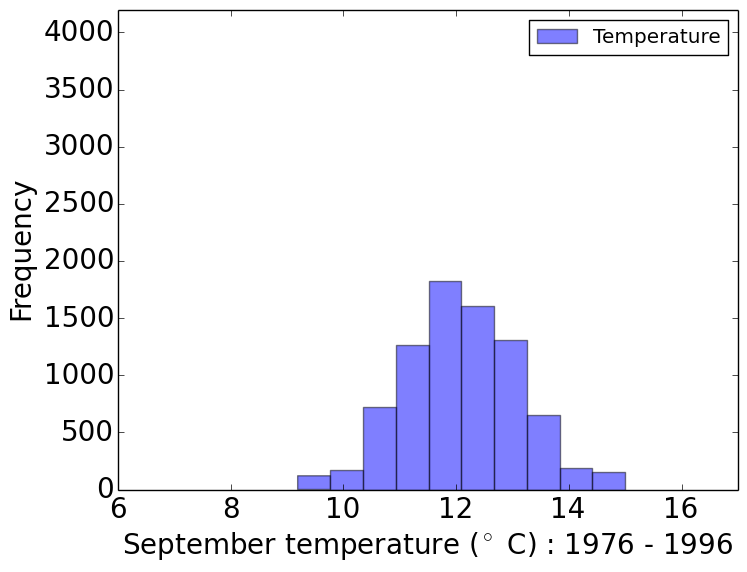}}\\
\subcaptionbox{September -- SMC: 1996 - 2016\label{fig:sep13}}{\includegraphics[width = 2.0in]{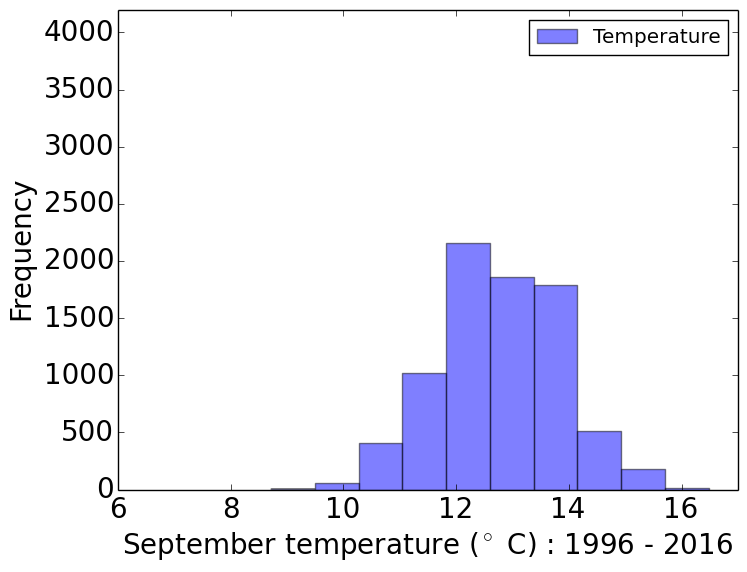}} &
\subcaptionbox{September -- SMC: Total\label{fig:sep14}}{\includegraphics[width = 2.0in]{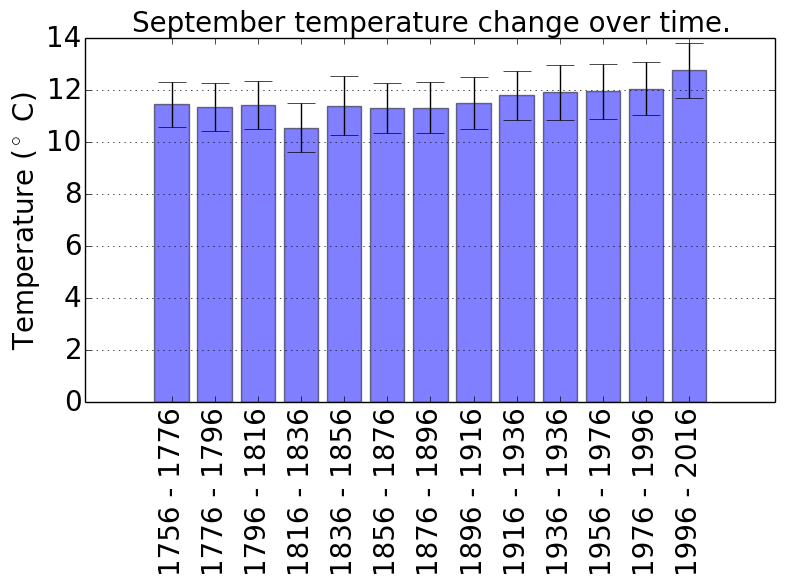}} &
\end{tabular}
\label{fig:sep1}
\end{adjustwidth}
\end{figure}

\begin{figure}
\begin{adjustwidth}{-6em}{0em}
\centering
\begin{tabular}{ccc}
\subcaptionbox{October -- SMC: 1756 - 1776\label{fig:oct1}}{\includegraphics[width = 2.0in]{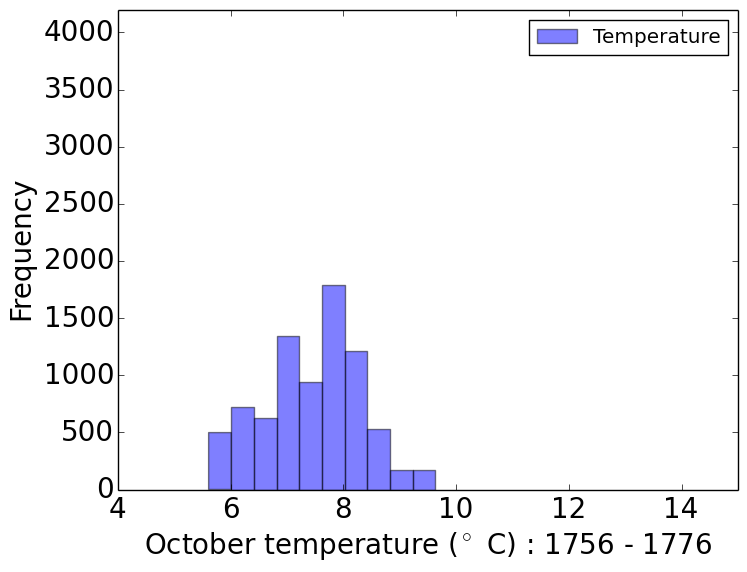}} &
\subcaptionbox{October -- SMC: 1776 - 1796\label{fig:oct2}}{\includegraphics[width = 2.0in]{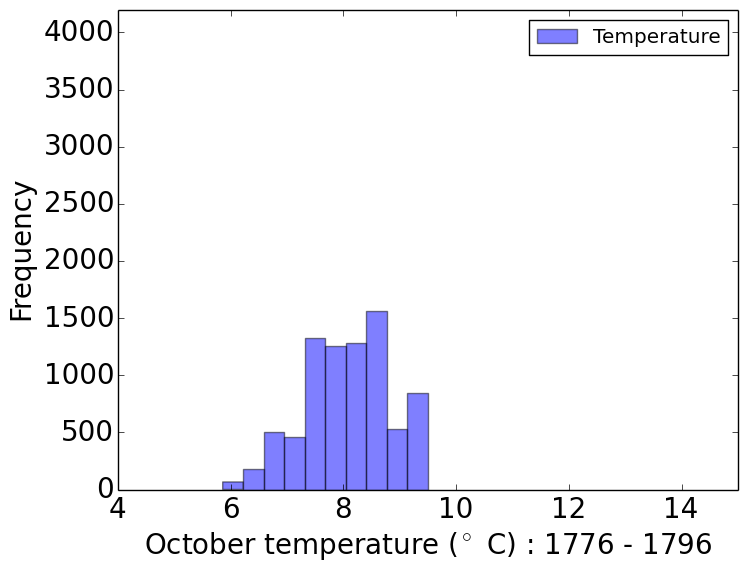}} &
\subcaptionbox{October -- SMC: 1796 - 1816\label{fig:oct3}}{\includegraphics[width = 2.0in]{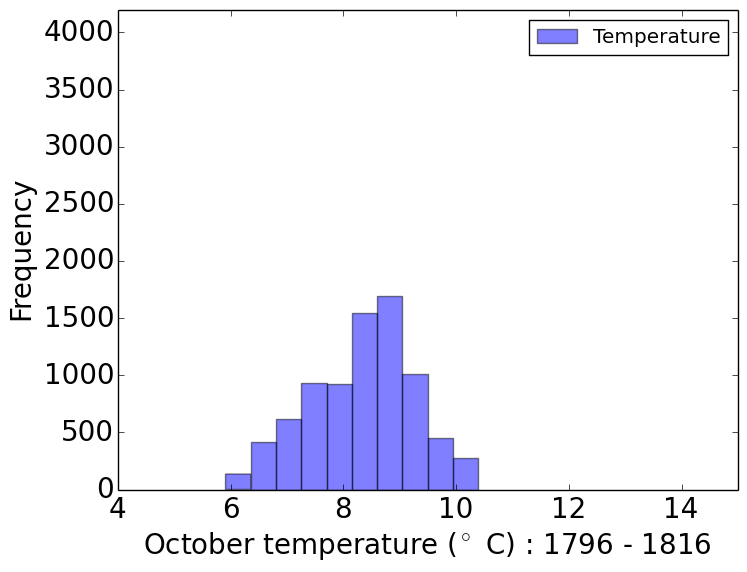}}\\
\subcaptionbox{October -- SMC: 1816 - 1836\label{fig:oct4}}{\includegraphics[width = 2.0in]{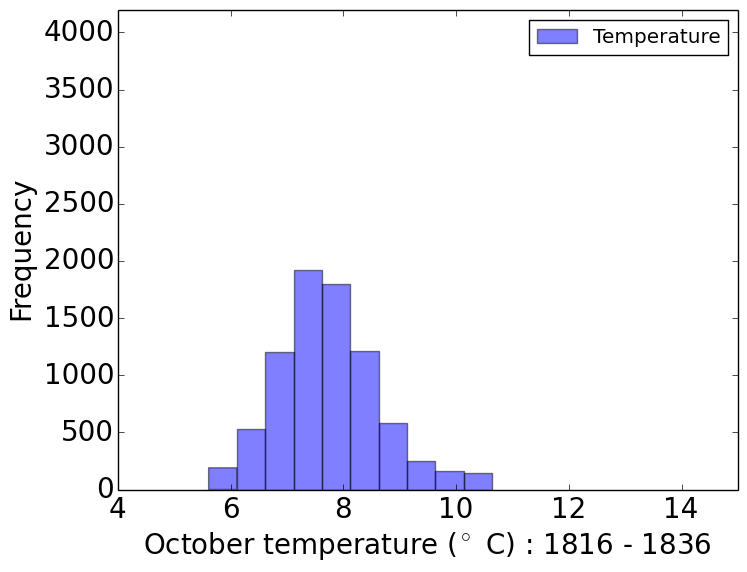}} &
\subcaptionbox{October -- SMC: 1836 - 1856\label{fig:oct5}}{\includegraphics[width = 2.0in]{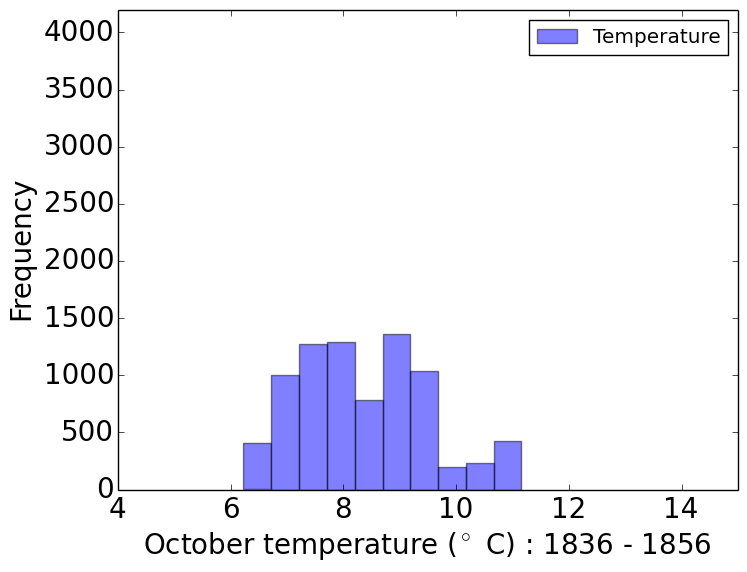}} &
\subcaptionbox{October -- SMC: 1856 - 1876\label{fig:oct6}}{\includegraphics[width = 2.0in]{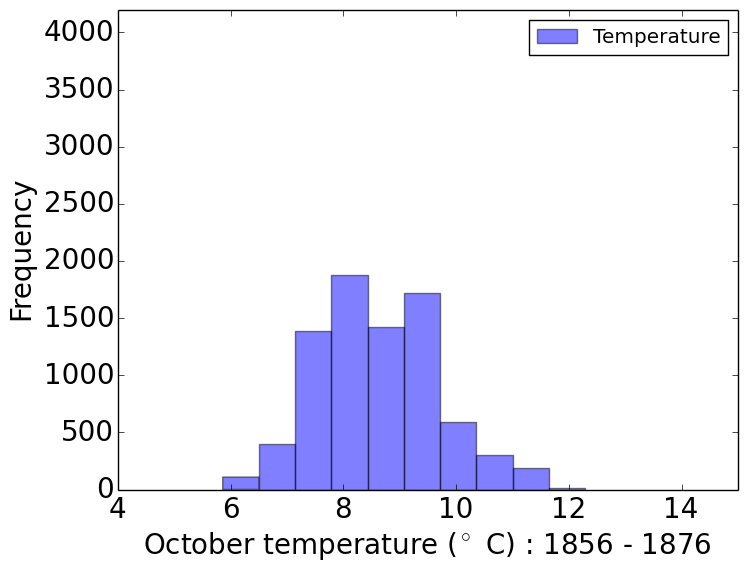}}\\
\subcaptionbox{October -- SMC: 1876 - 1896\label{fig:oct7}}{\includegraphics[width = 2.0in]{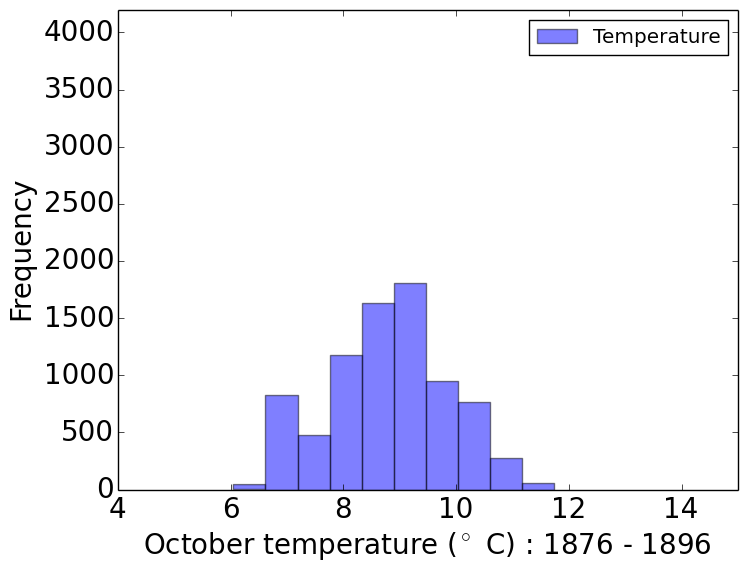}} &
\subcaptionbox{October -- SMC: 1896 - 1916\label{fig:oct8}}{\includegraphics[width = 2.0in]{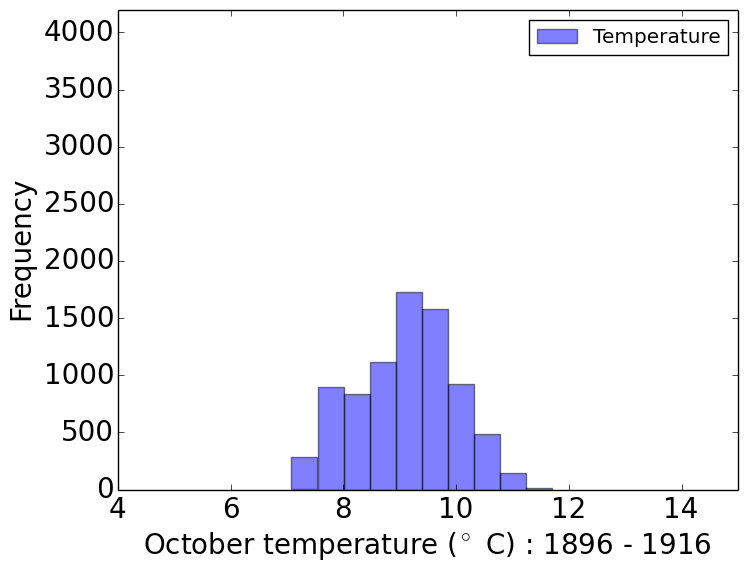}} &
\subcaptionbox{October -- SMC: 1916 - 1936\label{fig:oct9}}{\includegraphics[width = 2.0in]{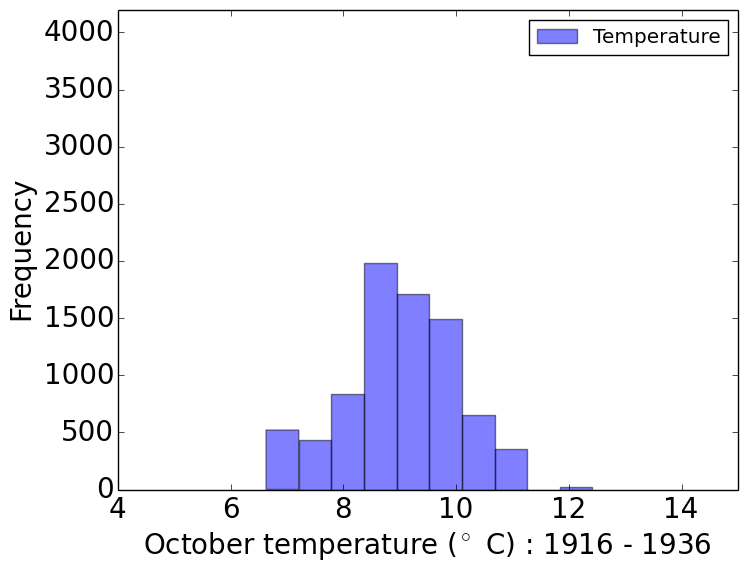}}\\
\subcaptionbox{October -- SMC: 1936 - 1936\label{fig:oct10}}{\includegraphics[width = 2.0in]{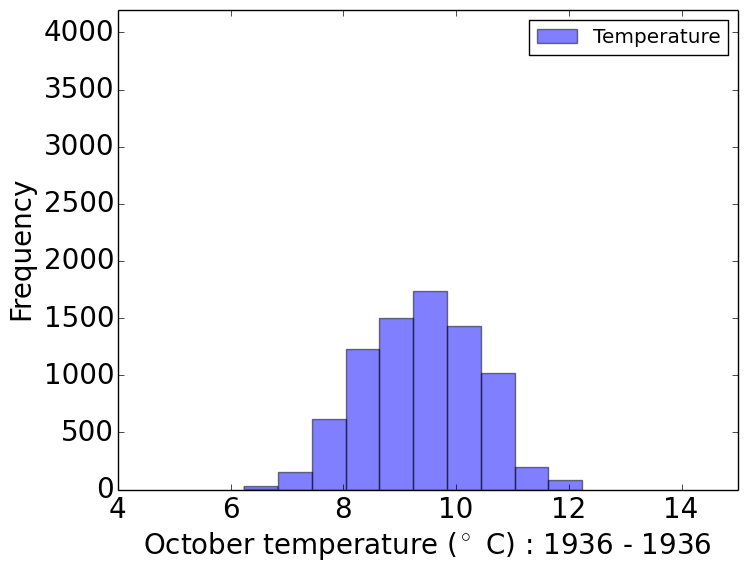}} &
\subcaptionbox{October -- SMC: 1956 - 1976\label{fig:oct11}}{\includegraphics[width = 2.0in]{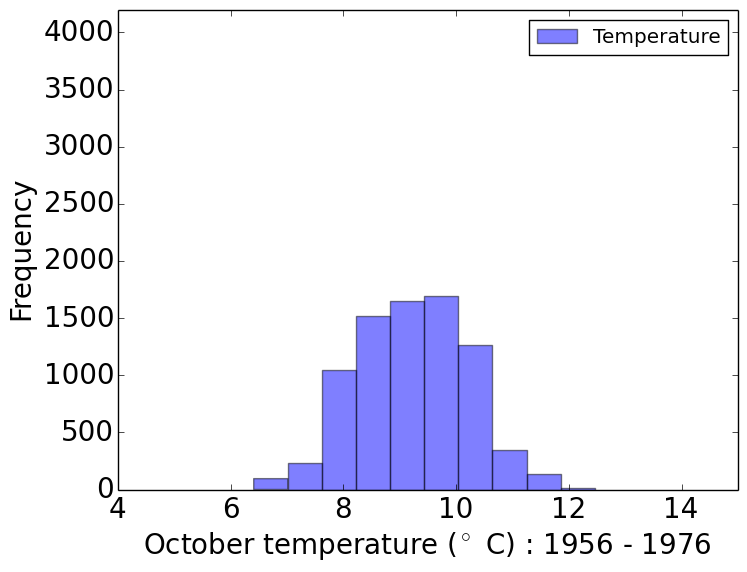}} &
\subcaptionbox{October -- SMC: 1976 - 1996\label{fig:oct12}}{\includegraphics[width = 2.0in]{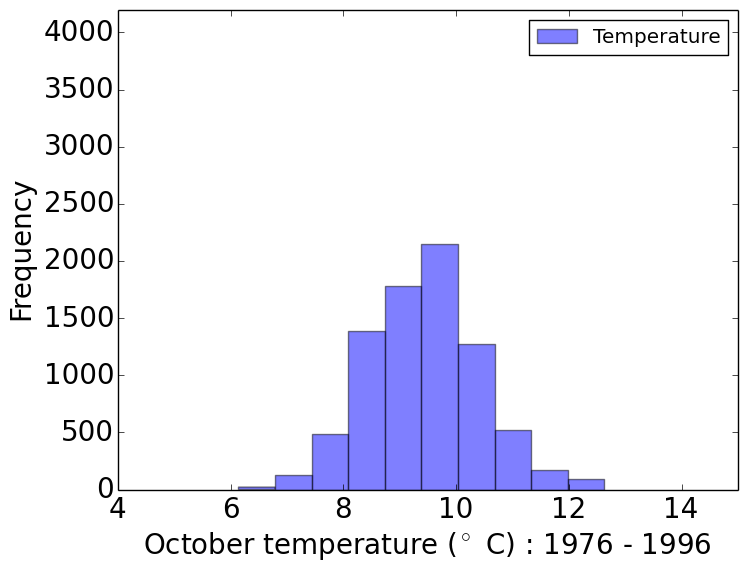}}\\
\subcaptionbox{October -- SMC: 1996 - 2016\label{fig:oct13}}{\includegraphics[width = 2.0in]{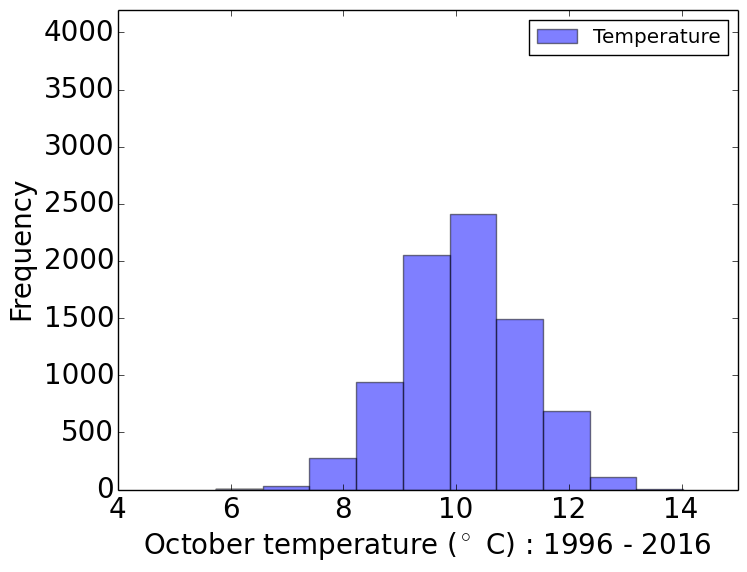}} &
\subcaptionbox{October -- SMC: Total\label{fig:oct14}}{\includegraphics[width = 2.0in]{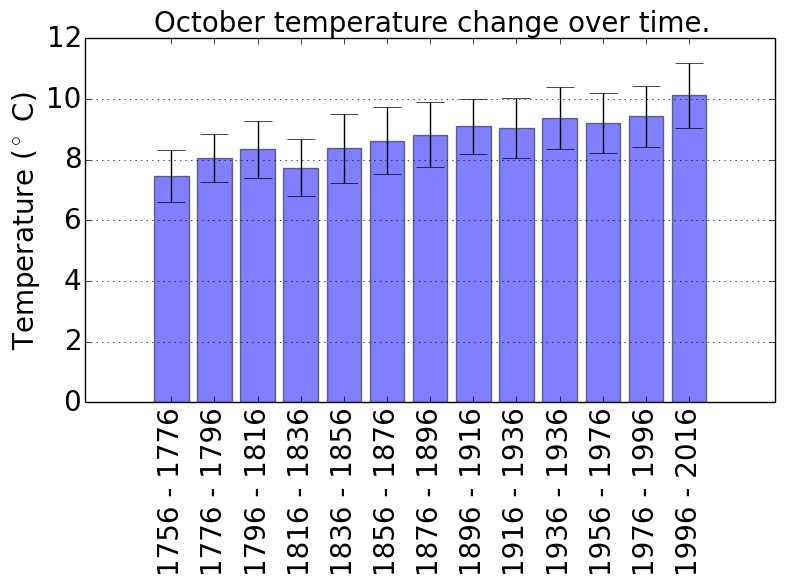}} &
\end{tabular}
\label{fig:oct1}
\end{adjustwidth}
\end{figure}

\begin{figure}
\begin{adjustwidth}{-6em}{0em}
\centering
\begin{tabular}{ccc}
\subcaptionbox{November -- SMC: 1756 - 1776\label{fig:nov1}}{\includegraphics[width = 2.0in]{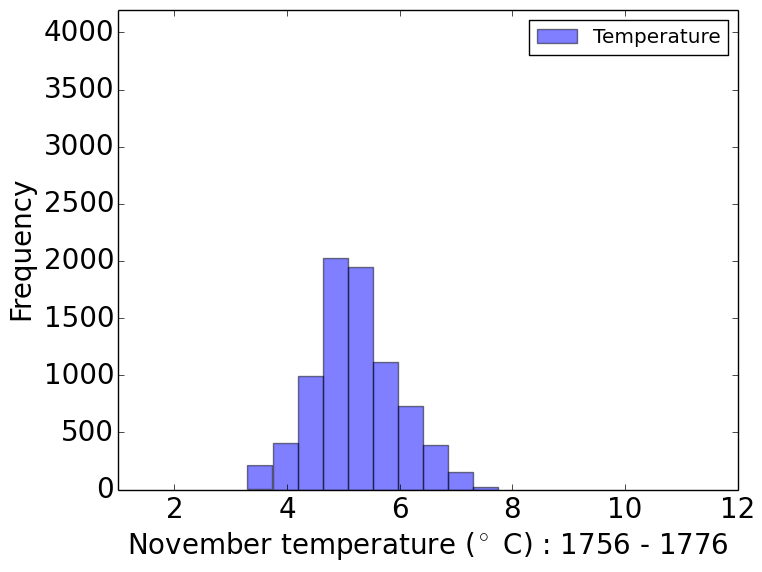}} &
\subcaptionbox{November -- SMC: 1776 - 1796\label{fig:nov2}}{\includegraphics[width = 2.0in]{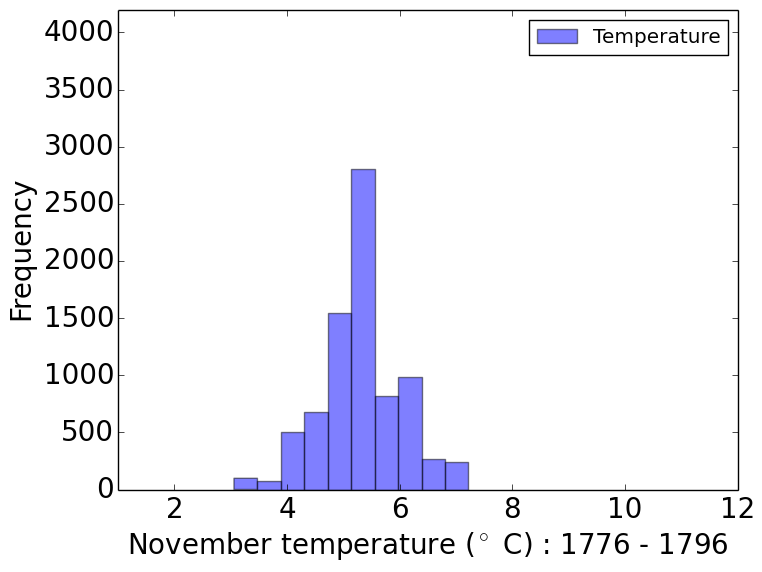}} &
\subcaptionbox{November -- SMC: 1796 - 1816\label{fig:nov3}}{\includegraphics[width = 2.0in]{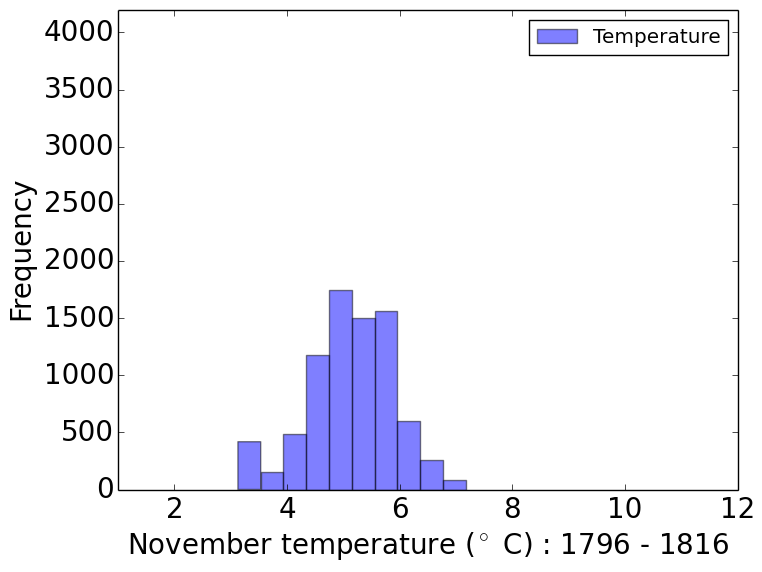}}\\
\subcaptionbox{November -- SMC: 1816 - 1836\label{fig:nov4}}{\includegraphics[width = 2.0in]{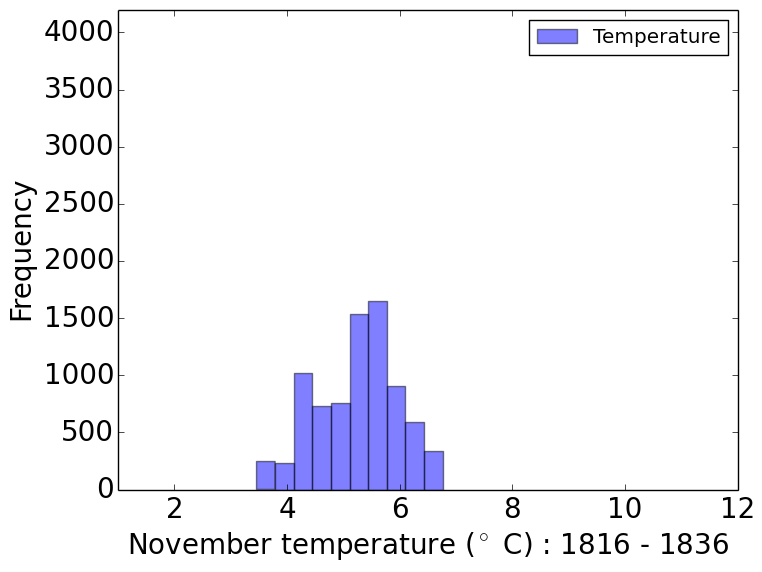}} &
\subcaptionbox{November -- SMC: 1836 - 1856\label{fig:nov5}}{\includegraphics[width = 2.0in]{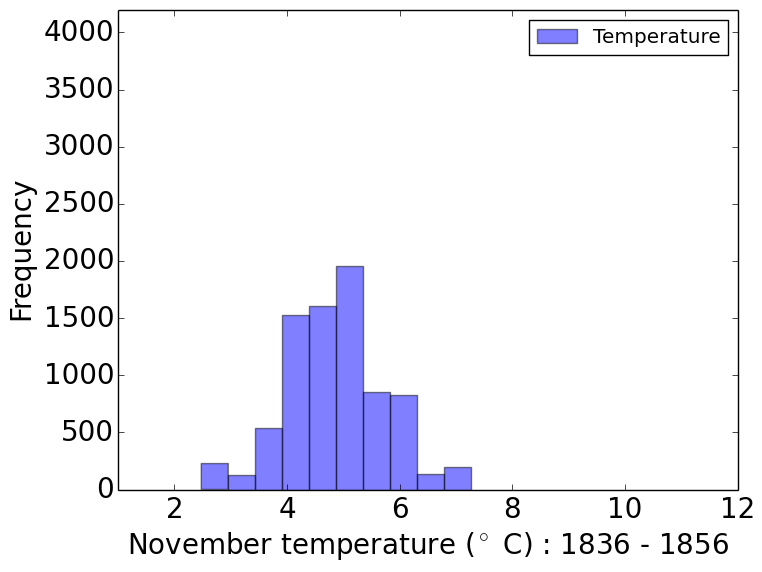}} &
\subcaptionbox{November -- SMC: 1856 - 1876\label{fig:nov6}}{\includegraphics[width = 2.0in]{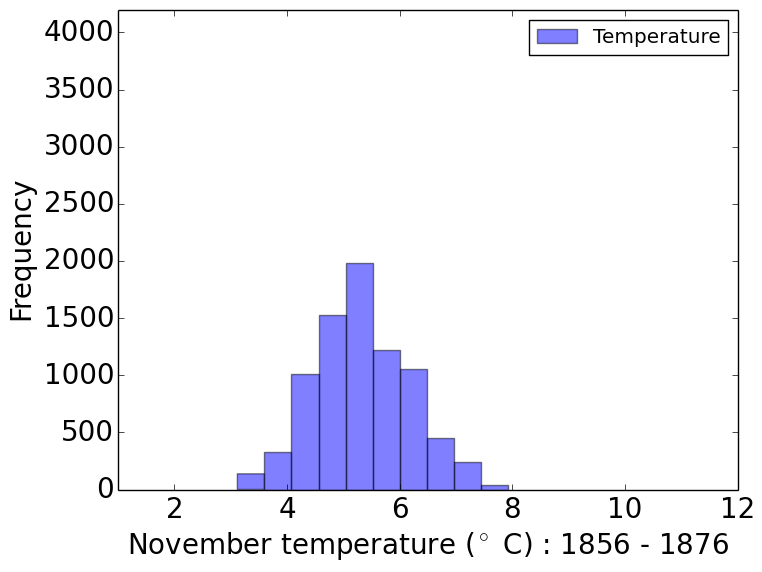}}\\
\subcaptionbox{November -- SMC: 1876 - 1896\label{fig:nov7}}{\includegraphics[width = 2.0in]{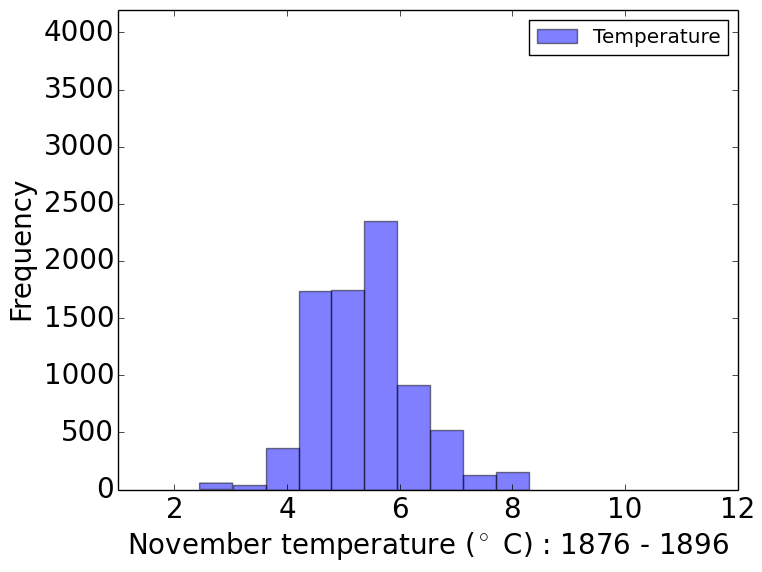}} &
\subcaptionbox{November -- SMC: 1896 - 1916\label{fig:nov8}}{\includegraphics[width = 2.0in]{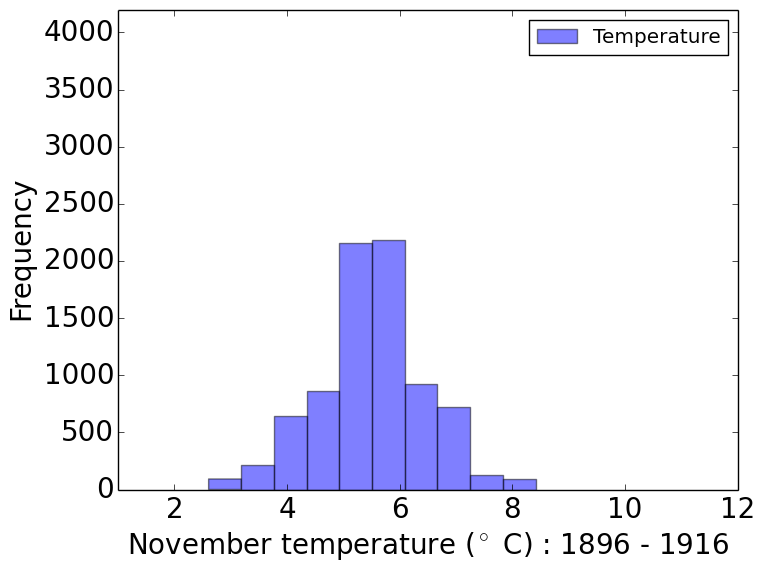}} &
\subcaptionbox{November -- SMC: 1916 - 1936\label{fig:nov9}}{\includegraphics[width = 2.0in]{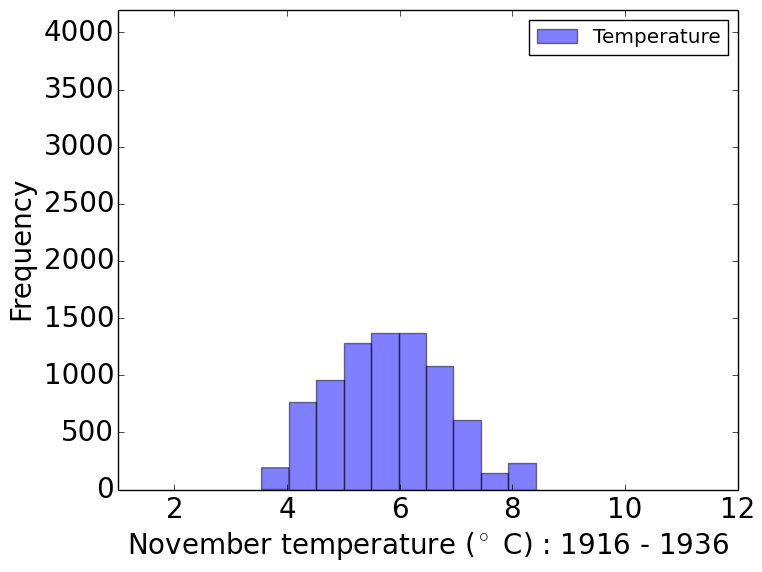}}\\
\subcaptionbox{November -- SMC: 1936 - 1936\label{fig:nov10}}{\includegraphics[width = 2.0in]{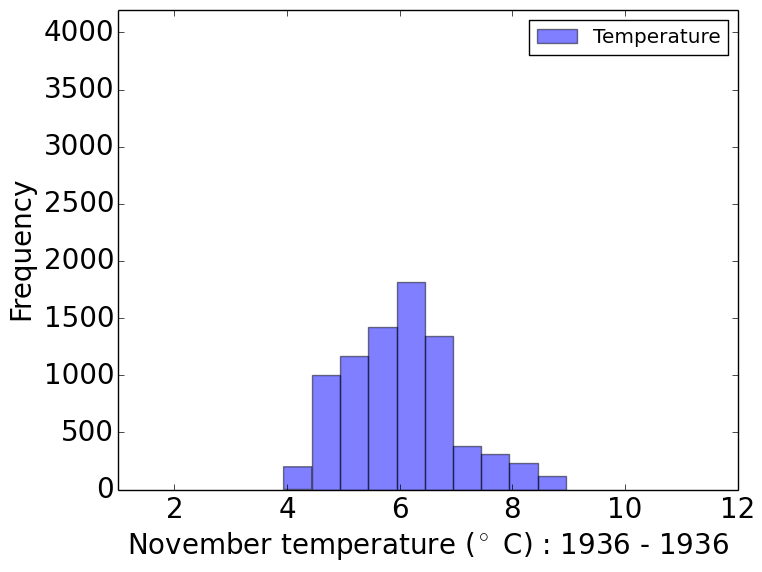}} &
\subcaptionbox{November -- SMC: 1956 - 1976\label{fig:nov11}}{\includegraphics[width = 2.0in]{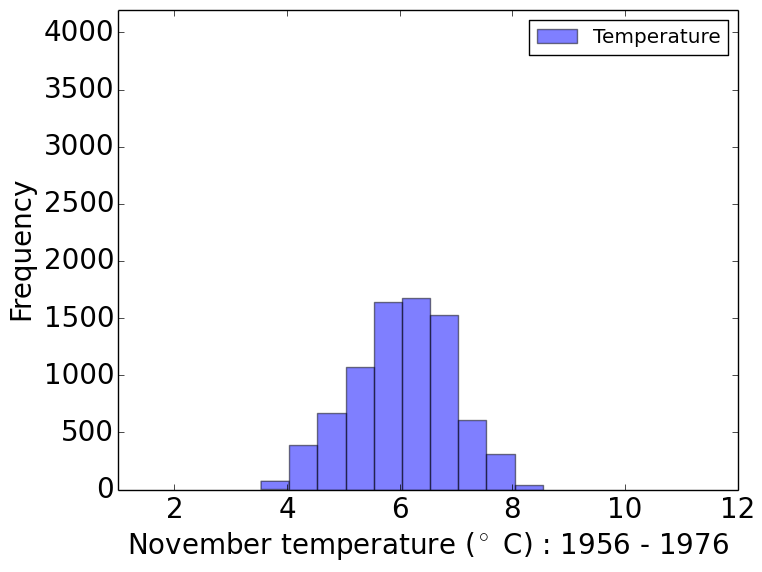}} &
\subcaptionbox{November -- SMC: 1976 - 1996\label{fig:nov12}}{\includegraphics[width = 2.0in]{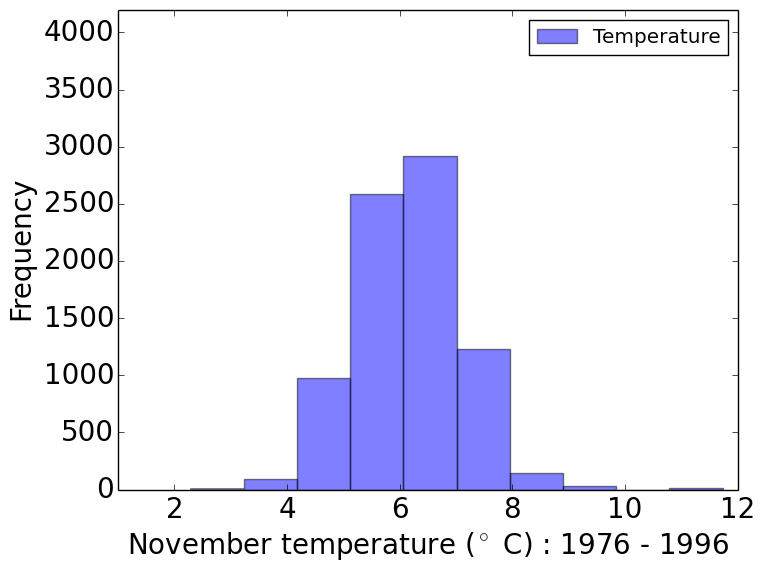}}\\
\subcaptionbox{November -- SMC: 1996 - 2016\label{fig:nov13}}{\includegraphics[width = 2.0in]{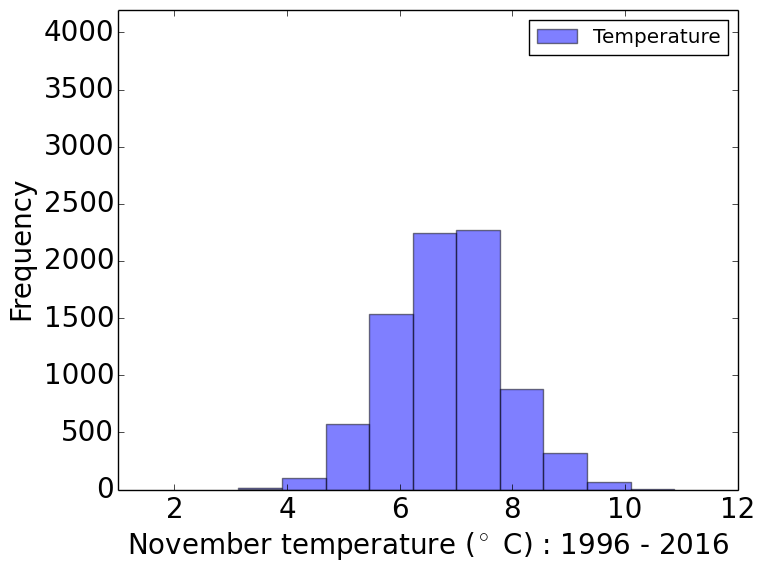}} &
\subcaptionbox{November -- SMC: Total\label{fig:nov14}}{\includegraphics[width = 2.0in]{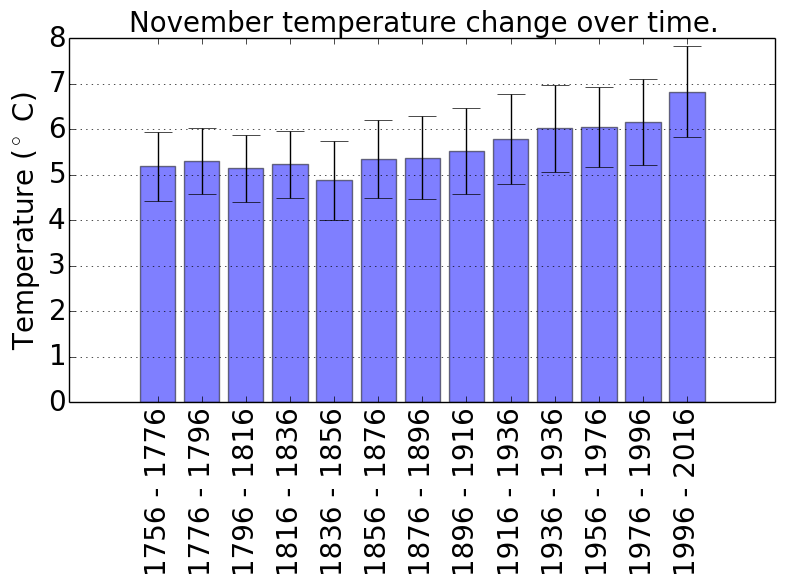}} &
\end{tabular}
\label{fig:nov1}
\end{adjustwidth}
\end{figure}

\begin{figure}
\begin{adjustwidth}{-6em}{0em}
\centering
\begin{tabular}{ccc}
\subcaptionbox{December -- SMC: 1756 - 1776\label{fig:dec1}}{\includegraphics[width = 2.0in]{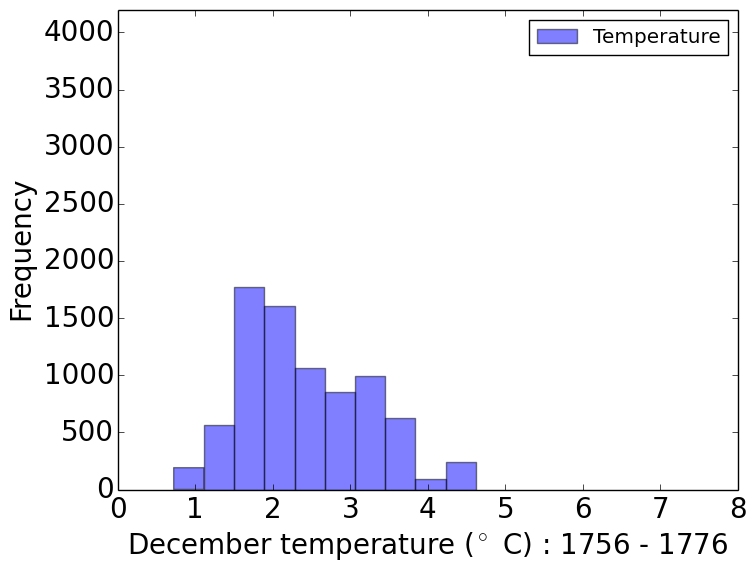}} &
\subcaptionbox{December -- SMC: 1776 - 1796\label{fig:dec2}}{\includegraphics[width = 2.0in]{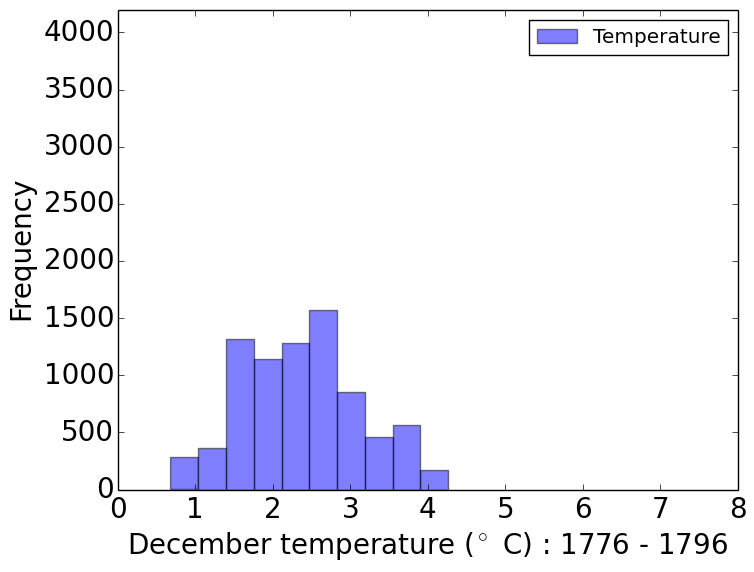}} &
\subcaptionbox{December -- SMC: 1796 - 1816\label{fig:dec3}}{\includegraphics[width = 2.0in]{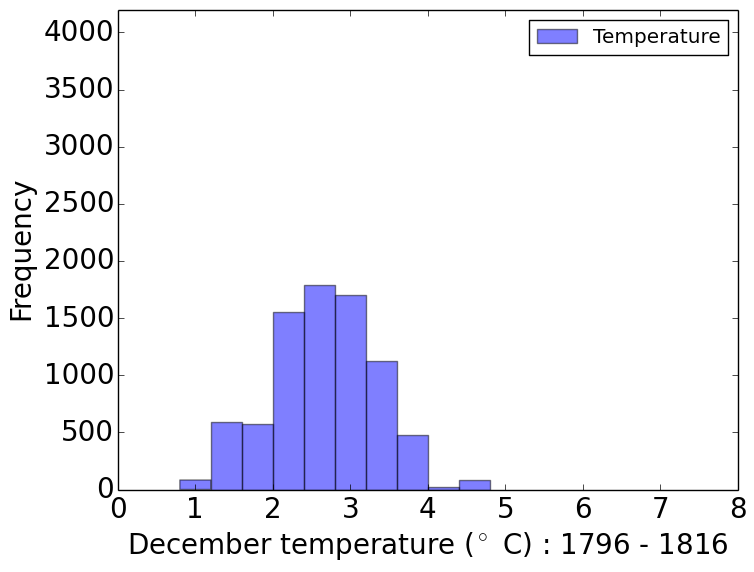}}\\
\subcaptionbox{December -- SMC: 1816 - 1836\label{fig:dec4}}{\includegraphics[width = 2.0in]{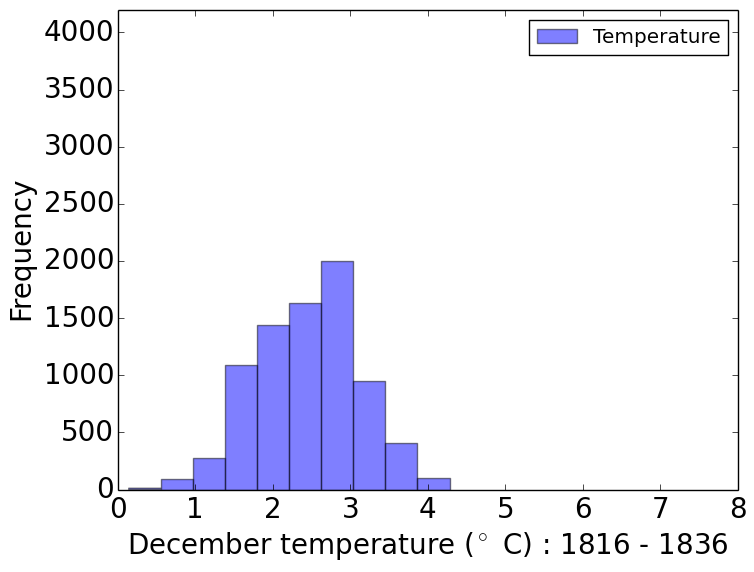}} &
\subcaptionbox{December -- SMC: 1836 - 1856\label{fig:dec5}}{\includegraphics[width = 2.0in]{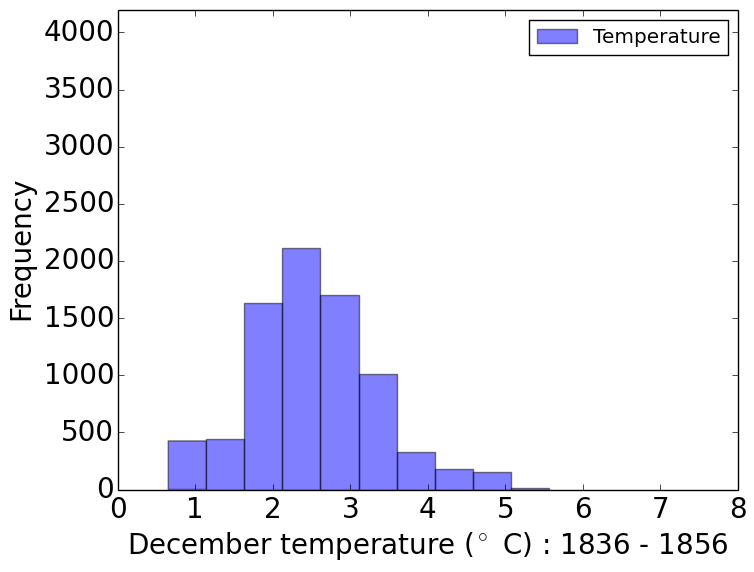}} &
\subcaptionbox{December -- SMC: 1856 - 1876\label{fig:dec6}}{\includegraphics[width = 2.0in]{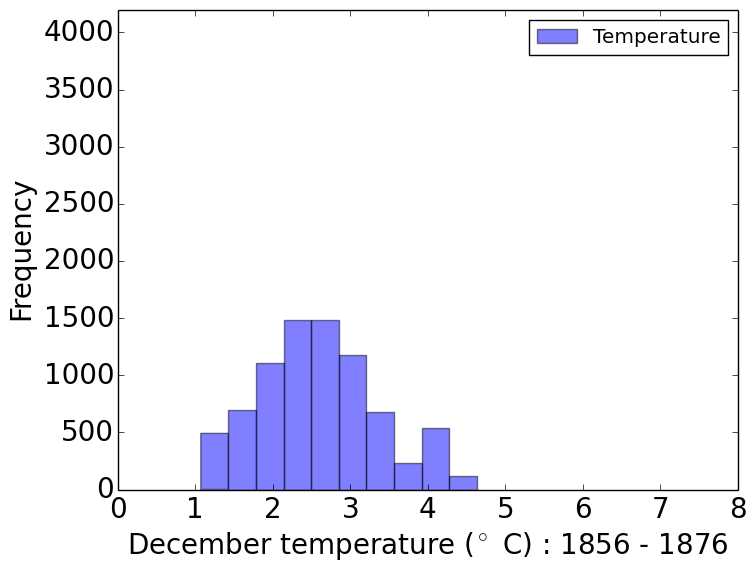}}\\
\subcaptionbox{December -- SMC: 1876 - 1896\label{fig:dec7}}{\includegraphics[width = 2.0in]{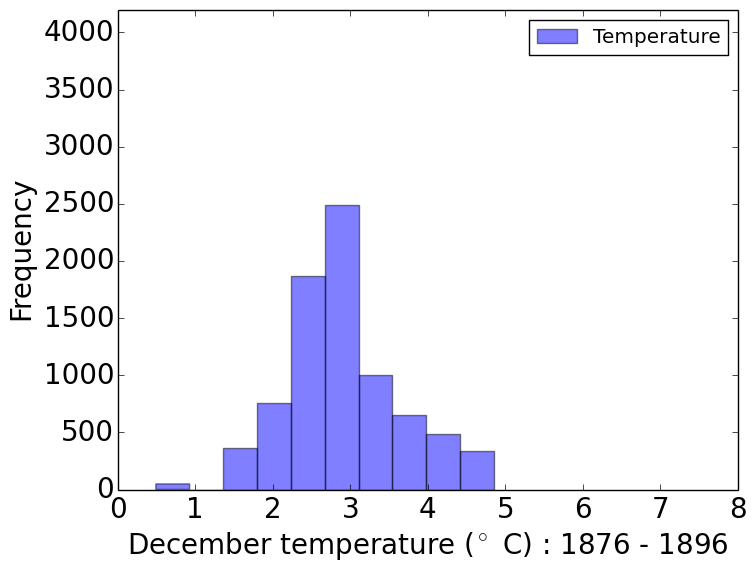}} &
\subcaptionbox{December -- SMC: 1896 - 1916\label{fig:dec8}}{\includegraphics[width = 2.0in]{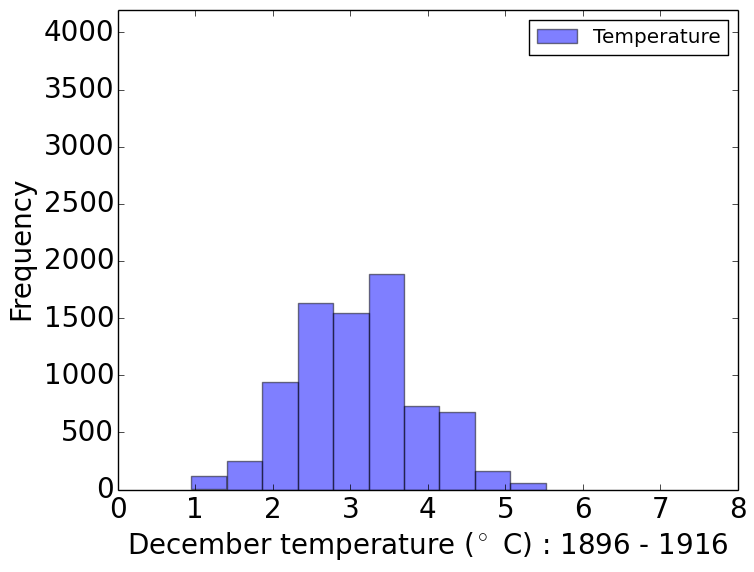}} &
\subcaptionbox{December -- SMC: 1916 - 1936\label{fig:dec9}}{\includegraphics[width = 2.0in]{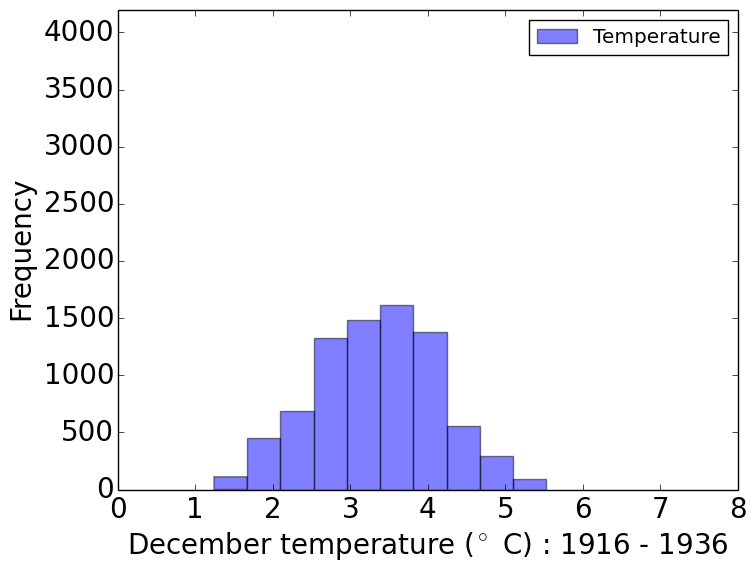}}\\
\subcaptionbox{December -- SMC: 1936 - 1936\label{fig:dec10}}{\includegraphics[width = 2.0in]{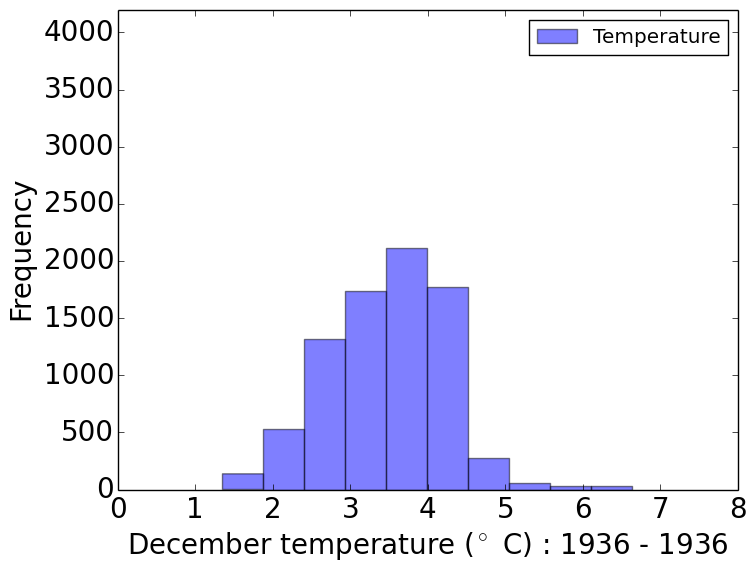}} &
\subcaptionbox{December -- SMC: 1956 - 1976\label{fig:dec11}}{\includegraphics[width = 2.0in]{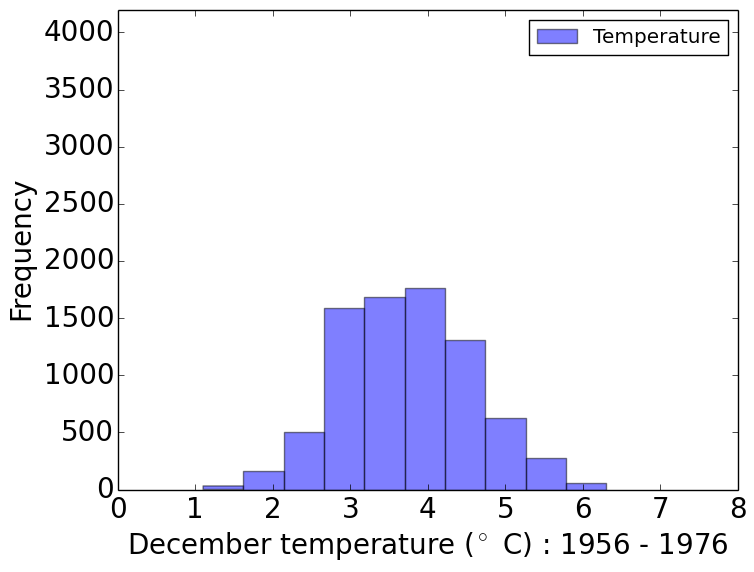}} &
\subcaptionbox{December -- SMC: 1976 - 1996\label{fig:dec12}}{\includegraphics[width = 2.0in]{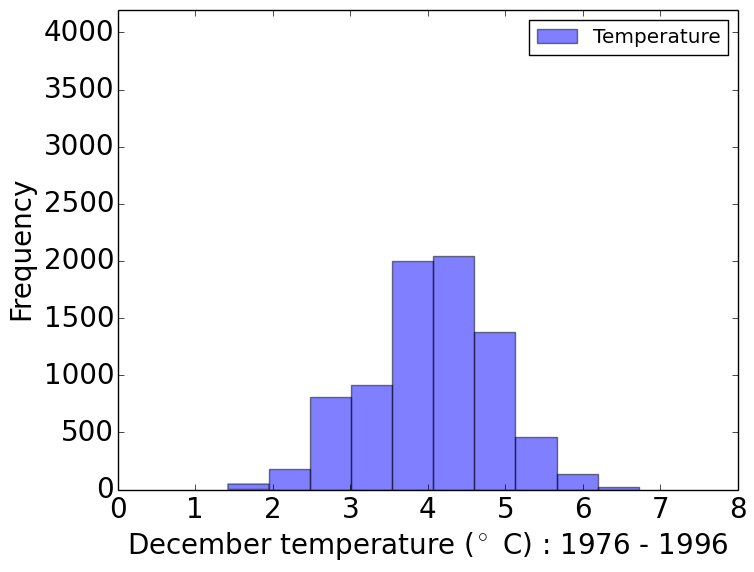}}\\
\subcaptionbox{December -- SMC: 1996 - 2016\label{fig:dec13}}{\includegraphics[width = 2.0in]{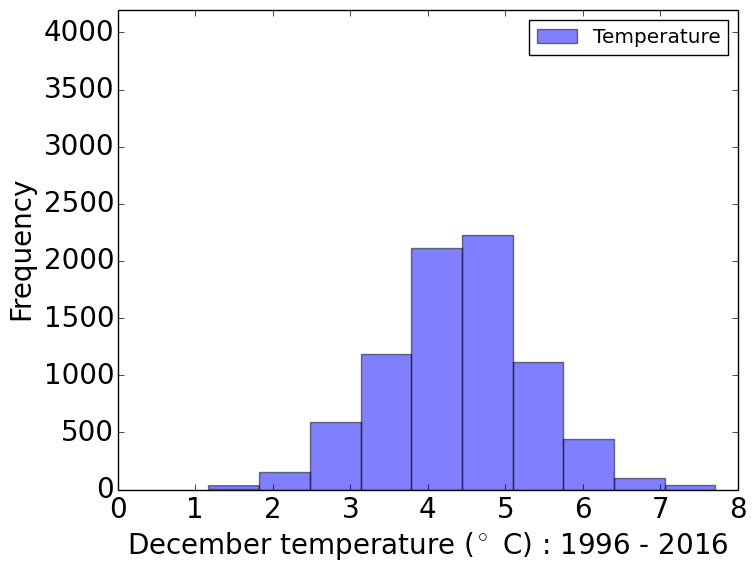}} &
\subcaptionbox{December -- SMC: Total\label{fig:dec14}}{\includegraphics[width = 2.0in]{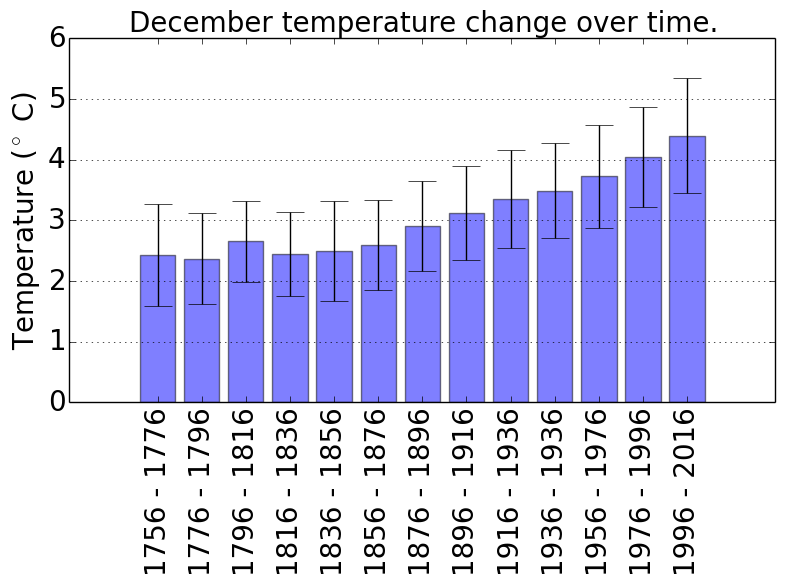}} &
\end{tabular}
\label{fig:dec1}
\end{adjustwidth}
\end{figure}

\begin{figure}
\begin{adjustwidth}{-6em}{0em}
\centering
\begin{tabular}{ccc}
\subcaptionbox{January -- TMCMC: 1756 - 1776\label{fig:jan_t1}}{\includegraphics[width = 2.0in]{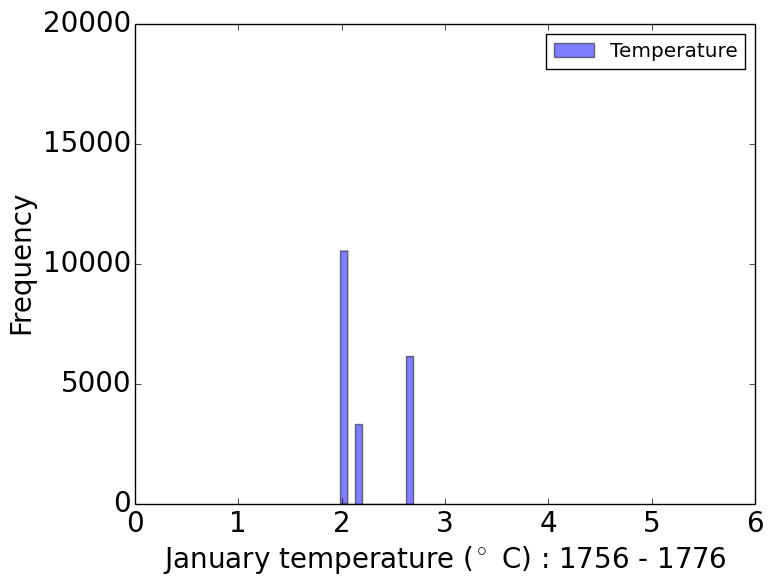}} &
\subcaptionbox{January -- TMCMC: 1776 - 1796\label{fig:jan_t2}}{\includegraphics[width = 2.0in]{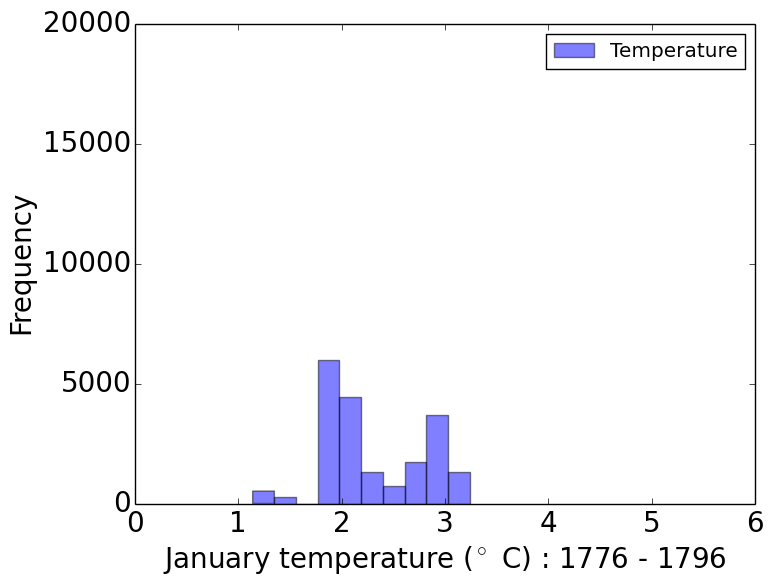}} &
\subcaptionbox{January -- TMCMC: 1796 - 1816\label{fig:jan_t3}}{\includegraphics[width = 2.0in]{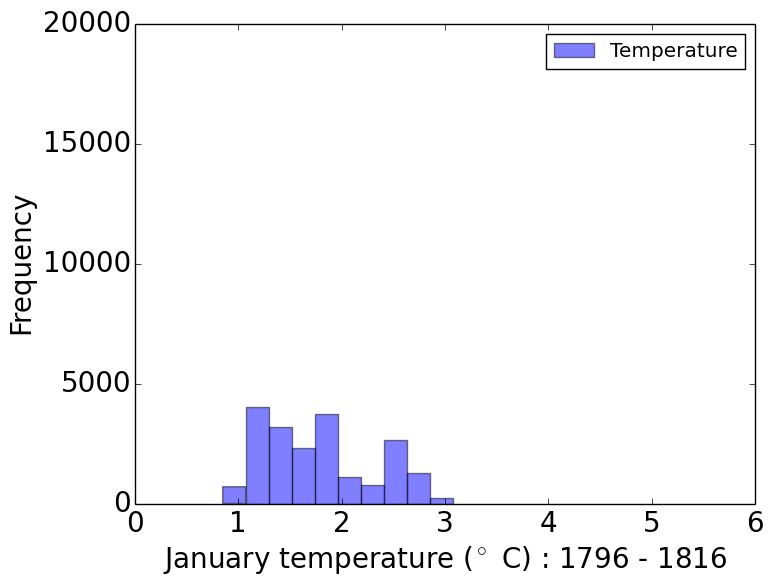}}\\
\subcaptionbox{January -- TMCMC: 1816 - 1836\label{fig:jan_t4}}{\includegraphics[width = 2.0in]{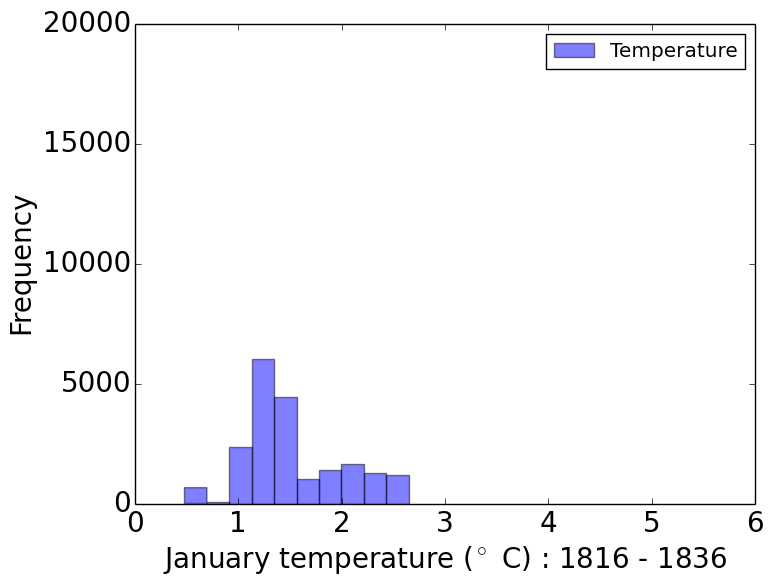}} &
\subcaptionbox{January -- TMCMC: 1836 - 1856\label{fig:jan_t5}}{\includegraphics[width = 2.0in]{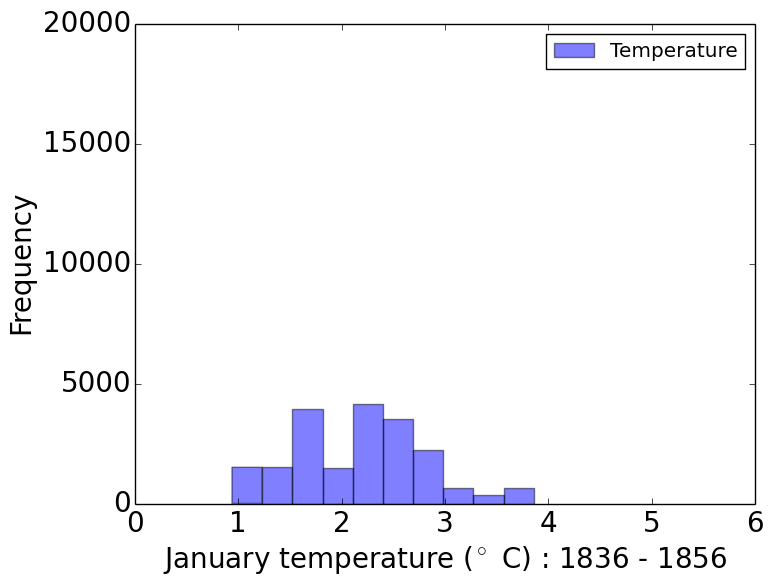}} &
\subcaptionbox{January -- TMCMC: 1856 - 1876\label{fig:jan_t6}}{\includegraphics[width = 2.0in]{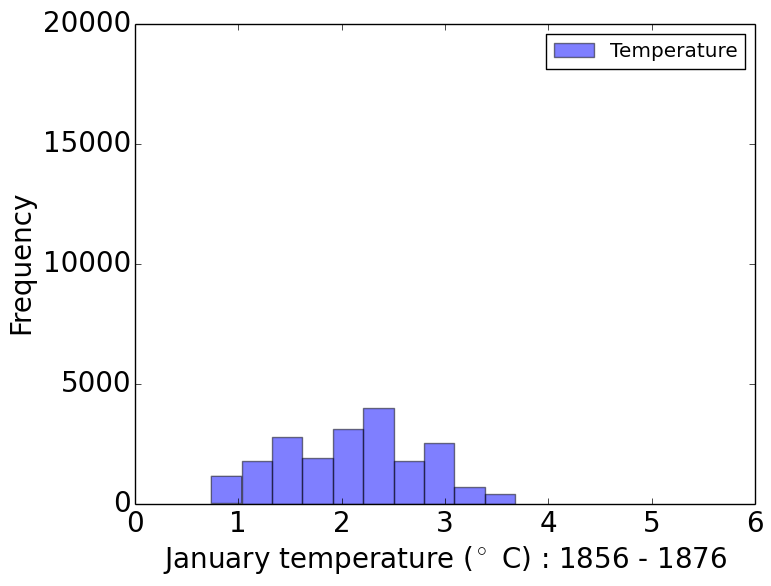}}\\
\subcaptionbox{January -- TMCMC: 1876 - 1896\label{fig:jan_t7}}{\includegraphics[width = 2.0in]{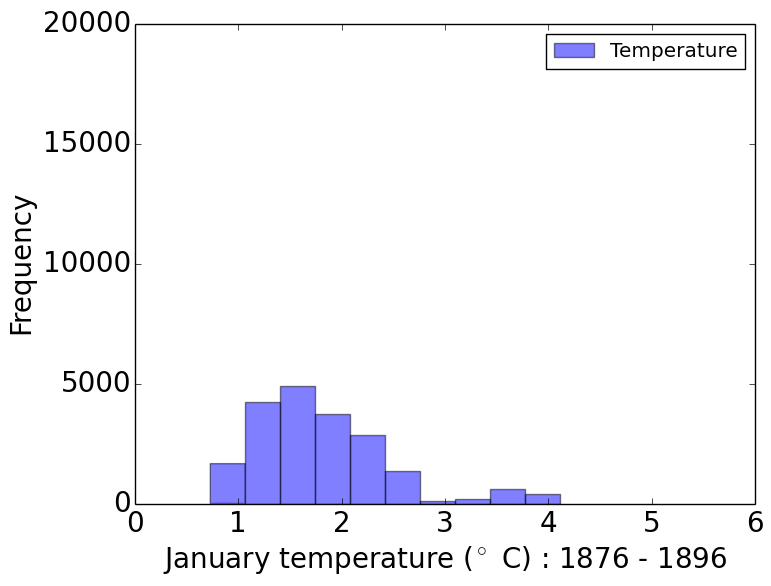}} &
\subcaptionbox{January -- TMCMC: 1896 - 1916\label{fig:jan_t8}}{\includegraphics[width = 2.0in]{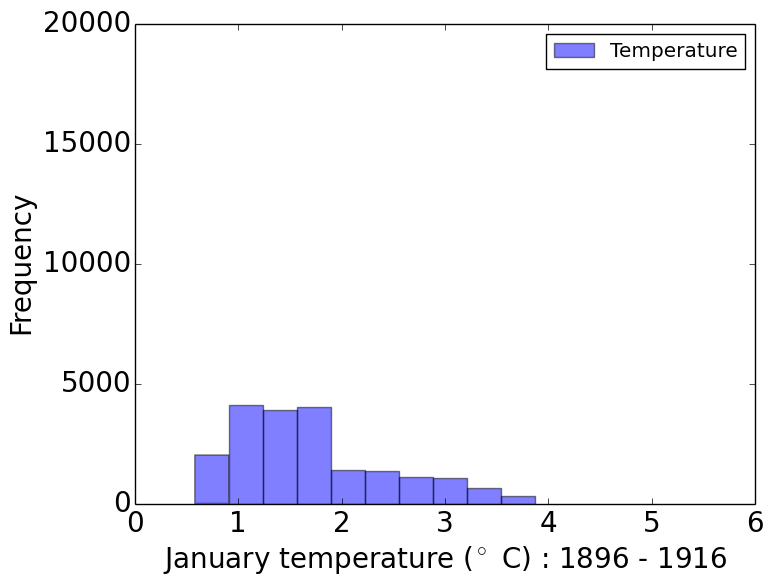}} &
\subcaptionbox{January -- TMCMC: 1916 - 1936\label{fig:jan_t9}}{\includegraphics[width = 2.0in]{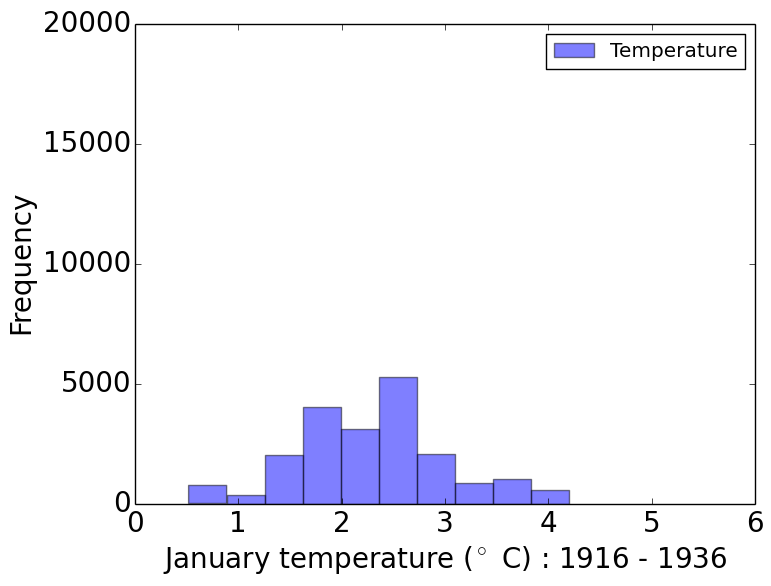}}\\
\subcaptionbox{January -- TMCMC: 1936 - 1936\label{fig:jan_t10}}{\includegraphics[width = 2.0in]{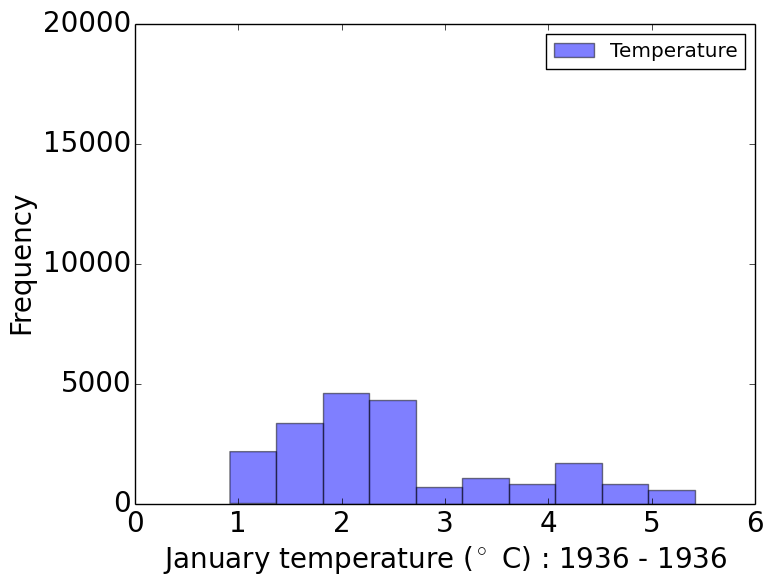}} &
\subcaptionbox{January -- TMCMC: 1956 - 1976\label{fig:jan_t11}}{\includegraphics[width = 2.0in]{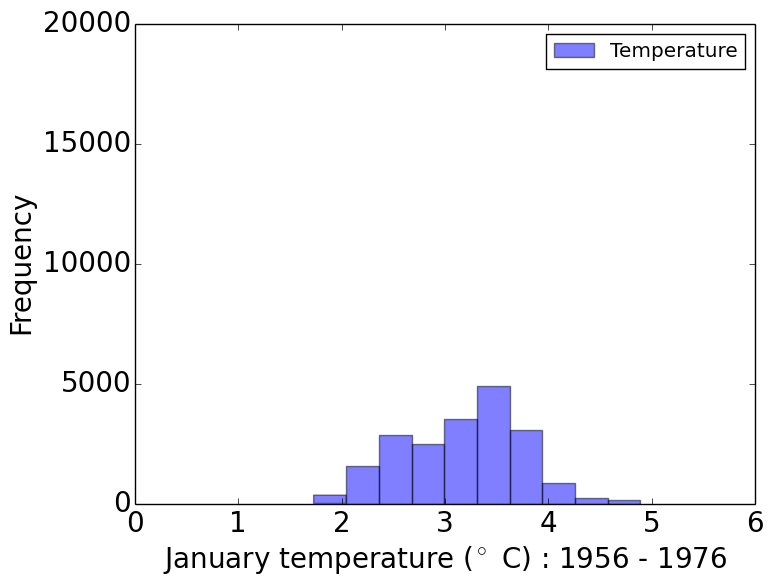}} &
\subcaptionbox{January -- TMCMC: 1976 - 1996\label{fig:jan_t12}}{\includegraphics[width = 2.0in]{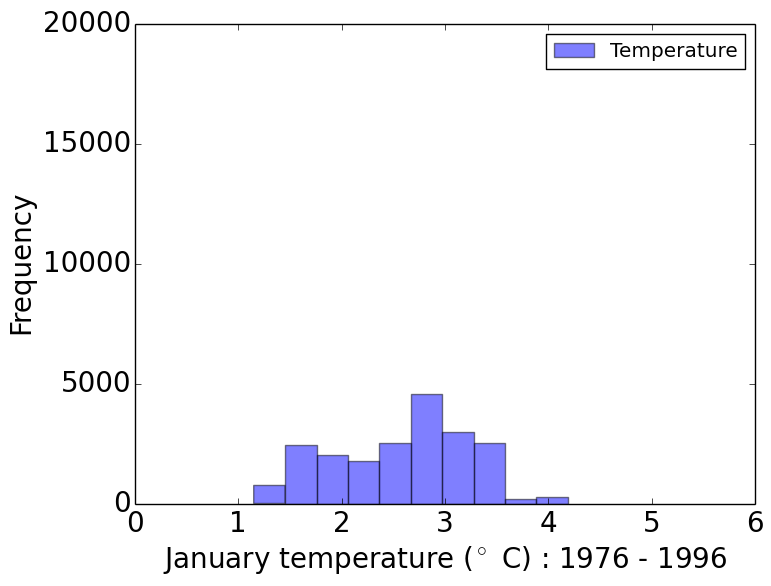}}\\
\subcaptionbox{January -- TMCMC: 1996 - 2016\label{fig:jan_t13}}{\includegraphics[width = 2.0in]{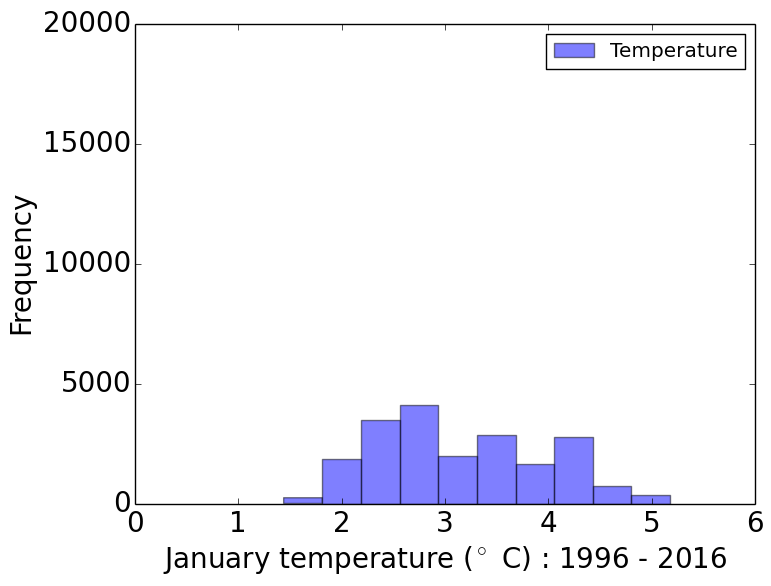}} &
\subcaptionbox{January -- TMCMC: Total\label{fig:jan_t14}}{\includegraphics[width = 2.0in]{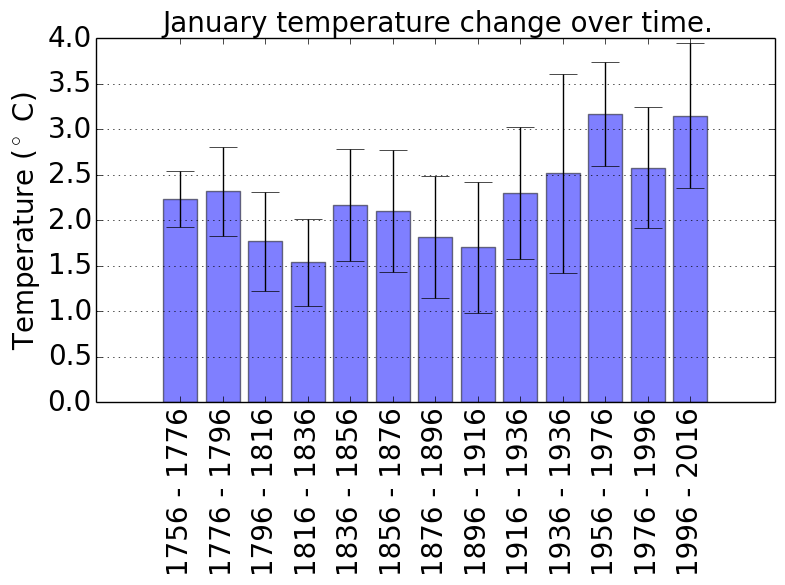}} &
\end{tabular}
\label{fig:jan_t1}
\end{adjustwidth}
\end{figure}

\begin{figure}
\begin{adjustwidth}{-6em}{0em}
\centering
\begin{tabular}{ccc}
\subcaptionbox{Feburary -- TMCMC: 1756 - 1776\label{fig:feb_t1}}{\includegraphics[width = 2.0in]{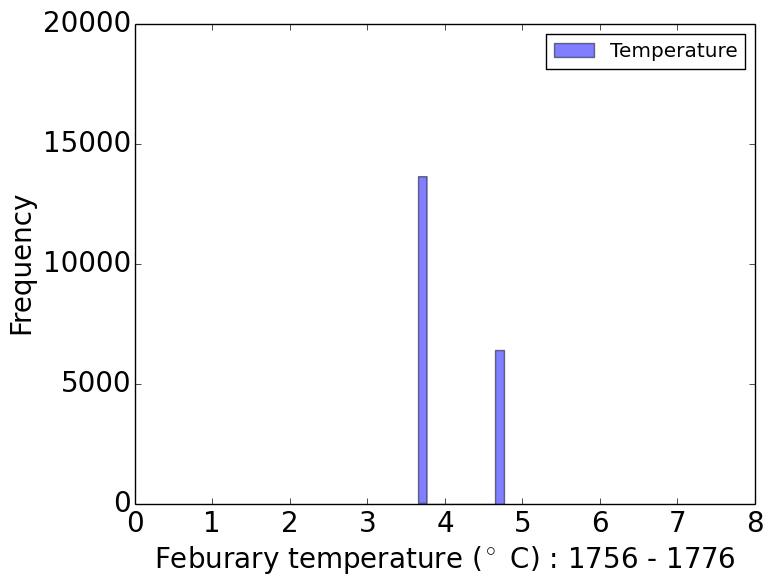}} &
\subcaptionbox{Feburary -- TMCMC: 1776 - 1796\label{fig:feb_t2}}{\includegraphics[width = 2.0in]{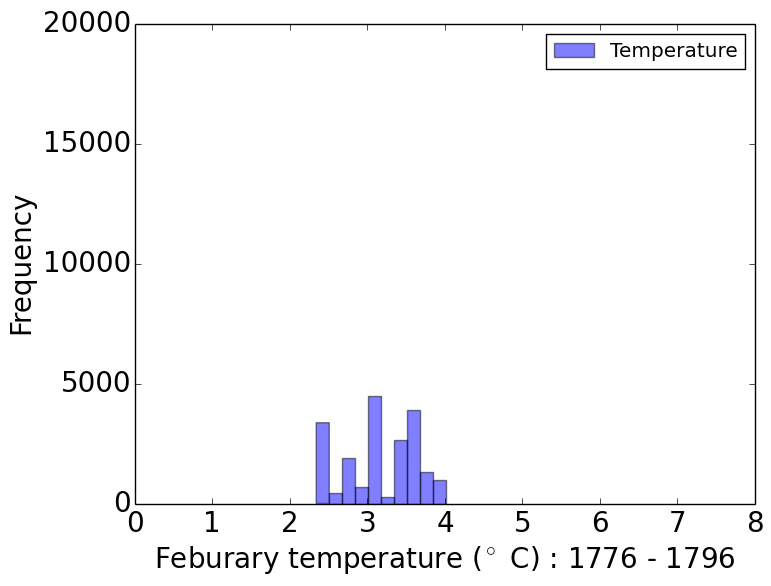}} &
\subcaptionbox{Feburary -- TMCMC: 1796 - 1816\label{fig:feb_t3}}{\includegraphics[width = 2.0in]{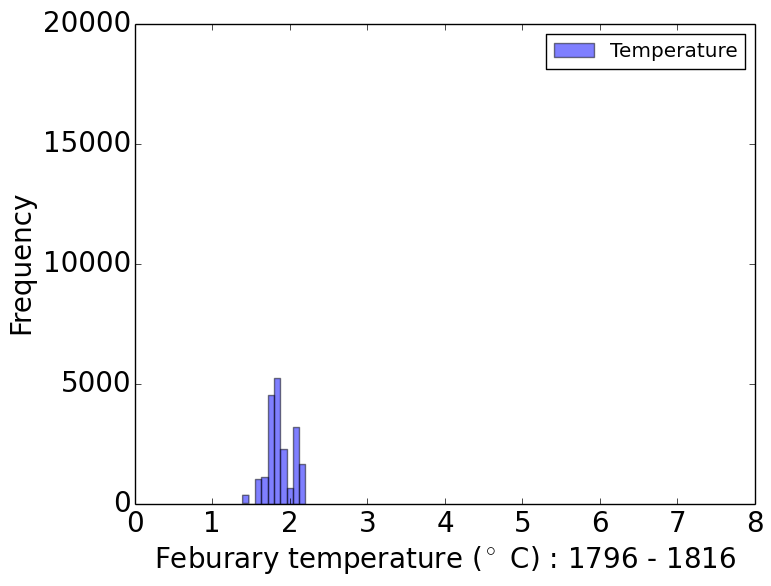}}\\
\subcaptionbox{Feburary -- TMCMC: 1816 - 1836\label{fig:feb_t4}}{\includegraphics[width = 2.0in]{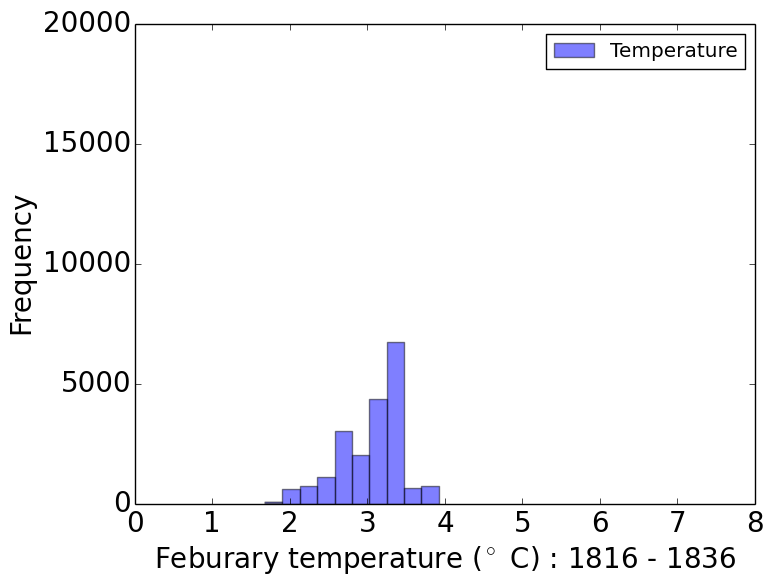}} &
\subcaptionbox{Feburary -- TMCMC: 1836 - 1856\label{fig:feb_t5}}{\includegraphics[width = 2.0in]{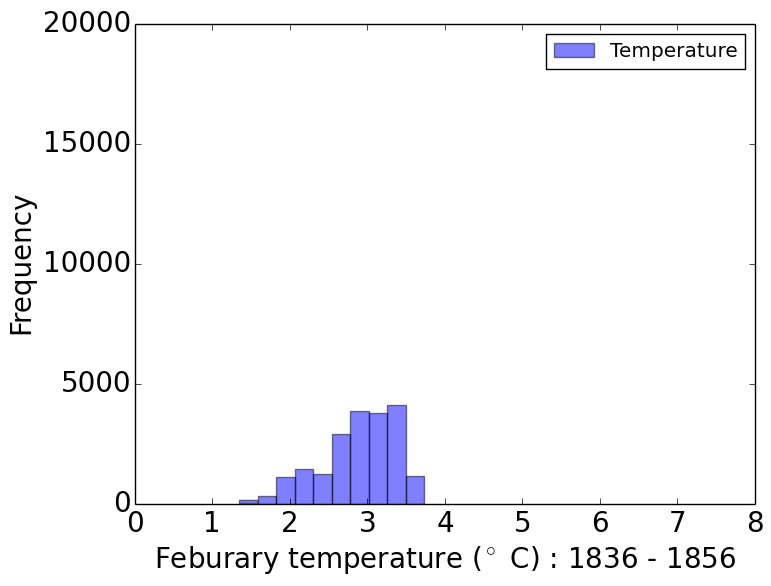}} &
\subcaptionbox{Feburary -- TMCMC: 1856 - 1876\label{fig:feb_t6}}{\includegraphics[width = 2.0in]{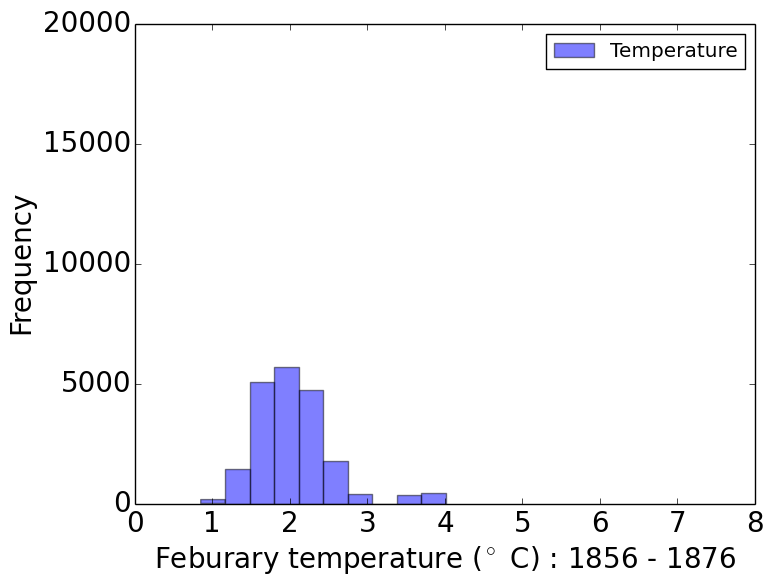}}\\
\subcaptionbox{Feburary -- TMCMC: 1876 - 1896\label{fig:feb_t7}}{\includegraphics[width = 2.0in]{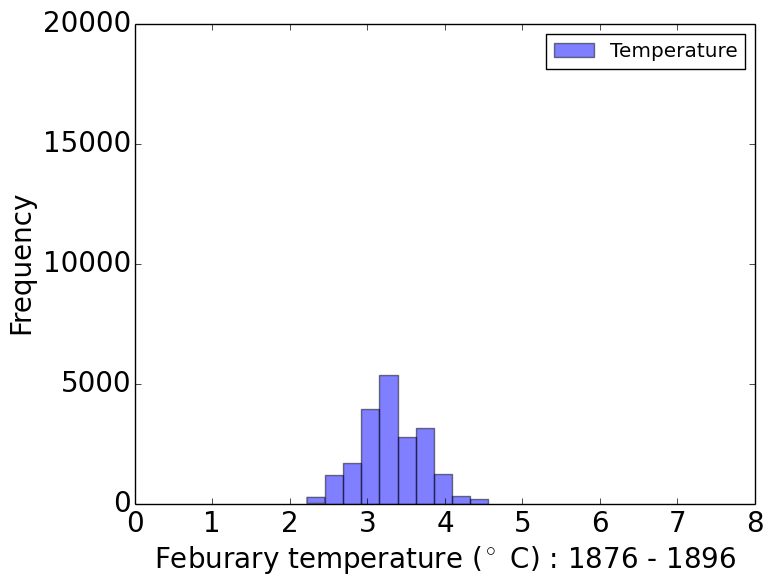}} &
\subcaptionbox{Feburary -- TMCMC: 1896 - 1916\label{fig:feb_t8}}{\includegraphics[width = 2.0in]{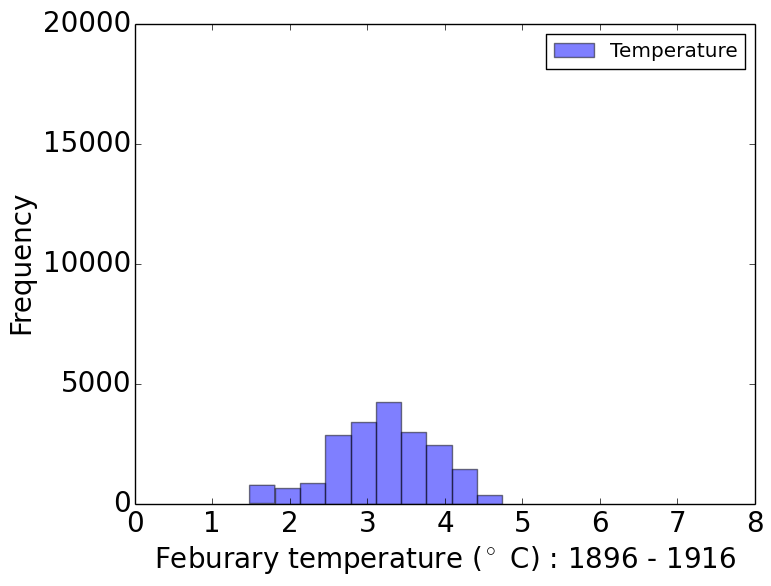}} &
\subcaptionbox{Feburary -- TMCMC: 1916 - 1936\label{fig:feb_t9}}{\includegraphics[width = 2.0in]{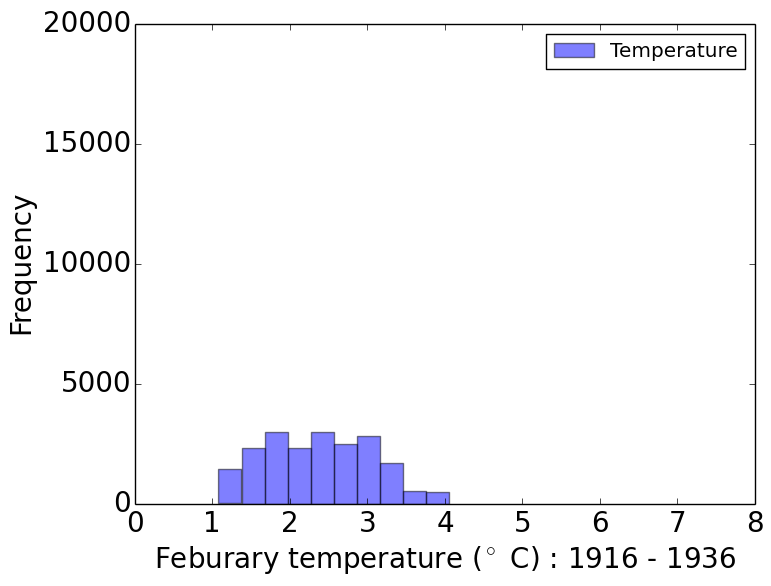}}\\
\subcaptionbox{Feburary -- TMCMC: 1936 - 1936\label{fig:feb_t10}}{\includegraphics[width = 2.0in]{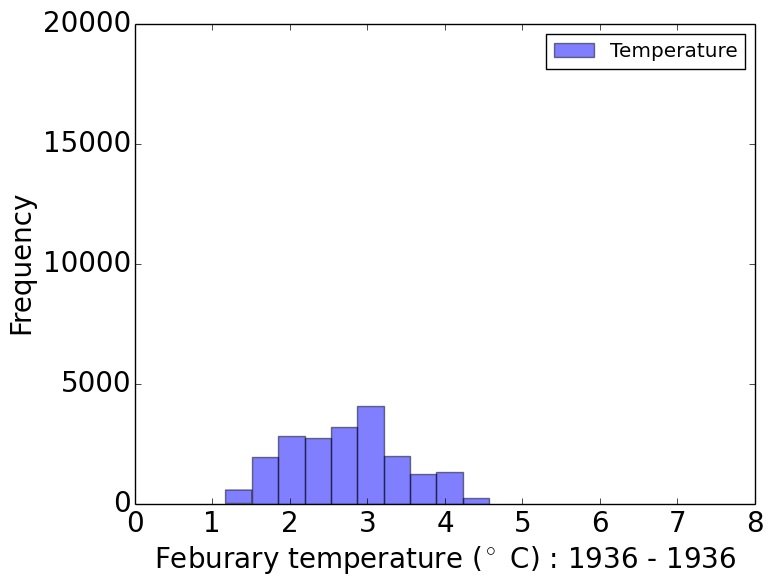}} &
\subcaptionbox{Feburary -- TMCMC: 1956 - 1976\label{fig:feb_t11}}{\includegraphics[width = 2.0in]{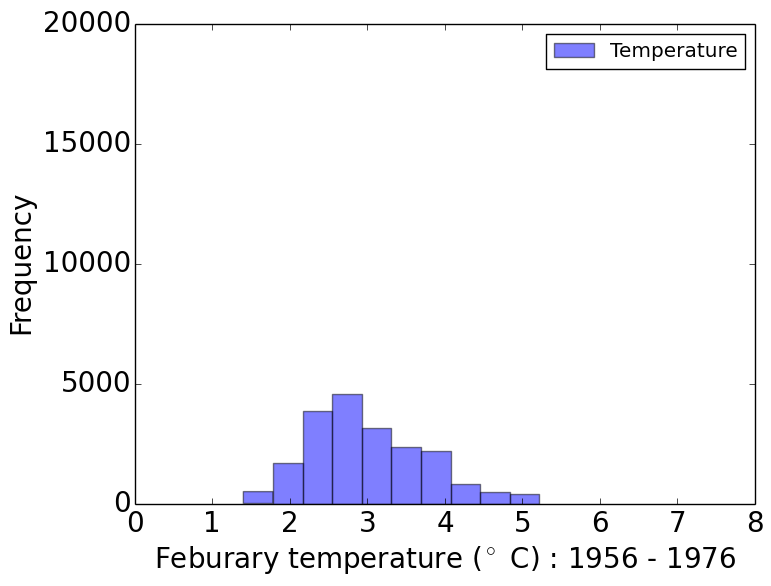}} &
\subcaptionbox{Feburary -- TMCMC: 1976 - 1996\label{fig:feb_t12}}{\includegraphics[width = 2.0in]{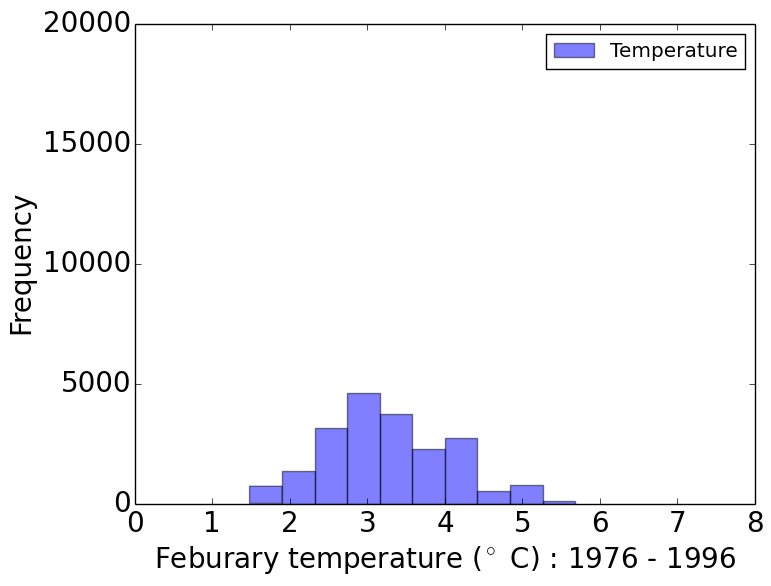}}\\
\subcaptionbox{Feburary -- TMCMC: 1996 - 2016\label{fig:feb_t13}}{\includegraphics[width = 2.0in]{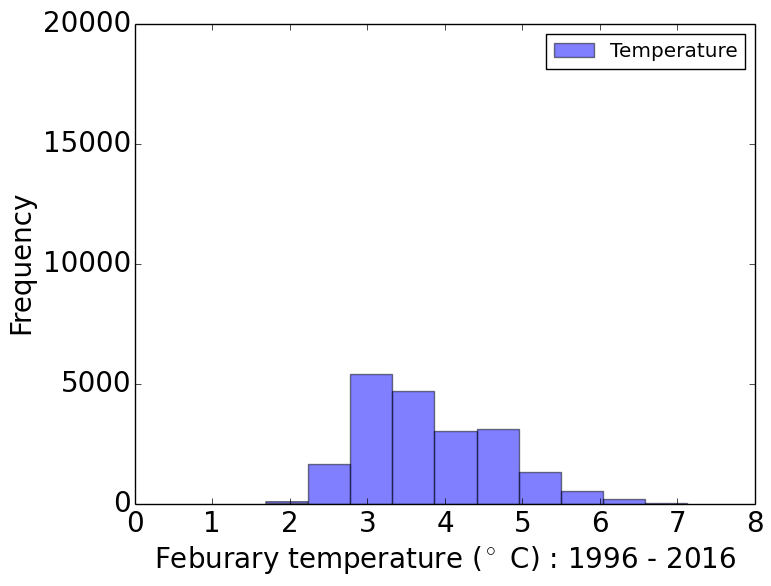}} &
\subcaptionbox{Feburary -- TMCMC: Total\label{fig:feb_t14}}{\includegraphics[width = 2.0in]{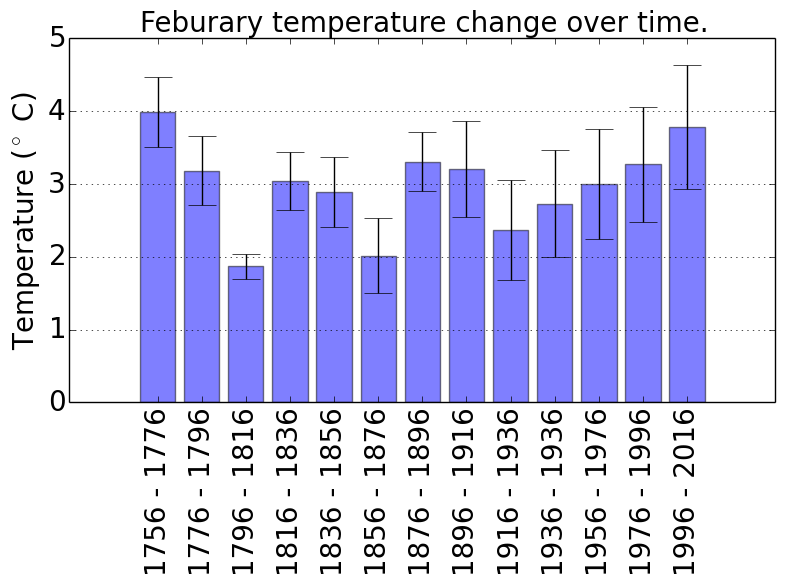}} &
\end{tabular}
\label{fig:feb_t1}
\end{adjustwidth}
\end{figure}

\begin{figure}
\begin{adjustwidth}{-6em}{0em}
\centering
\begin{tabular}{ccc}
\subcaptionbox{March -- TMCMC: 1756 - 1776\label{fig:mar_t1}}{\includegraphics[width = 2.0in]{images/tmcmc/mar/comb_tmcmc_hist_1.png}} &
\subcaptionbox{March -- TMCMC: 1776 - 1796\label{fig:mar_t2}}{\includegraphics[width = 2.0in]{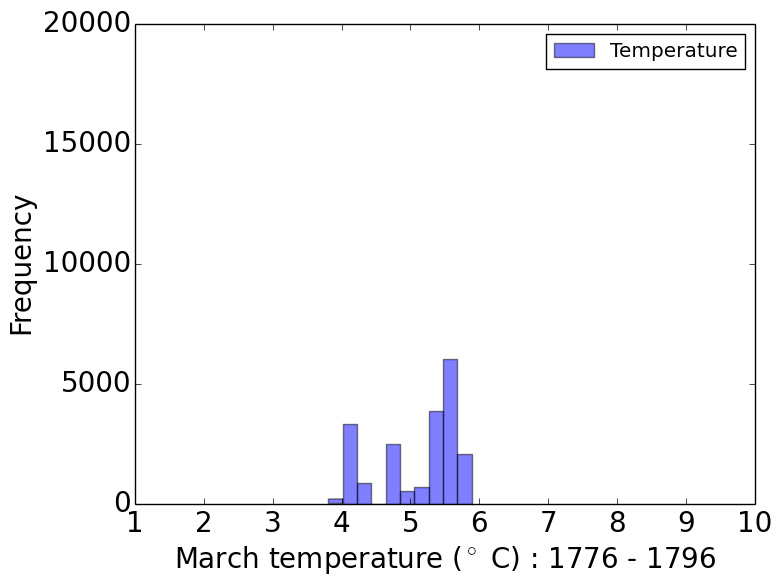}} &
\subcaptionbox{March -- TMCMC: 1796 - 1816\label{fig:mar_t3}}{\includegraphics[width = 2.0in]{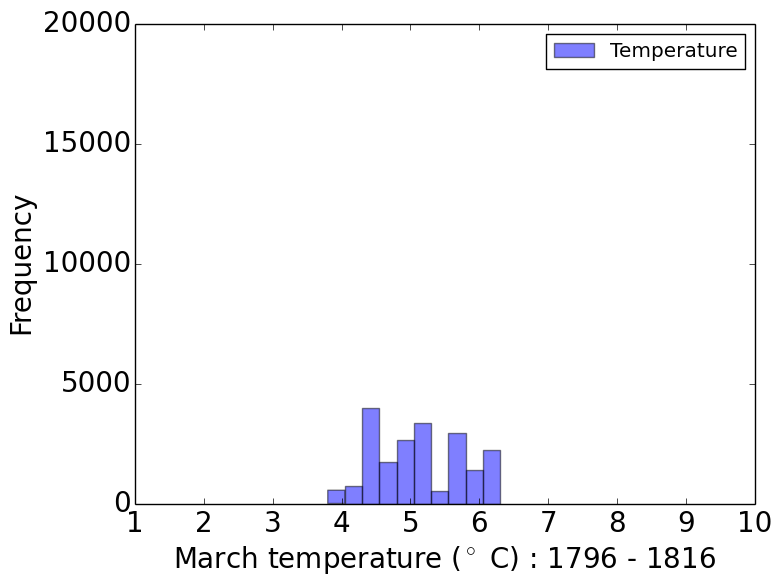}}\\
\subcaptionbox{March -- TMCMC: 1816 - 1836\label{fig:mar_t4}}{\includegraphics[width = 2.0in]{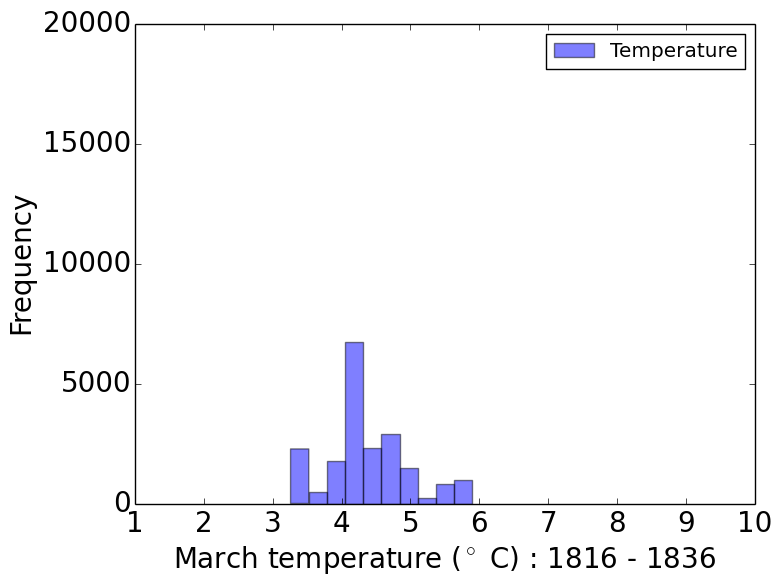}} &
\subcaptionbox{March -- TMCMC: 1836 - 1856\label{fig:mar_t5}}{\includegraphics[width = 2.0in]{images/tmcmc/mar/comb_tmcmc_hist_5.png}} &
\subcaptionbox{March -- TMCMC: 1856 - 1876\label{fig:mar_t6}}{\includegraphics[width = 2.0in]{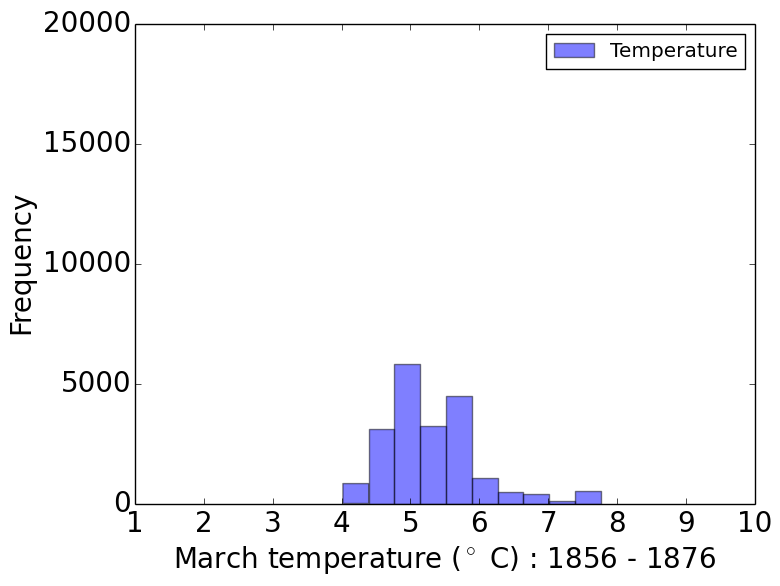}}\\
\subcaptionbox{March -- TMCMC: 1876 - 1896\label{fig:mar_t7}}{\includegraphics[width = 2.0in]{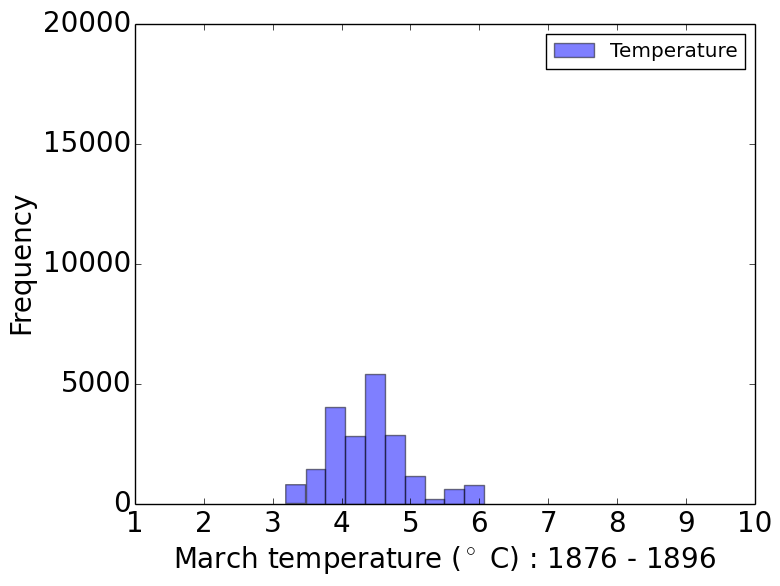}} &
\subcaptionbox{March -- TMCMC: 1896 - 1916\label{fig:mar_t8}}{\includegraphics[width = 2.0in]{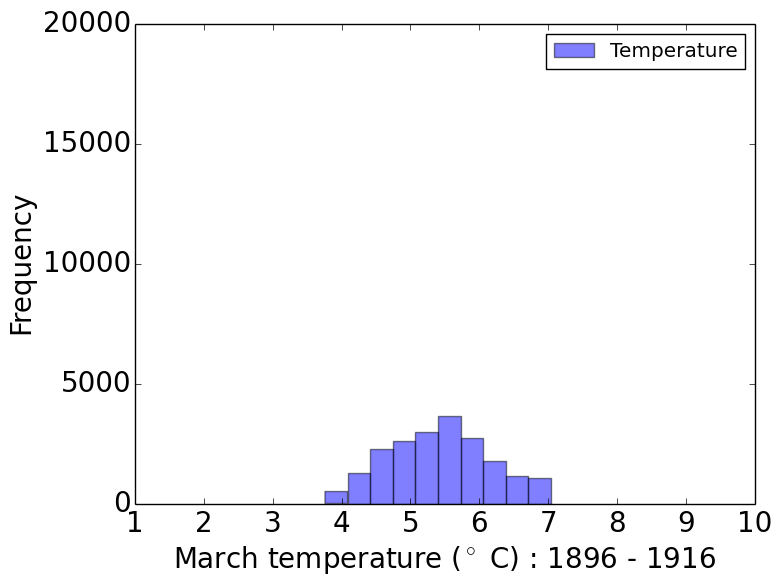}} &
\subcaptionbox{March -- TMCMC: 1916 - 1936\label{fig:mar_t9}}{\includegraphics[width = 2.0in]{images/tmcmc/mar/comb_tmcmc_hist_9.png}}\\
\subcaptionbox{March -- TMCMC: 1936 - 1936\label{fig:mar_t10}}{\includegraphics[width = 2.0in]{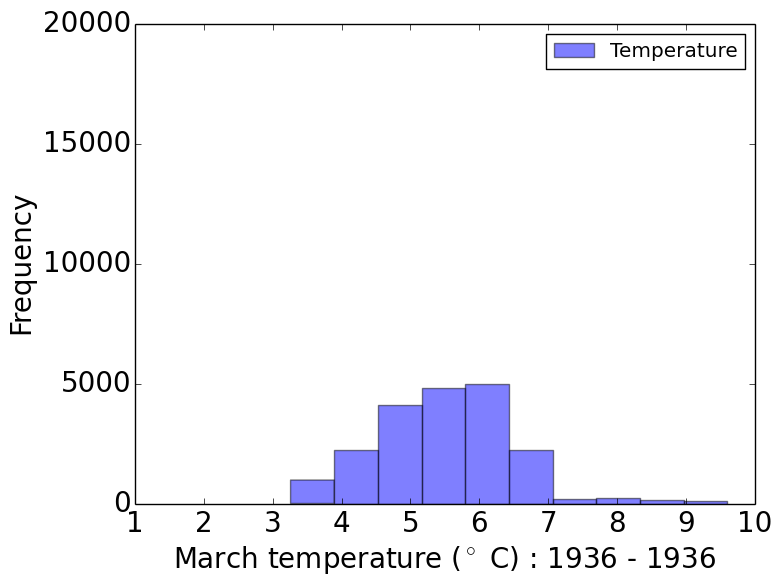}} &
\subcaptionbox{March -- TMCMC: 1956 - 1976\label{fig:mar_t11}}{\includegraphics[width = 2.0in]{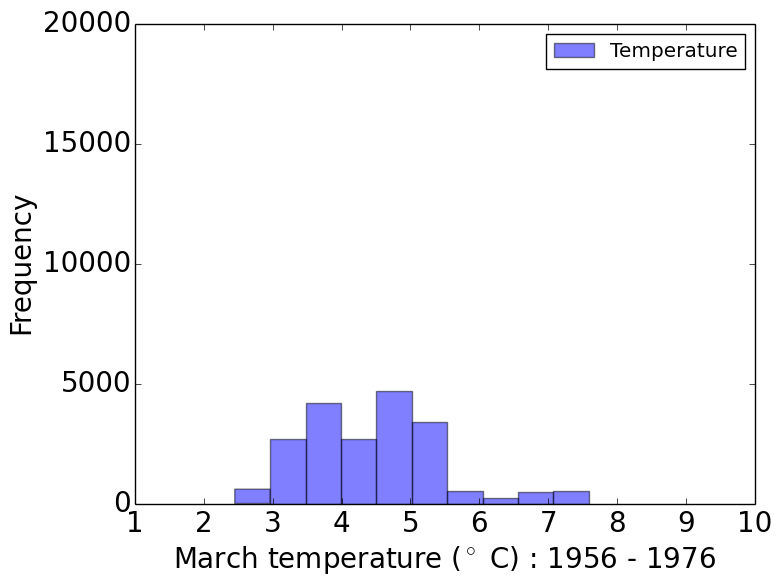}} &
\subcaptionbox{March -- TMCMC: 1976 - 1996\label{fig:mar_t12}}{\includegraphics[width = 2.0in]{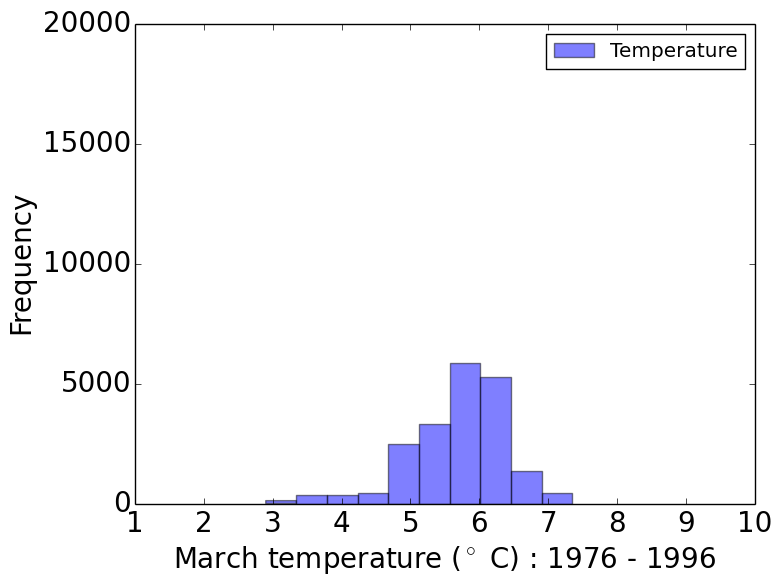}}\\
\subcaptionbox{March -- TMCMC: 1996 - 2016\label{fig:mar_t13}}{\includegraphics[width = 2.0in]{images/tmcmc/mar/comb_tmcmc_hist_13.png}} &
\subcaptionbox{March -- TMCMC: Total\label{fig:mar_t14}}{\includegraphics[width = 2.0in]{images/tmcmc/mar/tmcmc_aggregate.png}} &
\end{tabular}
\label{fig:mar_t1}
\end{adjustwidth}
\end{figure}

\begin{figure}
\begin{adjustwidth}{-6em}{0em}
\centering
\begin{tabular}{ccc}
\subcaptionbox{April -- TMCMC: 1756 - 1776\label{fig:apr_t1}}{\includegraphics[width = 2.0in]{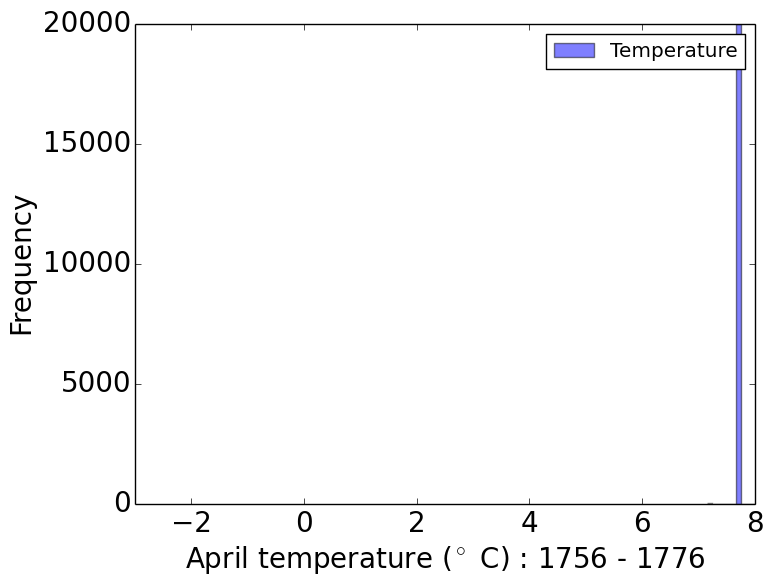}} &
\subcaptionbox{April -- TMCMC: 1776 - 1796\label{fig:apr_t2}}{\includegraphics[width = 2.0in]{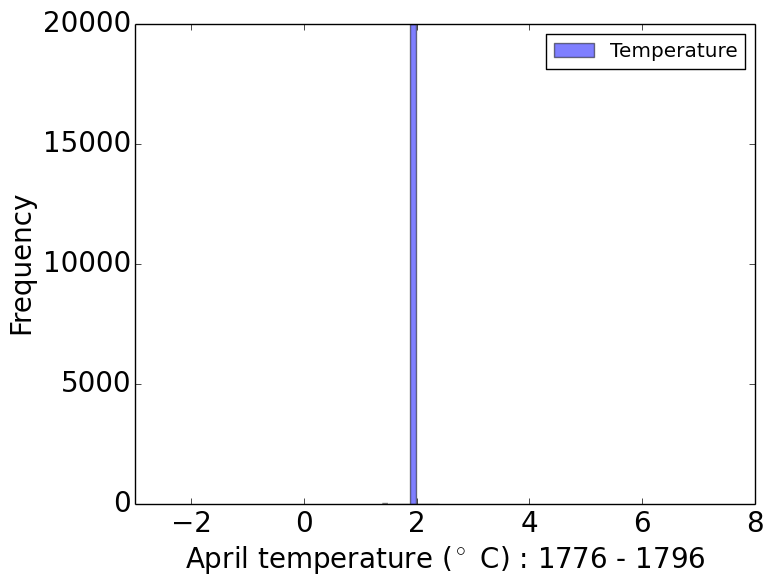}} &
\subcaptionbox{April -- TMCMC: 1796 - 1816\label{fig:apr_t3}}{\includegraphics[width = 2.0in]{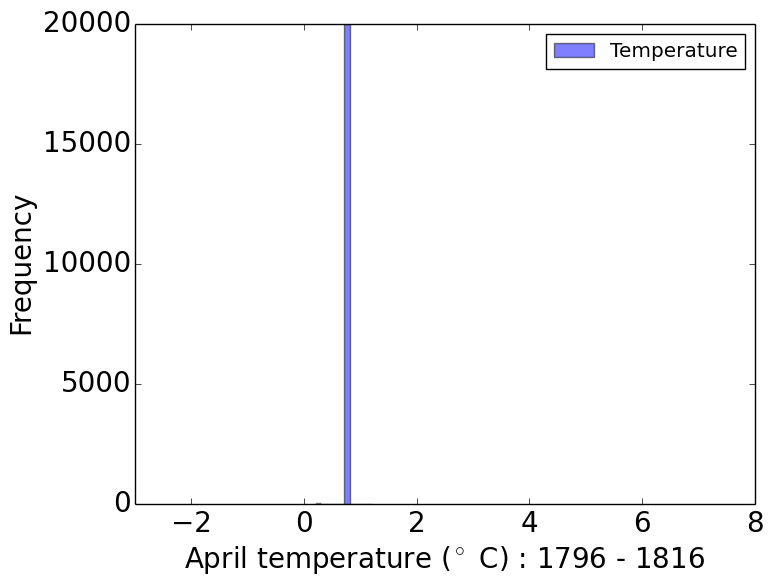}}\\
\subcaptionbox{April -- TMCMC: 1816 - 1836\label{fig:apr_t4}}{\includegraphics[width = 2.0in]{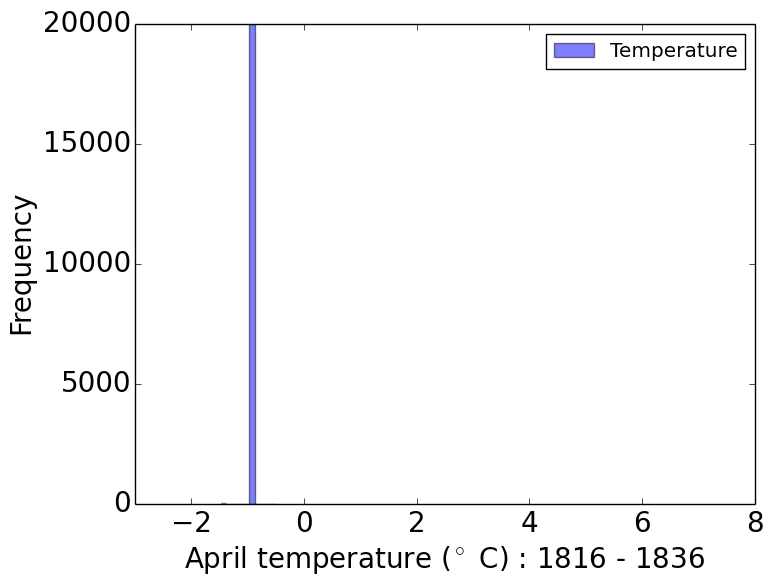}} &
\subcaptionbox{April -- TMCMC: 1836 - 1856\label{fig:apr_t5}}{\includegraphics[width = 2.0in]{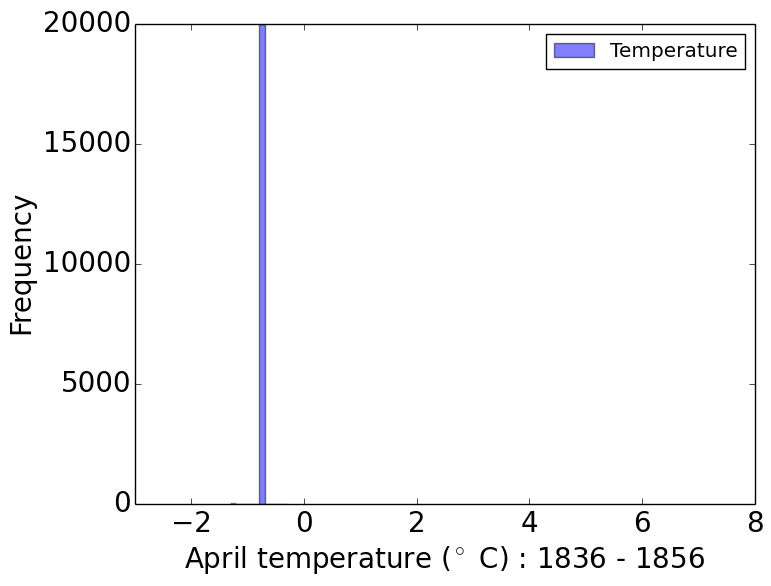}} &
\subcaptionbox{April -- TMCMC: 1856 - 1876\label{fig:apr_t6}}{\includegraphics[width = 2.0in]{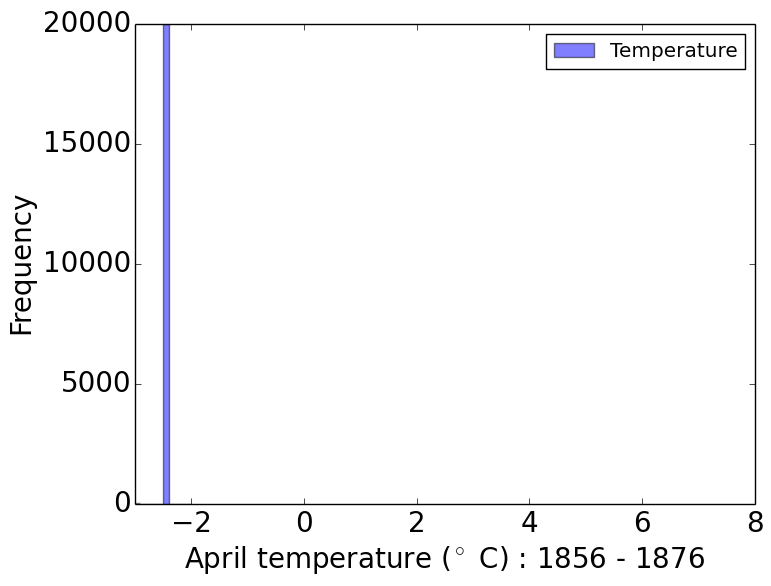}}\\
\subcaptionbox{April -- TMCMC: 1876 - 1896\label{fig:apr_t7}}{\includegraphics[width = 2.0in]{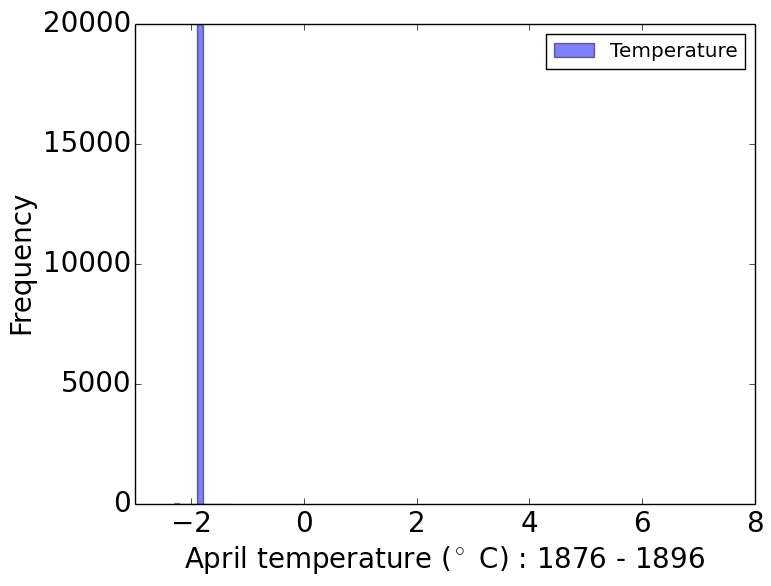}} &
\subcaptionbox{April -- TMCMC: 1896 - 1916\label{fig:apr_t8}}{\includegraphics[width = 2.0in]{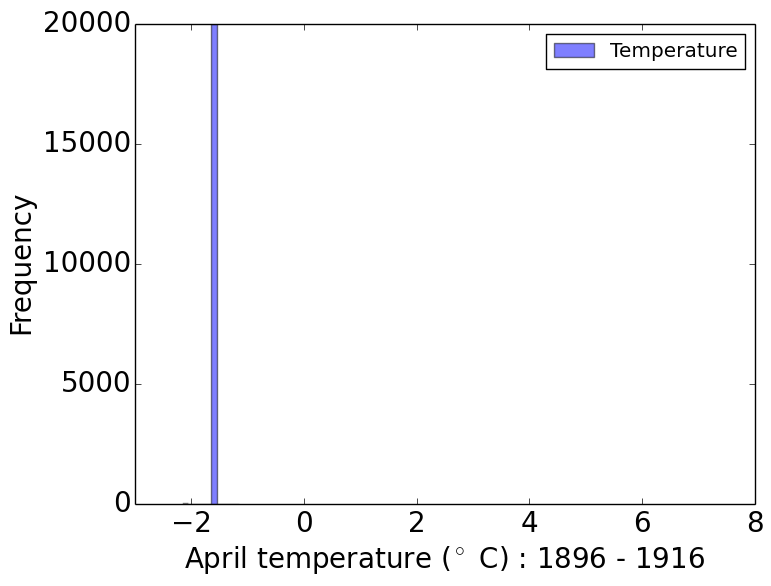}} &
\subcaptionbox{April -- TMCMC: 1916 - 1936\label{fig:apr_t9}}{\includegraphics[width = 2.0in]{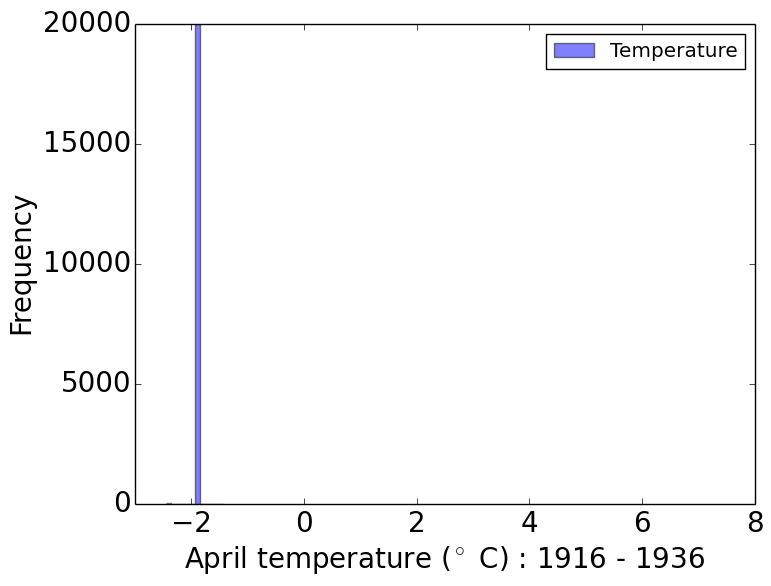}}\\
\subcaptionbox{April -- TMCMC: 1936 - 1936\label{fig:apr_t10}}{\includegraphics[width = 2.0in]{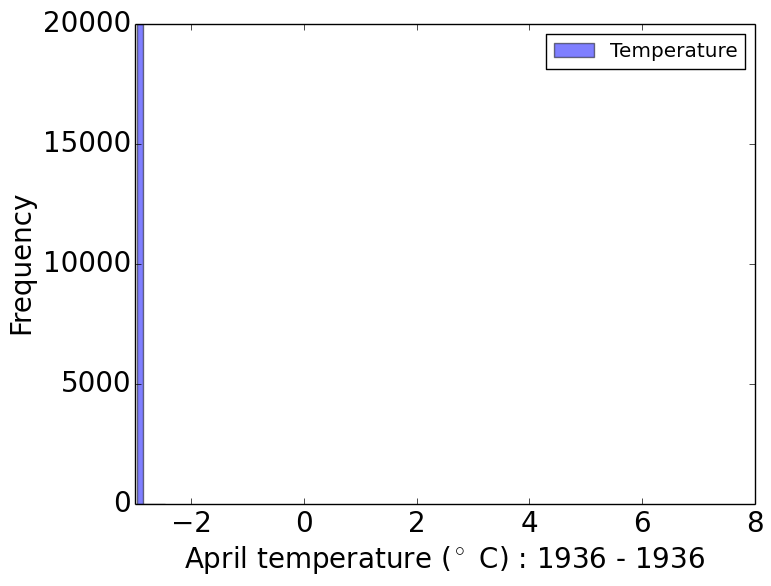}} &
\subcaptionbox{April -- TMCMC: 1956 - 1976\label{fig:apr_t11}}{\includegraphics[width = 2.0in]{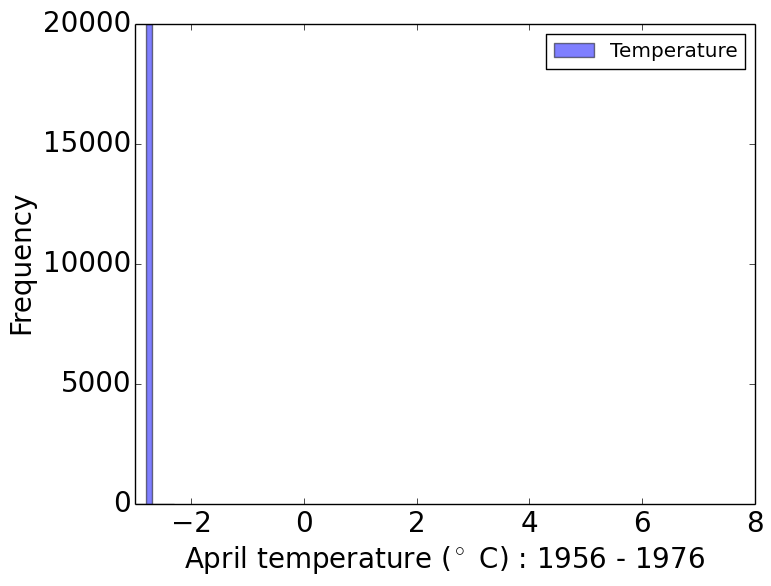}} &
\subcaptionbox{April -- TMCMC: 1976 - 1996\label{fig:apr_t12}}{\includegraphics[width = 2.0in]{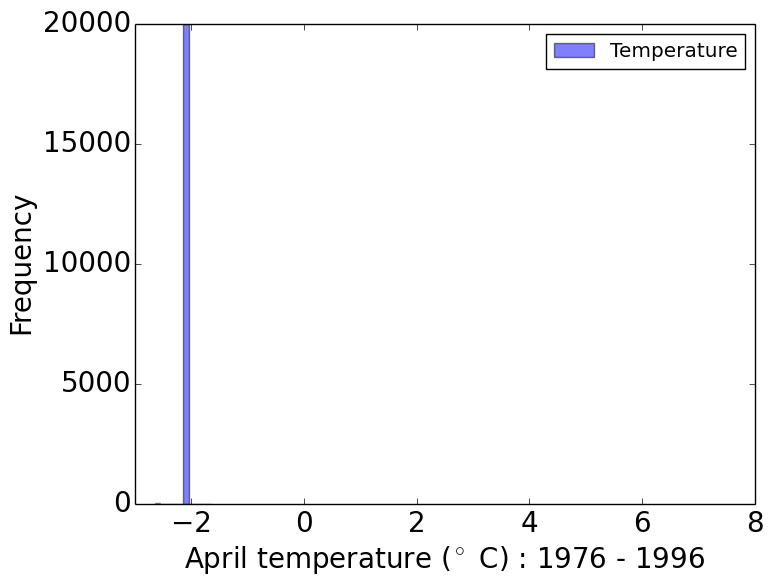}}\\
\subcaptionbox{April -- TMCMC: 1996 - 2016\label{fig:apr_t13}}{\includegraphics[width = 2.0in]{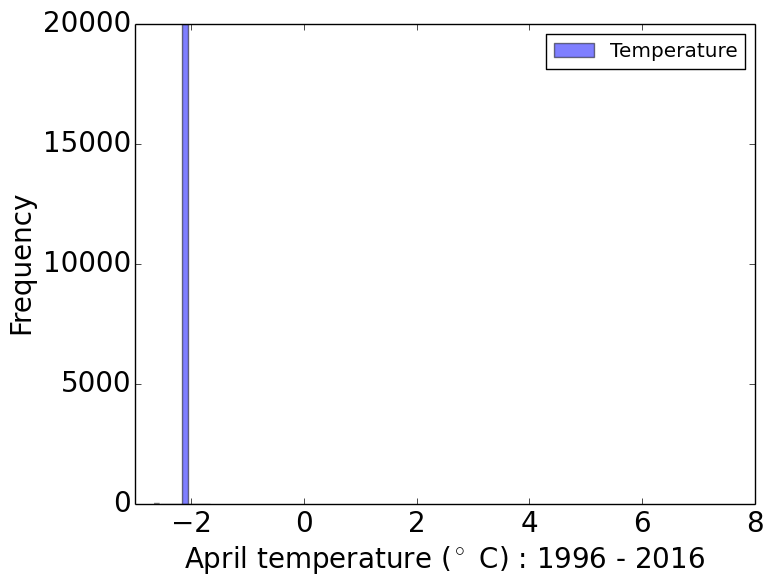}} &
\subcaptionbox{April -- TMCMC: Total\label{fig:apr_t14}}{\includegraphics[width = 2.0in]{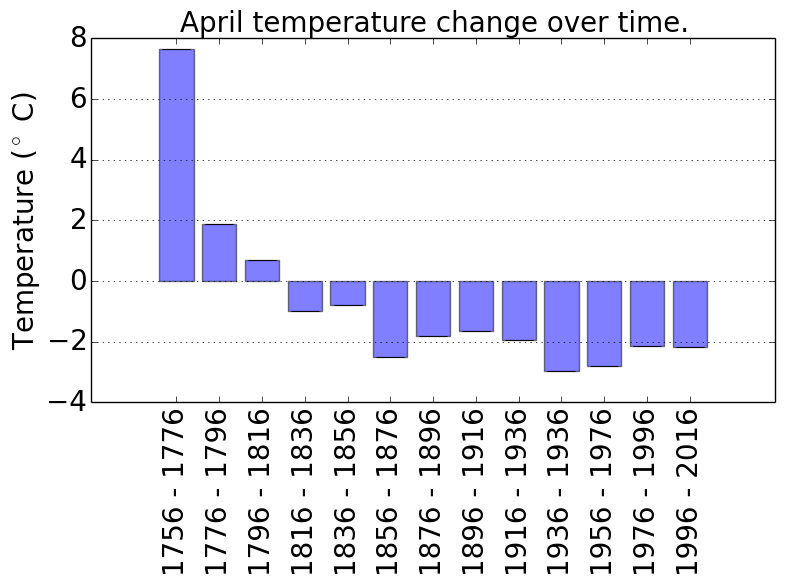}} &
\end{tabular}
\label{fig:apr_t1}
\end{adjustwidth}
\end{figure}

\begin{figure}
\begin{adjustwidth}{-6em}{0em}
\centering
\begin{tabular}{ccc}
\subcaptionbox{May -- TMCMC: 1756 - 1776\label{fig:may_t1}}{\includegraphics[width = 2.0in]{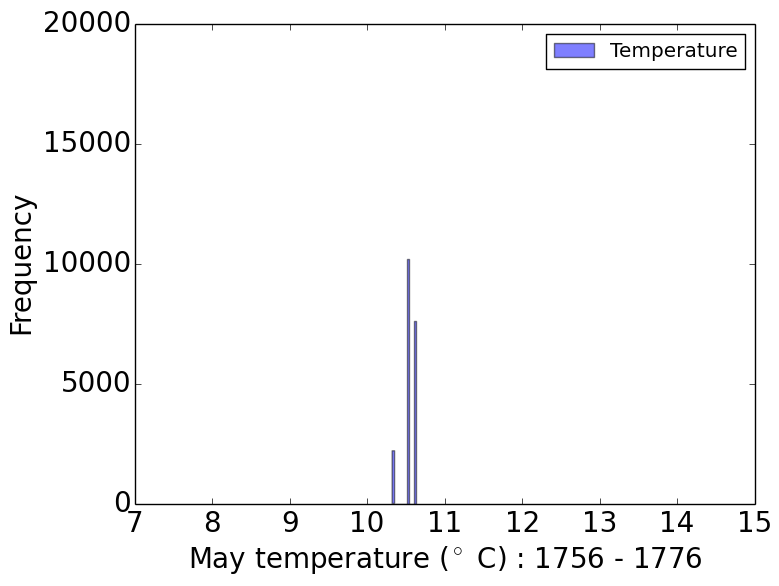}} &
\subcaptionbox{May -- TMCMC: 1776 - 1796\label{fig:may_t2}}{\includegraphics[width = 2.0in]{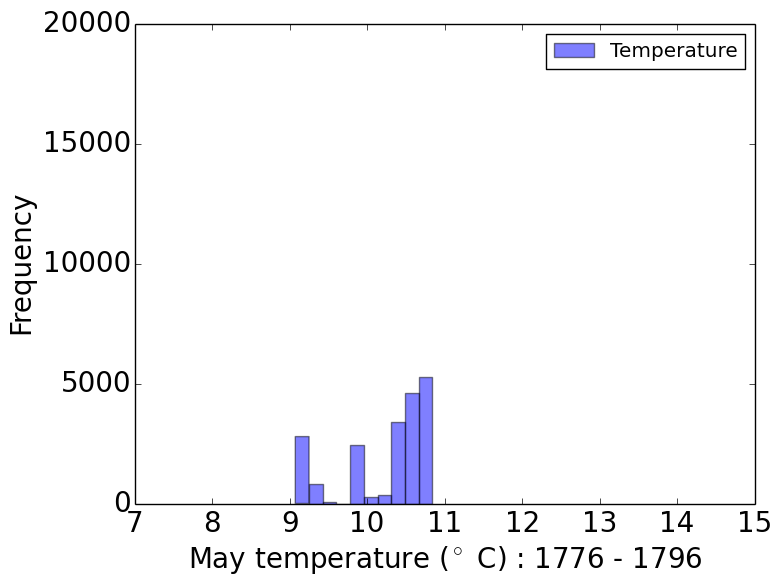}} &
\subcaptionbox{May -- TMCMC: 1796 - 1816\label{fig:may_t3}}{\includegraphics[width = 2.0in]{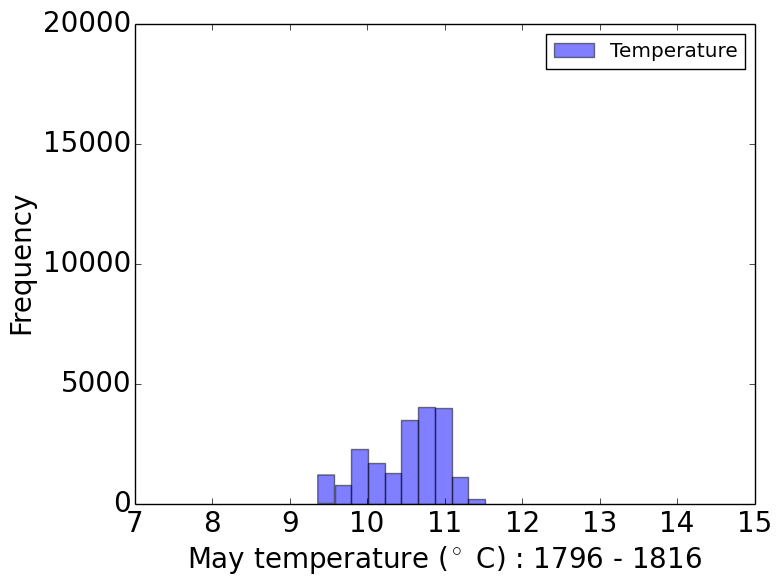}}\\
\subcaptionbox{May -- TMCMC: 1816 - 1836\label{fig:may_t4}}{\includegraphics[width = 2.0in]{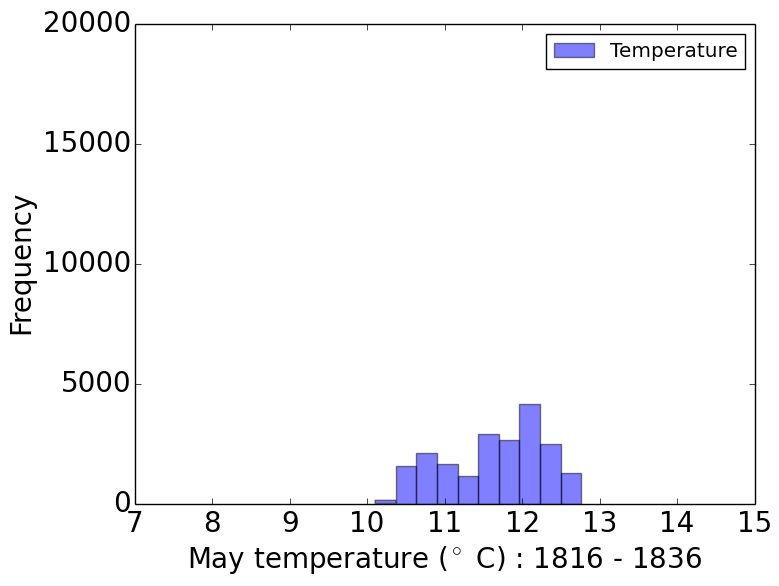}} &
\subcaptionbox{May -- TMCMC: 1836 - 1856\label{fig:may_t5}}{\includegraphics[width = 2.0in]{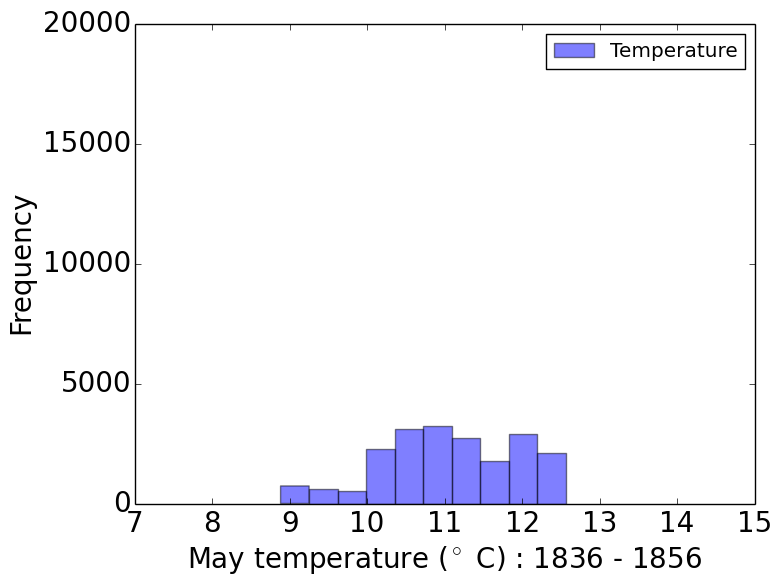}} &
\subcaptionbox{May -- TMCMC: 1856 - 1876\label{fig:may_t6}}{\includegraphics[width = 2.0in]{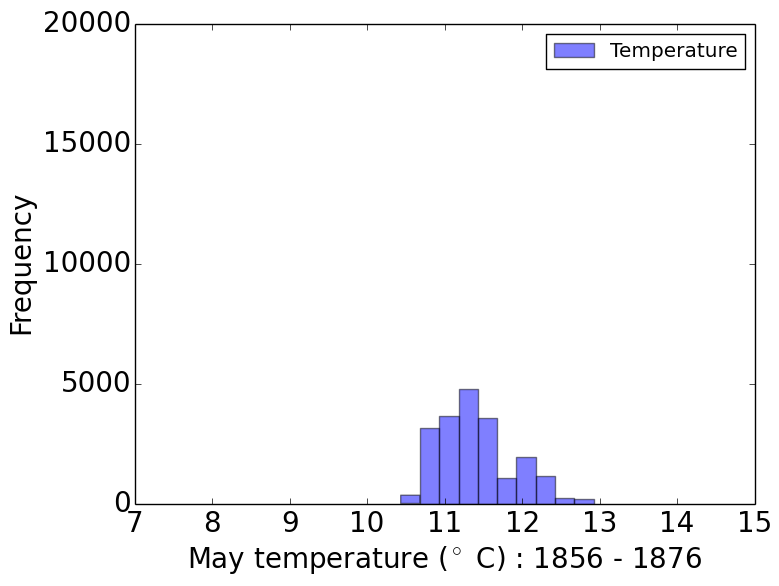}}\\
\subcaptionbox{May -- TMCMC: 1876 - 1896\label{fig:may_t7}}{\includegraphics[width = 2.0in]{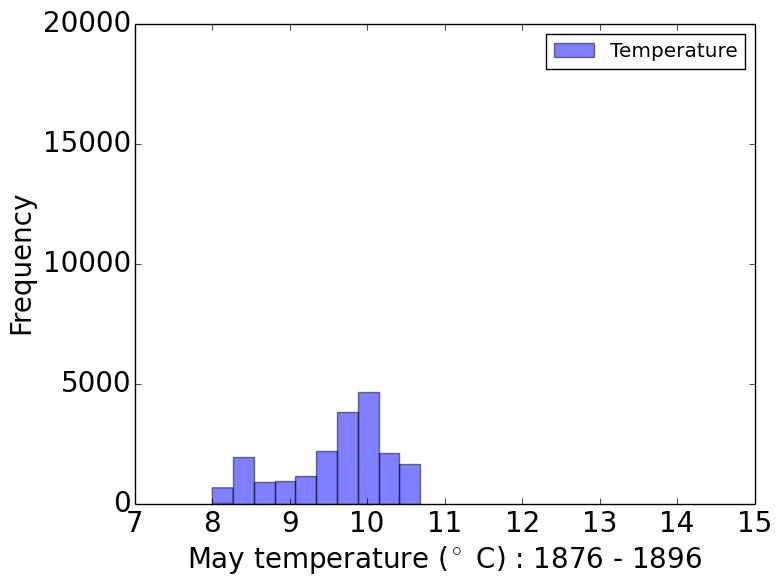}} &
\subcaptionbox{May -- TMCMC: 1896 - 1916\label{fig:may_t8}}{\includegraphics[width = 2.0in]{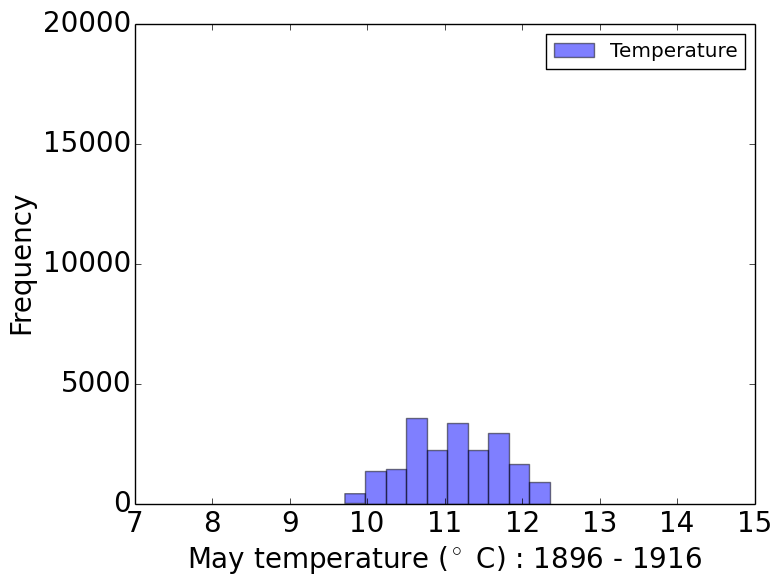}} &
\subcaptionbox{May -- TMCMC: 1916 - 1936\label{fig:may_t9}}{\includegraphics[width = 2.0in]{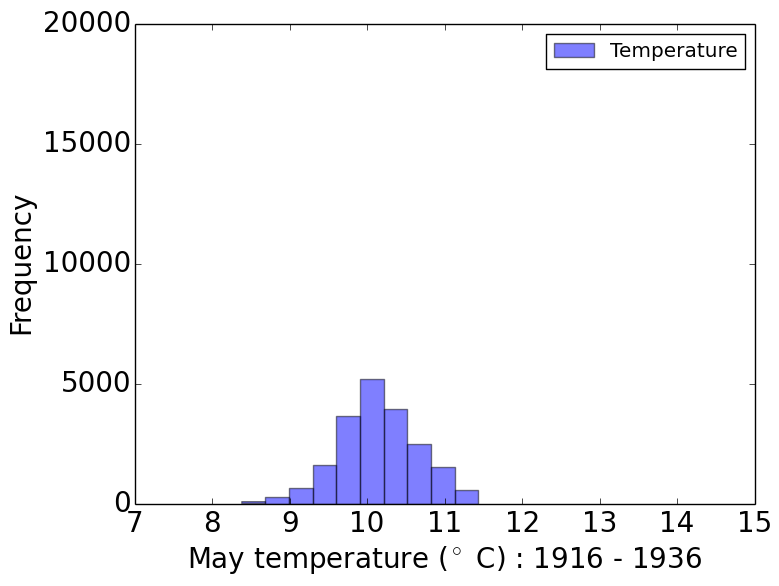}}\\
\subcaptionbox{May -- TMCMC: 1936 - 1936\label{fig:may_t10}}{\includegraphics[width = 2.0in]{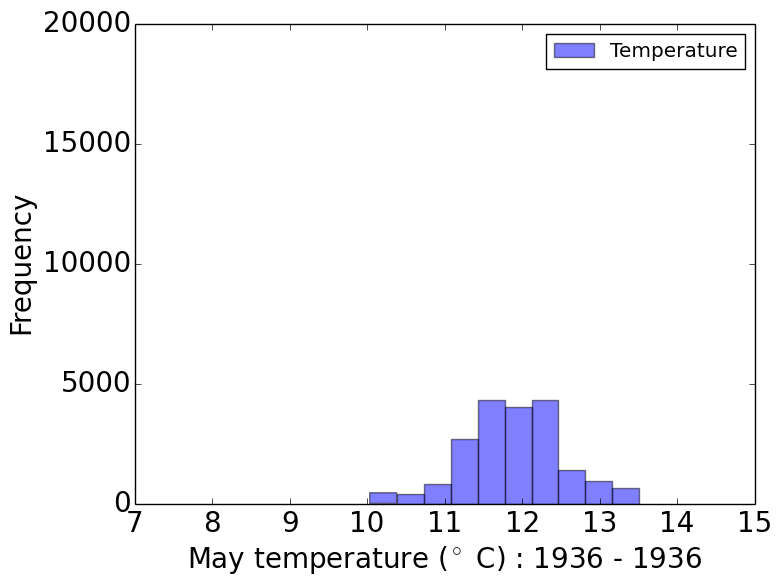}} &
\subcaptionbox{May -- TMCMC: 1956 - 1976\label{fig:may_t11}}{\includegraphics[width = 2.0in]{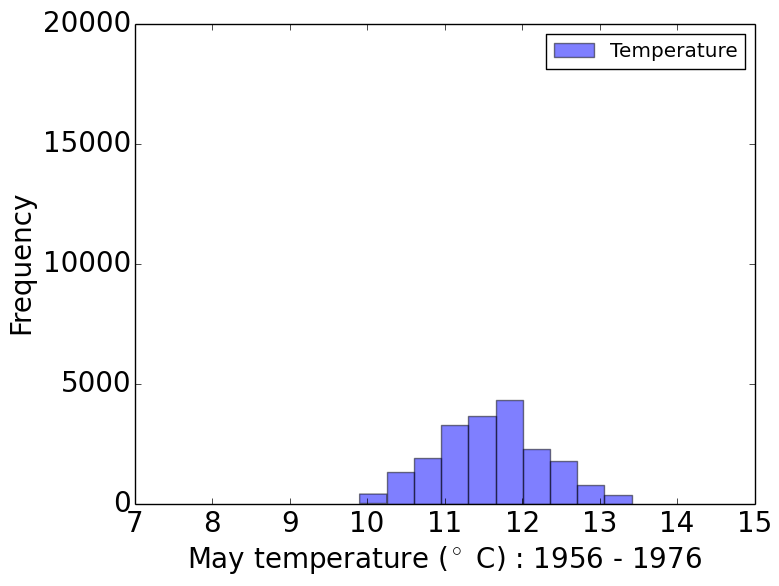}} &
\subcaptionbox{May -- TMCMC: 1976 - 1996\label{fig:may_t12}}{\includegraphics[width = 2.0in]{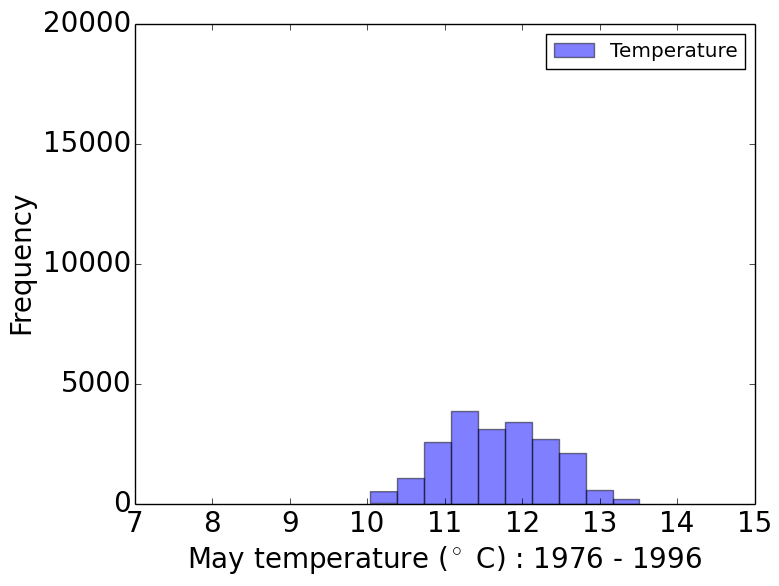}}\\
\subcaptionbox{May -- TMCMC: 1996 - 2016\label{fig:may_t13}}{\includegraphics[width = 2.0in]{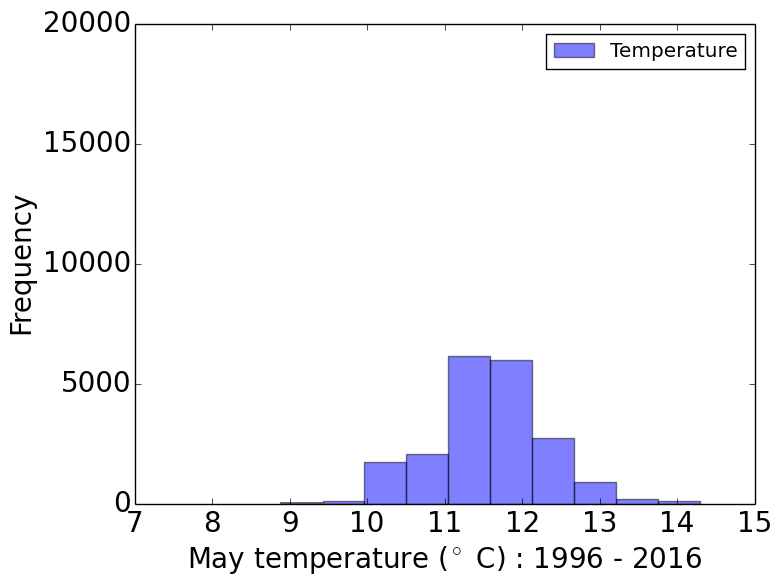}} &
\subcaptionbox{May -- TMCMC: Total\label{fig:may_t14}}{\includegraphics[width = 2.0in]{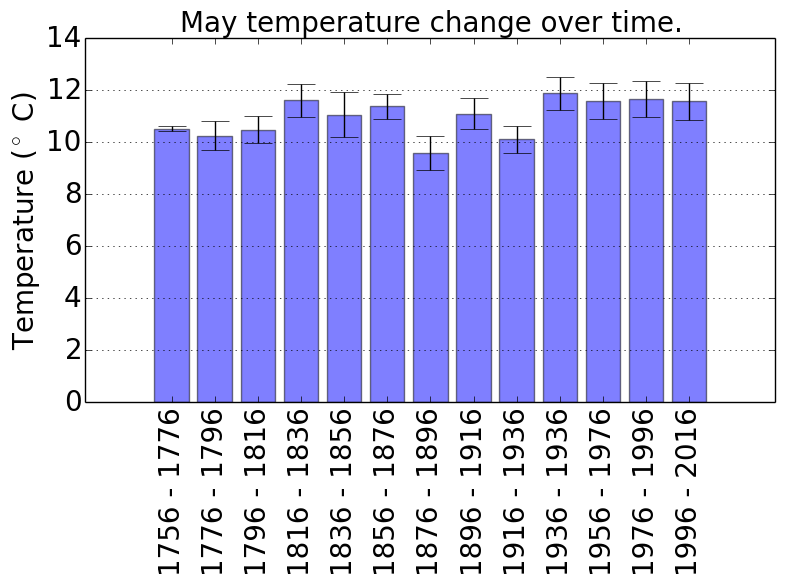}} &
\end{tabular}
\label{fig:may_t1}
\end{adjustwidth}
\end{figure}

\begin{figure}
\begin{adjustwidth}{-6em}{0em}
\centering
\begin{tabular}{ccc}
\subcaptionbox{June -- TMCMC: 1756 - 1776\label{fig:jun_t1}}{\includegraphics[width = 2.0in]{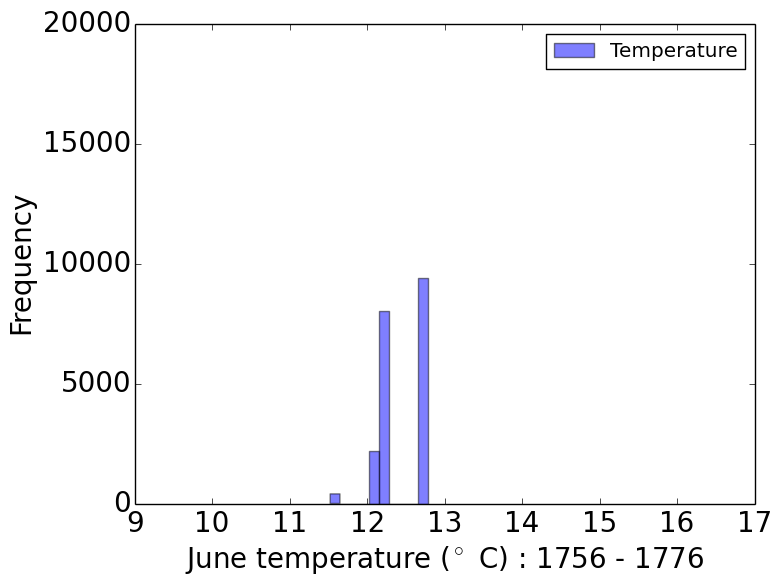}} &
\subcaptionbox{June -- TMCMC: 1776 - 1796\label{fig:jun_t2}}{\includegraphics[width = 2.0in]{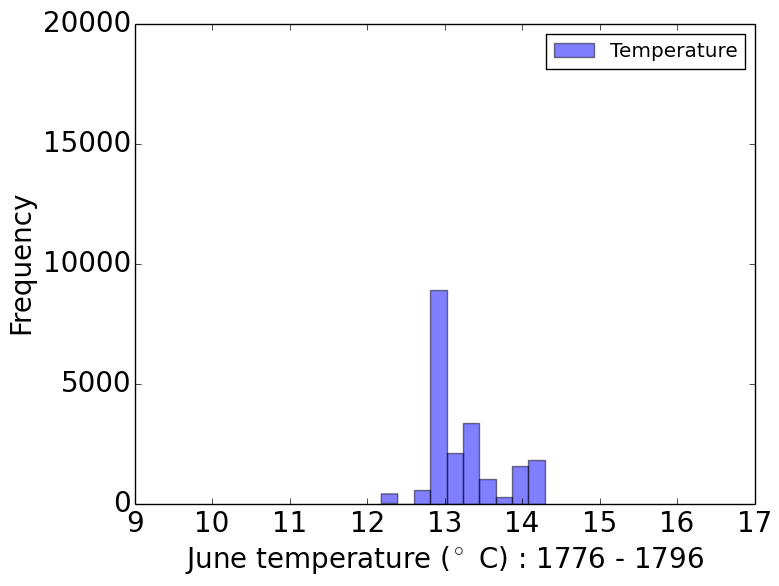}} &
\subcaptionbox{June -- TMCMC: 1796 - 1816\label{fig:jun_t3}}{\includegraphics[width = 2.0in]{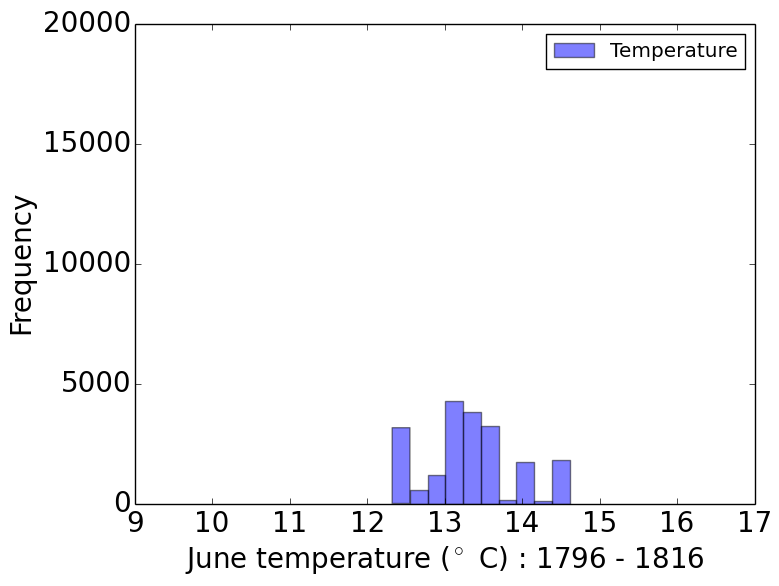}}\\
\subcaptionbox{June -- TMCMC: 1816 - 1836\label{fig:jun_t4}}{\includegraphics[width = 2.0in]{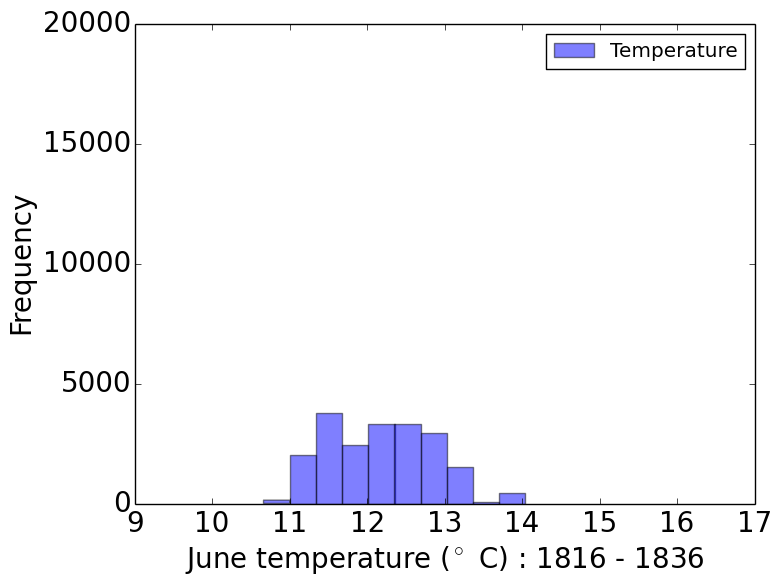}} &
\subcaptionbox{June -- TMCMC: 1836 - 1856\label{fig:jun_t5}}{\includegraphics[width = 2.0in]{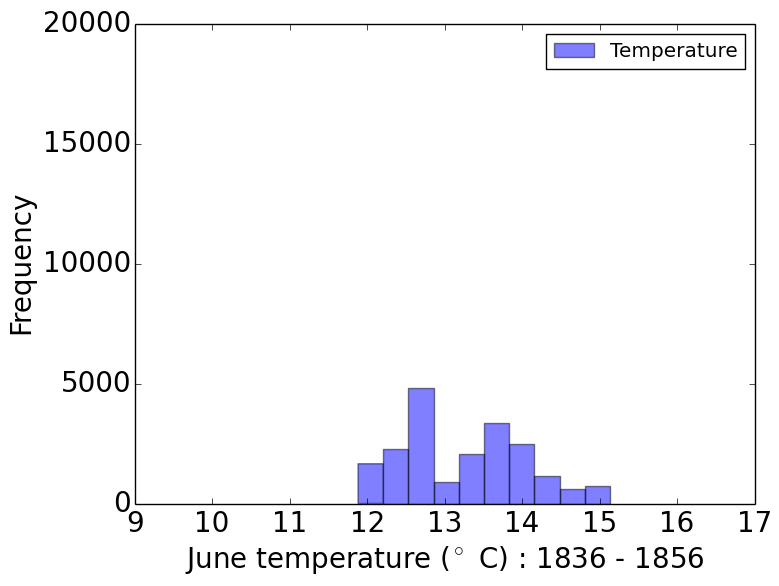}} &
\subcaptionbox{June -- TMCMC: 1856 - 1876\label{fig:jun_t6}}{\includegraphics[width = 2.0in]{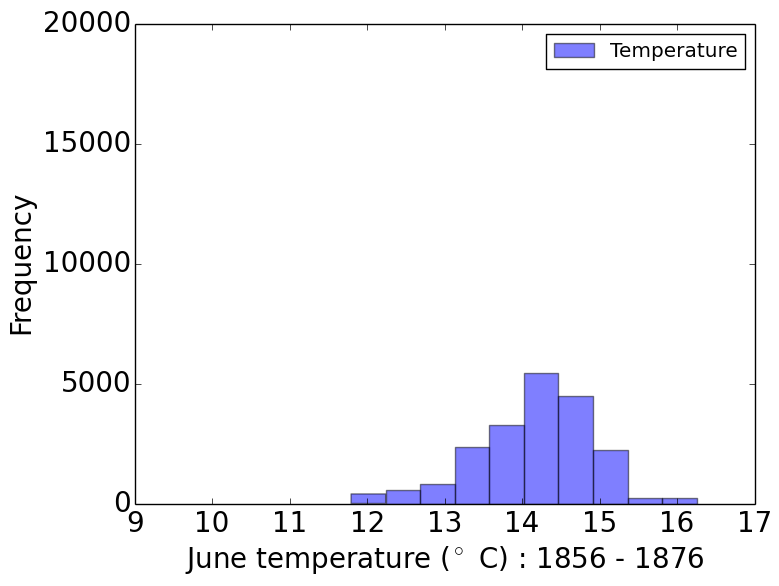}}\\
\subcaptionbox{June -- TMCMC: 1876 - 1896\label{fig:jun_t7}}{\includegraphics[width = 2.0in]{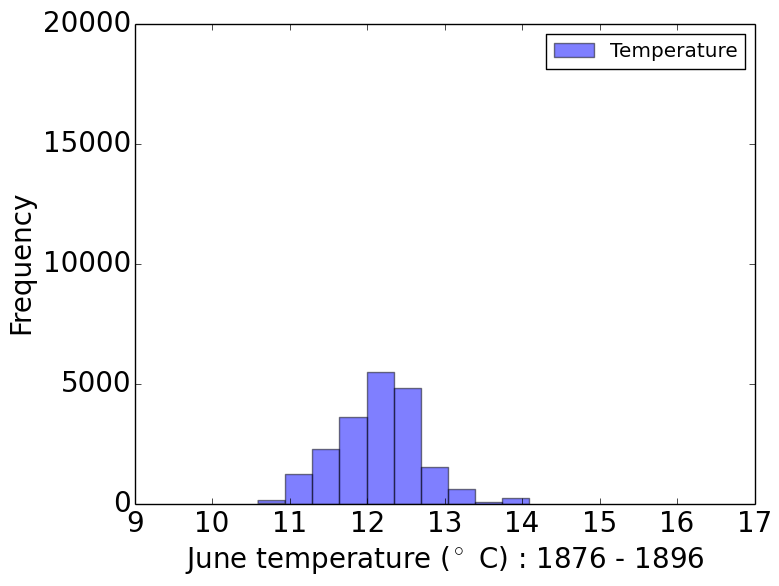}} &
\subcaptionbox{June -- TMCMC: 1896 - 1916\label{fig:jun_t8}}{\includegraphics[width = 2.0in]{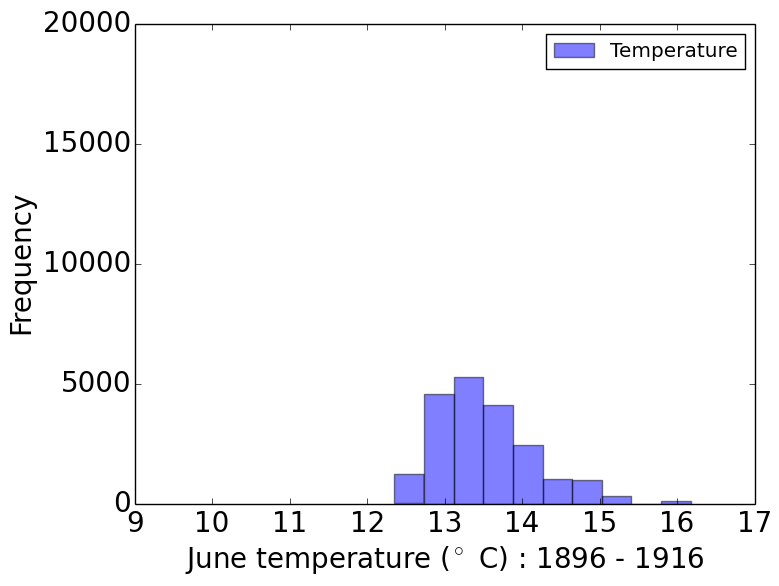}} &
\subcaptionbox{June -- TMCMC: 1916 - 1936\label{fig:jun_t9}}{\includegraphics[width = 2.0in]{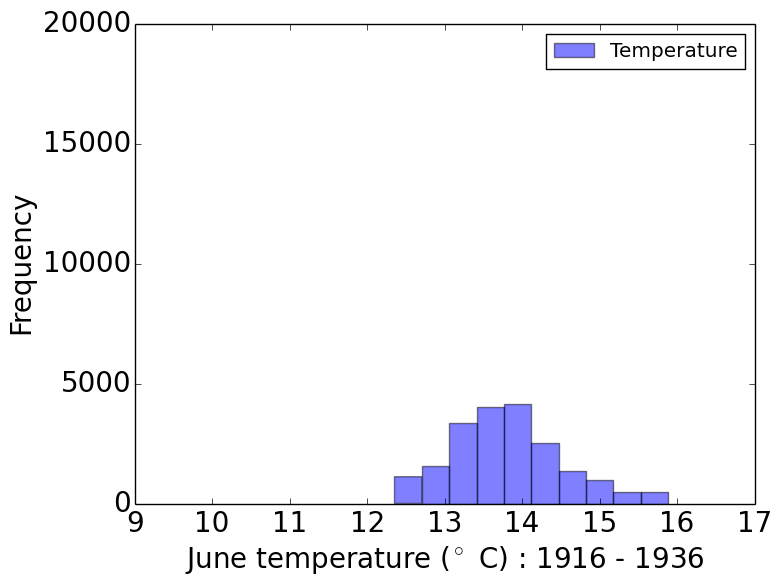}}\\
\subcaptionbox{June -- TMCMC: 1936 - 1936\label{fig:jun_t10}}{\includegraphics[width = 2.0in]{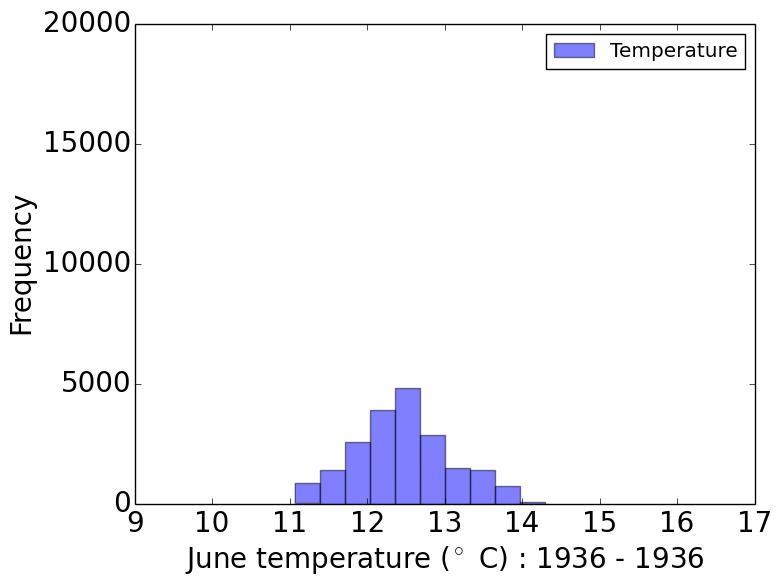}} &
\subcaptionbox{June -- TMCMC: 1956 - 1976\label{fig:jun_t11}}{\includegraphics[width = 2.0in]{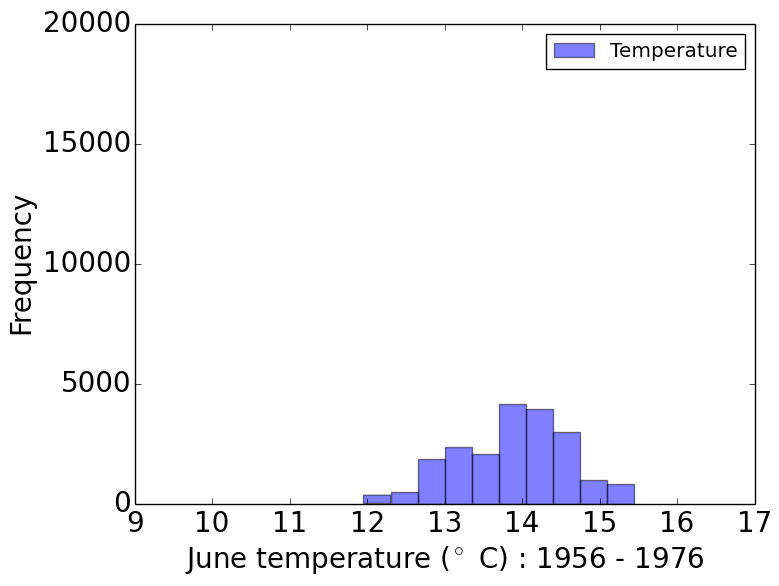}} &
\subcaptionbox{June -- TMCMC: 1976 - 1996\label{fig:jun_t12}}{\includegraphics[width = 2.0in]{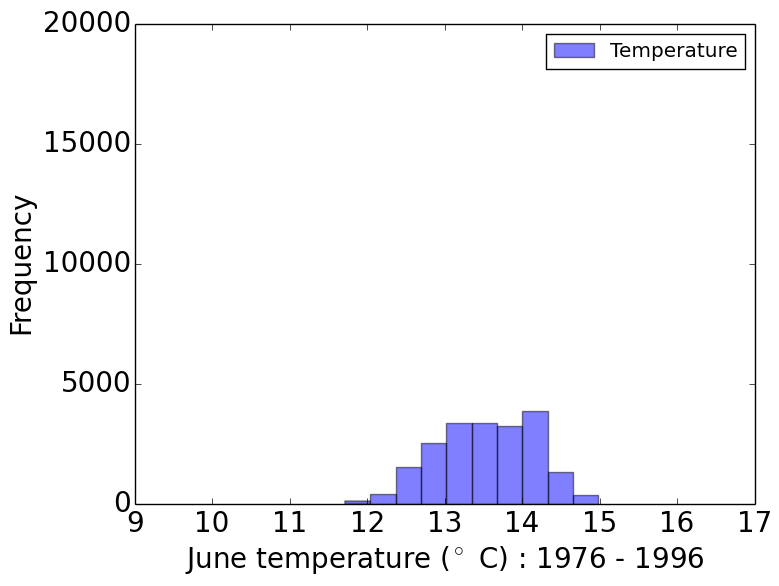}}\\
\subcaptionbox{June -- TMCMC: 1996 - 2016\label{fig:jun_t13}}{\includegraphics[width = 2.0in]{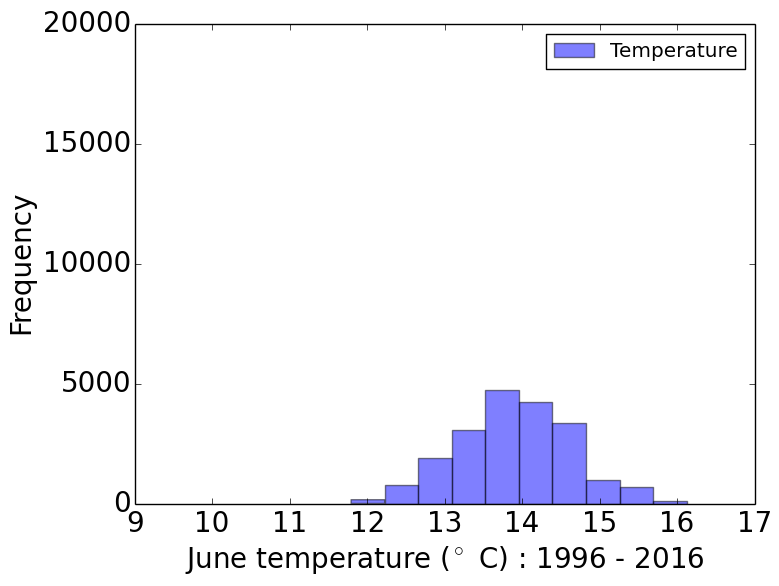}} &
\subcaptionbox{June -- TMCMC: Total\label{fig:jun_t14}}{\includegraphics[width = 2.0in]{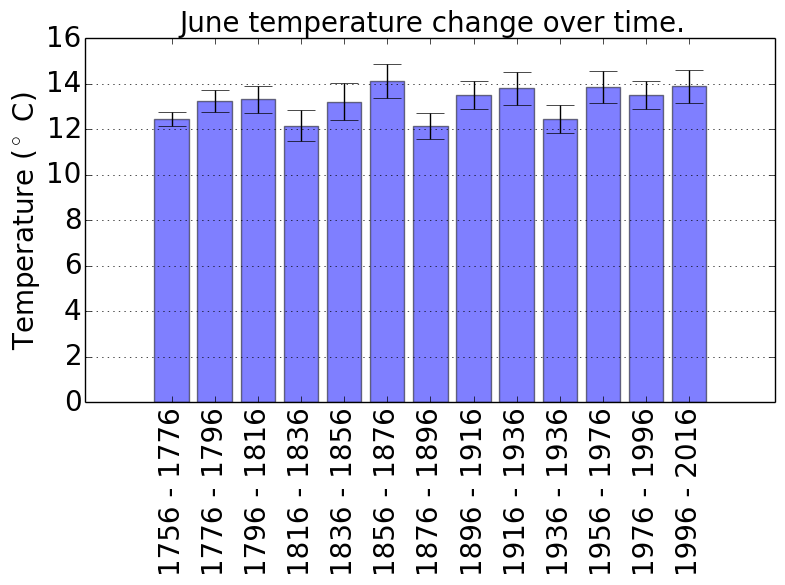}} &
\end{tabular}
\label{fig:jun_t1}
\end{adjustwidth}
\end{figure}

\begin{figure}
\begin{adjustwidth}{-6em}{0em}
\centering
\begin{tabular}{ccc}
\subcaptionbox{July -- TMCMC: 1756 - 1776\label{fig:jul_t1}}{\includegraphics[width = 2.0in]{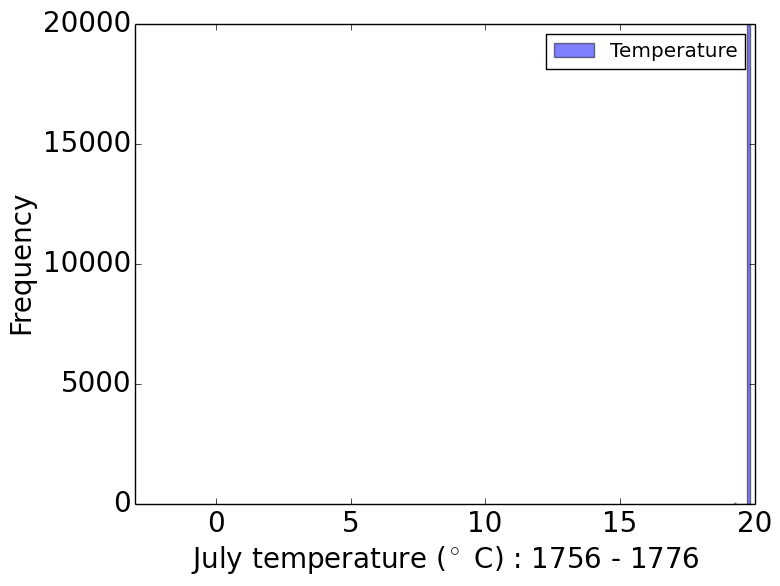}} &
\subcaptionbox{July -- TMCMC: 1776 - 1796\label{fig:jul_t2}}{\includegraphics[width = 2.0in]{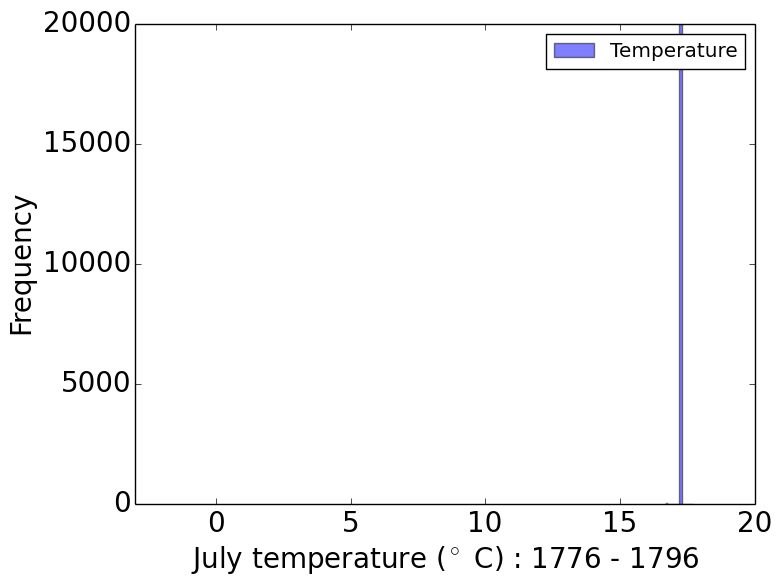}} &
\subcaptionbox{July -- TMCMC: 1796 - 1816\label{fig:jul_t3}}{\includegraphics[width = 2.0in]{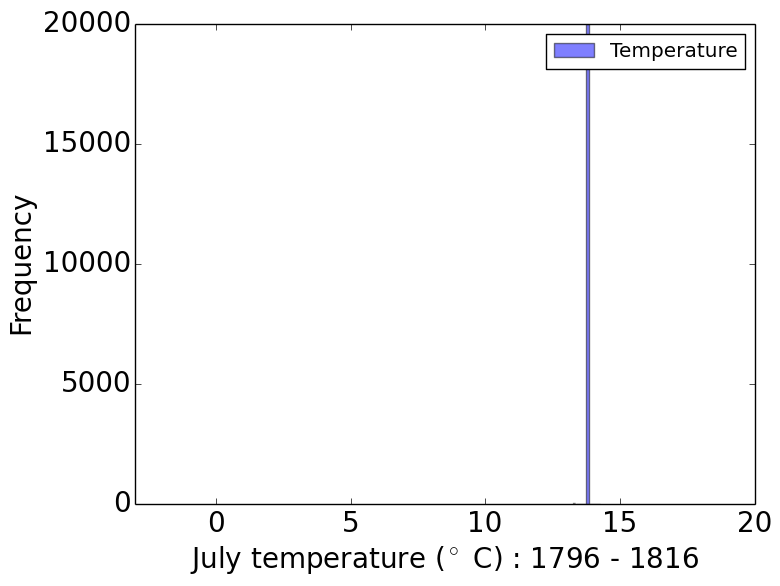}}\\
\subcaptionbox{July -- TMCMC: 1816 - 1836\label{fig:jul_t4}}{\includegraphics[width = 2.0in]{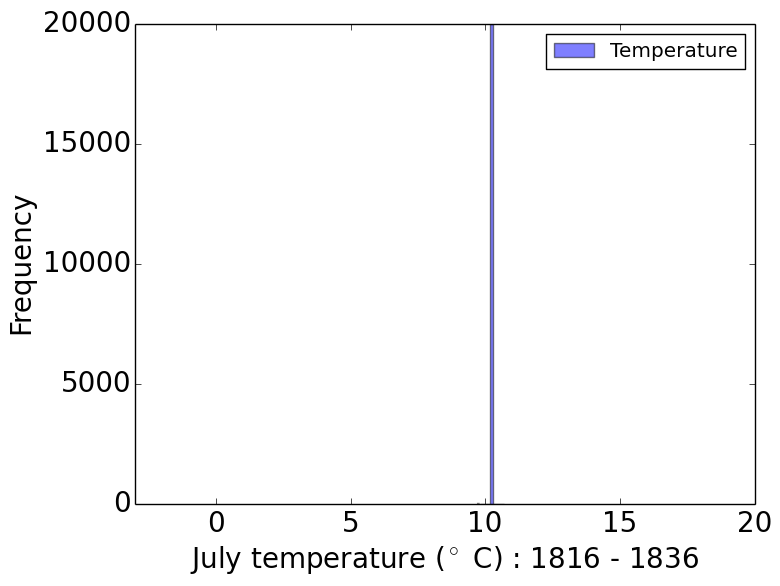}} &
\subcaptionbox{July -- TMCMC: 1836 - 1856\label{fig:jul_t5}}{\includegraphics[width = 2.0in]{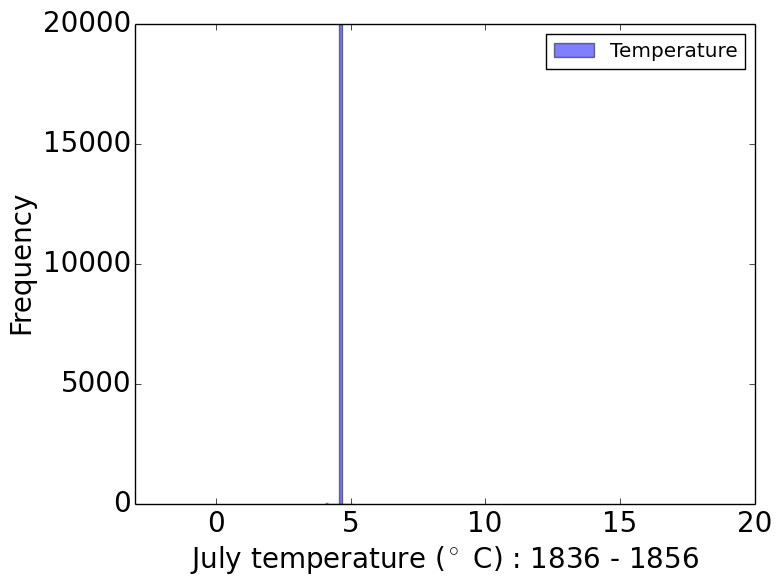}} &
\subcaptionbox{July -- TMCMC: 1856 - 1876\label{fig:jul_t6}}{\includegraphics[width = 2.0in]{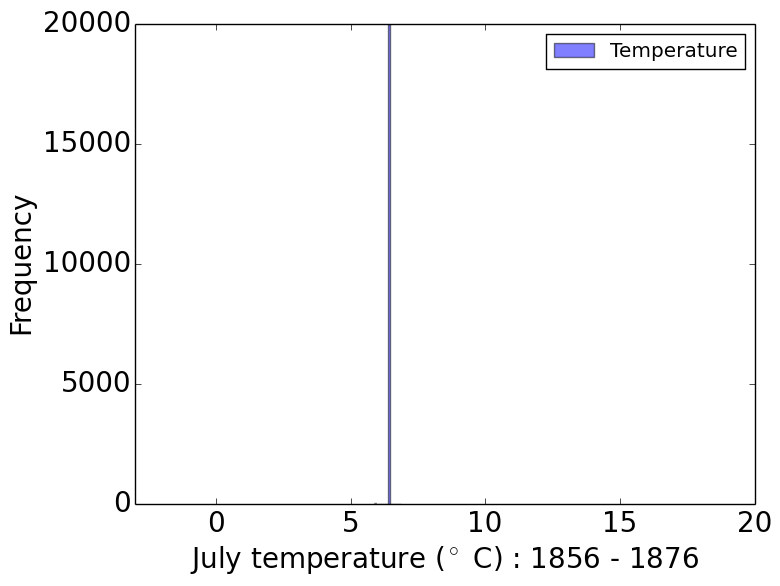}}\\
\subcaptionbox{July -- TMCMC: 1876 - 1896\label{fig:jul_t7}}{\includegraphics[width = 2.0in]{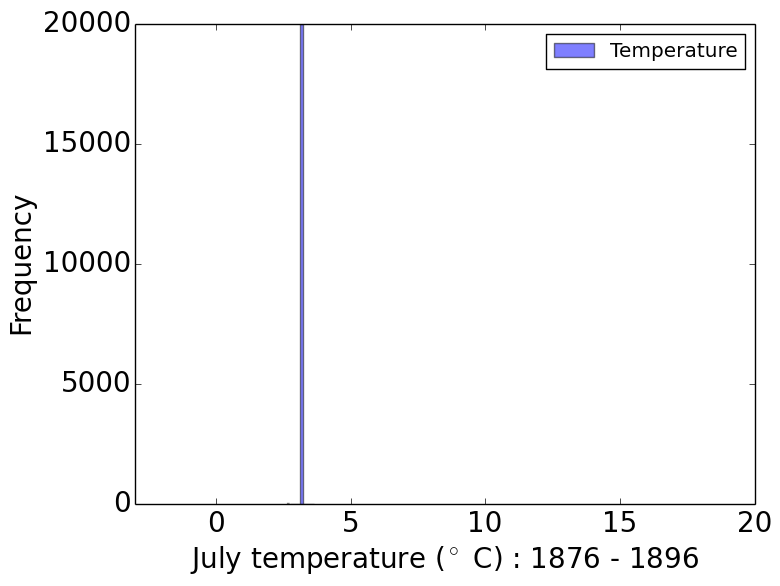}} &
\subcaptionbox{July -- TMCMC: 1896 - 1916\label{fig:jul_t8}}{\includegraphics[width = 2.0in]{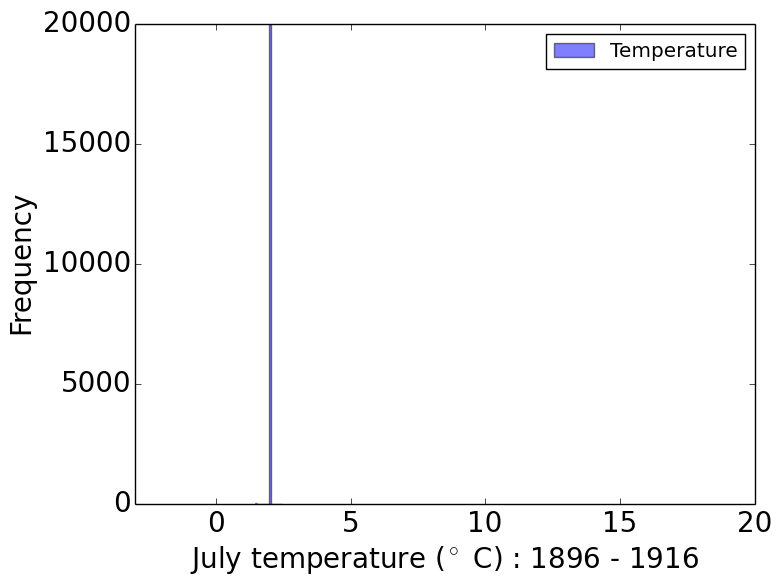}} &
\subcaptionbox{July -- TMCMC: 1916 - 1936\label{fig:jul_t9}}{\includegraphics[width = 2.0in]{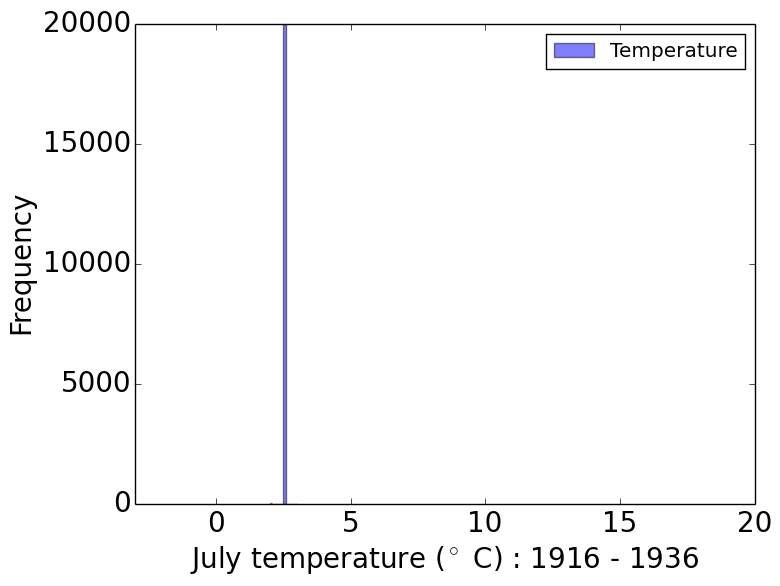}}\\
\subcaptionbox{July -- TMCMC: 1936 - 1936\label{fig:jul_t10}}{\includegraphics[width = 2.0in]{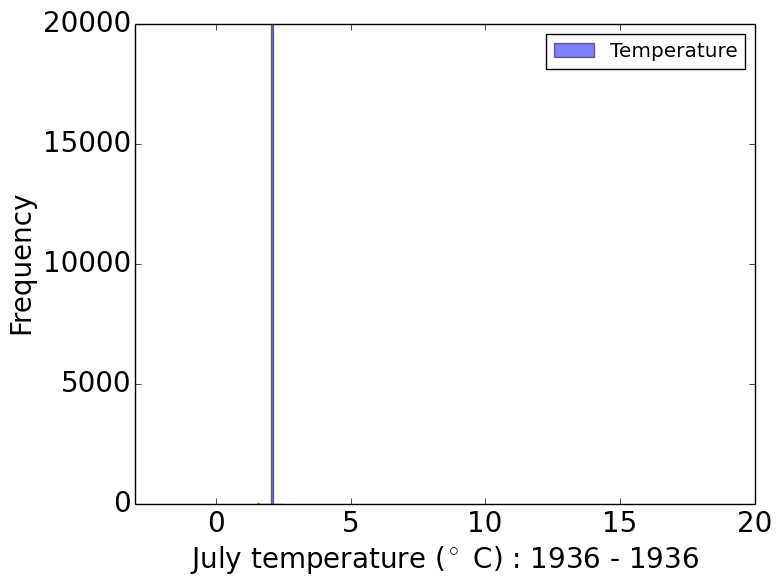}} &
\subcaptionbox{July -- TMCMC: 1956 - 1976\label{fig:jul_t11}}{\includegraphics[width = 2.0in]{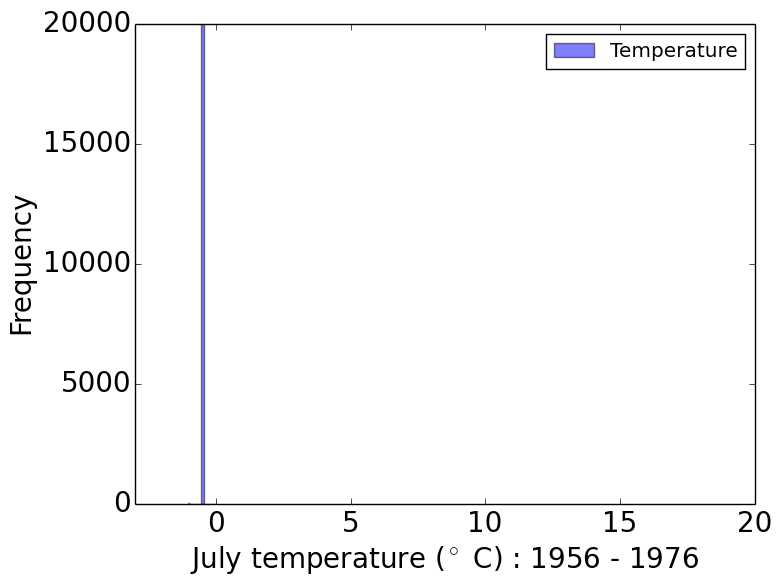}} &
\subcaptionbox{July -- TMCMC: 1976 - 1996\label{fig:jul_t12}}{\includegraphics[width = 2.0in]{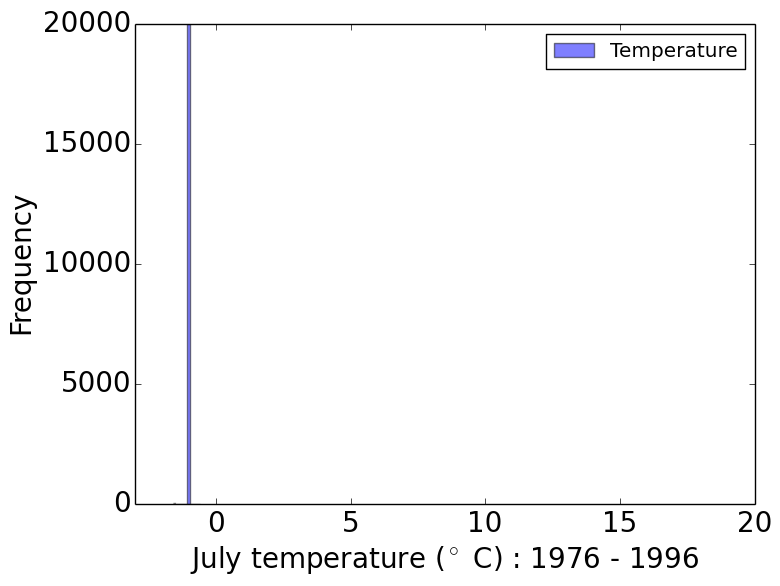}}\\
\subcaptionbox{July -- TMCMC: 1996 - 2016\label{fig:jul_t13}}{\includegraphics[width = 2.0in]{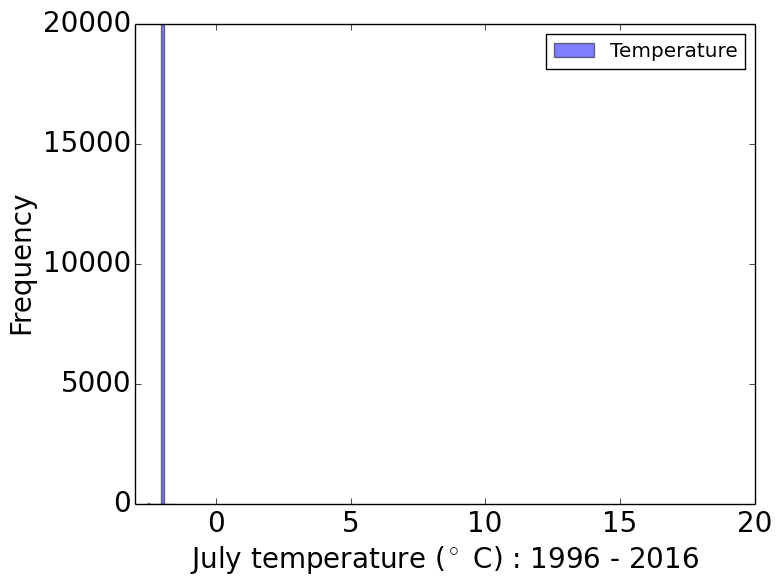}} &
\subcaptionbox{July -- TMCMC: Total\label{fig:jul_t14}}{\includegraphics[width = 2.0in]{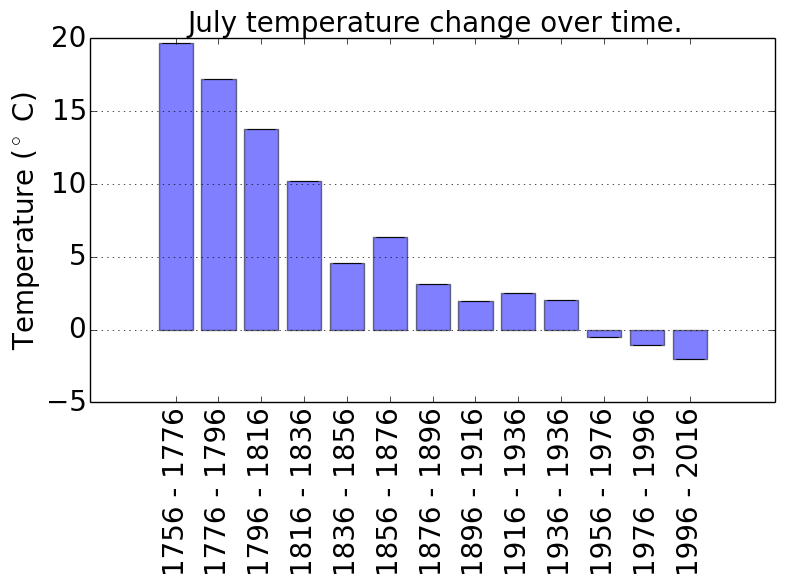}} &
\end{tabular}
\label{fig:jul_t1}
\end{adjustwidth}
\end{figure}

\begin{figure}
\begin{adjustwidth}{-6em}{0em}
\centering
\begin{tabular}{ccc}
\subcaptionbox{August -- TMCMC: 1756 - 1776\label{fig:aug_t1}}{\includegraphics[width = 2.0in]{images/tmcmc/aug/comb_tmcmc_hist_1.png}} &
\subcaptionbox{August -- TMCMC: 1776 - 1796\label{fig:aug_t2}}{\includegraphics[width = 2.0in]{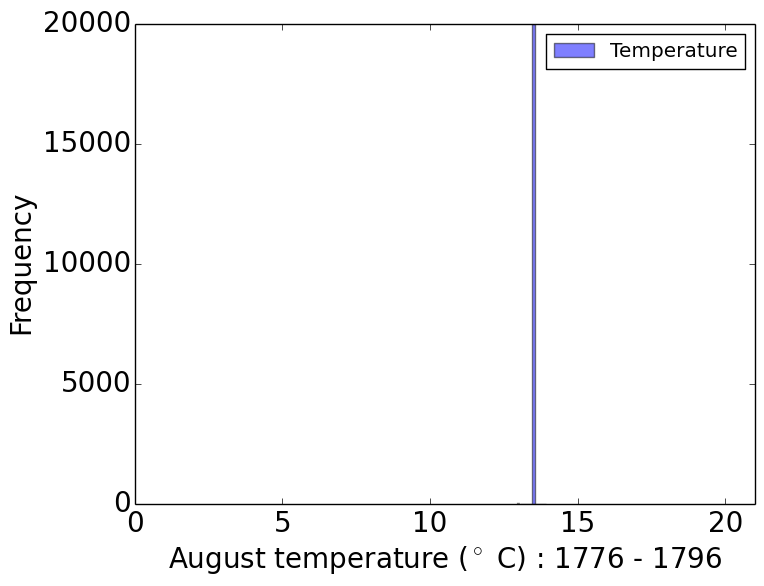}} &
\subcaptionbox{August -- TMCMC: 1796 - 1816\label{fig:aug_t3}}{\includegraphics[width = 2.0in]{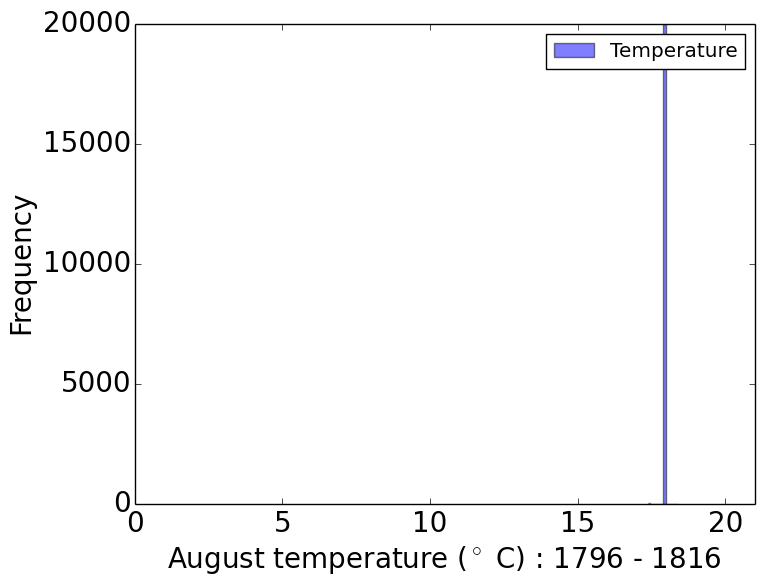}}\\
\subcaptionbox{August -- TMCMC: 1816 - 1836\label{fig:aug_t4}}{\includegraphics[width = 2.0in]{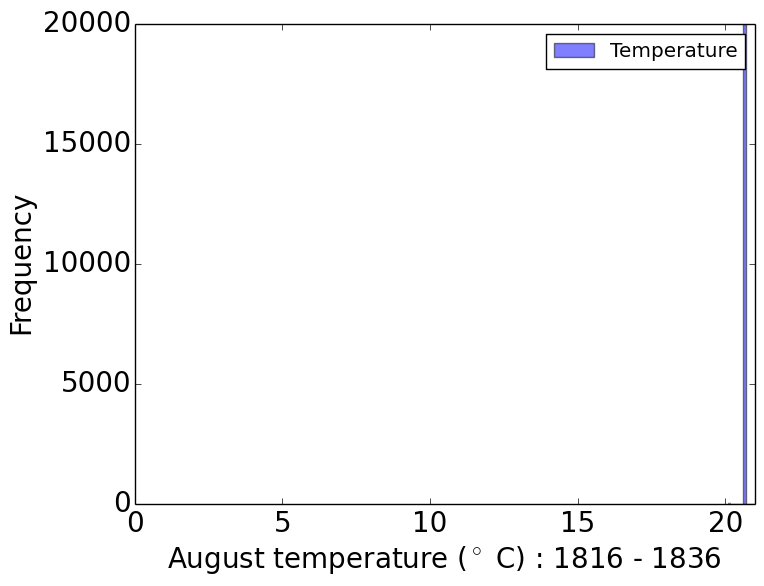}} &
\subcaptionbox{August -- TMCMC: 1836 - 1856\label{fig:aug_t5}}{\includegraphics[width = 2.0in]{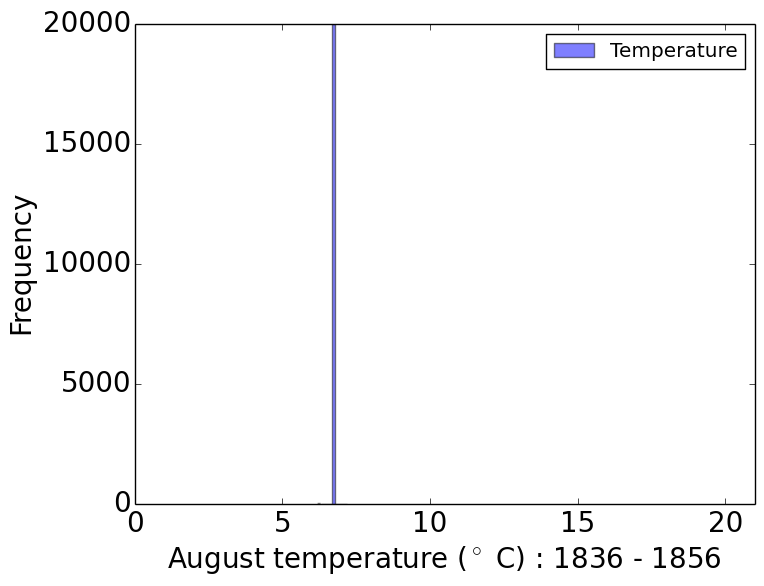}} &
\subcaptionbox{August -- TMCMC: 1856 - 1876\label{fig:aug_t6}}{\includegraphics[width = 2.0in]{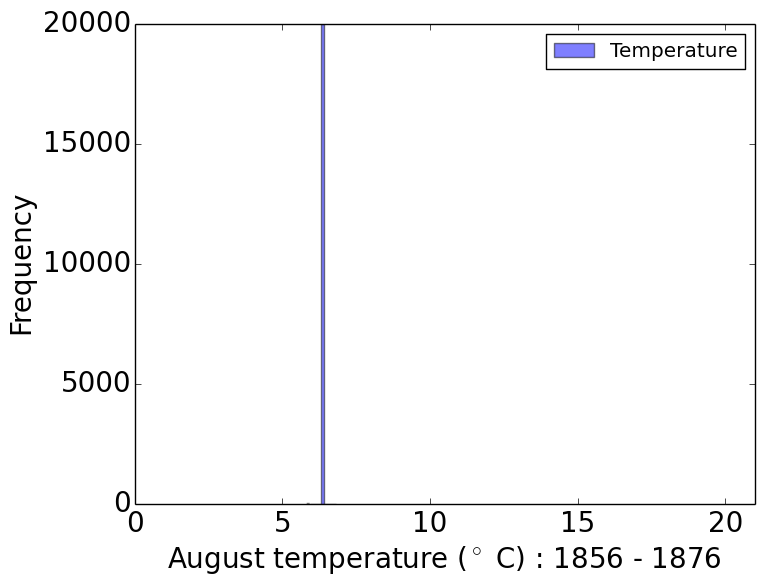}}\\
\subcaptionbox{August -- TMCMC: 1876 - 1896\label{fig:aug_t7}}{\includegraphics[width = 2.0in]{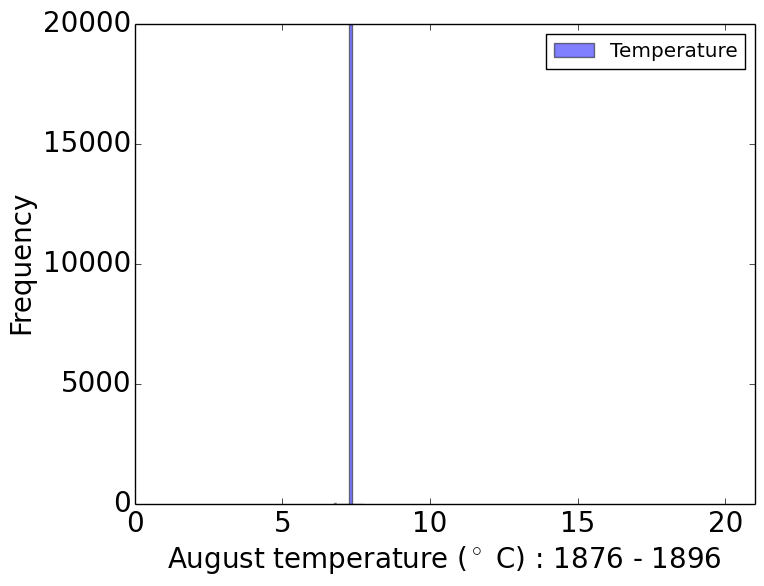}} &
\subcaptionbox{August -- TMCMC: 1896 - 1916\label{fig:aug_t8}}{\includegraphics[width = 2.0in]{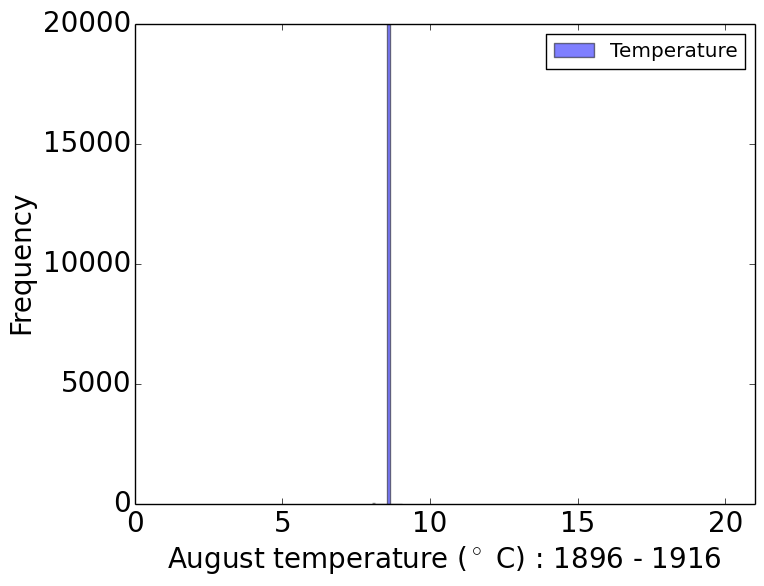}} &
\subcaptionbox{August -- TMCMC: 1916 - 1936\label{fig:aug_t9}}{\includegraphics[width = 2.0in]{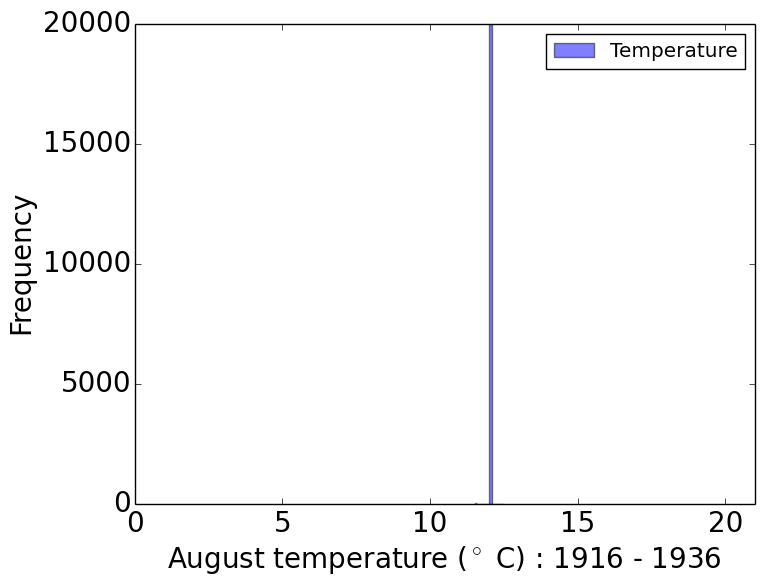}}\\
\subcaptionbox{August -- TMCMC: 1936 - 1936\label{fig:aug_t10}}{\includegraphics[width = 2.0in]{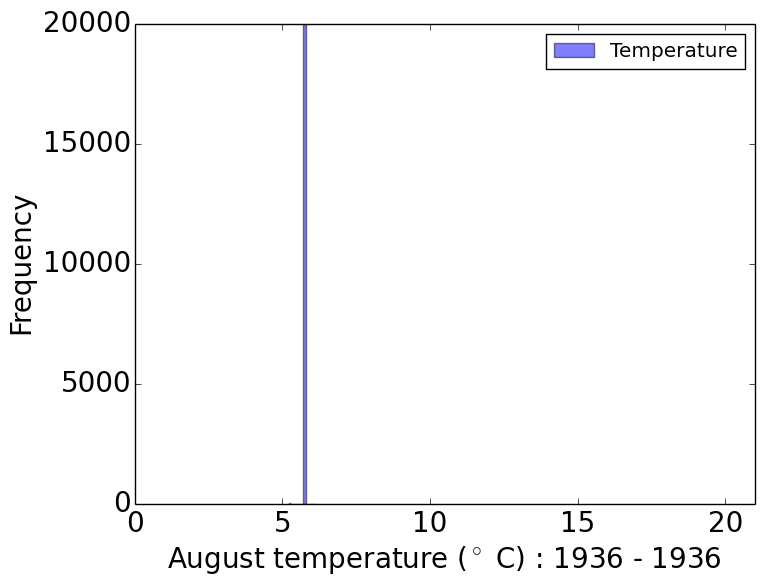}} &
\subcaptionbox{August -- TMCMC: 1956 - 1976\label{fig:aug_t11}}{\includegraphics[width = 2.0in]{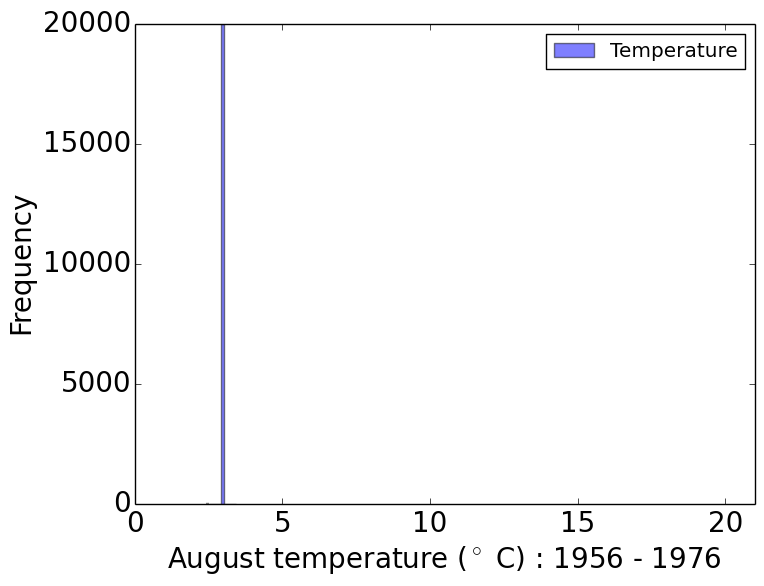}} &
\subcaptionbox{August -- TMCMC: 1976 - 1996\label{fig:aug_t12}}{\includegraphics[width = 2.0in]{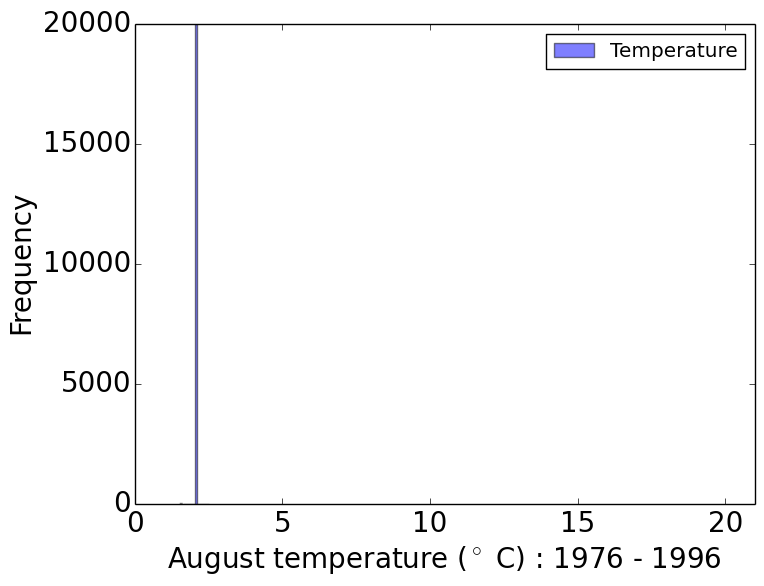}}\\
\subcaptionbox{August -- TMCMC: 1996 - 2016\label{fig:aug_t13}}{\includegraphics[width = 2.0in]{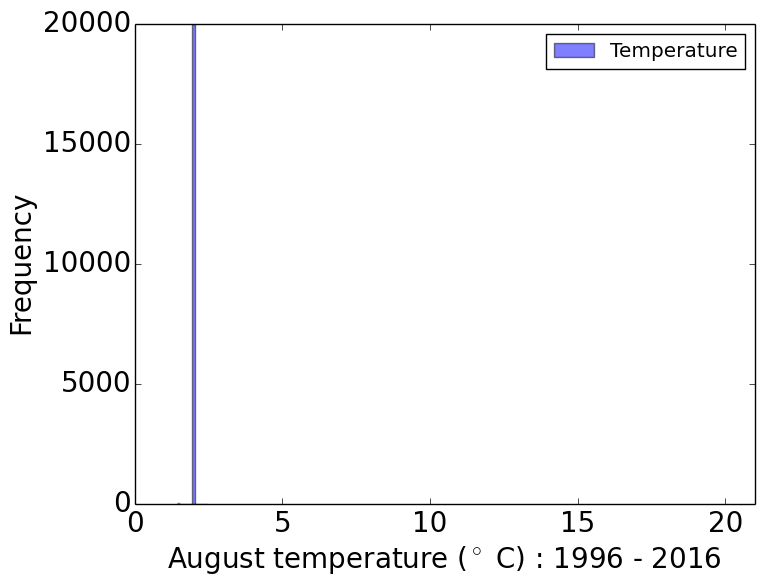}} &
\subcaptionbox{August -- TMCMC: Total\label{fig:aug_t14}}{\includegraphics[width = 2.0in]{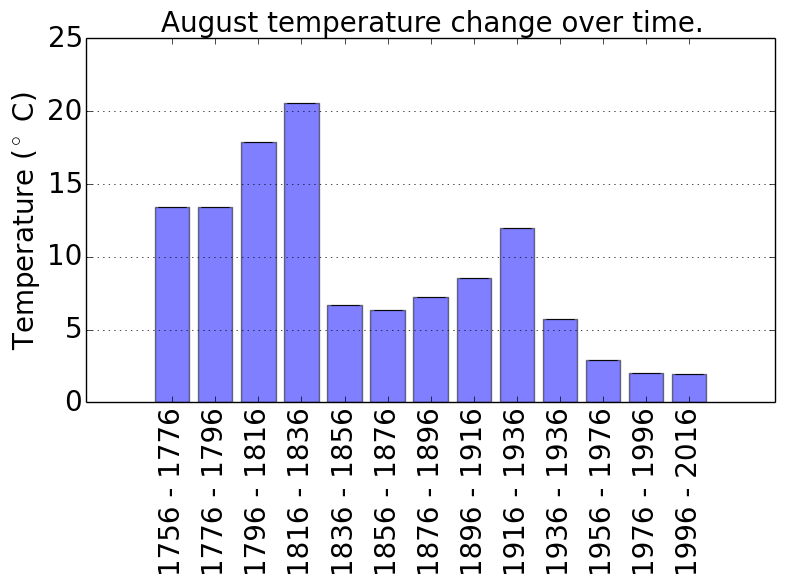}} &
\end{tabular}
\label{fig:aug_t1}
\end{adjustwidth}
\end{figure}

\begin{figure}
\begin{adjustwidth}{-6em}{0em}
\centering
\begin{tabular}{ccc}
\subcaptionbox{September -- TMCMC: 1756 - 1776\label{fig:sep_t1}}{\includegraphics[width = 2.0in]{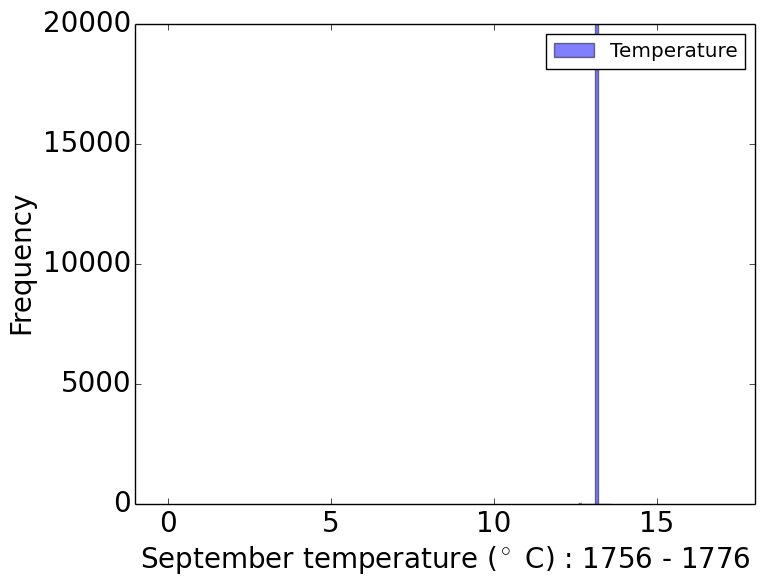}} &
\subcaptionbox{September -- TMCMC: 1776 - 1796\label{fig:sep_t2}}{\includegraphics[width = 2.0in]{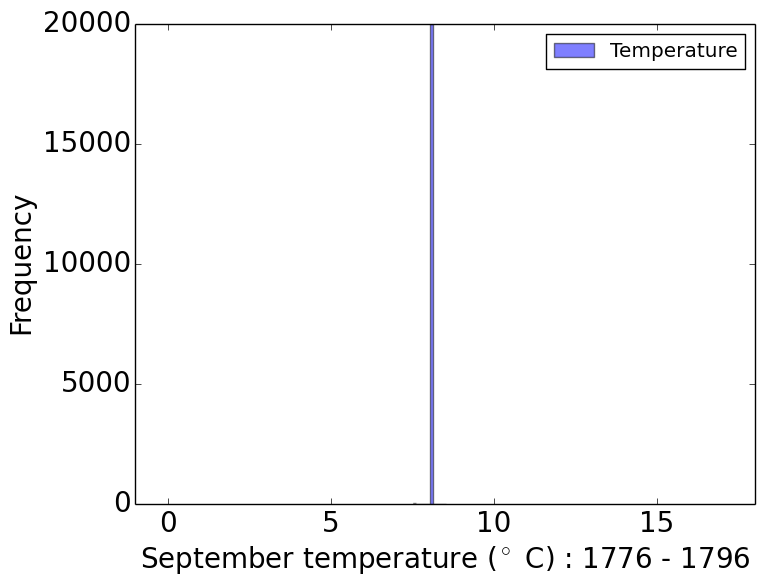}} &
\subcaptionbox{September -- TMCMC: 1796 - 1816\label{fig:sep_t3}}{\includegraphics[width = 2.0in]{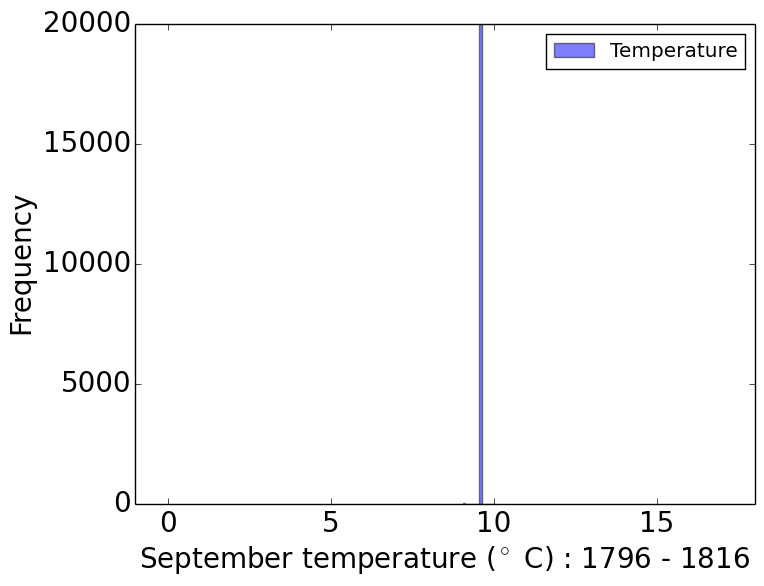}}\\
\subcaptionbox{September -- TMCMC: 1816 - 1836\label{fig:sep_t4}}{\includegraphics[width = 2.0in]{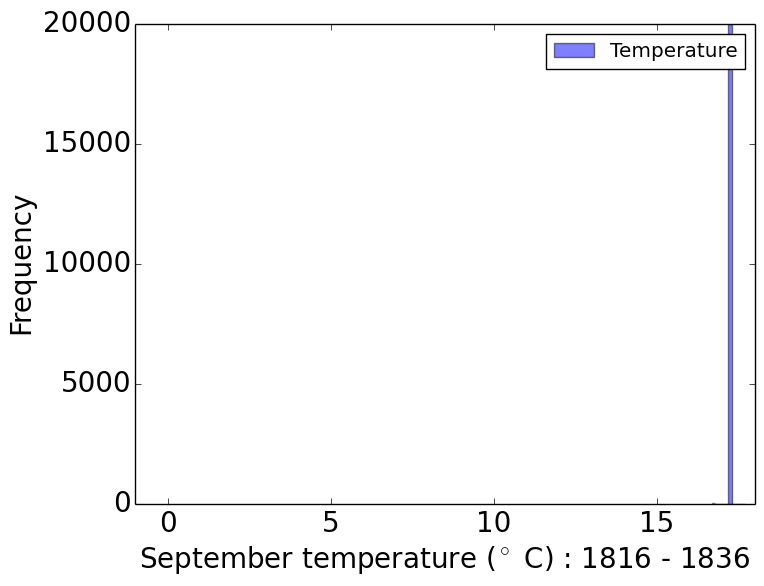}} &
\subcaptionbox{September -- TMCMC: 1836 - 1856\label{fig:sep_t5}}{\includegraphics[width = 2.0in]{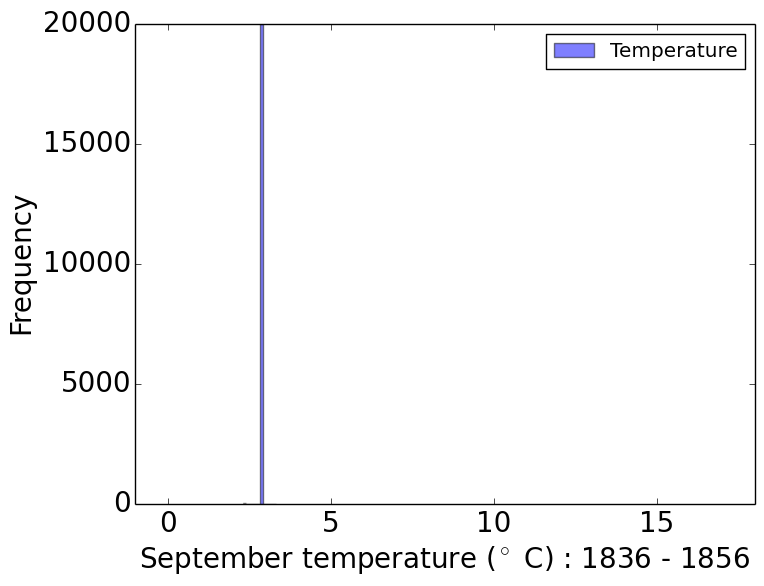}} &
\subcaptionbox{September -- TMCMC: 1856 - 1876\label{fig:sep_t6}}{\includegraphics[width = 2.0in]{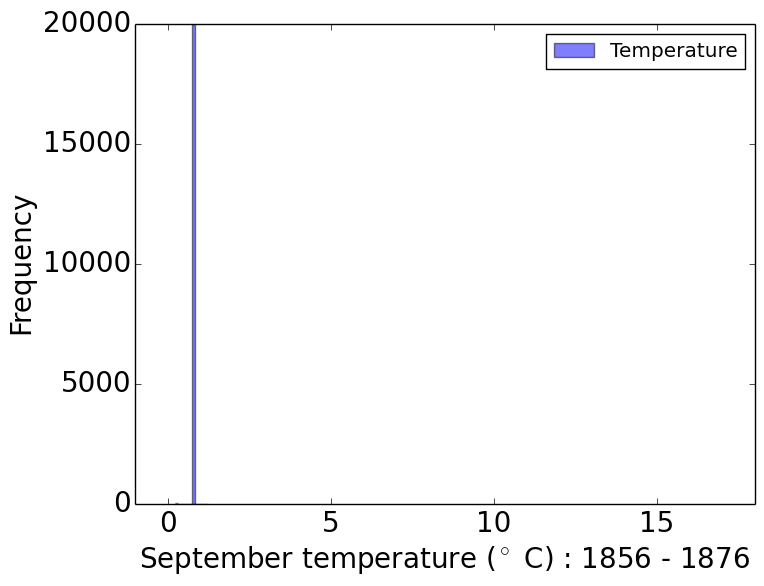}}\\
\subcaptionbox{September -- TMCMC: 1876 - 1896\label{fig:sep_t7}}{\includegraphics[width = 2.0in]{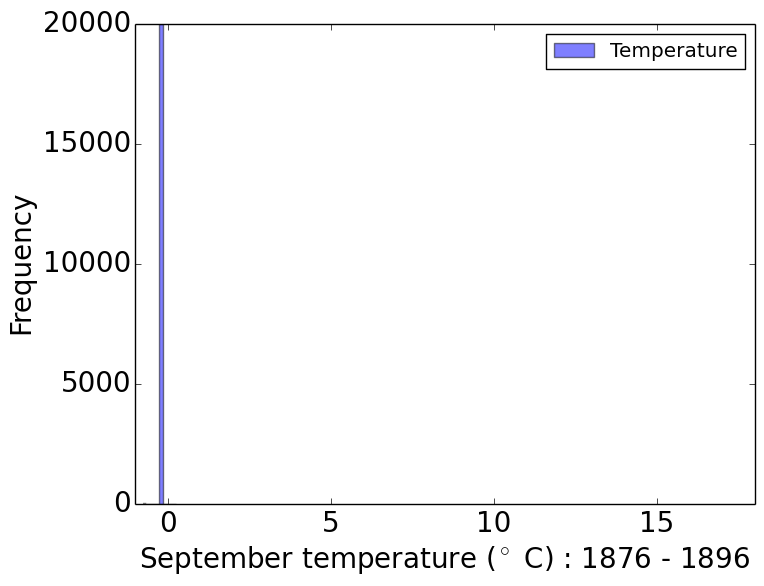}} &
\subcaptionbox{September -- TMCMC: 1896 - 1916\label{fig:sep_t8}}{\includegraphics[width = 2.0in]{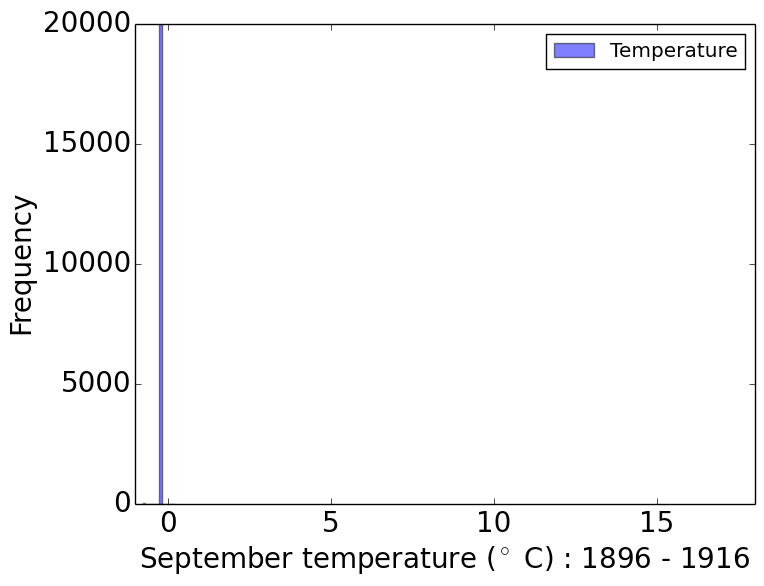}} &
\subcaptionbox{September -- TMCMC: 1916 - 1936\label{fig:sep_t9}}{\includegraphics[width = 2.0in]{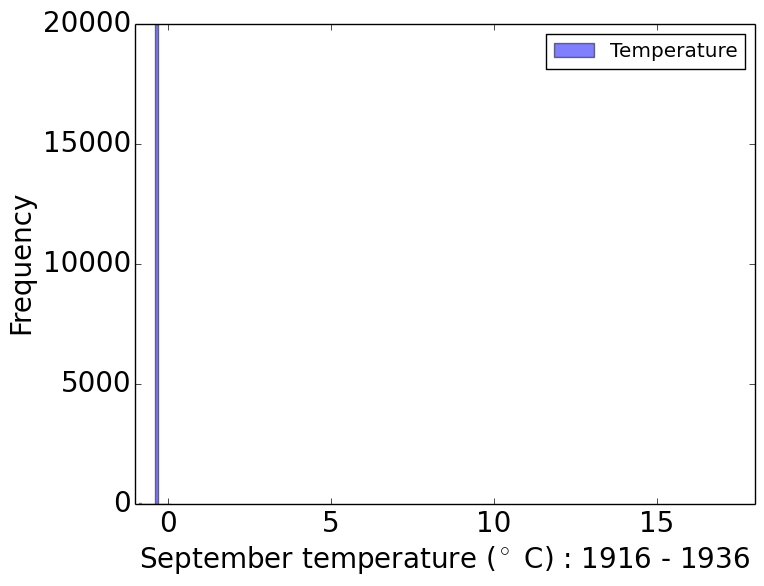}}\\
\subcaptionbox{September -- TMCMC: 1936 - 1936\label{fig:sep_t10}}{\includegraphics[width = 2.0in]{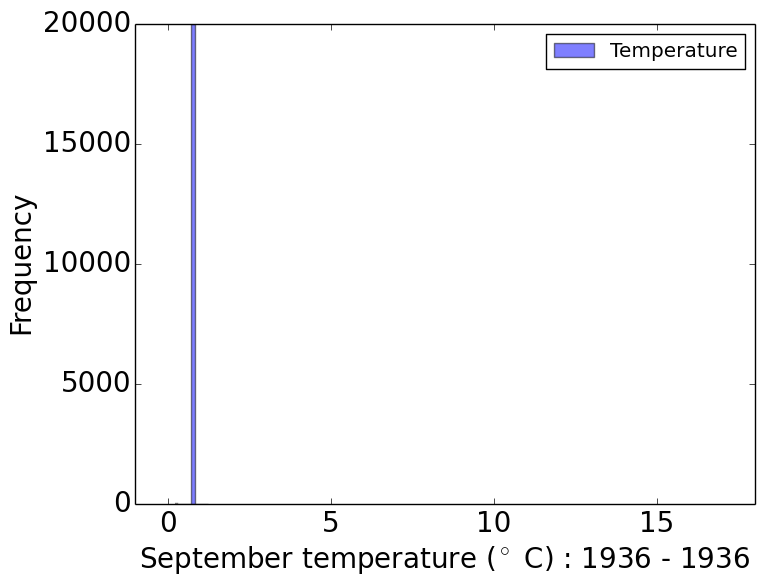}} &
\subcaptionbox{September -- TMCMC: 1956 - 1976\label{fig:sep_t11}}{\includegraphics[width = 2.0in]{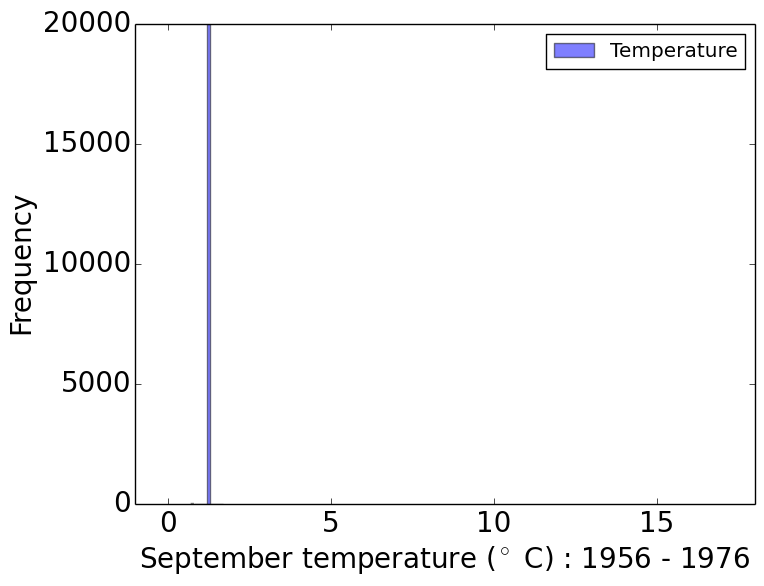}} &
\subcaptionbox{September -- TMCMC: 1976 - 1996\label{fig:sep_t12}}{\includegraphics[width = 2.0in]{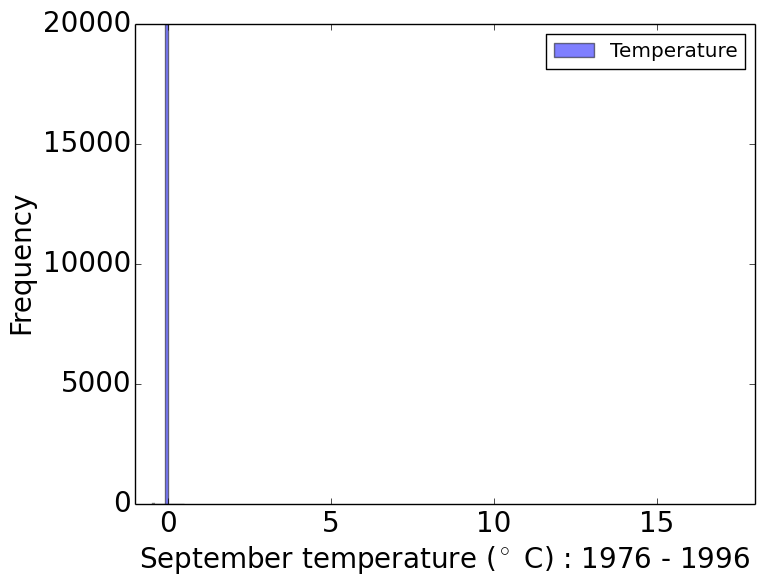}}\\
\subcaptionbox{September -- TMCMC: 1996 - 2016\label{fig:sep_t13}}{\includegraphics[width = 2.0in]{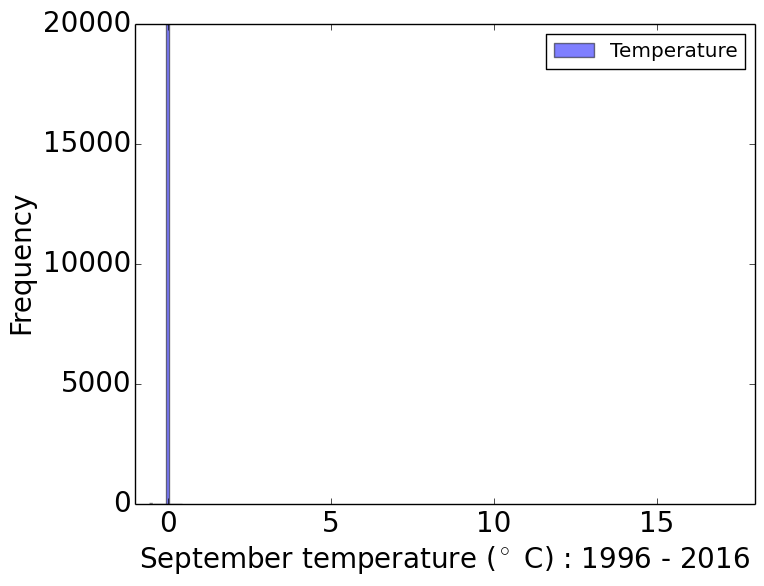}} &
\subcaptionbox{September -- TMCMC: Total\label{fig:sep_t14}}{\includegraphics[width = 2.0in]{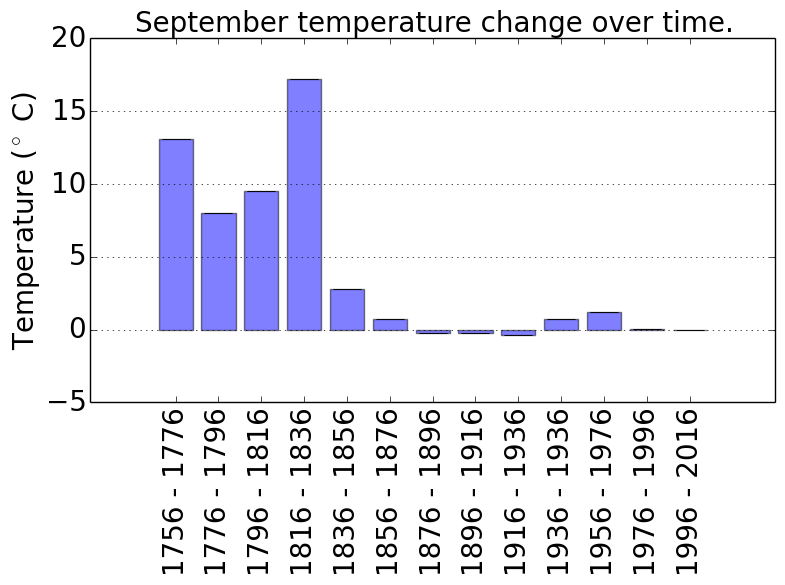}} &
\end{tabular}
\label{fig:sep_t1}
\end{adjustwidth}
\end{figure}

\begin{figure}
\begin{adjustwidth}{-6em}{0em}
\centering
\begin{tabular}{ccc}
\subcaptionbox{October -- TMCMC: 1756 - 1776\label{fig:oct_t1}}{\includegraphics[width = 2.0in]{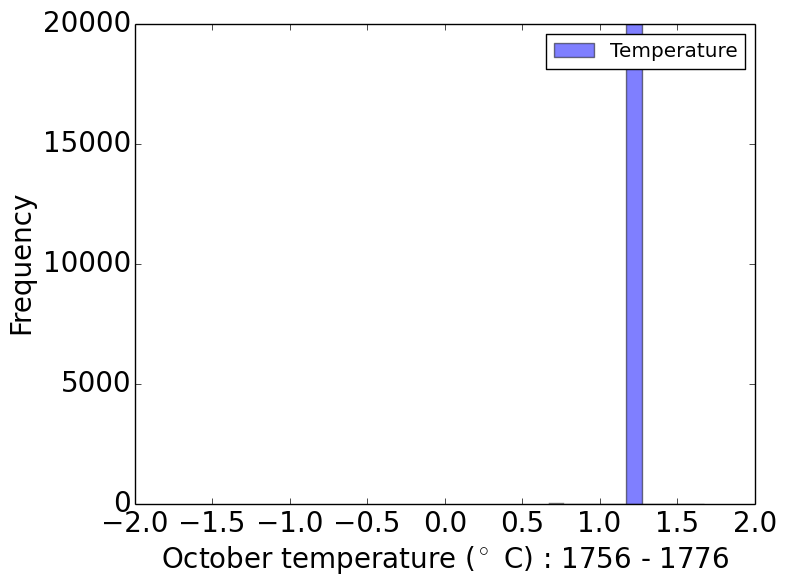}} &
\subcaptionbox{October -- TMCMC: 1776 - 1796\label{fig:oct_t2}}{\includegraphics[width = 2.0in]{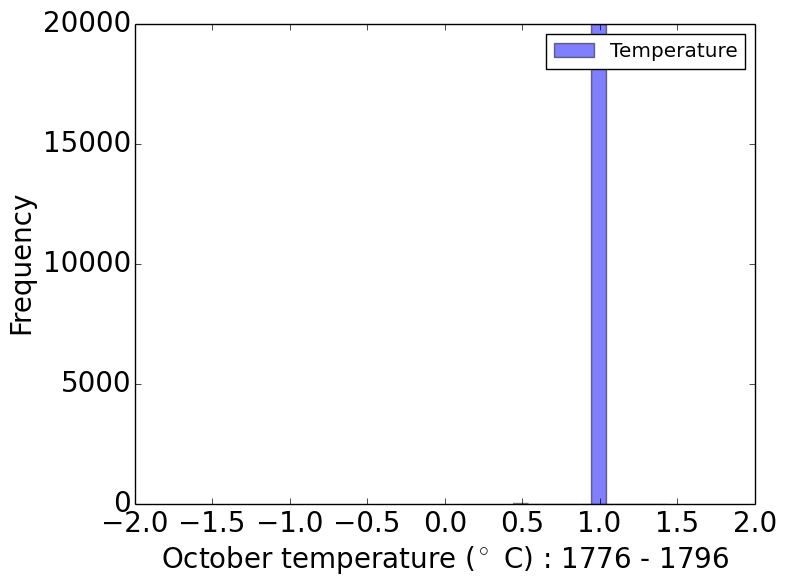}} &
\subcaptionbox{October -- TMCMC: 1796 - 1816\label{fig:oct_t3}}{\includegraphics[width = 2.0in]{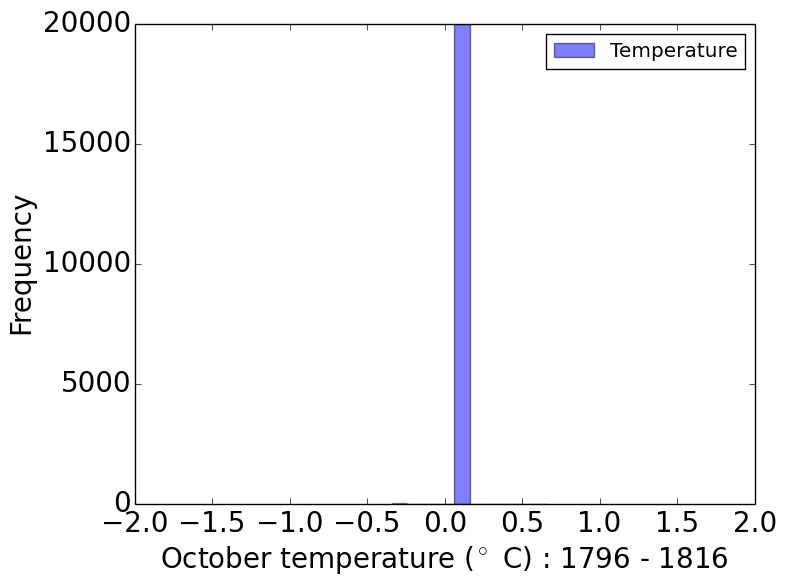}}\\
\subcaptionbox{October -- TMCMC: 1816 - 1836\label{fig:oct_t4}}{\includegraphics[width = 2.0in]{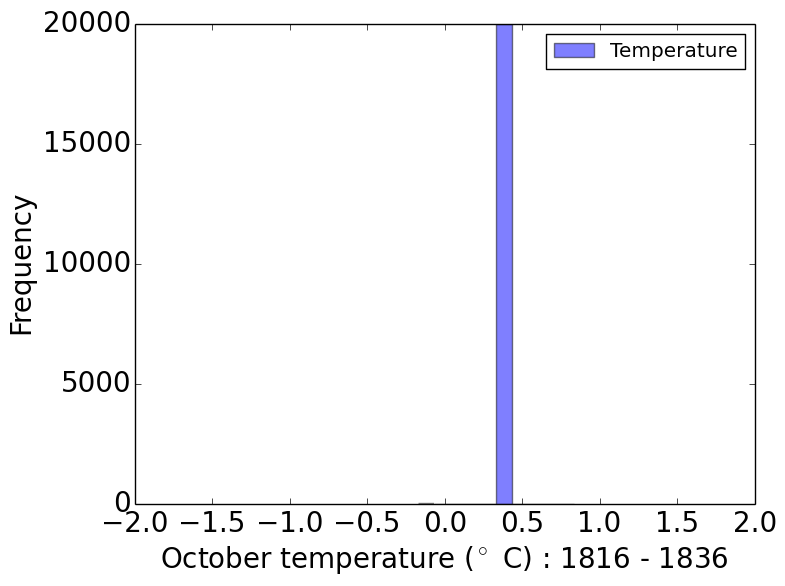}} &
\subcaptionbox{October -- TMCMC: 1836 - 1856\label{fig:oct_t5}}{\includegraphics[width = 2.0in]{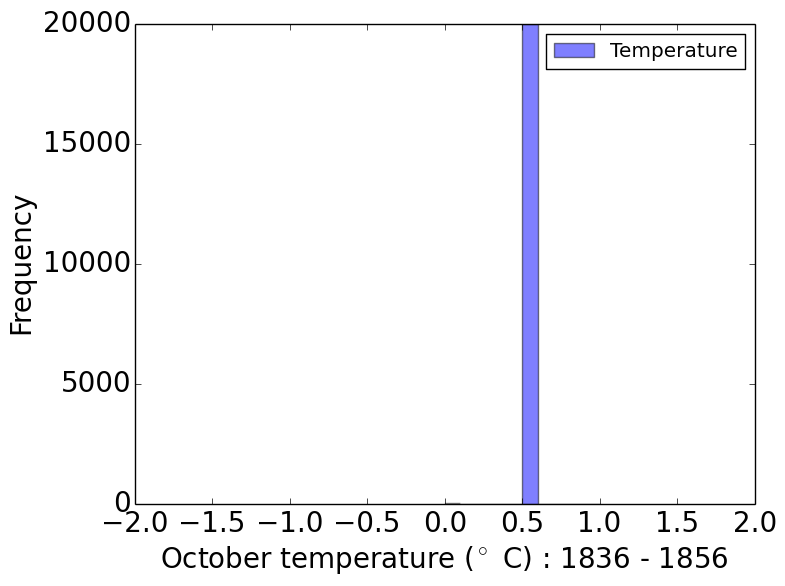}} &
\subcaptionbox{October -- TMCMC: 1856 - 1876\label{fig:oct_t6}}{\includegraphics[width = 2.0in]{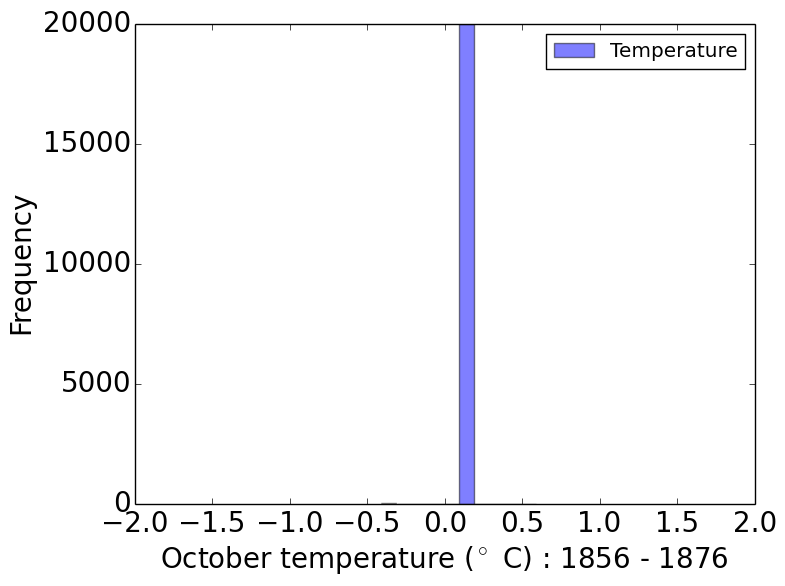}}\\
\subcaptionbox{October -- TMCMC: 1876 - 1896\label{fig:oct_t7}}{\includegraphics[width = 2.0in]{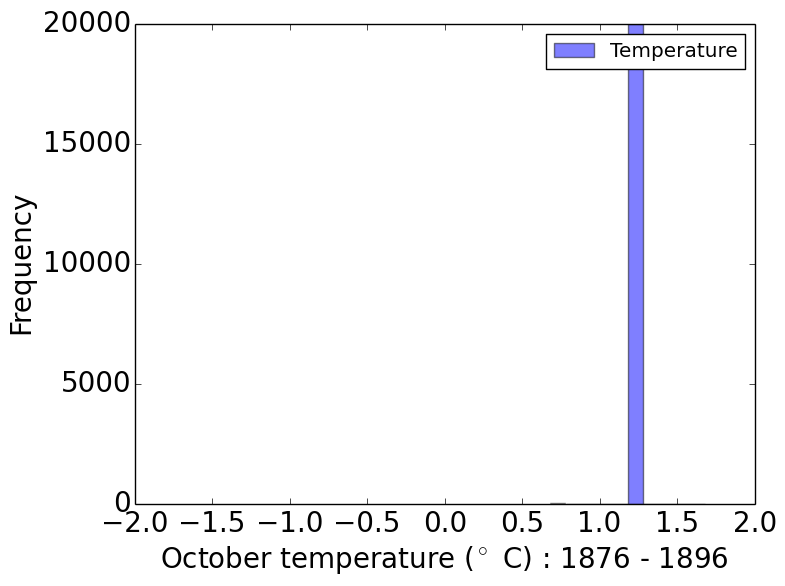}} &
\subcaptionbox{October -- TMCMC: 1896 - 1916\label{fig:oct_t8}}{\includegraphics[width = 2.0in]{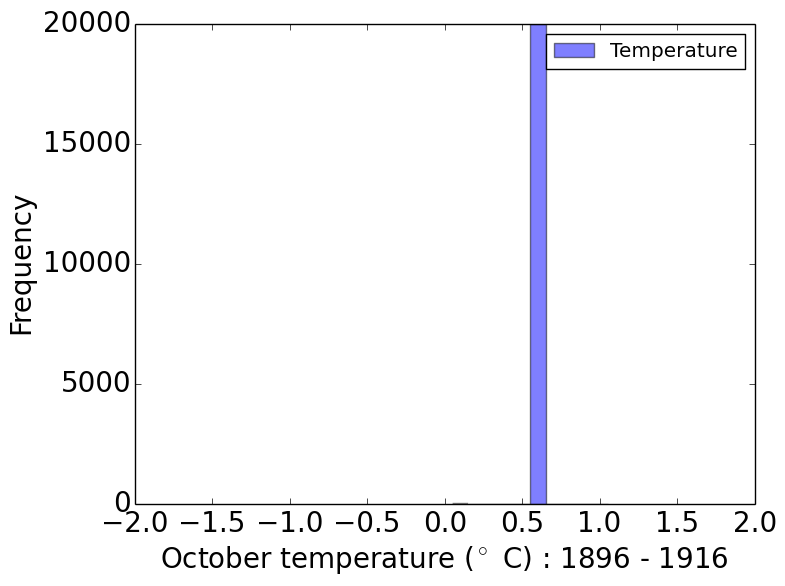}} &
\subcaptionbox{October -- TMCMC: 1916 - 1936\label{fig:oct_t9}}{\includegraphics[width = 2.0in]{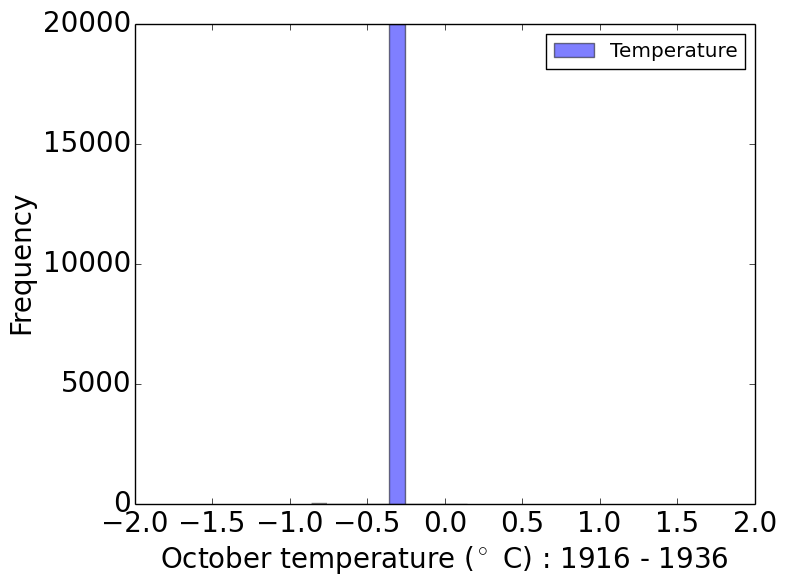}}\\
\subcaptionbox{October -- TMCMC: 1936 - 1936\label{fig:oct_t10}}{\includegraphics[width = 2.0in]{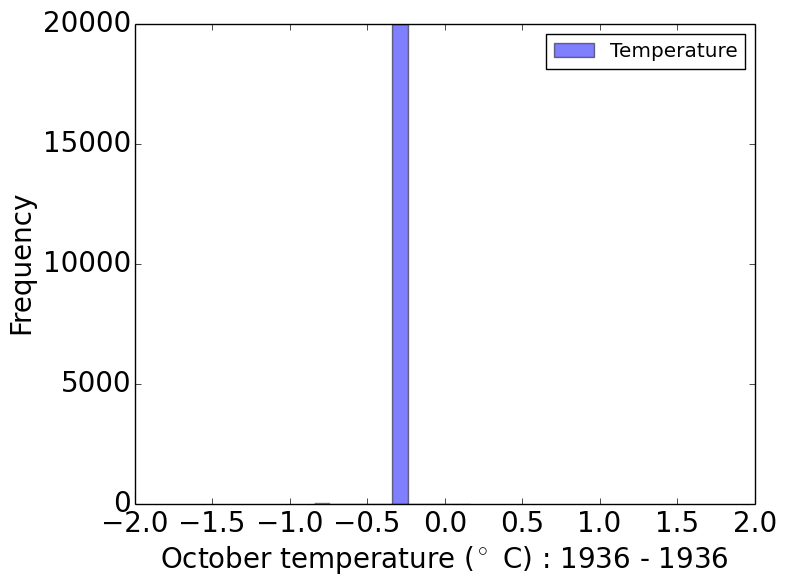}} &
\subcaptionbox{October -- TMCMC: 1956 - 1976\label{fig:oct_t11}}{\includegraphics[width = 2.0in]{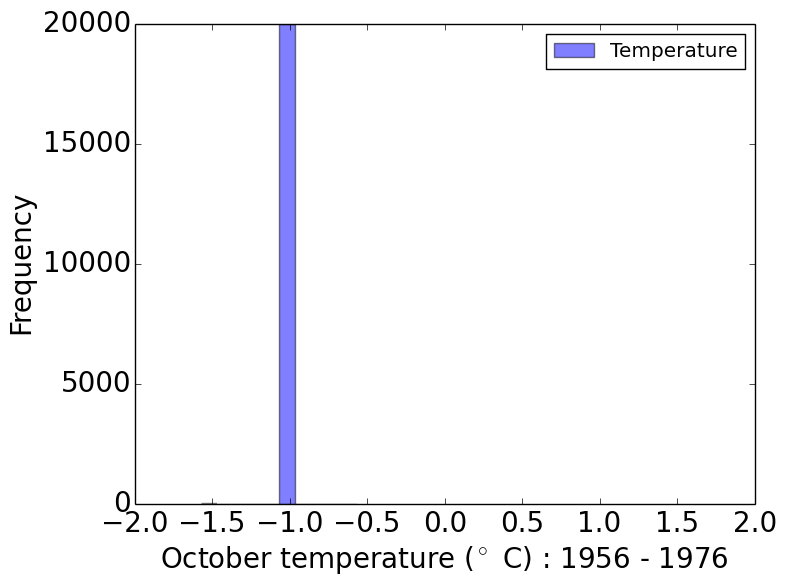}} &
\subcaptionbox{October -- TMCMC: 1976 - 1996\label{fig:oct_t12}}{\includegraphics[width = 2.0in]{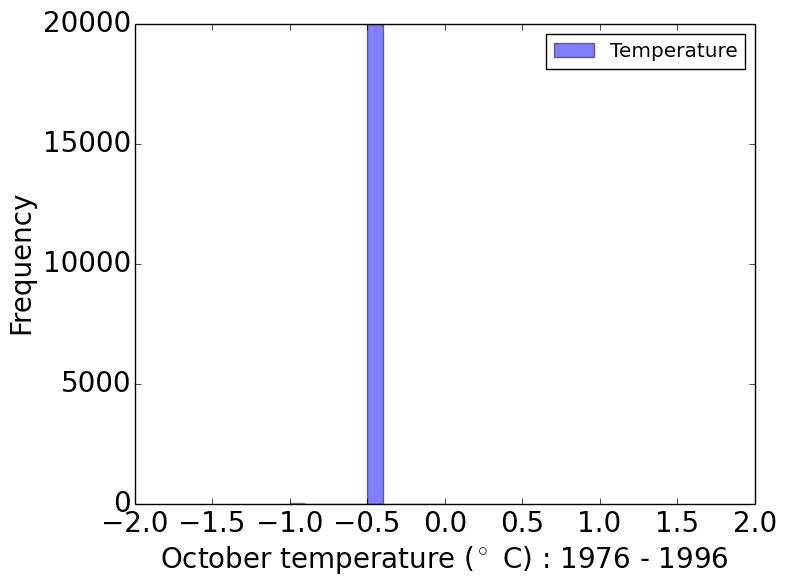}}\\
\subcaptionbox{October -- TMCMC: 1996 - 2016\label{fig:oct_t13}}{\includegraphics[width = 2.0in]{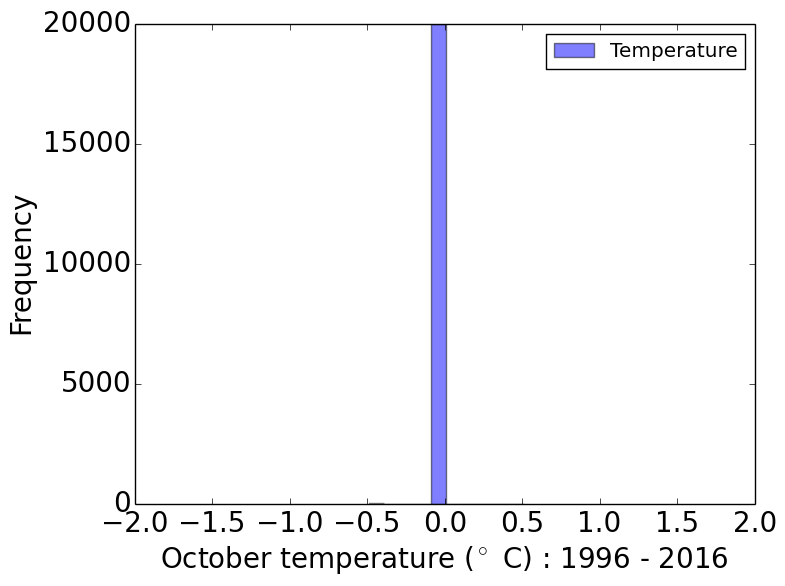}} &
\subcaptionbox{October -- TMCMC: Total\label{fig:oct_t14}}{\includegraphics[width = 2.0in]{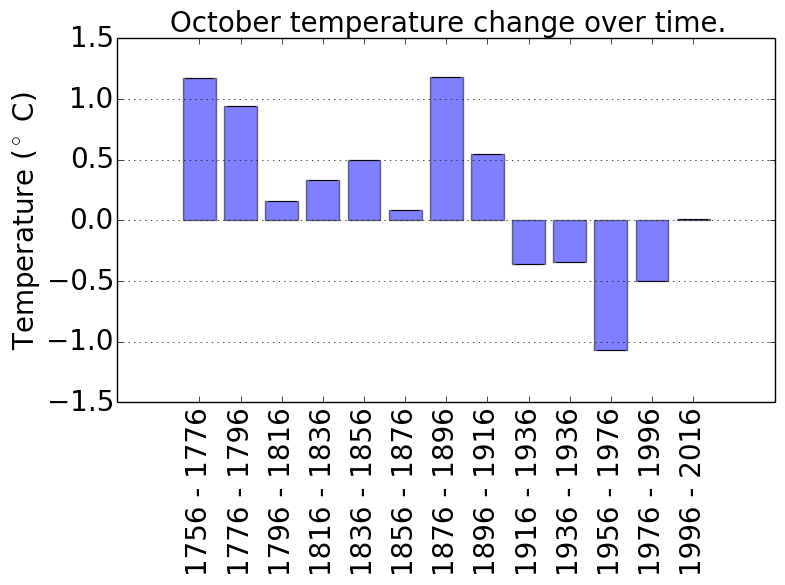}} &
\end{tabular}
\label{fig:oct_t1}
\end{adjustwidth}
\end{figure}

\begin{figure}
\begin{adjustwidth}{-6em}{0em}
\centering
\begin{tabular}{ccc}
\subcaptionbox{November -- TMCMC: 1756 - 1776\label{fig:nov_t1}}{\includegraphics[width = 2.0in]{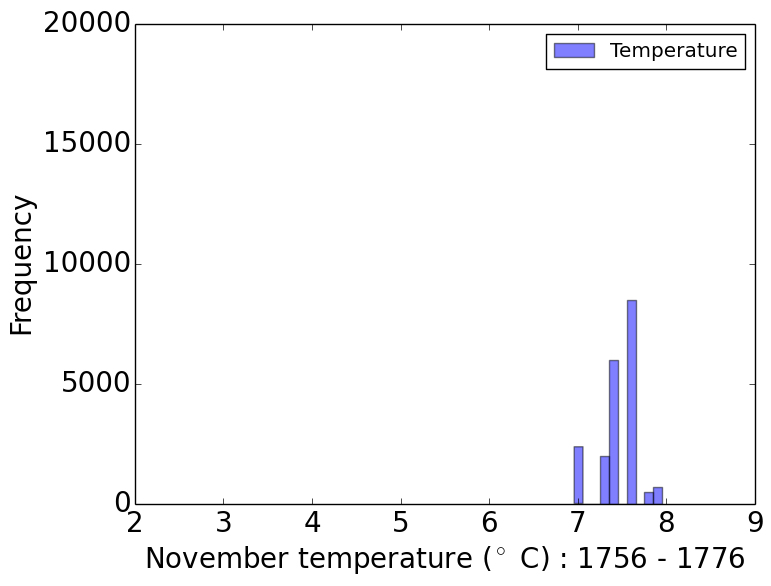}} &
\subcaptionbox{November -- TMCMC: 1776 - 1796\label{fig:nov_t2}}{\includegraphics[width = 2.0in]{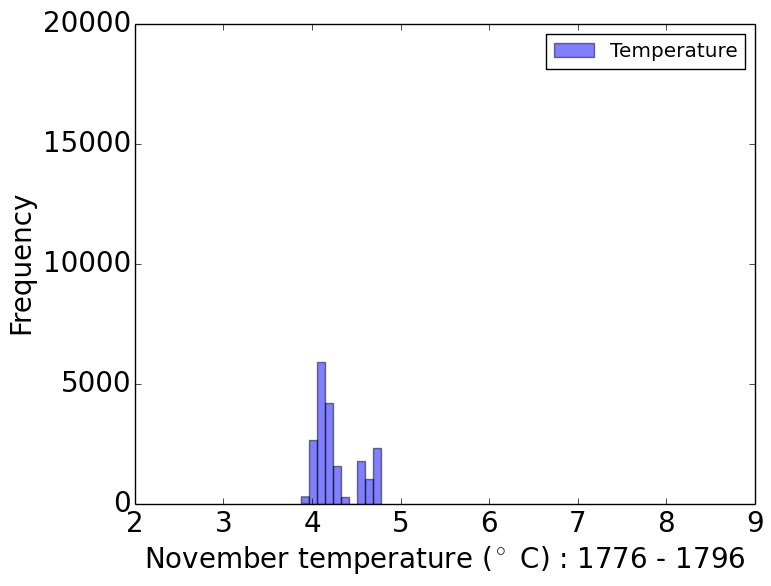}} &
\subcaptionbox{November -- TMCMC: 1796 - 1816\label{fig:nov_t3}}{\includegraphics[width = 2.0in]{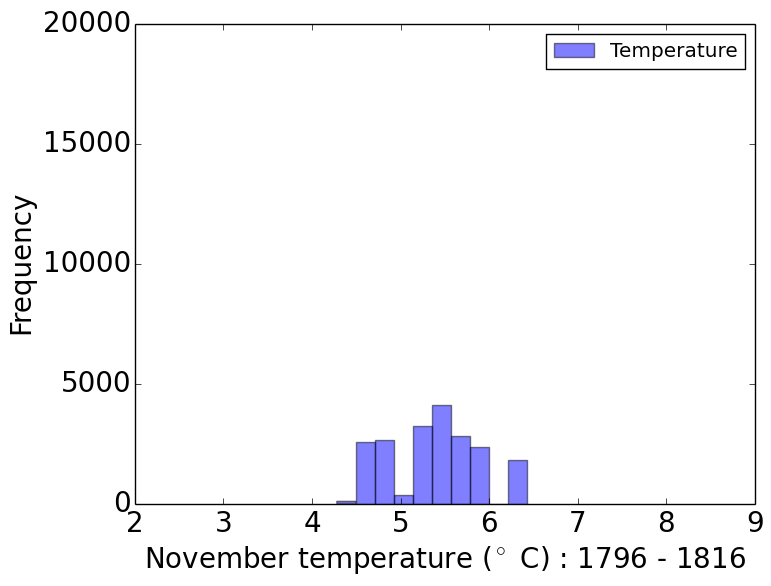}}\\
\subcaptionbox{November -- TMCMC: 1816 - 1836\label{fig:nov_t4}}{\includegraphics[width = 2.0in]{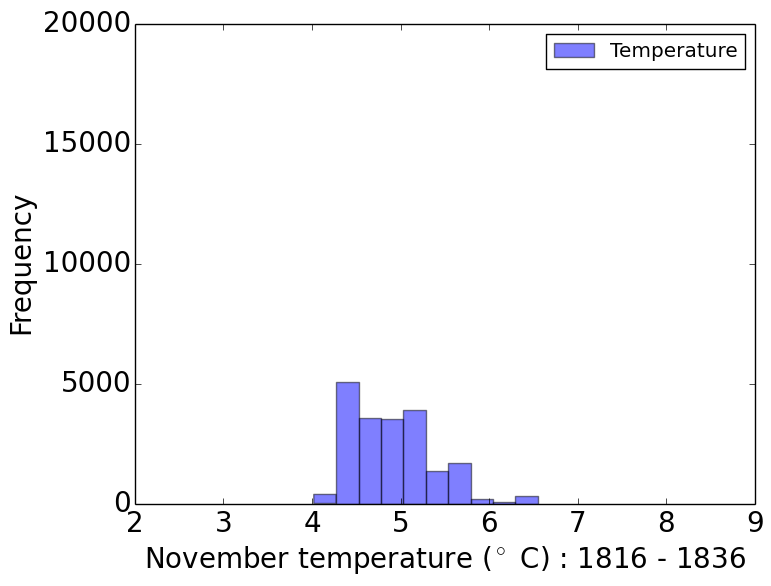}} &
\subcaptionbox{November -- TMCMC: 1836 - 1856\label{fig:nov_t5}}{\includegraphics[width = 2.0in]{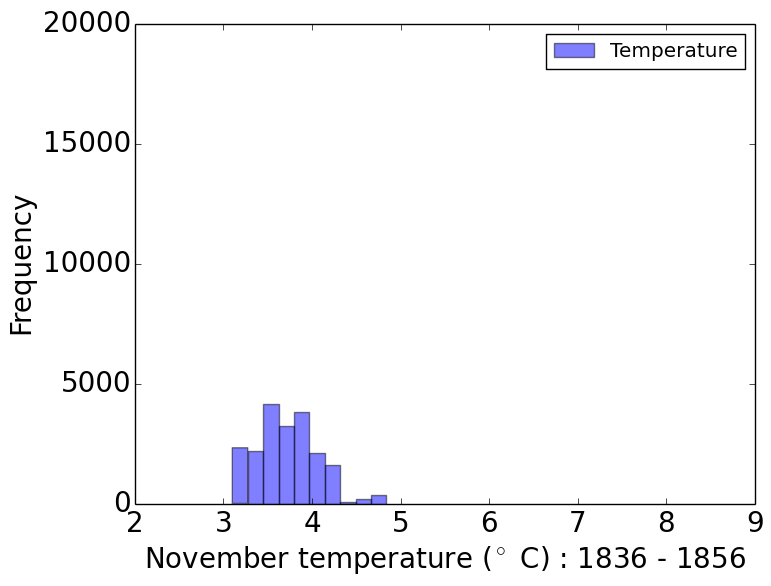}} &
\subcaptionbox{November -- TMCMC: 1856 - 1876\label{fig:nov_t6}}{\includegraphics[width = 2.0in]{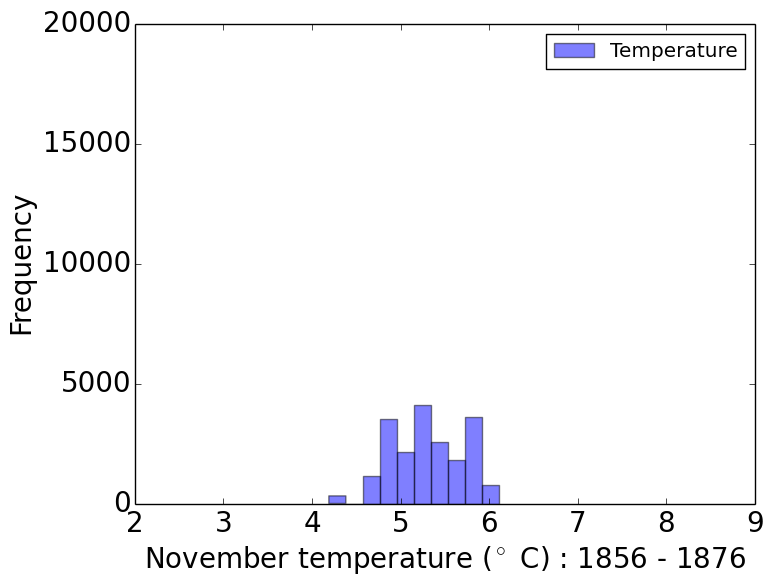}}\\
\subcaptionbox{November -- TMCMC: 1876 - 1896\label{fig:nov_t7}}{\includegraphics[width = 2.0in]{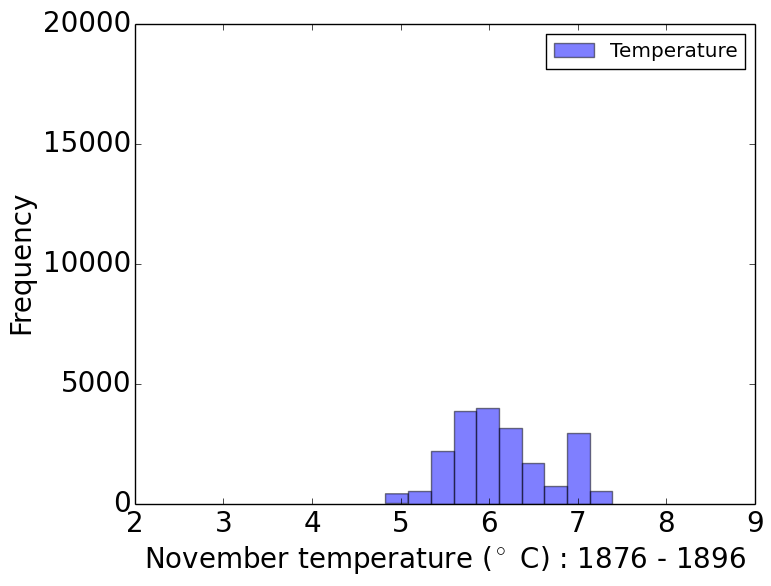}} &
\subcaptionbox{November -- TMCMC: 1896 - 1916\label{fig:nov_t8}}{\includegraphics[width = 2.0in]{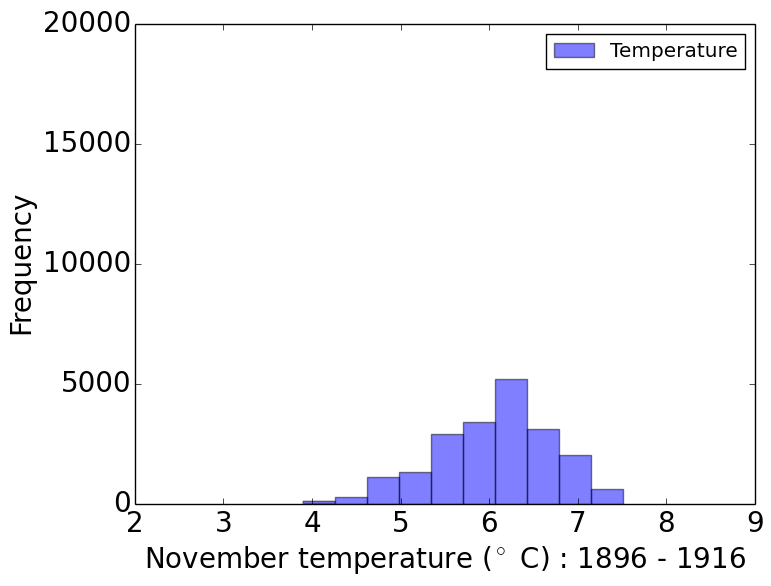}} &
\subcaptionbox{November -- TMCMC: 1916 - 1936\label{fig:nov_t9}}{\includegraphics[width = 2.0in]{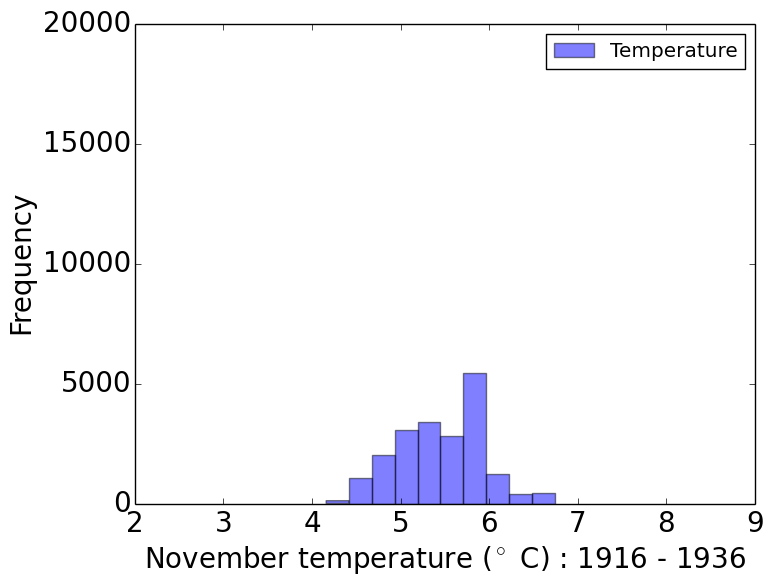}}\\
\subcaptionbox{November -- TMCMC: 1936 - 1936\label{fig:nov_t10}}{\includegraphics[width = 2.0in]{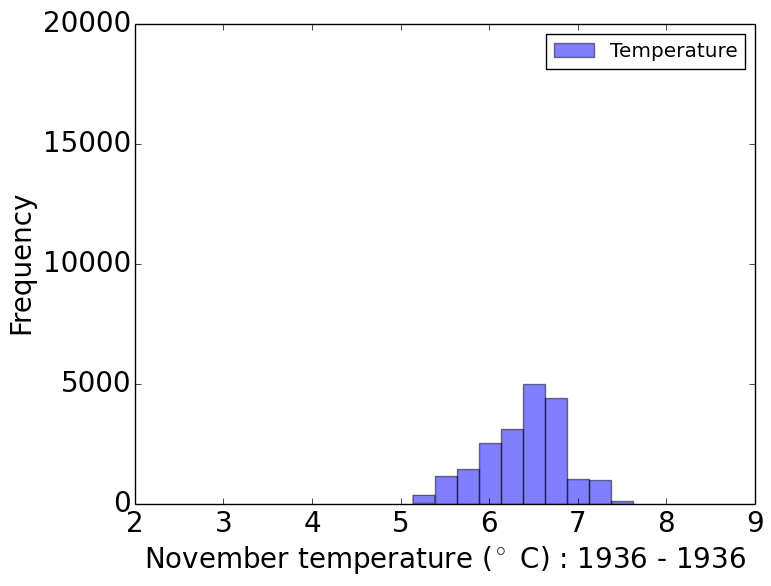}} &
\subcaptionbox{November -- TMCMC: 1956 - 1976\label{fig:nov_t11}}{\includegraphics[width = 2.0in]{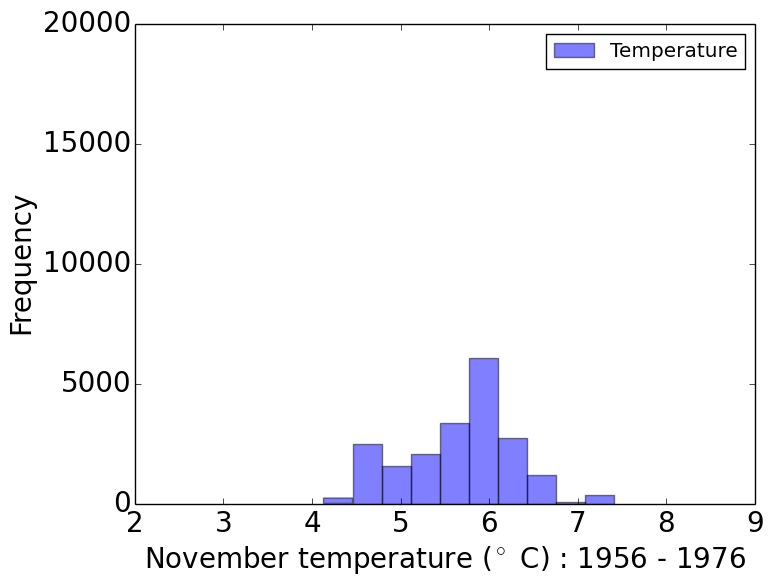}} &
\subcaptionbox{November -- TMCMC: 1976 - 1996\label{fig:nov_t12}}{\includegraphics[width = 2.0in]{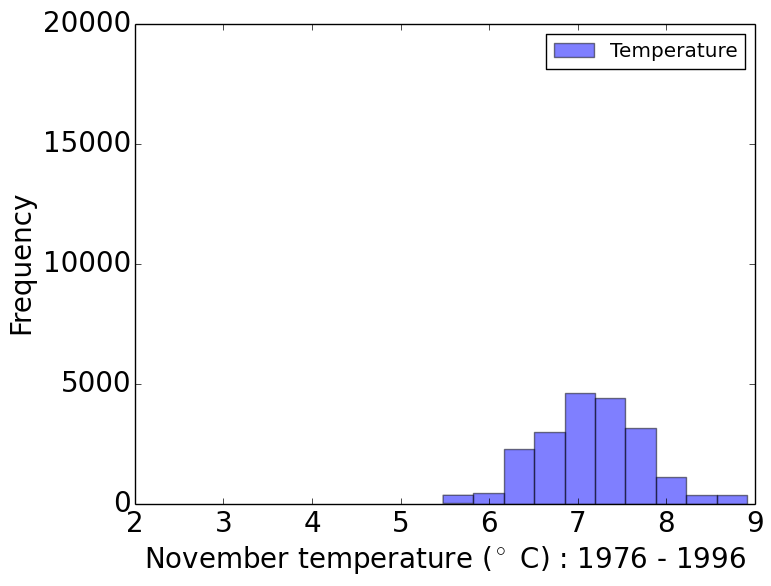}}\\
\subcaptionbox{November -- TMCMC: 1996 - 2016\label{fig:nov_t13}}{\includegraphics[width = 2.0in]{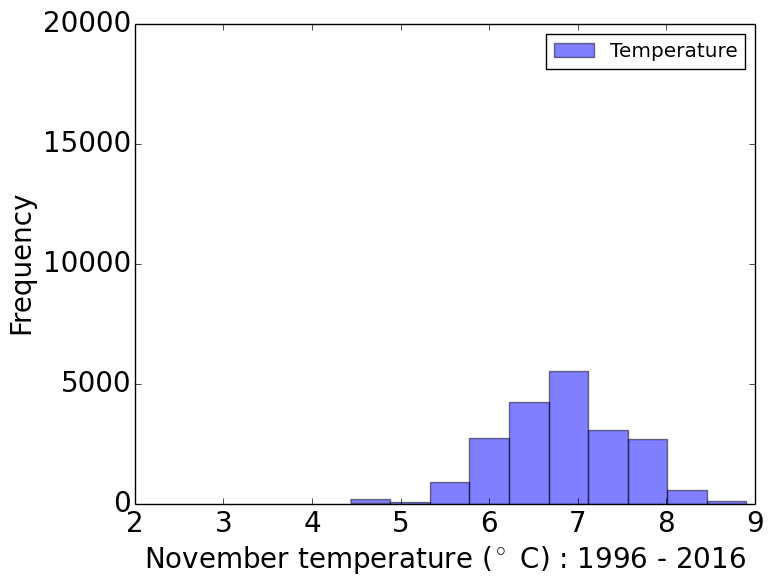}} &
\subcaptionbox{November -- TMCMC: Total\label{fig:nov_t14}}{\includegraphics[width = 2.0in]{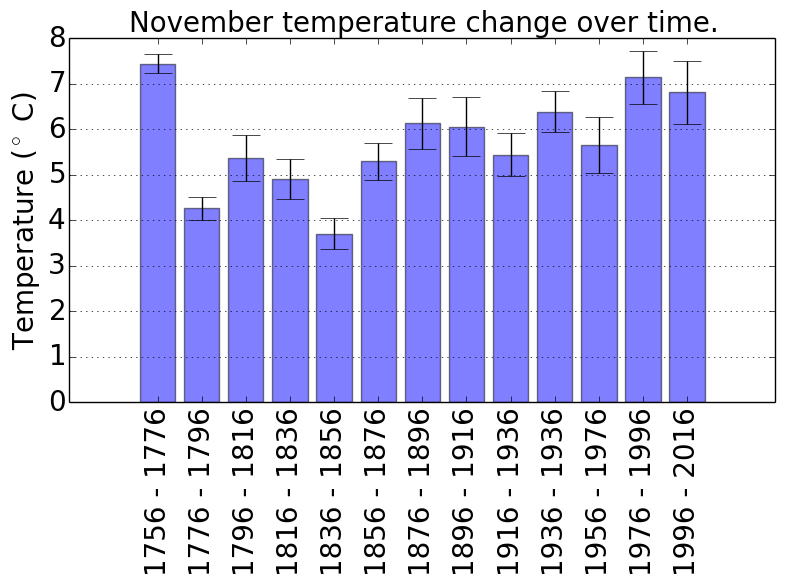}} &
\end{tabular}
\label{fig:nov_t1}
\end{adjustwidth}
\end{figure}

\begin{figure}
\begin{adjustwidth}{-6em}{0em}
\centering
\begin{tabular}{ccc}
\subcaptionbox{December -- TMCMC: 1756 - 1776\label{fig:dec_t1}}{\includegraphics[width = 2.0in]{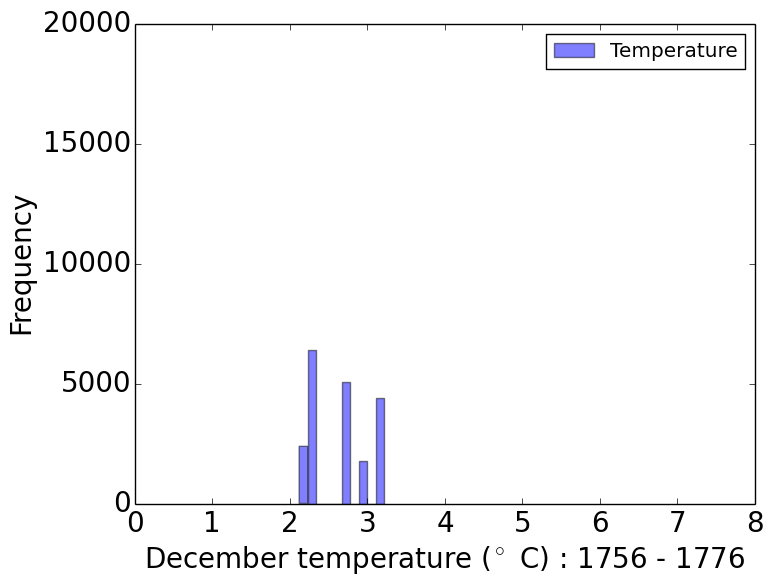}} &
\subcaptionbox{December -- TMCMC: 1776 - 1796\label{fig:dec_t2}}{\includegraphics[width = 2.0in]{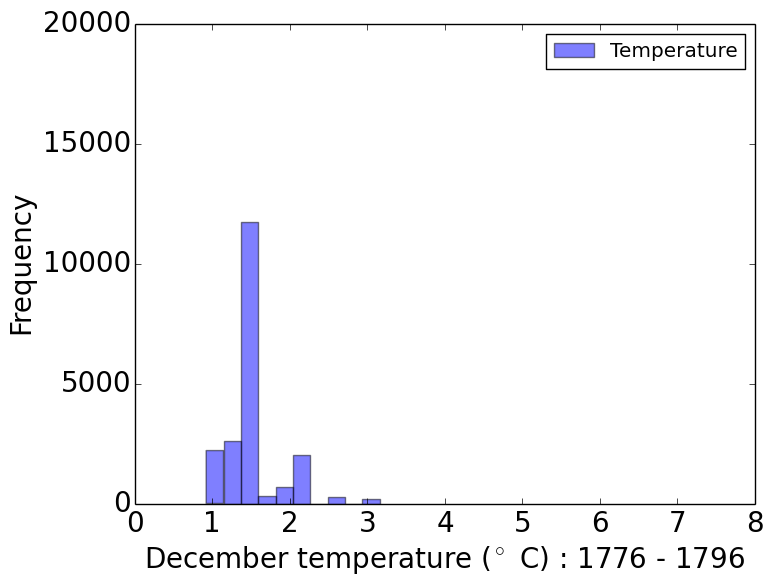}} &
\subcaptionbox{December -- TMCMC: 1796 - 1816\label{fig:dec_t3}}{\includegraphics[width = 2.0in]{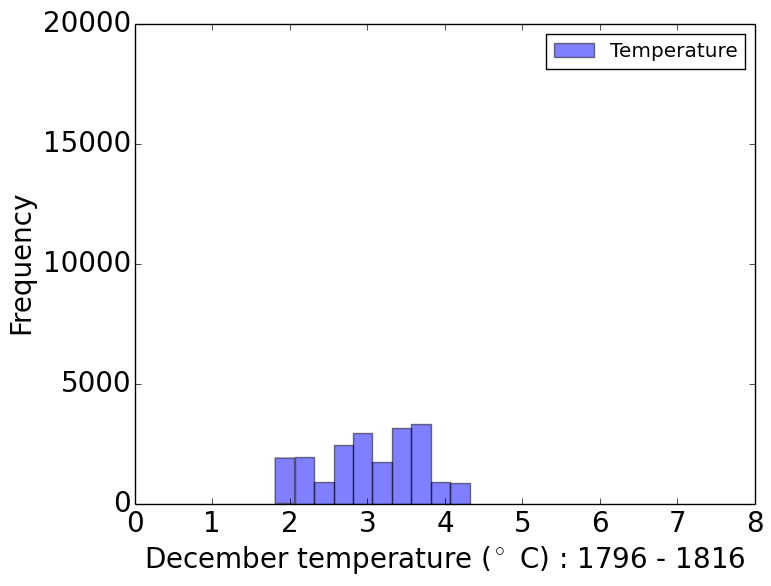}}\\
\subcaptionbox{December -- TMCMC: 1816 - 1836\label{fig:dec_t4}}{\includegraphics[width = 2.0in]{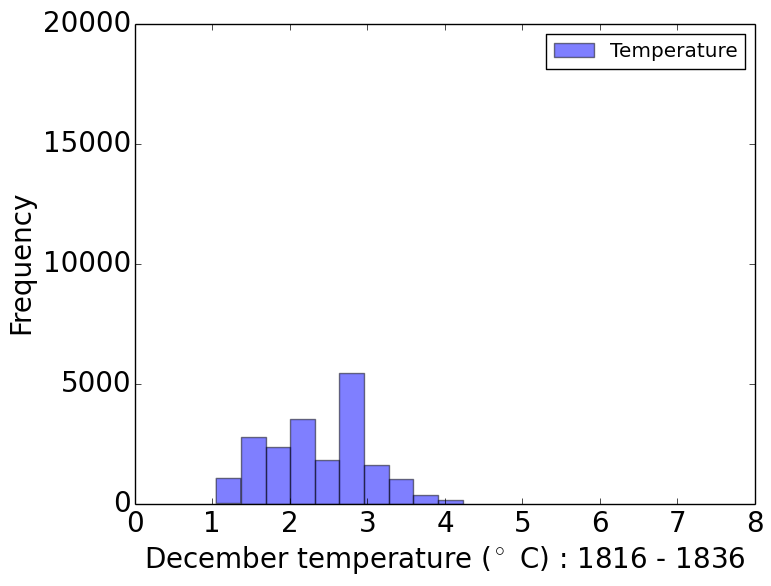}} &
\subcaptionbox{December -- TMCMC: 1836 - 1856\label{fig:dec_t5}}{\includegraphics[width = 2.0in]{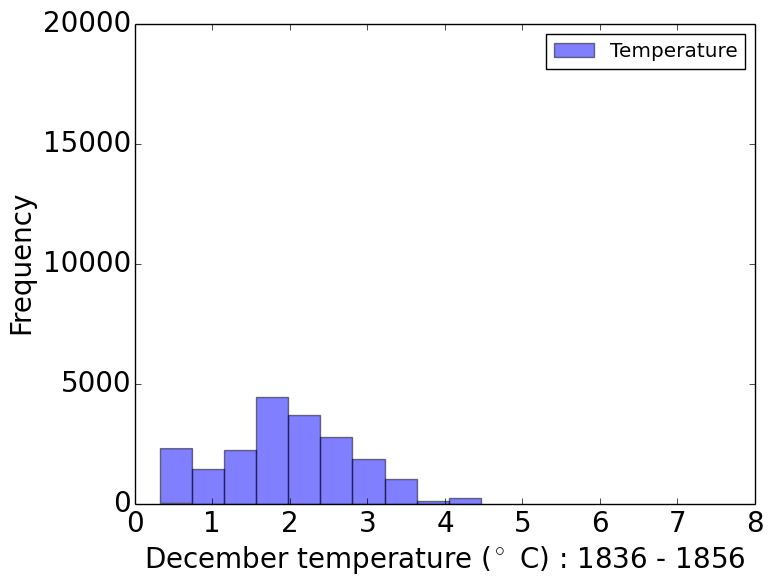}} &
\subcaptionbox{December -- TMCMC: 1856 - 1876\label{fig:dec_t6}}{\includegraphics[width = 2.0in]{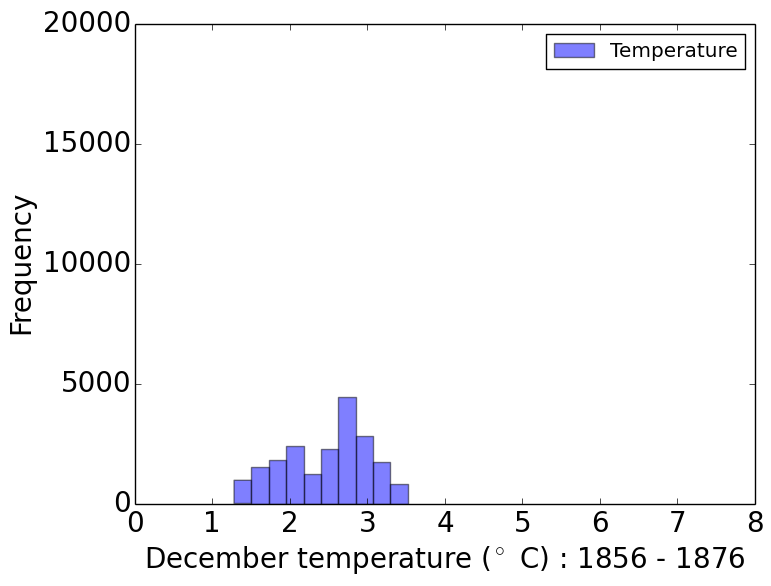}}\\
\subcaptionbox{December -- TMCMC: 1876 - 1896\label{fig:dec_t7}}{\includegraphics[width = 2.0in]{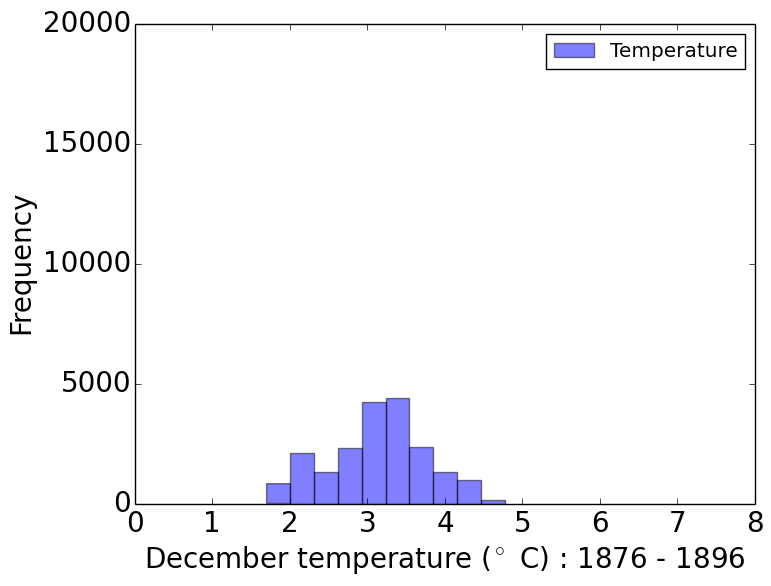}} &
\subcaptionbox{December -- TMCMC: 1896 - 1916\label{fig:dec_t8}}{\includegraphics[width = 2.0in]{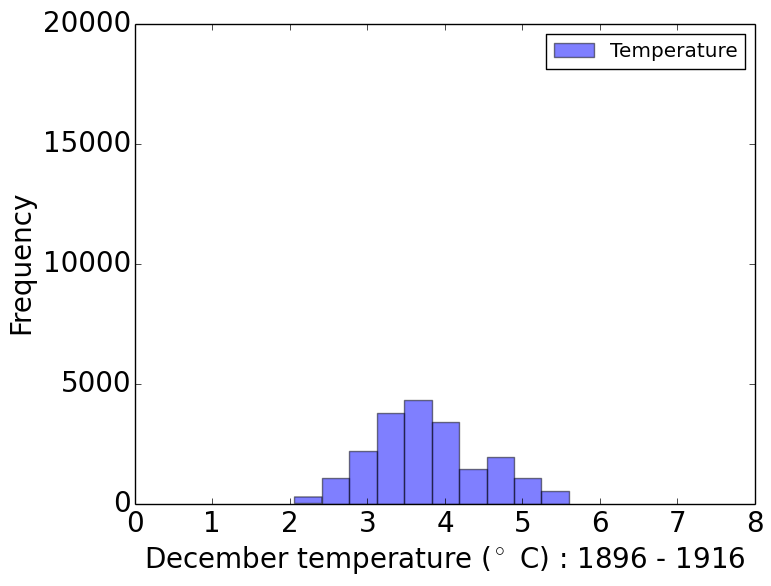}} &
\subcaptionbox{December -- TMCMC: 1916 - 1936\label{fig:dec_t9}}{\includegraphics[width = 2.0in]{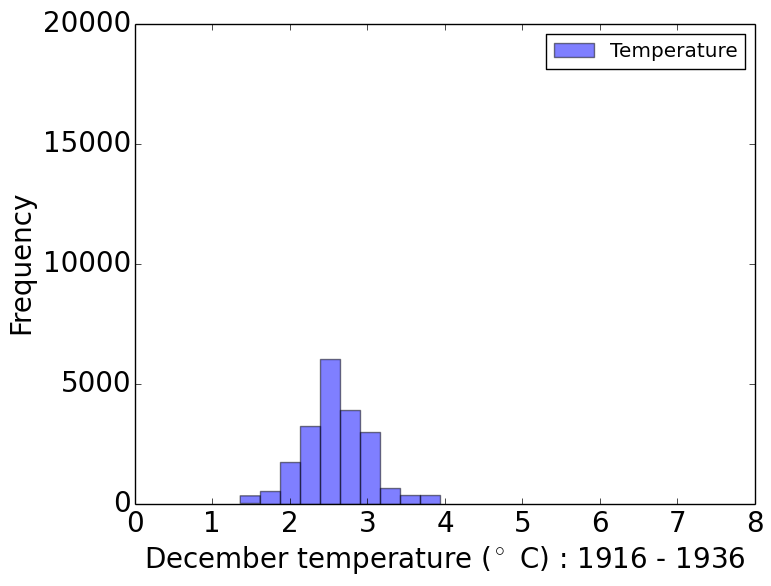}}\\
\subcaptionbox{December -- TMCMC: 1936 - 1936\label{fig:dec_t10}}{\includegraphics[width = 2.0in]{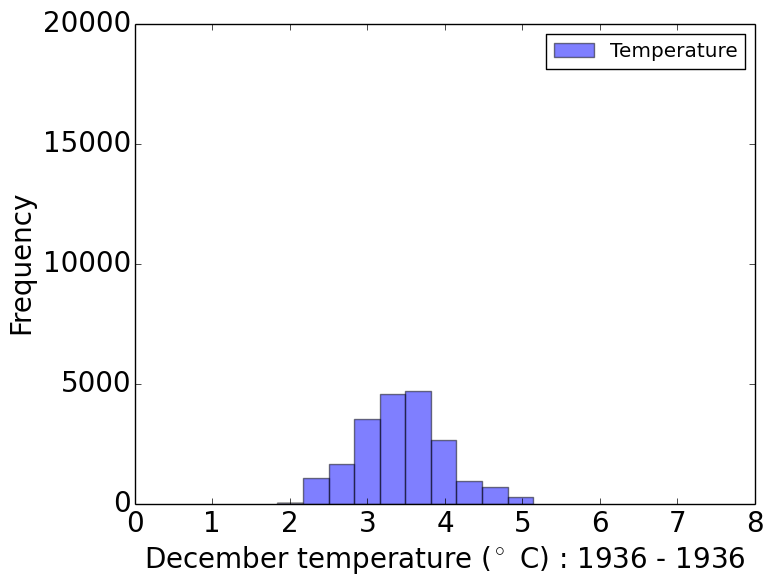}} &
\subcaptionbox{December -- TMCMC: 1956 - 1976\label{fig:dec_t11}}{\includegraphics[width = 2.0in]{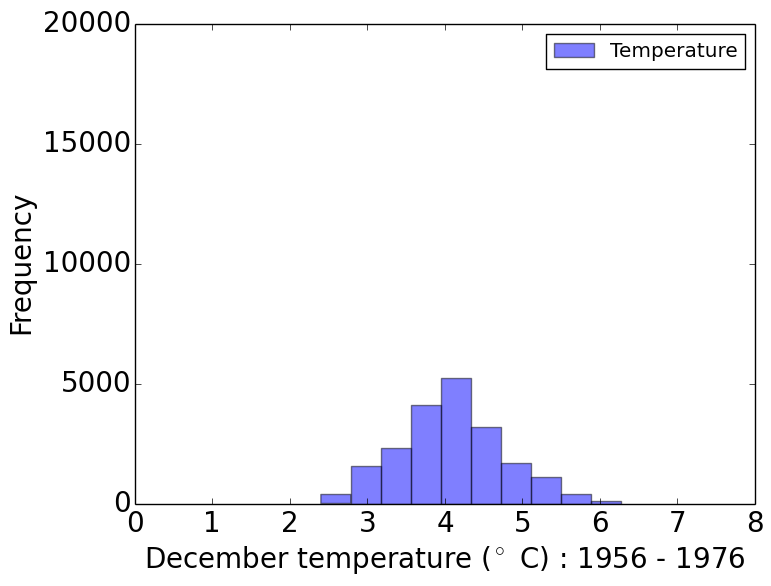}} &
\subcaptionbox{December -- TMCMC: 1976 - 1996\label{fig:dec_t12}}{\includegraphics[width = 2.0in]{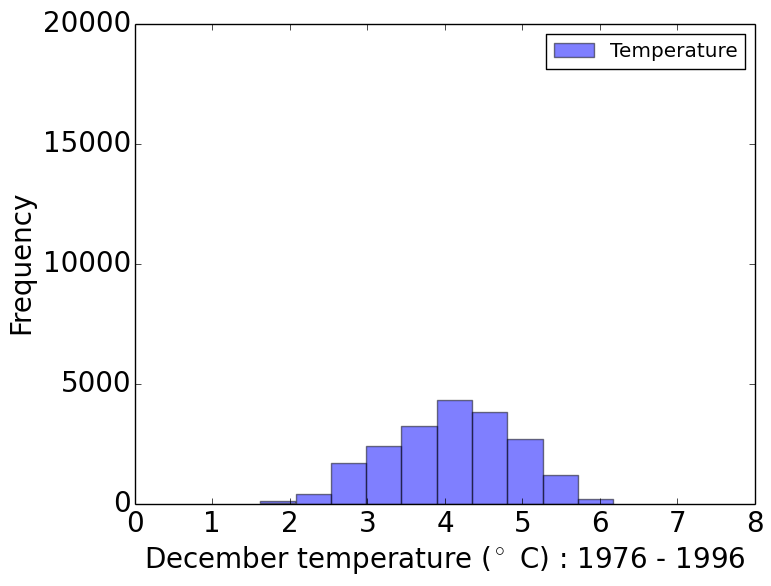}}\\
\subcaptionbox{December -- TMCMC: 1996 - 2016\label{fig:dec_t13}}{\includegraphics[width = 2.0in]{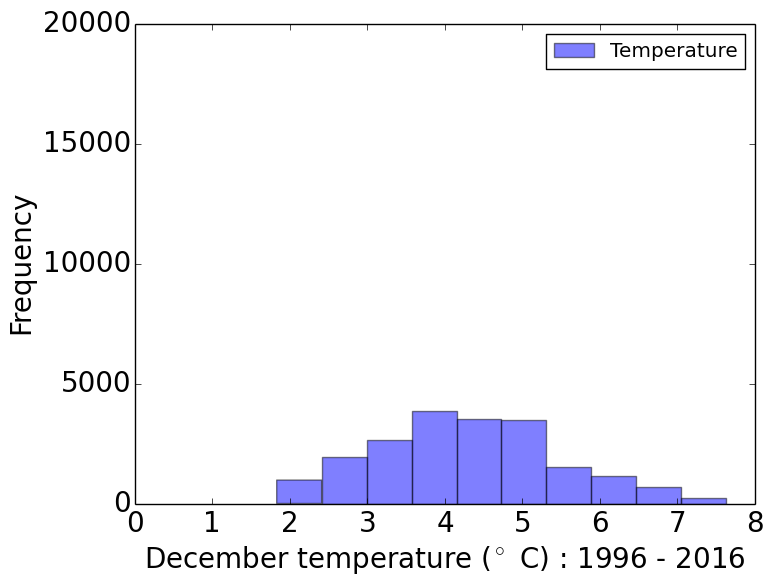}} &
\subcaptionbox{December -- TMCMC: Total\label{fig:dec_t14}}{\includegraphics[width = 2.0in]{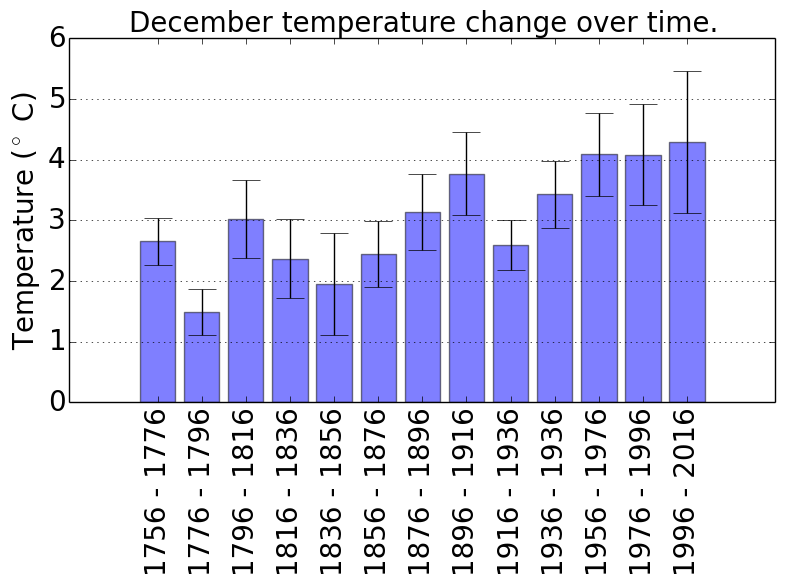}} &
\end{tabular}
\label{fig:dec_t1}
\end{adjustwidth}
\end{figure}

\end{document}